\newcommand{\ifprep}{\iftrue}   
\newcommand{\ifnotprep}{\iffalse}   
\newcommand{\ifall}{\iftrue}    

\documentclass[aps,prc,superscriptaddress,showpacs,floatfix]{revtex4}
\usepackage{placeins}
\usepackage{xspace}
\usepackage{longtable}
\usepackage{graphicx}
\usepackage[figuresright]{rotating}
\usepackage{afterpage}
\def\Journal#1#2#3#4{{#1} {\bf #2}, (#3) #4}
\def\NCA{\rm Nuovo Cimento}
\def\NIM{\rm Nucl. Instrum. Methods}
\def\NIMA{{\rm Nucl. Instrum. Methods} \bf A}
\def\NPB{{\rm Nucl. Phys.} \bf B}
\def\PLB{{\rm Phys. Lett.}  \bf B}
\def\PRL{\rm Phys. Rev. Lett.}
\def\PRD{{\rm Phys. Rev.} \bf D}
\def\ZPC{{\rm Z. Phys.} \bf C}

\newcommand{\kg}{\ensuremath{\mbox{kg}}\xspace}
\newcommand{\GeV}{\ensuremath{\mbox{GeV}}\xspace}
\newcommand{\MeV}{\ensuremath{\mbox{MeV}}\xspace}
\newcommand{\GeVc}{\ensuremath{\mbox{GeV}/c}\xspace}
\newcommand{\MeVc}{\ensuremath{\mbox{MeV}/c}\xspace}
\newcommand{\T}{\ensuremath{\mbox{T}}\xspace}
\newcommand{\cmsq}{\ensuremath{\mbox{cm}^2}\xspace}
\newcommand{\msq}{\ensuremath{\mbox{m}^2}\xspace}
\newcommand{\cm}{\ensuremath{\mbox{cm}}\xspace}
\newcommand{\mm}{\ensuremath{\mbox{mm}}\xspace}
\newcommand{\micron}{\ensuremath{\mu \mbox{m}}\xspace}
\newcommand{\mrad}{\ensuremath{\mbox{mrad}}\xspace}
\newcommand{\rad}{\ensuremath{\mbox{rad}}\xspace}
\newcommand{\ns}{\ensuremath{\mbox{ns}}\xspace}
\newcommand{\m}{\ensuremath{\mbox{m}}\xspace}
\newcommand{\s}{\ensuremath{\mbox{s}}\xspace}
\newcommand{\ms}{\ensuremath{\mbox{ms}}\xspace}
\newcommand{\micros}{\ensuremath{\mu \mbox{s}}\xspace}
\newcommand{\ps}{\ensuremath{\mbox{ps}}\xspace}
\newcommand{\barn}{\ensuremath{\mbox{barn}}\xspace}

\newcommand{\dedx}{\ensuremath{\mbox{d}E/\mbox{d}x}\xspace}
\newcommand{\pip}{\ensuremath{\pi^+}\xspace}
\newcommand{\pim}{\ensuremath{\pi^-}\xspace}
\newcommand{\pipm}{\ensuremath{\pi^{\pm}}\xspace}
\newcommand{\piz}{\ensuremath{\pi^0}\xspace}
\newcommand{\bfpip}{\ensuremath{\mathbf {\pi^+}}\xspace}
\newcommand{\bfpim}{\ensuremath{\mathbf {\pi^-}}\xspace}

\newcommand{\unknown}{\mbox{\bf XXXX}\xspace}
\newcommand{\dzeroprime}{\ensuremath{d'_0}\xspace}
\newcommand{\zzeroprime}{\ensuremath{z'_0}\xspace}
\newcommand{\evtspill}{\ensuremath{N_{\mathrm{evt}}}\xspace}
\newcommand{\pt}{\ensuremath{p_{\mathrm{T}}}\xspace}
\newcommand{\tht}{\ensuremath{\theta}\xspace}

\def\st{\scriptstyle}
\def\sst{\scriptscriptstyle}
\def\mco{\multicolumn}
\def\epp{\epsilon^{\prime}}
\def\vep{\varepsilon}
\def\ra{\rightarrow}
\def\ppg{\pi^+\pi^-\gamma}
\def\vp{{\bf p}}
\def\ko{K^0}
\def\kb{\bar{K^0}}
\def\al{\alpha}
\def\ab{\bar{\alpha}}
\def\be{\begin{equation}}
\def\ee{\end{equation}}
\def\bea{\begin{eqnarray}}
\def\eea{\end{eqnarray}}
\def\CPbar{\hbox{{\rm CP}\hskip-1.80em{/}}}

\renewcommand{\topfraction}{.8}
\renewcommand{\bottomfraction}{.8}
\renewcommand{\textfraction}{.15}
\renewcommand{\floatpagefraction}{.1}

\newcommand{\bfGeVc}{\ensuremath{\mathbf {\mbox{\bf GeV}/c}}\xspace}
\begin{document}
\title{{\Large EUROPEAN ORGANIZATION FOR NUCLEAR RESEARCH} \\
\ifprep
\vskip 1cm
\begin{flushright}
{\rm CERN-PH-EP/2009-021} 
\end{flushright}
\fi
\vskip 2cm

{\LARGE \bf Large-angle production of  charged pions 
with incident pion beams on nuclear targets} \\

\vskip 2cm
{\bf HARP Collaboration}
\vskip 2cm
\begin{flushleft}
{\rm  

 Measurements of the  double-differential \pipm production
 cross-section in the range of momentum $100~\MeVc \leq p \le 800~\MeVc$ 
 and angle $0.35~\rad \leq \theta  \le 2.15~\rad$
 using \pipm  beams incident on
 beryllium, aluminium, carbon, copper, tin, tantalum and lead targets are presented.
 The data were taken  with the large-acceptance HARP detector in the T9 beam
 line of the CERN Proton Synchrotron.
 The secondary pions were produced by beams in a momentum range from
 3~\GeVc to  12.9~\GeVc hitting a solid target with a thickness of
 5\% of a nuclear interaction length.  
 The tracking and identification of the
 produced particles was performed using a small-radius
 cylindrical time projection chamber placed inside a solenoidal
 magnet. 
 Incident particles were identified by an elaborate system of beam
 detectors.
 Results are obtained for the double-differential cross-sections 
  $
  {{\mathrm{d}^2 \sigma}}/{{\mathrm{d}p\mathrm{d}\theta }}
  $
 at six incident-beam momenta. 
 Data at 3~\GeVc, 5~\GeVc, 8~\GeVc, and 12~\GeVc are available for all
 targets while additional data at 8.9~\GeVc and 12.9 \GeVc were taken in
 positive particle  beams on Be and Al targets, respectively.
 The measurements are compared with several generators of GEANT4 and the
 MARS Monte Carlo simulation.
}
\end{flushleft}

\ifprep
\vskip 5cm
\centerline{\em{(to be published in Physical Review C)}}
\fi
\clearpage
}

\author{M.~Apollonio} 
\altaffiliation{Now at Imperial College, University of London, UK.}
\affiliation{Universit\`{a} degli Studi e Sezione INFN, Trieste, Italy}
\author{A.~Artamonov}   
\altaffiliation{ITEP, Moscow, Russian Federation.}
\affiliation{ CERN, Geneva, Switzerland}
\author{A. Bagulya} 
\affiliation{P. N. Lebedev Institute of Physics (FIAN), Russian Academy of
Sciences, Moscow, Russia}
\author{G.~Barr}
\affiliation{Nuclear and Astrophysics Laboratory, University of Oxford, UK} 
\author{A.~Blondel}
\affiliation{Section de Physique, Universit\'{e} de Gen\`{e}ve, Switzerland} 
\author{F.~Bobisut} 
\affiliation{Sezione INFN$^{(a)}$ and Universit\'a degli Studi$^{(b)}$, 
Padova, Italy}
\author{M.~Bogomilov}
\affiliation{ Faculty of Physics, St. Kliment Ohridski University, Sofia,
  Bulgaria}
\author{M.~Bonesini}
\affiliation{Sezione INFN Milano Bicocca, Milano, Italy} 
\author{C.~Booth} 
\affiliation{ Dept. of Physics, University of Sheffield, UK}
\author{S.~Borghi}  
\altaffiliation{now at the University of Glasgow, UK}
\affiliation{Section de Physique, Universit\'{e} de Gen\`{e}ve, Switzerland}
\author{S.~Bunyatov}
\affiliation{Joint Institute for Nuclear Research, JINR Dubna, Russia} 
\author{J.~Burguet--Castell}
\affiliation{Instituto de F\'{i}sica Corpuscular, IFIC, CSIC and Universidad de Valencia, Spain}
\author{M.G.~Catanesi}
\affiliation{Sezione INFN, Bari, Italy} 
\author{A.~Cervera--Villanueva}
\affiliation{Instituto de F\'{i}sica Corpuscular, IFIC, CSIC and Universidad de Valencia, Spain}
\author{P.~Chimenti}  
\affiliation{Universit\`{a} degli Studi e Sezione INFN, Trieste, Italy}
\author{L.~Coney} 
\affiliation{Columbia University, New York, USA}
\author{E.~Di~Capua}
\affiliation{Universit\`{a} degli Studi e Sezione INFN, Ferrara, Italy} 
\author{U.~Dore}
\affiliation{ Universit\`{a} ``La Sapienza'' e Sezione INFN Roma I, Roma,
  Italy}
\author{J.~Dumarchez}
\affiliation{ LPNHE, Universit\'{e}s de Paris VI et VII, Paris, France}
\author{R.~Edgecock}
\affiliation{Rutherford Appleton Laboratory, Chilton, Didcot, UK} 
\author{M.~Ellis}     
\altaffiliation{Now at FNAL, Batavia, Illinois, USA.} 
\affiliation{Rutherford Appleton Laboratory, Chilton, Didcot, UK}
\author{F.~Ferri} 
\affiliation{Sezione INFN Milano Bicocca, Milano, Italy}
\author{U.~Gastaldi}
\affiliation{Laboratori Nazionali di Legnaro dell' INFN, Legnaro, Italy}
\author{S.~Giani} 
\affiliation{ CERN, Geneva, Switzerland}
\author{G.~Giannini} 
\affiliation{Universit\`{a} degli Studi e Sezione INFN, Trieste, Italy}
\author{D.~Gibin}
\affiliation{Sezione INFN$^{(a)}$ and Universit\'a degli Studi$^{(b)}$, 
Padova, Italy}
\author{S.~Gilardoni}       
\affiliation{ CERN, Geneva, Switzerland} 
\author{P.~Gorbunov}  
\altaffiliation{ITEP, Moscow, Russian Federation.}
\affiliation{ CERN, Geneva, Switzerland}
\author{C.~G\"{o}\ss ling}
\affiliation{ Institut f\"{u}r Physik, Universit\"{a}t Dortmund, Germany}
\author{J.J.~G\'{o}mez--Cadenas} 
\affiliation{Instituto de F\'{i}sica Corpuscular, IFIC, CSIC and Universidad de Valencia, Spain}
\author{A.~Grant}  
\affiliation{ CERN, Geneva, Switzerland}
\author{J.S.~Graulich}
\altaffiliation{Now at Section de Physique, Universit\'{e} de Gen\`{e}ve, Switzerland.}
\affiliation{Institut de Physique Nucl\'{e}aire, UCL, Louvain-la-Neuve,
  Belgium} 
\author{G.~Gr\'{e}goire}
\affiliation{Institut de Physique Nucl\'{e}aire, UCL, Louvain-la-Neuve,
  Belgium} 
\author{V.~Grichine}  
\affiliation{P. N. Lebedev Institute of Physics (FIAN), Russian Academy of
Sciences, Moscow, Russia}
\author{A.~Grossheim} 
\altaffiliation{Now at TRIUMF, Vancouver, Canada.}
\affiliation{ CERN, Geneva, Switzerland} 
\author{A.~Guglielmi$^{(a)}$}
\affiliation{Sezione INFN$^{(a)}$ and Universit\'a degli Studi$^{(b)}$, 
Padova, Italy}
\author{L.~Howlett}
\affiliation{ Dept. of Physics, University of Sheffield, UK}
\author{A.~Ivanchenko}
\altaffiliation{ On leave from Novosibirsk University,  Russia.}
\affiliation{ CERN, Geneva, Switzerland}
\author{V.~Ivanchenko} 
\altaffiliation{On leave  from Ecoanalitica, Moscow State University,
Moscow, Russia}
\affiliation{ CERN, Geneva, Switzerland}
\author{A.~Kayis-Topaksu}
\altaffiliation{Now at \c{C}ukurova University, Adana, Turkey.}
\affiliation{ CERN, Geneva, Switzerland}
\author{M.~Kirsanov}
\affiliation{Institute for Nuclear Research, Moscow, Russia}
\author{D.~Kolev} 
\affiliation{ Faculty of Physics, St. Kliment Ohridski University, Sofia,
  Bulgaria}
\author{A.~Krasnoperov} 
\affiliation{Joint Institute for Nuclear Research, JINR Dubna, Russia}
\author{J. Mart\'{i}n--Albo}
\affiliation{Instituto de F\'{i}sica Corpuscular, IFIC, CSIC and Universidad de Valencia, Spain}
\author{C.~Meurer}
\affiliation{Institut f\"{u}r Physik, Forschungszentrum Karlsruhe, Germany}
\noaffiliation{}
\author{M.~Mezzetto$^{(a)}$}
\affiliation{Sezione INFN$^{(a)}$ and Universit\'a degli Studi$^{(b)}$, 
Padova, Italy}
\author{G.~B.~Mills}
\affiliation{Los Alamos National Laboratory, Los Alamos, USA}
\author{M.C.~Morone}
\altaffiliation{Now at University of Rome Tor Vergata, Italy.}   
\affiliation{Section de Physique, Universit\'{e} de Gen\`{e}ve, Switzerland}
\author{P.~Novella} 
\affiliation{Instituto de F\'{i}sica Corpuscular, IFIC, CSIC and Universidad de Valencia, Spain}
\author{D.~Orestano}
\affiliation{Sezione INFN$^{(c)}$ and Universit\'a$^{(d)}$  Roma Tre, 
Roma, Italy}
\author{V.~Palladino}
\affiliation{Universit\`{a} ``Federico II'' e Sezione INFN, Napoli, Italy}
\author{J.~Panman}
\thanks{Corresponding author (J.~Panman).~E-mail: 
jaap.panman@cern.ch}
\affiliation{ CERN, Geneva, Switzerland}
 \author{I.~Papadopoulos}  
\affiliation{ CERN, Geneva, Switzerland}
\author{F.~Pastore} 
\affiliation{Sezione INFN$^{(c)}$ and Universit\'a$^{(d)}$  Roma Tre, 
Roma, Italy}
\author{S.~Piperov}
\affiliation{ Institute for Nuclear Research and Nuclear Energy,
Academy of Sciences, Sofia, Bulgaria}
\author{N.~Polukhina}
\affiliation{P. N. Lebedev Institute of Physics (FIAN), Russian Academy of
Sciences, Moscow, Russia}
\author{B.~Popov} 
\altaffiliation{Also supported by LPNHE, Paris, France.}
\affiliation{Joint Institute for Nuclear Research, JINR Dubna, Russia}
\author{G.~Prior}  
\altaffiliation{Now at CERN}
\affiliation{Section de Physique, Universit\'{e} de Gen\`{e}ve, Switzerland}
\author{E.~Radicioni}
\affiliation{Sezione INFN, Bari, Italy}
\author{D.~Schmitz}
\affiliation{Columbia University, New York, USA}
\author{R.~Schroeter}
\affiliation{Section de Physique, Universit\'{e} de Gen\`{e}ve, Switzerland}
\author{G.~Skoro}
\affiliation{ Dept. of Physics, University of Sheffield, UK}
\author{M.~Sorel}
\affiliation{Instituto de F\'{i}sica Corpuscular, IFIC, CSIC and Universidad de Valencia, Spain}
\author{E.~Tcherniaev}
\affiliation{ CERN, Geneva, Switzerland}
 \author{P.~Temnikov}
\affiliation{ Institute for Nuclear Research and Nuclear Energy,
Academy of Sciences, Sofia, Bulgaria}
\author{V.~Tereschenko}  
\affiliation{Joint Institute for Nuclear Research, JINR Dubna, Russia}
\author{A.~Tonazzo}
\affiliation{Sezione INFN$^{(c)}$ and Universit\'a$^{(d)}$  Roma Tre, 
Roma, Italy}
\author{L.~Tortora$^{(c)}$}
\affiliation{Sezione INFN$^{(c)}$ and Universit\'a$^{(d)}$  Roma Tre, 
Roma, Italy}
\author{R.~Tsenov}
\affiliation{ Faculty of Physics, St. Kliment Ohridski University, Sofia,
  Bulgaria}
\author{I.~Tsukerman}  
\altaffiliation{ITEP, Moscow, Russian Federation.}
\affiliation{ CERN, Geneva, Switzerland}
\author{G.~Vidal--Sitjes}  
\altaffiliation{Now at Imperial College, University of London, UK.}
\affiliation{Universit\`{a} degli Studi e Sezione INFN, Ferrara, Italy}
\author{C.~Wiebusch}   
\altaffiliation{Now at III Phys. Inst. B, RWTH Aachen, Germany.}
\affiliation{ CERN, Geneva, Switzerland}
\author{P.~Zucchelli}
\altaffiliation{Now at SpinX Technologies, Geneva, Switzerland; On leave
from INFN, Sezione di Ferrara, Italy.} 
\affiliation{ CERN, Geneva, Switzerland}
%
%
 \collaboration{\bf HARP Collaboration}
 \noaffiliation

 \maketitle

\clearpage

\section{Introduction}

In many particle and astroparticle physics experiments external 
knowledge of hadron production is required to make optimal use of
the recorded data and to help design the experimental facilities.
The HARP experiment~\cite{harp-prop} is motivated by this need for
precise hadron production measurements.
It has taken data with beams of pions and protons
with momenta from 1.5~\GeVc to 15~\GeVc hitting targets made of a large
range of materials.
To provide a large angular and momentum coverage of the produced charged
particles the experiment comprises two spectrometers, a forward
spectrometer built around a dipole magnet covering the angular range up
to 250~mrad and a large-angle spectrometer constructed in a solenoidal
magnet with an angular acceptance of 
$0.35~\rad \leq \theta  \le 2.15~\rad$.

The main objectives are to measure pion yields for a quantitative
design of the proton driver of future superbeams (high-intensity
conventional beams) and a neutrino factory~\cite{ref:nufact}, 
to provide measurements to improve calculations of the atmospheric neutrino
flux~\cite{Battistoni,Stanev,Gaisser,Engel}
and to measure particle yields as input for the flux
calculation of accelerator neutrino experiments~\cite{ref:physrep}, 
such as K2K~\cite{ref:k2k,ref:k2kfinal},
MiniBooNE~\cite{ref:miniboone} and SciBooNE~\cite{ref:sciboone}. 
In addition to these specific aims, the data provided by HARP are
valuable for validating hadron production models used in simulation
programs. 
These simulations are playing an important role in the interpretation
and design of modern large particle-physics experiments.
In particular, the simulation of calorimeter response and secondary
interactions in tracking systems needs to be supported by experimental
hadron production data.

This paper presents the final  
measurements of the double-differential production cross-section, 
$
{{\mathrm{d}^2 \sigma^{\pi}}}/{{\mathrm{d}p\mathrm{d}\theta }}
$
for \pipm production at large angles by positively and negatively
charged pions of 3~\GeVc, 5~\GeVc, 8~\GeVc, 8.9~\GeVc (\pip--Be),
12~\GeVc and 12.9~\GeVc (\pip--Al) momenta impinging
on a thin beryllium, carbon, aluminium, copper, tin, tantalum or lead
target of 5\% nuclear interaction length.  
The data are taken with the large-angle spectrometer of the HARP detector.
The results of a similar analysis of \pipm production data taken
with proton beams on the same set of nuclear targets 
can be found in Ref.~\cite{ref:harp:la}.

The results presented in this paper, covering an extended range of solid
targets in the same experiment, make it possible to perform systematic 
comparisons of hadron production models with measurements 
at different incoming beam momenta over a large range 
of target atomic number $A$.
The performance of these models can be further tested by extending the
comparisons with our results on pion production in p--$A$ interactions
obtained with the forward spectrometer.
These results are the subject of other HARP publications
\cite{ref:harp:alPaper,ref:harp:bePaper,ref:harp:carbonfw,ref:harp:cnofw,ref:harp:protonforward}.  
The analysis of results taken with charged pion beams on the full range
of targets presented in this paper, but measured with the forward
spectrometer, can be found in Ref.~\cite{ref:harp:pionforward}. 
Pion production data in this energy region are extremely
scarce, especially in pion beams,  and HARP
is the first experiment to provide a large data set taken with many 
different targets, full particle identification and a large-acceptance
detector~\footnote{%
Cross-sections for some of the data-sets based on the same raw data as
the ones used in this paper have been published by a different
group~\cite{ref:cdp:la}. 
We disagree with the analysis of Ref.~\cite{ref:cdp:la} (see
Ref.~\cite{ref:tpcmom} for differences in detector calibration).
}.

Data were taken in the T9 beam of the CERN PS. The collected statistics,
for the different nuclear targets, are reported in
Tables~\ref{tab:events-pip} and \ref{tab:events-pim}.  
The analysis proceeds by selecting tracks in the time projection
chamber (TPC) in events with incident beams of charged pions of both
polarities.  
Momentum and polar angle measurements and particle identification are
based on the measurements of track position and energy deposition in
the TPC.
An unfolding method is used to correct for experimental resolution,
efficiency and acceptance and to obtain the double-differential pion
production cross-sections.  
The analysis follows the same methods as used for the determination of
\pipm production by protons.  
These analysis methods are documented in Ref.~\cite{ref:harp:tantalum} with
improvements described in Ref.~\cite{ref:harp:la} and will be only
briefly outlined here.

\section{Experimental apparatus and data selection.}
\label{sec:apparatus}
\begin{figure}[tb]
 \begin{center}
  \includegraphics[width=12cm]{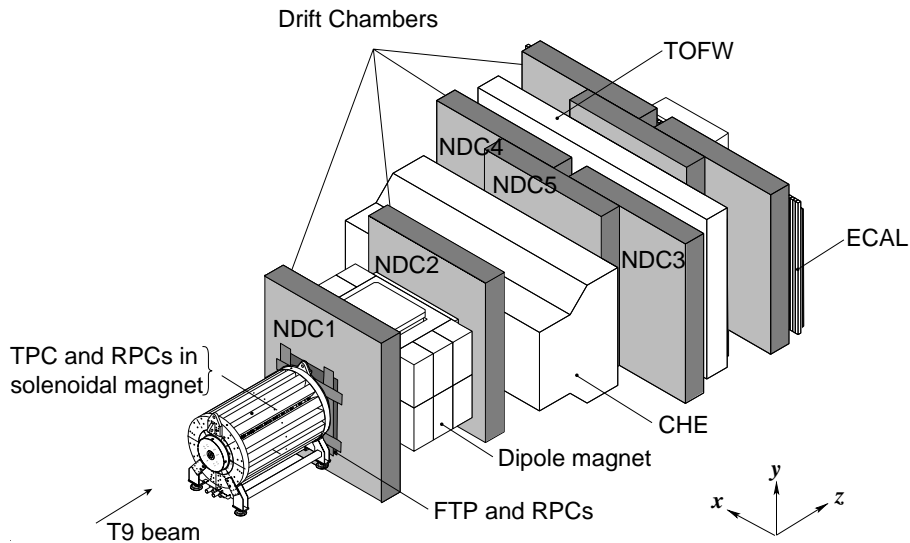}
 \end{center}
\caption{Schematic layout of the HARP detector. 
The convention for the coordinate system is shown in the lower-right
corner. 
The three most downstream (unlabelled) drift chamber modules are only partly
equipped with electronics and are not used for tracking. 
The detector covers a total length of 13.5 m along the beam axis 
and has a maximum width of 6.5 m perpendicular to the beam.
The beam muon identifier is visible as the most downstream detector
 (white box).
}
\label{fig:harp}
\end{figure}

The HARP detector is shown in Fig.~\ref{fig:harp} and is 
 described in detail in Ref.~\cite{ref:harpTech}.
 The forward spectrometer, mainly used in the analysis for the conventional
 neutrino beams and atmospheric neutrino flux, consists of a dipole magnet,
 large planar drift chambers 
 (NDCs)~\cite{NOMAD_NIM_DC} , a time-of-flight wall (TOFW) \cite{ref:tofPaper}, 
 a threshold Cherenkov counter
 (CHE), and an electromagnetic calorimeter (ECAL).

In the large-angle region a cylindrical TPC with a radius of 408~\mm
is  positioned inside a solenoidal magnet with a field of 0.7~\T. 
The TPC detector was designed to measure and identify tracks in the
angular region from 0.25~\rad to 2.5~\rad with respect to the beam axis.
The target is placed inside the inner field cage (IFC) of the TPC such that,
in addition to particles produced in the forward direction, 
backward-going tracks can be measured.
The TPC is used
for tracking, momentum determination and measurement of the
energy deposition \dedx for particle identification~\cite{ref:tpc:ieee}.
A set of resistive plate chambers (RPCs) form a barrel inside the solenoid 
around the TPC to measure the arrival time of the secondary
particles~\cite{ref:rpc}. 
Charged particle identification (PID) is achieved with the TPC by measuring the 
ionization per unit length in the gas (\dedx) as a function of the total
momentum of the particle. 
Additional PID can be performed through a time-of-flight measurement
with the RPCs; this method is used to calculate the efficiency and to
provide an independent calibration of the PID based on \dedx. 
The method used for the PID and its calibration are discussed in more
detail in Section~\ref{sec:pid}.

In addition to the data taken with the solid targets of
5\% nuclear interaction length ($\lambda_{\mathrm{I}}$),
runs were also taken with an empty target holder to check backgrounds. 
Data taken with a liquid hydrogen target at 3~\GeVc, 5~\GeVc and
8~\GeVc incident-beam momentum together with cosmic-ray data were used 
to provide an absolute calibration of the efficiency, momentum scale and
resolution of the detector~\cite{ref:tpcmom}. 

The momentum of the T9 beam is known with a precision of
the order of 1\%~\cite{ref:t9}. 
The absolute normalization of the number of incident pions is
performed using incident-pion triggers. 
These are triggers where the same selection on the beam particle is
applied but no selection on the interaction is performed.
The rate of this trigger was down-scaled by a factor 64.
The total number of these incident-beam triggers is typically 50,000 per
data set thus introducing a negligible statistical error. 
A cross-check of the absolute normalization is provided by counting
tracks in the forward spectrometer.
The dimensions and masses of the solid targets were carefully measured.
The purity of the target materials exceeded 99.9\%.
The uncertainties in thickness and density of the targets are well below
1\%. 

Beam instrumentation provides identification of the incoming
particle, the determination of the time when it hits the target, 
and the impact point and direction of the beam particle
on the target. 
It is based on a set of four multi-wire proportional chambers (MWPCs)
to measure position and direction of the incoming beam particles 
and time-of-flight (TOF) detectors and two
N$_2$-filled Cherenkov counters to identify incoming particles.  
Several trigger detectors are installed to select events with an
interaction and to define the normalization.

Besides the usual need for calibration of the detector, a number of
hardware shortfalls, discovered mainly after the end of data-taking,
had to be overcome to use the TPC data reliably in the analysis.
The TPC is affected by a relatively large number of dead or noisy 
pads and static and dynamic distortions of the reconstructed
trajectories.
A first set of results on the production by protons of pions at large angles
has been published by the HARP Collaboration in
Refs.~\cite{ref:harp:tantalum,ref:harp:cacotin,
ref:harp:bealpb}, based on the analysis of the data in the beginning of
each accelerator spill. 
The reduction of the data set was necessary to avoid problems in the TPC
detector responsible for dynamic distortions to the image of the
particle trajectories as the ion charge was building up during each
spill.   
Corrections that allow the use of the full statistics to be made have
been developed (see Ref.~\cite{ref:dyn}) and are applied in this analysis
in the same manner as in Ref.~\cite{ref:harp:la}. 
The size of the corrections for dynamic distortions grows as function of
the time within each accelerator spill and for each data set the part of
the spill that can be reliably corrected is checked.
The fraction of events usable for the analysis is typically 80\%, but
varies for the different data sets (see Tables~\ref{tab:events-pip} and
\ref{tab:events-pim}). 
The presence of a possible residual momentum bias in the TPC measurement 
due to the dynamic distortions was investigated using a large set of
calibration methods.  
A dedicated paper~\cite{ref:tpcmom}
addresses this point and shows that our estimation of momentum bias
is below $3\%$.

The beam of positive particles used for this measurement contains mainly 
positrons, muons, pions and protons, with small components of kaons and
deuterons and heavier ions.
Its composition depends on the selected beam momentum.
The proton fraction in the incoming positive-particle beam varies from
35\% at 3~\GeVc to 92\% at 12~\GeVc.  
This explains the relatively low statistics for some of the 12~\GeVc
data sets taken with the \pip beam.
The negatively charged particle beam is mainly composed of pions with
small background components of muons and electrons.

At the first stage of the analysis a favoured beam particle type is selected
using the beam time-of-flight system and the two Cherenkov
counters.
A value of the pulse height consistent with the absence of a signal in both beam
Cherenkov detectors distinguishes protons (and kaons) from electrons and pions.
We also ask for time measurements to be present which are needed for calculating 
the arrival time of the beam proton at the target. 
The beam TOF system is used to reject ions, such as deuterons, and to
separate protons from pions at low momenta.
At 3~\GeVc, the TOF measurement allows the selection of
pions from protons to be made at more than 5{$\sigma$}.
In most beam settings the nitrogen pressure in the beam Cherenkov
counters was too low for kaons to be above the threshold.
Kaons are thus counted in the proton sample.
However, the fraction of kaons has been measured in the 12.9~\GeVc beam
configuration and are found to contribute less than 0.5\%, and hence are
negligible in the pion beam sample.
Electrons radiate in the Cherenkov counters and would be counted as
pions. 
In the 3~\GeVc beam, electrons are identified by both Cherenkov counters,
since the pressure was such that pions remained below threshold.
In the 5~\GeVc beam electrons could be tagged by one Cherenkov counter
only, while the other Cherenkov counter was used to tag pions.
The $e/\pi$ fraction was measured to be 1\% in the 3~\GeVc beam and 
$<10^{-3}$ in the  5~\GeVc beam.
By extrapolation from the lower-energy beam settings this electron
contamination can be estimated to be negligible ($<10^{-3}$) for the
beams where it cannot be measured directly.
More details on the beam particle selection can be found in 
Refs.~\cite{ref:harpTech} and \cite{ref:harp:alPaper}.

In addition to the momentum-selected beam of protons and pions originating
from the T9 production target one expects also the presence of muons
from pion decay both downstream and upstream of the beam momentum selection.  
Therefore, precise absolute knowledge of the muon rate incident on the
HARP targets is required when measurements 
of particle production with incident pions are performed. 
The particle identification detectors in the beam do not distinguish
muons from pions. 
A separate measurement of the muon component has been performed
using data-sets without target (``empty-target data-sets'').
Since the empty-target data were taken with the same beam parameter
settings as the data taken with targets, the beam composition can be
measured in the empty-target runs using the forward spectrometer  
and then used as an overall correction
for the counting of pions in the runs with targets. 
Muons are recognized by their longer range in the beam muon identifier
(BMI).
The BMI is a small instrumented stack of iron absorbers at the
downstream end of the spectrometer.
The punch-through background in the BMI is measured counting the
protons (identified with the beam detectors) thus mis-identified as muons
by the BMI. 
A comparison of the punch-through rate between simulated incoming pions
and protons was used to determine a correction for the difference
between pions and protons and to determine the systematic error.
This difference is the dominant systematic error in the beam composition
measurement. 
The aim was to determine the composition of the beam as it strikes the
target, thus muons produced in pion decays after the HARP target
should be considered as a background to the measurement of muons in the
beam. 
The rate of these latter background muons, which depends mainly on the total 
inelastic cross-section and pion decay,  was calculated by a Monte Carlo 
simulation using GEANT4~\cite{ref:geant4}.
The muon fraction in the beam (at the target) is obtained taking
into account the efficiency of the BMI selection criteria as well as the
punch-through and decay backgrounds.
The analyses for the various beam settings give results for
$R=\mu/(\mu+\pi)$ of (4.2$\pm$1)\% and (5.2$\pm$1)\% for the low-momentum beams (3~\GeVc
and 5~\GeVc) and between (4.1$\pm$1)\% and (2.8$\pm$1)\% for the highest momenta (from
8~\GeVc to 12.9~\GeVc).
The uncertainty in these fractions is dominated by the systematic
uncertainty in the punch-through background.
The fact that the background does not scale with the decay probability
for pions is due to the limited acceptance of the beam-line to transport
the decay muons. 
The muon contamination is taken into account in the normalization of the
pion beam and adds a systematic error of 1\% to the overall
normalization. 

A set of MWPCs is used to retain events with only one
beam particle for which the trajectory extrapolates to the target.
An identical beam particle selection was performed for events
triggered with the incident-beam trigger in order to provide an
absolute normalization of the incoming pions.
This trigger selected every 64$^{th}$ beam particle coincidence
outside the dead-time of the data acquisition system.

\begin{table}[tbp] 
\caption{Total number of events and tracks used in the various nuclear 
  5\%~$\lambda_{\mathrm{I}}$ target data sets taken with the \pip beam and the number of
  pions on target as calculated from the pre-scaled incident pion triggers.} 
\label{tab:events-pip}
{\small
\begin{center}
\begin{tabular}{llrrrrrr} \hline
\bf{Data set (\bfGeVc)}          &         &\bf{3}&\bf{5}&\bf{8}&\bf{8.9} &\bf{12} &\bf{12.9}\\ \hline
    Total DAQ events     & (Be)     & 1399714 & 1473815 & 1102415 &  7236396 &  1211220 &   --      \\
                         &  (C)     & 1345461 & 2628362 & 1878590 &  --      &  1855615 &   --      \\
                         & (Al)     & 1586331 & 1787620 & 1706919 &   --     &  619021  &  5401701  \\
                         & (Cu)     & 623965  & 2089292 & 2613229 &   --     &  748443  &  --       \\
                         & (Sn)     & 1652751 & 2827934 & 2422110 &   --     &  1803035 &  --       \\
                         & (Ta)     & 2202760 & 2094286 & 2045631 &   --     &  886307  &  --       \\   
                         & (Pb)     & 1299264 & 2110904 & 2314552 &   --     &  486875  &  --       \\   
 Accepted pions with LAI &  (Be)    & 162739   & 202279 &  87076  &  750776  & 37056   &   --      \\
                         &  (C)     & 130343   & 345591 & 167675  &   --     & 42694   &  --       \\
                         &  (Al)    & 168224   & 274449 & 179039  &   --     & 14604   &  184480   \\
                         &  (Cu)    &  93049   & 349035 & 285211  &   --     & 20360   &  --       \\
                         &  (Sn)    & 210640   & 487780 & 308137  &   --     & 51603   &  --       \\
                         &  (Ta)    & 241334   & 342810 & 265647  &   --     & 25029   &  --       \\
                         &  (Pb)    & 199482   & 314916 & 298547  &   --     & 13584   &  --       \\
  Fraction of triggers used   & (Be)  & 79\%  &  75\% &   83\% &  94\% &    79\%   &  -- \\
  in the analysis             & (C)   &  95\% &  90\% &   83\% &  --   &    84\%   &  -- \\   
                              & (Al)  & 78\%  &  80\% &   63\% &  --   &   96\%  &  72\% \\
                              & (Cu)  & 91\%  &  76\% &   66\% &  --   &   76\%  &   --  \\
                              & (Sn)  & 97\%  &  73\% &   67\% &  --   &   76\%  &   --  \\
                              & (Ta)  & 86\%  &  81\% &   69\% &  --   &   76\%  &   --  \\
                              & (Pb)  & 74\%  &  56\% &   69\% &  --   &   50\%  &   --  \\
  \bf{$\bfpim$ selected with PID} & (Be)  & 20343  & 27018  & 12716   &  124869 &  5569 &   --     \\
                                  & (C)   & 16350  & 48343  & 23640   &  --     &  5375 &   --     \\
                                  & (Al)  & 18424  & 38759  & 22115   &  --     &  2909 &   29234  \\ 
                                  & (Cu)  & 10546  & 46845  & 38646   &  --     &  3607 &   --     \\
                                  & (Sn)  & 24864  & 66259  & 47208   &  --     & 10412 &   --     \\
                                  & (Ta)  & 23933  & 47515  & 39317   &  --     &  5128 &   --     \\
                                  & (Pb)  & 15661  & 32318  & 42878   &  --     &  2121 &   --     \\
  \bf{$\bfpip$ selected with PID} & (Be)  & 36568 & 44588 & 20033 &  193302 &  7780 &   --    \\
                                  & (C)   & 31807 & 85745 & 39621 &  --     &  8030 &   --    \\
                                  & (Al)  & 32843 & 62491 & 33036 &  --     &  4147 &   41799 \\
                                  & (Cu)  & 16933 & 70251 & 55031 &  --     &  5089 &   --    \\
                                  & (Sn)  & 36697 & 90915 & 59842 &  --     & 13099 &   --    \\
                                  & (Ta)  & 33122 & 63390 & 50176 &  --     &  6565 &   --    \\
                                  & (Pb)  & 21592 & 40937 & 52738 &  --     &  2438 &   --    \\
\end{tabular}
\end{center}
}
\end{table}

The length of the accelerator spill is 400~ms with a typical intensity
of 15~000 beam particles per spill in the positive beam and lower for
the negative beam.
The average number of events recorded by the data acquisition ranges
from 300 to 350 per spill for the different beam momenta and beam
polarities. 
The analysis proceeds by first selecting a beam pion hitting the
target, not accompanied by other beam tracks. 
Then an event is required to give a large-angle interaction (LAI) trigger  to be
retained. 
After the event selection the sample of tracks to be used for analysis
is defined.
Tables~\ref{tab:events-pip} and \ref{tab:events-pim} show the number of
events and the number of \pipm selected in the analysis for the
\pip and \pim data, respectively.
The large difference between the first and second set of rows (``Total
DAQ events'' and ``Accepted pions with LAI'') is due to the relatively
large fraction of protons in the 
beam and to the larger number of triggers taken for the measurements
with the forward dipole spectrometer (``forward triggers'').
The entry ``fraction of triggers used'' shows the part of the data for
which the dynamic distortions in the TPC could be calibrated reliably.

\begin{table}[tbp] 
\caption{Total number of events and tracks used in the various nuclear 
  5\%~$\lambda_{\mathrm{I}}$ target data sets taken with the \pim beam and the number of
  pions on target as calculated from the pre-scaled incident pion triggers.} 
\label{tab:events-pim}
{\small
\begin{center}
\begin{tabular}{llrrrr} \hline
\bf{Data set (\bfGeVc)}  &         &\bf{3}&\bf{5}&\bf{8}&\bf{12} \\ \hline
    Total DAQ events     & (Be)     & 1526198 &  748266 & 1425157 &  641631  \\
                         &  (C)     & 1994971 & 1434316 & 1454223 &  640065  \\
                         & (Al)     & 1835559 & 1094546 & 1217161 &  741230  \\
                         & (Cu)     & 1765764 & 1583102 & 1625665 &  304937  \\
                         & (Sn)     & 1981587 & 1544197 & 1403508 & 1154081  \\
                         & (Ta)     & 1193731 & 1444547 & 1218251 &  957654  \\   
                         & (Pb)     & 1242128 & 2027048 & 1486681 & 1386090  \\   
 Accepted pions with LAI &  (Be)    & 283861 & 196671 & 376363 & 196812   \\
                         &  (C)     & 344399 & 310321 & 408962 & 190306   \\
                         &  (Al)    & 376908 & 299988 & 413953 & 271824   \\
                         &  (Cu)    & 361839 & 497580 & 627258 & 119295   \\
                         &  (Sn)    & 375766 & 484138 & 567525 & 523802   \\
                         &  (Ta)    & 181035 & 456243 & 553778 & 404771   \\
                         &  (Pb)    & 147493 & 643923 & 686721 & 513335   \\
  Fraction of triggers used   & (Be)  & 97\% & 75\% & 65\% & 80\%   \\
  in the analysis             & (C)   & 90\% & 75\% & 80\% & 92\%   \\   
                              & (Al)  & 98\% & 58\% & 73\% & 96\%   \\
                              & (Cu)  & 70\% & 52\% & 72\% & 77\%   \\
                              & (Sn)  & 87\% & 74\% & 69\% & 80\%   \\
                              & (Ta)  & 49\% & 44\% & 69\% & 85\%   \\
                              & (Pb)  & 62\% & 45\% & 64\% & 67\%   \\
  \bf{$\bfpim$ selected with PID} & (Be)  & 79779 & 42817 &  65739 &  44174  \\
                                  & (C)   & 71616 & 63415 &  80815 &  39518  \\
                                  & (Al)  & 90961 & 50171 &  81210 &  72663  \\ 
                                  & (Cu)  & 57728 & 73853 & 122404 &  28409  \\
                                  & (Sn)  & 63391 & 92619 &  98667 & 129086  \\
                                  & (Ta)  & 16314 & 51497 &  85026 & 108590  \\
                                  & (Pb)  & 15797 & 73549 & 102324 & 112895  \\
  \bf{$\bfpip$ selected with PID} & (Be)  & 43386 & 26606 &  48419 &  34605  \\
                                  & (C)   & 43926 & 43862 &  63498 &  34040  \\
                                  & (Al)  & 60205 & 36936 &  66073 &  64202  \\
                                  & (Cu)  & 39533 & 56143 & 103464 &  25374  \\
                                  & (Sn)  & 42991 & 69343 &  80807 & 113925  \\
                                  & (Ta)  & 10325 & 39055 &  69544 &  94774  \\
                                  & (Pb)  &  9820 & 54855 &  84766 &  98637  \\
\end{tabular}
\end{center}
}
\end{table}

\section{Particle identification}
\label{sec:pid}

The particle identification in the large-angle region uses 
the \dedx information provided by  the TPC. 
The electron, pion and proton populations are well separated at most
momentum values. 
Simple momentum-dependent cuts are used to
separate the different populations. 
The pions are identified by removing electrons and protons.
The kaon population is negligible.
The cuts were optimized to maximize the purity of the pion sample,
accepting a lower efficiency in the selection.  
More details are given in Ref.~\cite{ref:harp:tantalum}.

The measurement of the velocity $\beta$ of secondary particles by the
time-of-flight determination with the RPC detectors using the BTOF as
starting-time reference provides complementary particle identification.
It allows the efficiency and purity of the PID algorithm using \dedx
to be studied for a large subset of the TPC tracks.  
A statistical accuracy of the order of 0.2\% can be
obtained in the PID efficiency determination. 

\begin{figure}[tbp]
 \begin{center}
  \includegraphics[width=0.46\textwidth]{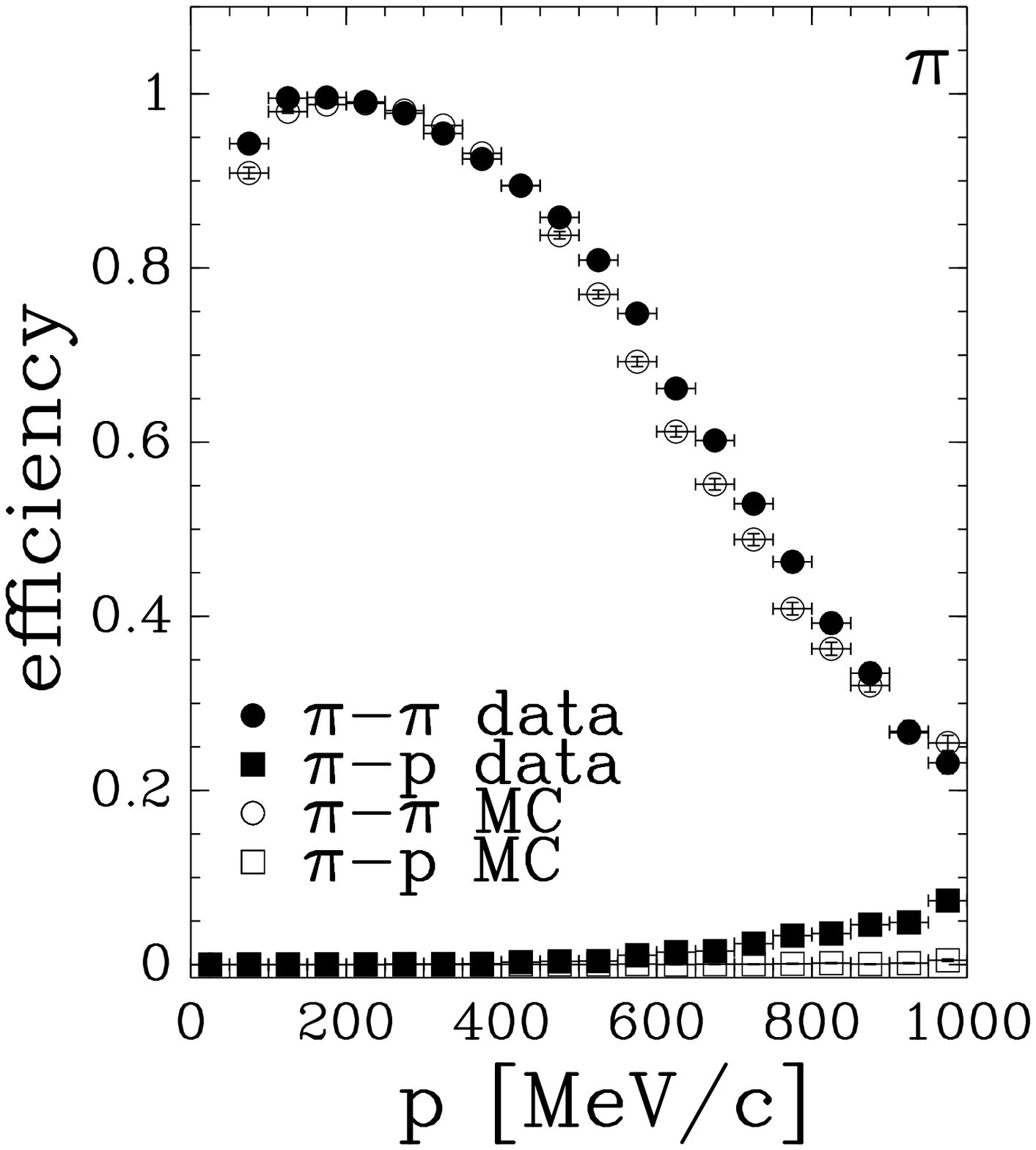}
  ~~
  \includegraphics[width=0.46\textwidth]{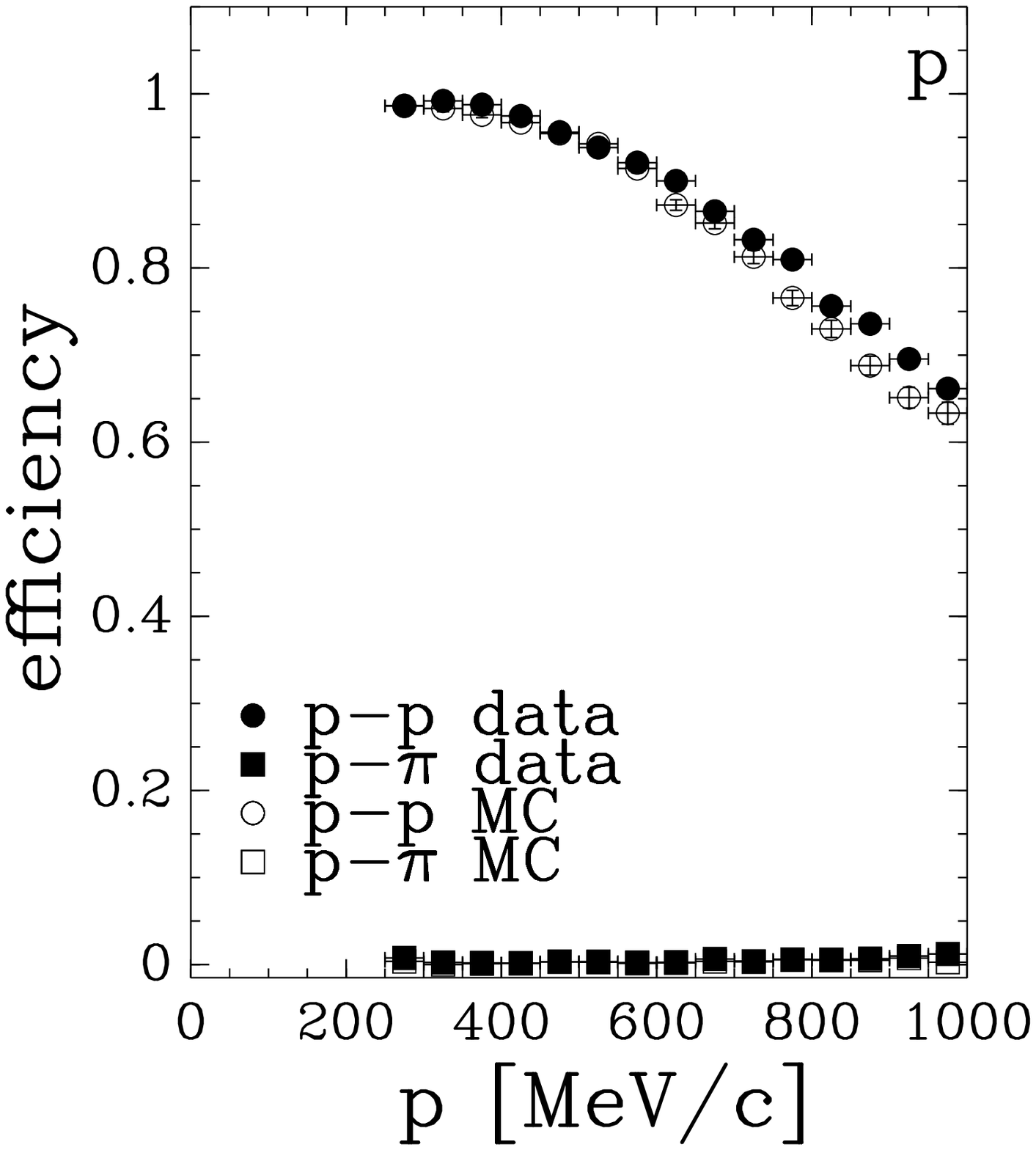}
 \end{center}
 \caption{Performance of the PID using the \dedx as a function of the
  measured momentum of the particle.  The particles are selected using
  TOF.  Left: for negative pions produced in a positive-pion beam;
  Right: for protons produced in a proton beam.
  The filled (open) circles show the efficiency measured with the data
  (Monte Carlo), the filled (open) squares represent the fraction of
  particles misidentified as antiprotons (left) and pions (right) in the
  data (Monte Carlo).
}
\label{fig:dedx:pid}
\end{figure}

The choice to use \dedx as the principal PID estimator is motivated by
two facts. 
The first argument is given by the fact that \dedx is obtained as a
property of the same points which constitute the TPC track, while the
TOF is obtained by matching the track to an external device.
It is observed that the background in the matching is not negligible. 
Converted photons from \piz production can hit the same -- rather large
-- RPC pad as the one pointed to by the track.  
This background depends on the pad position in the RPC barrel and is
different for every momentum setting. 
Thus a different background subtraction would have to be determined for
each momentum-target dataset.
The second argument is the increased complexity of the analysis that
would be introduced by having to combine two PID detectors of which the
response is highly non-Gaussian.
The probability density functions of both the response of the \dedx and
the TOF would have to be determined as function of all relevant
parameters. 
The gain in efficiency one would obtain with such a procedure would be
rather limited and would not balance the additional systematics
introduced. 
On the contrary, the availability of an independent PID device makes it
possible to determine the efficiency and purity of the selection with
the main device in a straightforward manner, without the need to know
the efficiency of the independent auxiliary PID device.

The measurement of $\beta$ allows an almost independent selection of a
very pure proton sample to be made in the momentum range
300~\MeVc--800~\MeVc  with a purity better than 99.8\%. 
The purity of the sample was checked using negative particles and
verifying that no particles identified as antiprotons are present.
While a proton sample was obtained using interactions of incoming
protons, a pure pion sample was prepared by using negative pions
selected by TOF produced by incident positive pions.
The behaviour of positive pions was also checked for momenta below
500~\MeVc (where they can be selected without proton contamination)
and was found to be equal to that of negative pions.

Protons are selected by requiring a high \dedx, while at higher momenta
pions are selected with low \dedx.  
To ensure purity of both samples there are ``unidentified'' particles
between the two samples.
At low momenta electrons are rejected by selecting low \dedx, while pions
are accepted with a higher \dedx.
This separation is not pure above 125~\MeVc, so an electron
subtraction is needed in the analysis.
This procedure will be explained below.

The result in terms of efficiency and 
of the fraction of misidentified particles  
is shown in Fig.~\ref{fig:dedx:pid}.
For the pions, the drop in efficiency toward higher momenta is caused by
the need to make a hard cut to remove protons. 
The migration of pions and protons into the wrong sample is kept below
the percent level in the momentum range of this analysis ($p <
800~\MeVc$). 
This is important for the measurement of the  \pip production rate 
since the proton production rate is significantly larger in some
of the bins.
The small differences in efficiency (up to $\approx5$\%)  which are
visible between the data and the simulation are dealt with in the
analysis by an {\em ad hoc} correction to the cross-sections.
It has been checked that the angular dependences of the PID efficiency
and purity are negligible. 

As already stated, electrons and positrons with a momentum above
125~\MeVc cannot be separated cleanly from pions.
A simulation study shows that the dominant source of electron background
is due to \piz production.
The approach chosen is not to attempt to identify these electrons but
to consider these as background to be subtracted. 
The assumption is made that the \piz spectrum is similar to the
spectrum of charged pions.
With this assumption, the electron background is negligible above
250~\MeVc--300~\MeVc.
The subtraction uses an iterative approach.
Initial \pim and \pip spectra are obtained in an analysis without \piz
subtraction. 
The spectra of pions with opposite charge compared to the beam are then used in
the Monte-Carlo (MC) simulation for the \piz distributions.  
A full simulation of the production and decay into $\gamma$'s with
subsequent conversion in the detector materials is used to predict the
background electron and positron tracks.
Most of these tracks have a momentum below the threshold for this
analysis or low enough to be recognized by \dedx.
The tracks with a PID below the expected value for pions can be
rejected as background.
In the region below 125~\MeVc, a large fraction of the electrons can be
unambiguously identified.
These tracks are used as relative normalization between
data and MC.
The remaining background is then estimated from the distributions of
the simulated electron and positron tracks which are accepted as pion
tracks with the same criteria as used to select the data.
These normalized distributions are subtracted from the data
before the unfolding procedure is applied.
The initial pion spectra obtained in the first pass are not subtracted
for this background.  
This is not a large problem, since this overestimation occurs in a
momentum region where the majority of electrons from these \piz's are
too soft to disturb the analysis.
Uncertainties in the assumption of the \piz spectrum are taken into
account by an alternative assumption that their spectrum follows the
distribution of pions with the same charge as the beam.
An additional systematic error of 10\% is assigned to the
normalization of the \piz subtraction using the identified electrons
and positrons.
At low momenta and small angles the \piz subtraction introduces the
largest systematic uncertainty. 
It is in principle possible to reject more electrons and positrons by
constructing a combined PID estimator based on \dedx and TOF.
Indeed, such an analysis was performed and gave consistent results.
However, its systematic errors are more difficult to estimate.

\section{Data analysis}
\label{sec:analysis}

Only a short outline of the data analysis procedure is presented here:
for further details see Refs. \cite{ref:harp:la,ref:harp:tantalum}. 
The double-differential cross-section for the production of a particle of 
type $\alpha$ can be expressed in the laboratory system as:

\begin{equation}
{\frac{{\mathrm{d}^2 \sigma_{\alpha}}}{{\mathrm{d}p_i \mathrm{d}\theta_j }}} =
\frac{1}{{N_{\mathrm{pot}} }}\frac{A}{{N_A \rho t}}
 \sum_{i',j',\alpha'} M_{ij\alpha i'j' \alpha'}^{-1} \cdot
{N_{i'j'}^{\alpha'} } 
\ ,
\label{eq:cross}
\end{equation}

where $\frac{{\mathrm{d}^2 \sigma_{\alpha}}}{{\mathrm{d}p_i \mathrm{d}\theta_j }}$
is expressed in bins of true momentum ($p_i$), angle ($\theta_j$) and
particle type ($\alpha$).

The ``raw yield'' $N_{i'j'}^{\alpha'}$ 
is the number of particles of observed type $\alpha'$ in bins of reconstructed
momentum ($p_{i'}$) and  angle ($\theta_{j'}$). 
These particles must satisfy the event, track and PID 
selection criteria.
Although, owing to the stringent PID selection,  the background from
misidentified protons in the pion sample is small, the pion and proton
raw yields ($N_{i'j'}^{\alpha'}$, for 
$\alpha'=\pim, \pip, \mathrm{p}$) have been measured simultaneously. 
It is thus possible to correct for the small remaining proton
background in the pion data without prior assumptions concerning the
proton production cross-section.

Various techniques are described in the literature to obtain the
matrix   $ M_{ij\alpha i'j' \alpha'}^{-1}$.
In this analysis an unfolding technique is used.
It performs  
a simultaneous unfolding of $p$, $\theta$ and PID, with a correction
matrix $M^{-1}$ computed mainly using the Monte Carlo simulation.  

The matrix $ M_{ij\alpha i'j' \alpha'}^{-1}$ 
corrects for the  efficiency and the resolution of the detector. 
It unfolds the true variables $i,j,\alpha$ from the reconstructed
variables $i',j',\alpha'$  with a Bayesian technique~\cite{dagostini} 
and corrects  
the observed number of particles to take into account effects such as 
trigger efficiency, reconstruction efficiency, acceptance, absorption,
pion decay, tertiary production, 
PID efficiency, PID misidentification and electron background. 
The central assumption of the method is that the 
probability density function in the (``true'') physical parameters
(``physical distribution'') can be approximated by a histogram with bins
of sufficiently small width.
A population in the physical distribution of events in a
given cell $ij\alpha$ generates a distribution in the measured variables,
$M_{ij\alpha i'j'\alpha'}$. 
Thus the observed distribution in the measurements can be
represented by a linear superposition of such populations.
The task of the unfolding procedure consists then of finding the
number of events in the physical bins for which the predicted
superposition in the measurement space gives the best description of
the data.
The method used to correct for the various effects is  described in
more detail in Ref.~\cite{ref:harp:tantalum}.

To predict the population of the migration matrix element 
$M_{ij\alpha i'j'\alpha'}$, the resolution, efficiency
and acceptance of the detector are obtained from the Monte Carlo.
This is accurate provided the Monte Carlo
simulation describes these quantities correctly. 
Where some deviations
from the control samples measured from the data are found, 
the data are used to introduce (small) {\em ad hoc} corrections to the
Monte Carlo. 
Using the unfolding approach, possible known biases in the measurements
are taken into account automatically as long as they are described by
the Monte Carlo.
In the experiment simulation, which is based on the GEANT4
toolkit~\cite{ref:geant4}, the materials in the beam-line and the 
detector are accurately described as well as
the relevant features of the detector response and 
the digitization process.
The time-dependent properties of the TPC, such as pulse-height
calibration per channel and the presence of dead channels, were
reproduced for each individual data set by running a dedicated
set of high-statistics simulations corresponding to each data set.
In general, the Monte Carlo simulation compares well with the data, as
shown in Ref.~\cite{ref:harp:tantalum}. 
For all important issues, physical benchmarks have been used to validate
the analysis.
The absolute efficiency and the measurement of the angle and momentum
was determined with elastic scattering. 
The momentum and angular resolution was determined exploiting the two
halves of cosmic-ray tracks crossing the TPC volume.
The efficiency of the particle identification was checked using two
independent detector systems.
Only the latter needs a small {\em ad hoc} correction compared to the
simulation.  

The factor  $\frac{A}{{N_A \rho t}}$ in Eq.~\ref{eq:cross}
is the inverse of the number of target nuclei per unit area
($A$ is the atomic mass,
$N_A$ is the Avogadro number, $\rho$ and $t$ are the target density
and thickness).
We do not make a correction for the attenuation
of the beam in the target, so that strictly speaking the
cross-sections are valid for a $\lambda_{\mathrm{I}}=5\%$ target.
The result is normalized to the number of incident pions on the target
$N_{\mathrm{pot}}$. 
The absolute normalization of the result is calculated in the first
instance relative to the number of incident-beam particles accepted by
the selection. 
After unfolding, the factor  $\frac{A}{{N_A \rho t}}$ is applied.
The beam normalization using down-scaled incident pion triggers 
has uncertainties smaller than 2\% for all beam momentum settings.

The background due to interactions of the primary
pions outside the target (called the empty target background) is
measured using data taken without the target mounted in the target
holder.
Owing to the selection criteria which only accept events from the
target region and the good definition of the interaction point this
background is negligible ($< 10^{-5}$).
To subtract backgrounds generated by 
\piz's produced in hadronic interactions of the incident-beam particle,
the assumption is made that the \piz spectrum is similar to the
spectrum of charged pions.
In an iterative procedure the production spectra of pions with opposite
charge compared to the beam particles are used for the
subtraction, while the difference between \pip and \pim production is
used to estimate the systematic error as explained in
Section~\ref{sec:pid}. 
The absorption and decay of particles is simulated by the Monte Carlo.
The generated single particle can re-interact and produce background
particles by hadronic or electromagnetic processes.
These processes are simulated and  additional
particles reconstructed in the TPC in the same event are taken into
account in the unfolding procedure as background.

The effects of the systematic uncertainties on the final results are estimated
by repeating the analysis with the relevant input modified within the
estimated uncertainty intervals.
In many cases this procedure requires the construction of a set of
different migration matrices.
The correlations of the variations between the cross-section bins are
evaluated and expressed in the covariance matrix.
Each systematic error source is represented by its own covariance
matrix.
The sum of these matrices describes the total systematic error.
The magnitude of the systematic errors 
will be shown in Section \ref{sec:results}.

\section{Experimental results}
\label{sec:results}

The measured double-differential cross-sections for the 
production of \pip and \pim in the laboratory system as a function of
the momentum and the polar angle for each incident beam momentum 
\ifall
are shown in 
Figures~\ref{fig:xs-p-th-pbeam-be} to \ref{fig:xs-pim-th-pbeam-pb} for 
targets from Be to Pb.
The error bars  shown are the
square-roots of the diagonal elements in the covariance matrix,
where statistical and systematic uncertainties are combined
in quadrature.
Correlations cannot be shown in the figures.
The overall scale error ($< 2\%$) is not shown.
\else
are available in EPAPS Document No.~[E-PRVCAN-80-078912].
\fi
The correlation of the statistical errors (introduced by the unfolding
procedure) are typically smaller than 20\% for adjacent momentum bins and
even smaller for adjacent angular bins.
The correlations of the systematic errors are larger, typically 80\% for
adjacent bins.
\ifall
The results of this analysis are also tabulated in Appendix A.
\else
The results of this analysis are also tabulated in EPAPS Document
No.~[E-PRVCAN-80-078912]. 
\fi
\begin{table}[tbp] 
\small{
\begin{center}
\caption{Experimental uncertainties for the analysis of the data taken
 with beryllium, carbon, aluminium, copper, tin, tantalum  and lead targets in the
3~\GeVc and 5~\GeVc \pip beams. The
 numbers represent the uncertainty in percent of 
  the cross-section integrated over the angle and momentum region indicated. 
} 
\label{tab:errors-3-pip}
\vspace{2mm}
\begin{tabular}{ l l rrr | rrr | rr} \hline
\bf{p (\GeVc) }&\multicolumn{4}{c|}{0.1 -- 0.3}
                            &\multicolumn{3}{c|}{0.3 -- 0.5}
                            &\multicolumn{2}{c}{0.5 -- 0.7} \\
\hline
\bf{Angle}& &350--&950--&1550--
            &350--&950--&1550--
                       &350--&950-- \\
\bf{(\mrad)}  &  &950&1550&2150
                       &950&1550&2150
                       &950&1550 \\
\hline
\bf{3 \GeVc }&&&&&&&&\\
\hline
\bf{Total syst.} & (Be) &   8.6 &  5.1 &  4.4 &  4.3 &  3.7 &  7.5 &  7.2 & 12.1 \\
                 & (C)  &   9.2 &  4.9 &  3.6 &  4.2 &  4.0 &  6.9 &  6.7 & 12.3 \\
                 & (Al) &  10.7 &  5.7 &  3.8 &  3.9 &  4.2 &  6.0 &  7.0 & 11.9 \\
                 & (Cu) &  10.7 &  8.1 &  6.8 &  3.6 &  4.3 &  6.3 &  6.5 & 10.5 \\
                 & (Sn) &  13.8 &  7.0 &  5.0 &  3.6 &  4.4 &  5.1 &  6.8 & 11.3 \\
                 & (Ta) &  18.8 &  9.6 &  8.0 &  4.0 &  4.6 &  4.9 &  6.8 &  9.7 \\
                 & (Pb) &  13.2 &  7.5 &  6.5 &  3.7 &  4.4 &  5.8 &  7.4 &  9.9 \\
                       
\bf{Statistics}  & (Be) &   1.7 &  1.6 &  2.2 &  1.3 &  1.8 &  3.4 &  1.4 &  2.5 \\
                 & (C)  &   1.8 &  1.6 &  2.0 &  1.3 &  1.8 &  3.4 &  1.5 &  2.7 \\
                 & (Al) &   1.8 &  1.5 &  1.9 &  1.3 &  1.8 &  3.2 &  1.5 &  2.6 \\
                 & (Cu) &   2.5 &  2.2 &  2.6 &  1.9 &  2.5 &  4.3 &  2.2 &  3.6 \\
                 & (Sn) &   1.8 &  1.5 &  1.7 &  1.3 &  1.7 &  2.7 &  1.5 &  2.3 \\
                 & (Ta) &   2.0 &  1.5 &  1.7 &  1.4 &  1.7 &  2.6 &  1.6 &  2.4 \\
                 & (Pb) &   2.3 &  1.8 &  2.1 &  1.7 &  2.1 &  3.4 &  2.1 &  3.0 \\
                
\hline
\bf{5 \GeVc }&&&&&&&&\\
\hline
\bf{Total syst.} & (Be) &   9.7 &  5.1 &  3.4 &  4.6 &  4.4 &  7.3 &  7.1 & 12.2 \\
                 & (C)  &   9.9 &  5.1 &  3.6 &  4.3 &  4.4 &  6.9 &  6.3 & 10.7 \\
                 & (Al) &  11.3 &  5.6 &  3.6 &  4.3 &  4.1 &  6.7 &  7.0 & 11.8 \\
                 & (Cu) &  10.7 &  8.0 &  7.1 &  3.8 &  4.7 &  6.4 &  6.6 & 11.3 \\
                 & (Sn) &  10.0 &  6.4 &  5.1 &  3.4 &  4.4 &  6.6 &  6.6 & 11.6 \\
                 & (Ta) &  19.6 &  8.5 &  7.5 &  3.8 &  4.4 &  5.5 &  6.4 & 10.5 \\
                 & (Pb) &  12.5 &  7.4 &  6.6 &  3.8 &  4.9 &  5.5 &  6.9 & 11.3 \\
              
\bf{Statistics } & (Be) &   1.5 &  1.4 &  1.9 &  1.1 &  1.6 &  3.2 &  1.2 &  2.4 \\
                 & (C)  &   1.1 &  1.0 &  1.3 &  0.8 &  1.2 &  2.2 &  0.8 &  1.7 \\
                 & (Al) &   1.3 &  1.1 &  1.4 &  0.9 &  1.3 &  2.4 &  1.0 &  1.9 \\
                 & (Cu) &   1.2 &  1.1 &  1.4 &  0.9 &  1.2 &  2.1 &  1.0 &  1.8 \\
                 & (Sn) &   1.1 &  1.0 &  1.2 &  0.8 &  1.1 &  1.9 &  1.0 &  1.6 \\
                 & (Ta) &   1.4 &  1.1 &  1.3 &  0.9 &  1.2 &  2.0 &  1.1 &  1.7 \\
                 & (Pb) &   1.7 &  1.3 &  1.6 &  1.2 &  1.6 &  2.6 &  1.4 &  2.3 \\

\end{tabular}
\end{center}
}
\end{table}
\begin{table}[tbp]
\small{
\caption{Experimental uncertainties for the analysis of the data taken
 with beryllium, carbon, aluminium, copper, tin, tantalum  and lead targets in the
8~\GeVc, 8.9~\GeVc, 12~\GeVc and 12.9 \GeVc \pip beam. The
 numbers represent the uncertainty in percent of 
  the cross-section integrated over the angle and momentum region indicated. 
} 
\label{tab:errors-4-pip}

\begin{center}
\vspace{2mm}
\begin{tabular}{ l l rrr | rrr | rr} \hline
\bf{p (\GeVc) }&\multicolumn{4}{c|}{0.1 -- 0.3}
                            &\multicolumn{3}{c|}{0.3 -- 0.5}
                            &\multicolumn{2}{c}{0.5 -- 0.7} \\
\hline
\bf{Angle}& &350--&950--&1550--
            &350--&950--&1550--
                       &350--&950-- \\
\bf{(\mrad)}  &  &950&1550&2150
                       &950&1550&2150
                       &950&1550 \\
\hline
\bf{8 \GeVc }&&&&&&&&\\
\hline
\bf{Total syst.} & (Be) &   9.2 &  4.8 &  3.4 &  4.2 &  4.5 &  6.1 &  7.1 & 10.7 \\
                 & (C)  &  10.0 &  4.8 &  3.3 &  4.2 &  3.7 &  6.0 &  6.1 & 10.2 \\
                 & (Al) &  11.3 &  5.3 &  3.6 &  4.0 &  3.8 &  6.0 &  6.9 & 11.5 \\
                 & (Cu) &  10.3 &  7.8 &  7.0 &  3.7 &  4.0 &  5.5 &  6.1 & 10.2 \\
                 & (Sn) &   9.1 &  6.3 &  5.3 &  3.3 &  4.5 &  6.2 &  6.6 & 11.0 \\
                 & (Ta) &  16.6 &  8.4 &  7.7 &  3.9 &  4.5 &  5.5 &  6.7 & 10.3 \\
                 & (Pb) &  11.1 &  7.6 &  6.2 &  3.7 &  4.6 &  6.1 &  6.4 & 10.0 \\
              
\bf{Statistics}  & (Be) &   2.0 &  2.0 &  2.8 &  1.5 &  2.5 &  4.8 &  1.7 &  3.7 \\
                 & (C)  &   1.6 &  1.5 &  2.0 &  1.1 &  1.8 &  3.3 &  1.2 &  2.5 \\
                 & (Al) &   1.8 &  1.6 &  2.0 &  1.2 &  1.8 &  3.4 &  1.3 &  2.6 \\
                 & (Cu) &   1.4 &  1.3 &  1.6 &  1.0 &  1.4 &  2.5 &  1.1 &  2.0 \\
                 & (Sn) &   1.3 &  1.2 &  1.5 &  1.0 &  1.4 &  2.4 &  1.1 &  2.0 \\
                 & (Ta) &   1.6 &  1.3 &  1.6 &  1.0 &  1.4 &  2.4 &  1.2 &  2.0 \\
                 & (Pb) &   1.4 &  1.2 &  1.5 &  1.0 &  1.4 &  2.4 &  1.2 &  1.9 \\
\hline
\bf{8.9 \GeVc }&&&&&&&&\\
\hline
\bf{Total syst.}   & (Be) &   9.3 &  4.7 &  3.1 &  4.2 &  4.4 &  7.7 &  7.9 & 12.4 \\
\bf{Statistics}    & (Be) &   0.6 &  0.5 &  0.7 &  0.4 &  0.6 &  1.1 &  0.4 &  0.9 \\
\hline
\bf{12 \GeVc }&&&&&&&&\\
\hline
\bf{Total syst.} & (Be) &   9.1 &  4.8 &  3.4 &  4.1 &  3.3 &  6.3 &  7.1 & 13.2 \\
                 & (C)  &  10.2 &  4.6 &  3.4 &  4.0 &  3.7 &  6.6 &  6.5 &  9.4 \\
                 & (Al) &  12.7 &  5.7 &  3.9 &  3.4 &  4.0 &  5.8 &  7.3 & 13.0 \\
                 & (Cu) &  10.6 &  7.8 &  6.7 &  3.4 &  3.7 &  4.9 &  6.9 & 11.5 \\
                 & (Sn) &  10.1 &  6.7 &  5.9 &  3.0 &  4.2 &  6.5 &  6.0 & 10.3 \\
                 & (Ta) &  17.4 &  7.9 &  7.6 &  2.6 &  4.4 &  6.3 &  6.8 & 10.0 \\
                 & (Pb) &  11.1 &  7.6 &  6.7 &  2.8 &  4.7 &  4.8 &  6.7 &  9.9 \\

\bf{Statistics}  & (Be) &   3.2 &  3.4 &  4.5 &  2.4 &  4.0 &  8.5 &  2.7 &  6.1 \\
                 & (C)  &   3.3 &  3.3 &  4.4 &  2.4 &  4.0 &  7.9 &  2.6 &  6.2 \\
                 & (Al) &   4.7 &  4.6 &  6.1 &  3.3 &  5.6 & 12.3 &  3.8 &  8.8 \\
                 & (Cu) &   4.2 &  4.4 &  5.5 &  3.1 &  4.9 &  9.5 &  3.6 &  7.6 \\
                 & (Sn) &   2.6 &  2.6 &  3.2 &  2.0 &  2.9 &  5.4 &  2.2 &  4.1 \\
                 & (Ta) &   3.7 &  3.5 &  4.4 &  2.8 &  4.2 &  8.4 &  3.3 &  6.2 \\
                 & (Pb) &   6.1 &  5.7 &  7.6 &  4.9 &  7.0 & 15.7 &  5.7 & 11.0 \\
\hline
\bf{12.9 \GeVc }&&&&&&&&\\
\hline
\bf{Total syst.} & (Al) &  11.8 &  5.2 &  3.8 &  3.7 &  3.8 &  6.7 &  7.2 & 11.1 \\
\bf{Statistics}  & (Al) &   1.4 &  1.4 &  1.8 &  1.0 &  1.7 &  3.0 &  1.2 &  2.3 \\
\hline
\end{tabular}
\end{center}
}
\end{table}

\begin{table}[tbp] 
\small{
\begin{center}
\caption{Experimental uncertainties for the analysis of the data taken
 with beryllium, carbon, aluminium, copper, tin  and lead targets in the 
3~\GeVc and  5~\GeVc \pim beams.  The
 numbers represent the uncertainty in percent of 
  the cross-section integrated over the angle and momentum region indicated. 
} 
\label{tab:errors-3-pim}
\vspace{2mm}
\begin{tabular}{ l l rrr | rrr | rr} \hline
\bf{p (\GeVc) }&\multicolumn{4}{c|}{0.1 -- 0.3}
                            &\multicolumn{3}{c|}{0.3 -- 0.5}
                            &\multicolumn{2}{c}{0.5 -- 0.7} \\
\hline
\bf{Angle}& &350--&950--&1550--
            &350--&950--&1550--
                       &350--&950-- \\
\bf{(\mrad)}  &  &950&1550&2150
                       &950&1550&2150
                       &950&1550 \\
\hline
\bf{3 \GeVc }&&&&&&&&\\
\hline
\bf{Total syst.} & (Be)  &   8.8 &  4.9 &  3.3 &  4.3 &  4.0 &  5.8 &  6.8 & 10.7  \\
                 & (C)   &   8.8 &  4.9 &  3.4 &  4.1 &  4.0 &  6.0 &  7.1 & 10.7  \\
                 & (Al)  &  10.1 &  5.4 &  3.9 &  4.0 &  4.0 &  6.1 &  7.1 & 10.2  \\
                 & (Cu)  &  13.5 & 10.2 &  9.2 &  5.1 &  5.6 &  6.8 &  7.8 & 10.4  \\
                 & (Sn)  &  11.2 &  6.8 &  5.5 &  3.8 &  4.2 &  5.7 &  7.0 &  9.5  \\
                 & (Ta)  &  17.0 &  9.5 &  8.0 &  3.9 &  4.5 &  5.5 &  7.2 & 10.0  \\
                 & (Pb)  &  17.7 &  9.0 &  6.7 &  4.1 &  4.5 &  4.8 &  7.6 &  9.7  \\
		    	   	  	               	      
\bf{Statistics}    & (Be)  &   1.1 &  1.0 &  1.3 &  0.8 &  1.2 &  2.1 &  0.9 &  1.6  \\
                   & (C)   &   1.1 &  1.0 &  1.3 &  0.9 &  1.2 &  2.1 &  1.0 &  1.6  \\
                   & (Al)  &   1.1 &  0.9 &  1.1 &  0.8 &  1.0 &  1.8 &  0.9 &  1.4  \\
                   & (Cu)  &   1.4 &  1.2 &  1.4 &  1.0 &  1.3 &  2.2 &  1.2 &  1.8  \\
                   & (Sn)  &   1.3 &  1.1 &  1.3 &  1.1 &  1.3 &  2.1 &  1.2 &  1.7  \\
                   & (Ta)  &   2.9 &  2.3 &  2.7 &  2.1 &  2.5 &  4.0 &  2.4 &  3.4  \\
                   & (Pb)  &   3.0 &  2.2 &  2.6 &  2.2 &  2.7 &  4.1 &  2.5 &  3.6  \\
\hline
\bf{5 \GeVc }&&&&&&&&\\
\hline
\bf{Total syst.} & (Be)  &   9.5 &  4.8 &  3.2 &  4.3 &  4.0 &  6.5 &  6.7 & 10.9  \\
                 & (C)   &   9.4 &  5.1 &  3.4 &  4.2 &  4.3 &  6.4 &  6.7 & 10.6  \\
                 & (Al)  &  10.4 &  5.5 &  3.8 &  4.3 &  4.1 &  5.7 &  7.7 & 11.3  \\
                 & (Cu)  &  14.4 & 10.0 &  9.0 &  5.4 &  5.5 &  6.9 &  7.9 & 11.1  \\
                 & (Sn)  &  10.8 &  6.9 &  5.5 &  3.8 &  4.5 &  5.7 &  7.3 & 10.2  \\
                 & (Ta)  &  15.2 &  8.3 &  7.5 &  3.8 &  4.5 &  4.8 &  6.9 &  9.7  \\
                 & (Pb)  &  14.9 &  7.5 &  6.6 &  3.8 &  4.6 &  5.4 &  6.9 &  9.2  \\
  	     	    	   	  	               	      
\bf{Statistics}    & (Be)  &   1.5 &  1.4 &  1.8 &  1.1 &  1.6 &  3.0 &  1.2 &  2.3  \\
                   & (C)   &   1.2 &  1.1 &  1.5 &  0.9 &  1.3 &  2.4 &  1.0 &  1.8  \\
                   & (Al)  &   1.4 &  1.2 &  1.5 &  1.0 &  1.5 &  2.6 &  1.2 &  2.0  \\
                   & (Cu)  &   1.2 &  1.1 &  1.4 &  0.9 &  1.2 &  2.1 &  1.0 &  1.6  \\
                   & (Sn)  &   1.1 &  0.9 &  1.2 &  0.8 &  1.1 &  1.8 &  0.9 &  1.5  \\
                   & (Ta)  &   1.4 &  1.2 &  1.6 &  1.1 &  1.4 &  2.4 &  1.2 &  1.9  \\
                   & (Pb)  &   1.2 &  1.0 &  1.2 &  0.9 &  1.2 &  2.0 &  1.1 &  1.6  \\
\hline
\end{tabular}
\end{center}
}
\end{table}
\begin{table}[tbp]
\small{
\caption{Experimental uncertainties for the analysis of the data taken
 with beryllium, carbon, aluminium, copper, tin  and lead targets in the 
 8~\GeVc  and 12~\GeVc \pim beam. The
 numbers represent the uncertainty in percent of 
  the cross-section integrated over the angle and momentum region indicated. 
} 
\label{tab:errors-4-pim}

\begin{center}
\vspace{2mm}
\begin{tabular}{ l l rrr | rrr | rr} \hline
\bf{p (\GeVc) }&\multicolumn{4}{c|}{0.1 -- 0.3}
                            &\multicolumn{3}{c|}{0.3 -- 0.5}
                            &\multicolumn{2}{c}{0.5 -- 0.7} \\
\hline
\bf{Angle}& &350--&950--&1550--
            &350--&950--&1550--
                       &350--&950-- \\
\bf{(\mrad)}  &  &950&1550&2150
                       &950&1550&2150
                       &950&1550 \\
\hline
\bf{8 \GeVc }&&&&&&&&\\
\hline
\bf{Total syst.} & (Be)  &   9.6 &  4.9 &  3.6 &  4.3 &  3.8 &  6.2 &  6.8 & 10.9 \\
                 & (C)   &   9.9 &  4.9 &  3.4 &  4.2 &  3.9 &  6.4 &  6.9 & 11.1 \\
                 & (Al)  &  10.8 &  5.3 &  3.7 &  4.2 &  3.9 &  6.0 &  6.8 & 10.2 \\
                 & (Cu)  &  16.1 & 10.2 &  8.9 &  5.2 &  5.2 &  6.1 &  7.4 & 10.6 \\
                 & (Sn)  &  11.9 &  7.2 &  5.5 &  3.8 &  4.4 &  5.8 &  6.7 & 10.2 \\
                 & (Ta)  &  16.2 &  8.2 &  7.6 &  3.8 &  4.7 &  5.3 &  6.4 &  9.5 \\
                 & (Pb)  &  17.0 &  7.7 &  6.7 &  3.9 &  4.4 &  5.7 &  7.1 &  9.9 \\
  	     	    	   	  	               	      
\bf{Statistics}    & (Be)  &   1.2 &  1.2 &  1.5 &  0.8 &  1.3 &  2.4 &  0.9 &  1.9  \\
                   & (C)   &   1.1 &  1.0 &  1.3 &  0.7 &  1.1 &  2.1 &  0.8 &  1.6  \\
                   & (Al)  &   1.1 &  1.0 &  1.2 &  0.7 &  1.1 &  2.0 &  0.8 &  1.6  \\
                   & (Cu)  &   1.0 &  0.9 &  1.1 &  0.6 &  0.9 &  1.6 &  0.7 &  1.2  \\
                   & (Sn)  &   1.1 &  1.0 &  1.2 &  0.8 &  1.1 &  1.9 &  0.8 &  1.4  \\
                   & (Ta)  &   1.2 &  1.0 &  1.3 &  0.8 &  1.1 &  1.9 &  0.9 &  1.5  \\
                   & (Pb)  &   1.2 &  0.9 &  1.1 &  0.7 &  1.0 &  1.7 &  0.8 &  1.4  \\
\hline
\bf{12 \GeVc }&&&&&&&&\\
\hline
\bf{Total syst.}   & (Be)  &   9.7 &  5.1 &  3.5 &  4.6 &  4.0 &  6.0 &  6.9 & 10.0  \\
                   & (C)   &   9.6 &  4.6 &  3.3 &  4.1 &  3.8 &  6.0 &  7.0 & 10.9  \\
                   & (Al)  &   9.0 &  5.0 &  3.9 &  3.8 &  4.0 &  6.1 &  6.1 &  9.6  \\
                   & (Cu)  &  16.7 & 10.0 &  8.5 &  4.8 &  6.9 &  8.4 &  7.1 & 10.5  \\
                   & (Sn)  &  10.2 &  6.6 &  5.4 &  3.8 &  4.5 &  5.6 &  7.0 & 10.3  \\
                   & (Ta)  &  14.1 &  7.9 &  7.6 &  3.8 &  4.6 &  5.7 &  6.6 & 10.0  \\
                   & (Pb)  &  19.1 &  7.5 &  6.5 &  3.7 &  4.5 &  5.5 &  6.6 & 10.0  \\
  	     	    	   	  	               	      
\bf{Total syst.} & (Be)  &   1.4 &  1.4 &  1.9 &  1.0 &  1.6 &  3.0 &  1.1 &  2.2  \\
                 & (C)   &   1.4 &  1.4 &  1.8 &  1.0 &  1.7 &  3.1 &  1.1 &  2.3  \\
                 & (Al)  &   1.0 &  1.0 &  1.3 &  0.8 &  1.2 &  2.1 &  0.9 &  1.7  \\
                 & (Cu)  &   1.8 &  1.7 &  2.2 &  1.2 &  1.7 &  3.1 &  1.3 &  2.0  \\
                 & (Sn)  &   0.8 &  0.8 &  1.0 &  0.6 &  0.9 &  1.6 &  0.7 &  1.3  \\
                 & (Ta)  &   0.9 &  0.9 &  1.1 &  0.7 &  1.0 &  1.7 &  0.8 &  1.3  \\
                 & (Pb)  &   1.1 &  0.9 &  1.1 &  0.7 &  1.0 &  1.7 &  0.8 &  1.3  \\
\hline
\end{tabular}
\end{center}
}
\end{table}

One observes that only for the 3~\GeVc beam is the statistical error  
similar in magnitude to the systematic error, while the statistical
error is negligible for the 8~\GeVc and 12~\GeVc beam settings.
The statistical error is calculated by error propagation as part of the
unfolding procedure. 
It takes into account that the unfolding matrix is obtained from the
data themselves~\footnote{The migration matrix is calculated without
prior knowledge of the cross-sections, while the unfolding procedure
determines the unfolding matrix from the migration matrix and the
distributions found in the data.} and hence contributes also to the
statistical error. 
This procedure almost doubles the statistical error, but it avoids an important 
systematic error which would otherwise be introduced by assuming a
cross-section model {\em a priori} to calculate the corrections. 

The largest systematic error corresponds to the uncertainty in the
absolute momentum scale, which was estimated to be around 3\% using elastic
scattering~\cite{ref:harp:tantalum}.
At low momentum in the relatively small angle forward
direction, the uncertainty in the subtraction of the electron and
positron background due to \piz production is dominant ($\sim 6-10 \%$).
This uncertainty is split between the variation in the shape of the
\piz spectrum and the normalization using the recognized electrons. 
The target region definition  and
the uncertainty in the PID efficiency and background from tertiaries
(particles produced in secondary interactions)
are of similar size and are not negligible ($\sim 2-3 \%$).
Relatively small errors are introduced by the uncertainties in
the absorption correction, absolute knowledge of the angular and the
momentum resolution.
The correction for tertiaries is relatively large at low momenta and large 
angles ($\sim 3-5 \%$). 
As expected, this region is most affected by this component.

As already mentioned above, the overall normalization has an uncertainty
of 2\% and is not reported in the table.
It is mainly due to the uncertainty in the efficiency that beam pions
counted in the normalization actually hit the target and the muon
contamination in the beams, with smaller
components from the target density and beam particle counting procedure.

One observes the weak beam energy dependence of
pion production by incoming pions.
This is particularly striking for low $A$ targets and enhanced for
same-charge pion production.
The energy dependence is larger for the more forward bins in the range
covered in this analysis; backward production also shows little beam
energy dependence over the whole range in $A$.
The 3~\GeVc data tend to show a markedly different behaviour from all
higher beam energy data already at moderate $A$ (Cu) in the most forward
angular bins, a trend which extends to larger angles for larger $A$ targets.

\begin{figure}[tbp!]
\begin{center}
 \includegraphics[width=0.37\textwidth]{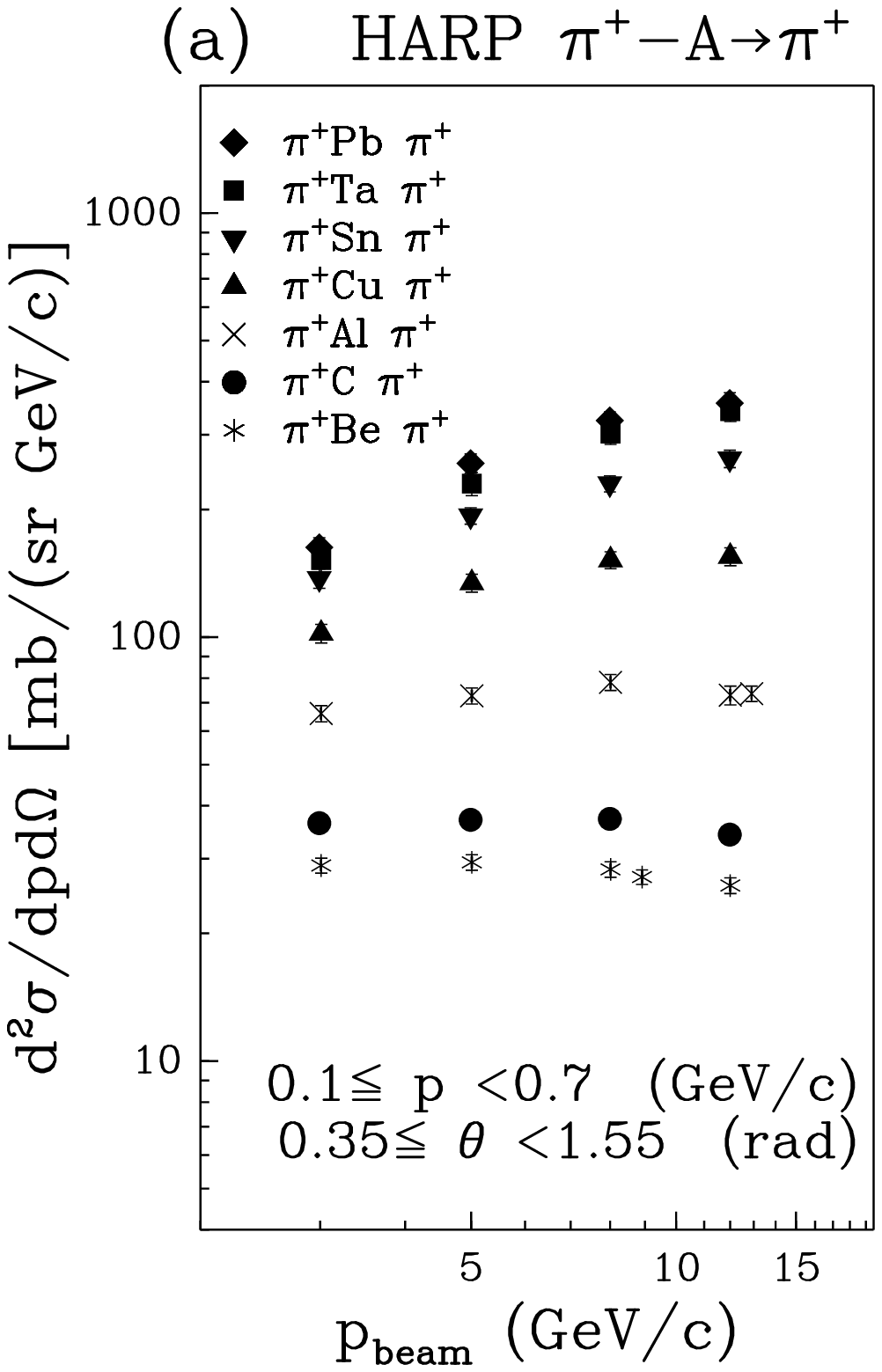}
 ~
 \includegraphics[width=0.37\textwidth]{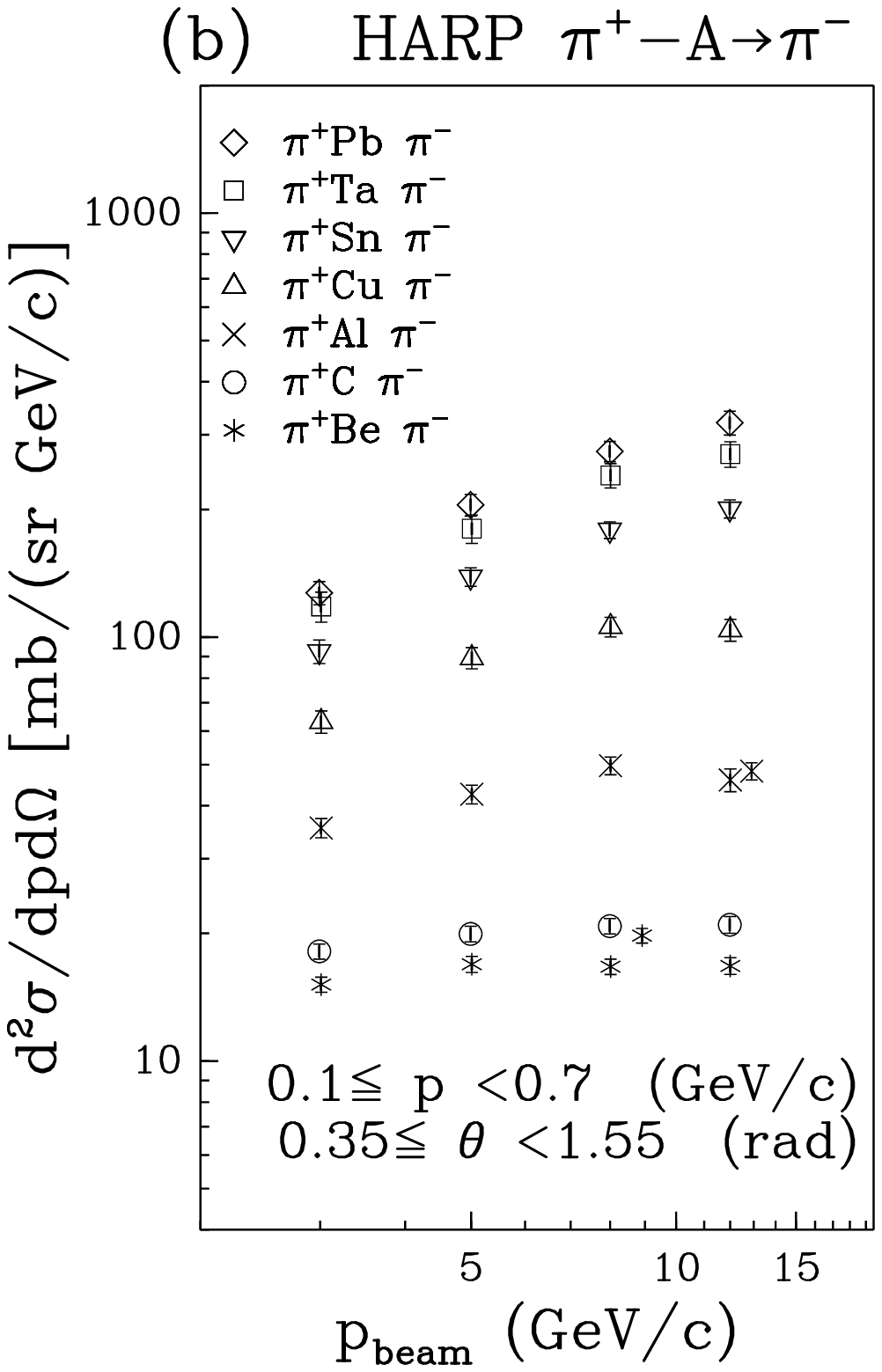}
 ~
 \includegraphics[width=0.37\textwidth]{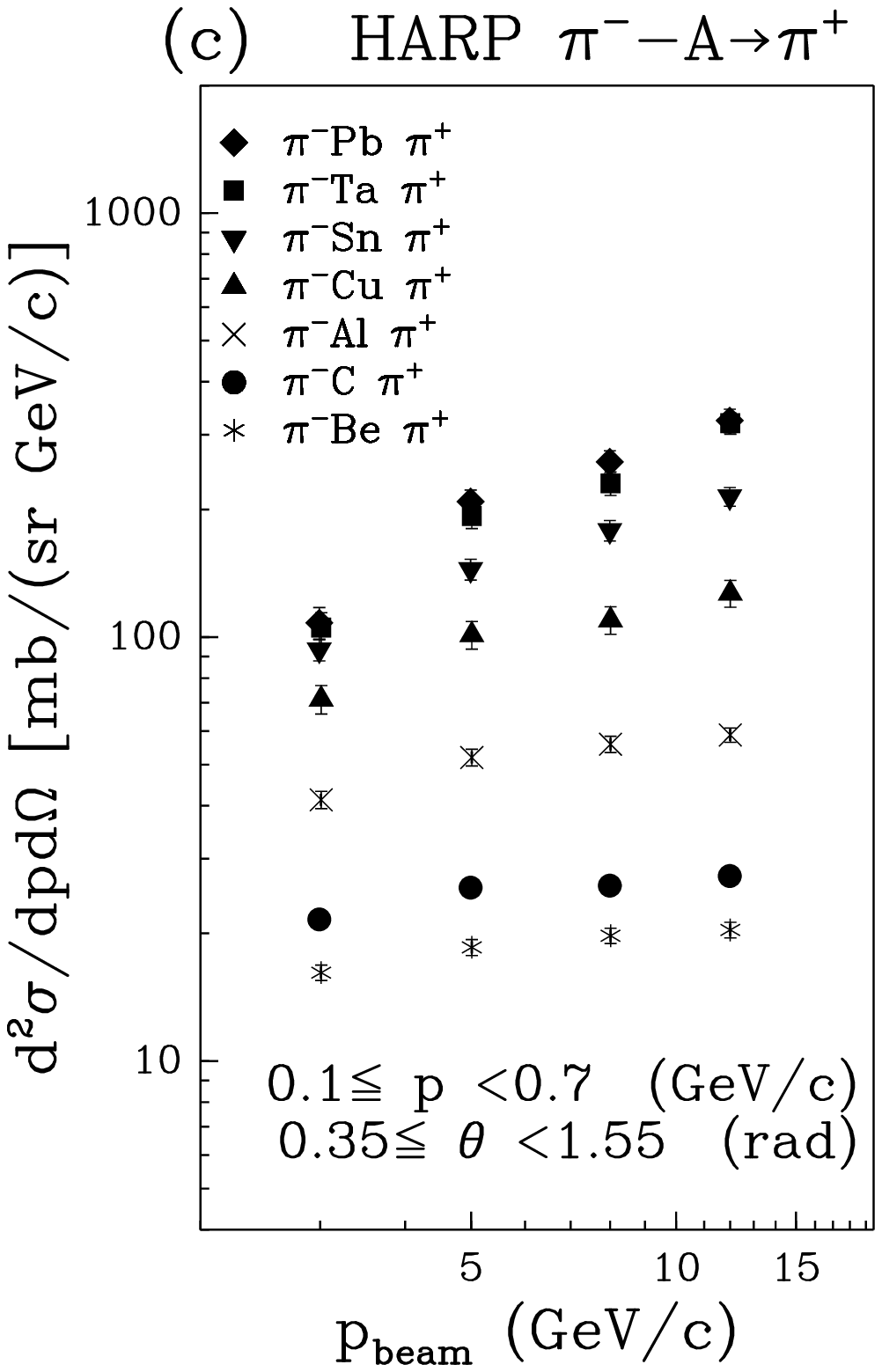}
 ~
 \includegraphics[width=0.37\textwidth]{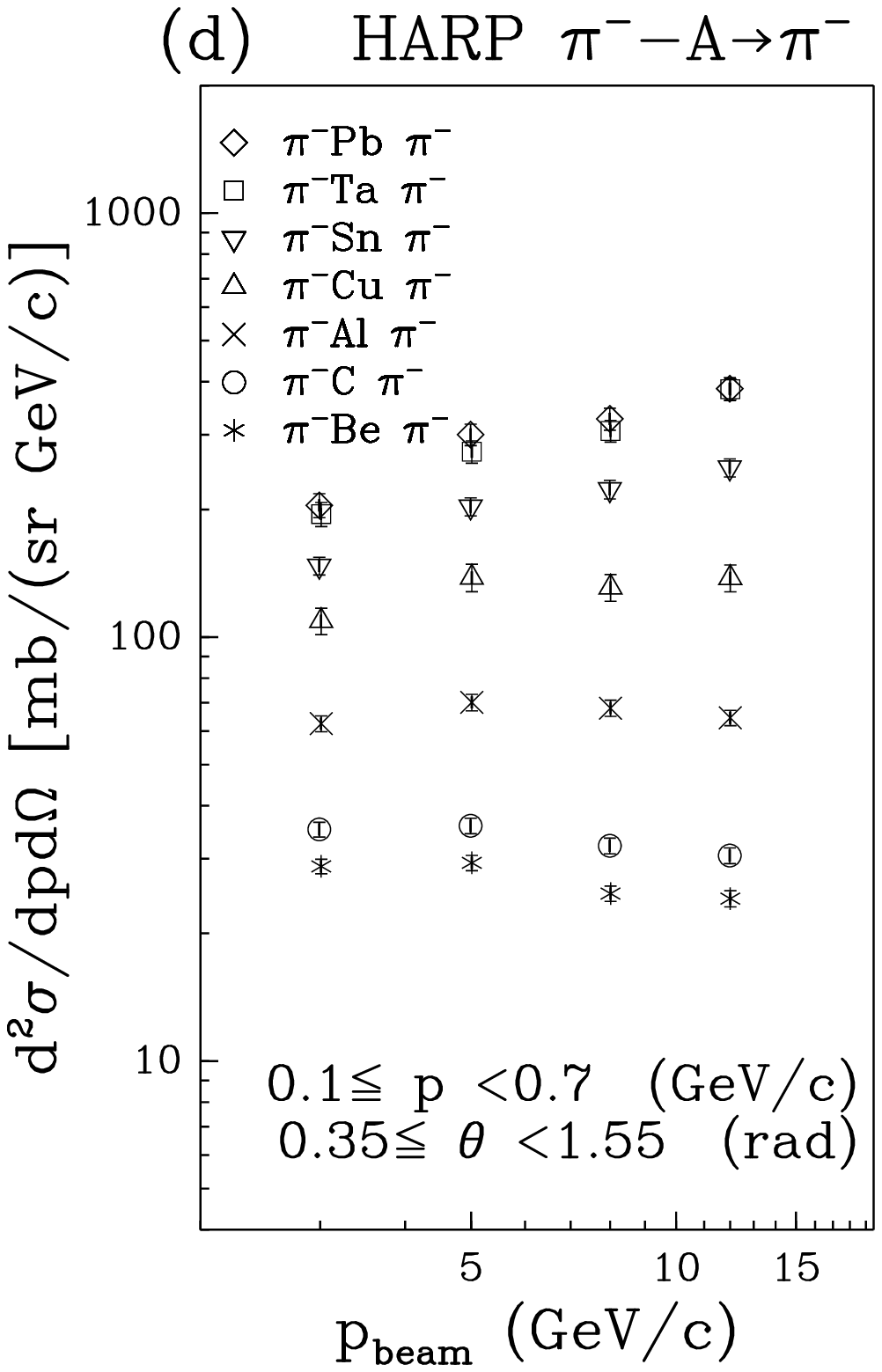}
 ~

\end{center}
\caption{
 Dependence on the beam momentum of the \pim (right) and \pip (left) 
  production yields
 in \pipm--Be, \pipm--C, \pipm--Al, \pipm--Cu, \pipm--Sn, \pipm--Ta, \pipm--Pb
 interactions averaged over the forward angular region 
 ($0.350~\rad \leq \theta < 1.550~\rad$) 
 and momentum region $100~\MeVc \leq p < 700~\MeVc$.
 The top-left panel (a) refers to \pip production in \pip beams, the
 top-right panel (b) to \pim production in \pip beams, while to
 bottom-left (c) and bottom-right (d) panels refer respectively to \pip
 and \pim production in \pim beams.
}
\label{fig:xs-trend}
\end{figure}

The dependence of the averaged pion yields on the incident-beam
momentum is shown in Fig.~\ref{fig:xs-trend}. 
The \pip and \pim yields are averaged over the region 
$0.350~\rad \leq \theta < 1.550~\rad$ and $100~\MeVc \leq p < 700~\MeVc$
(pions produced in the forward direction only).
Whereas the beam energy dependence of the yields in the
Be and C data differs clearly from the dependence in the Ta and Pb data
one can observe that the Al, Cu and Sn data display a
smooth transition between them.
The dependence in the Be and C data is much more flat with a saturation of
the yield between 8~\GeVc and 12~\GeVc
with the Al, Cu and Sn showing an intermediate behaviour.
The momentum dependence is larger for opposite-charge pion production
than for equal-charge production.

\begin{figure}[tbp!]
\begin{center}
 \includegraphics[width=0.45\textwidth]{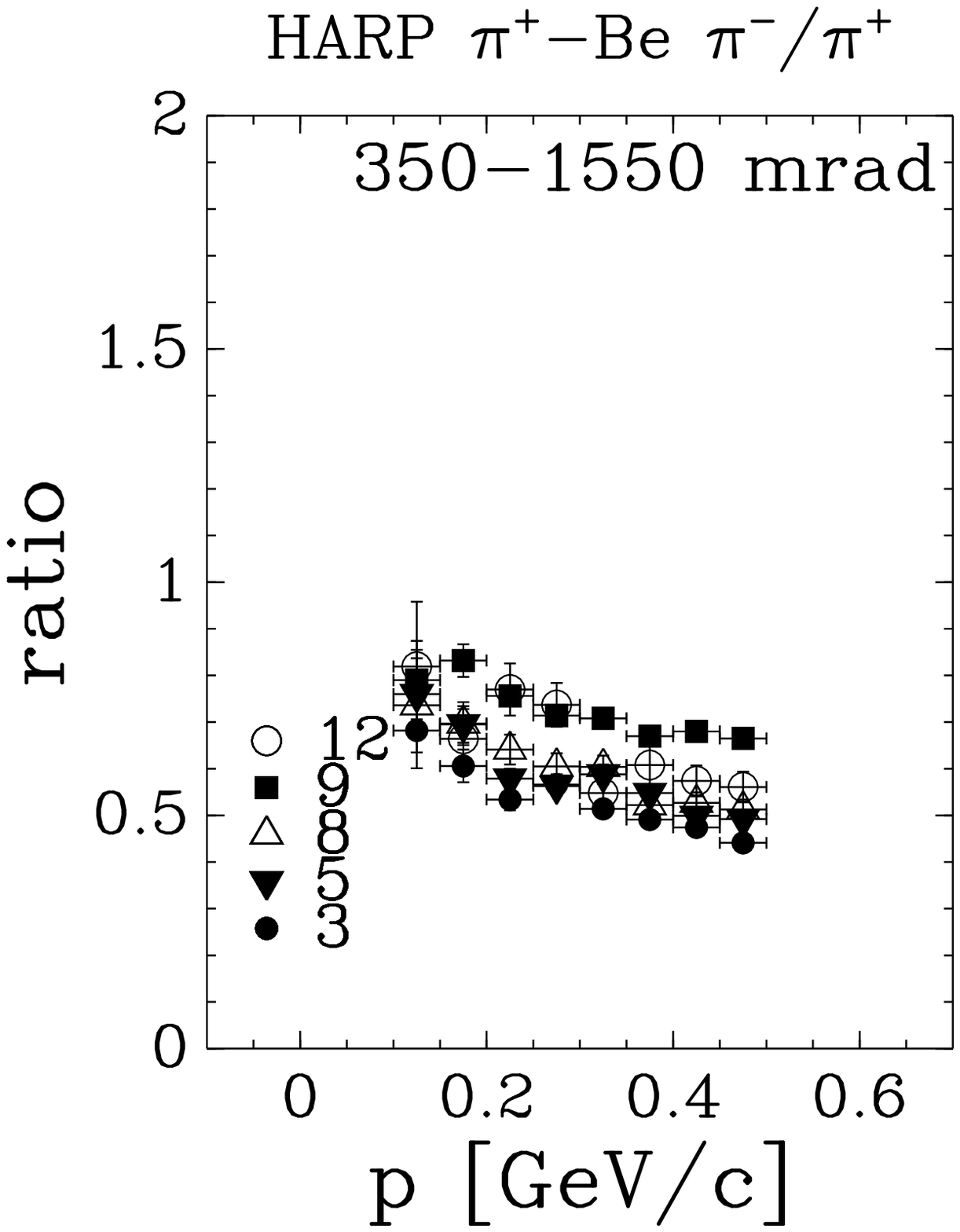}
 ~
 \includegraphics[width=0.45\textwidth]{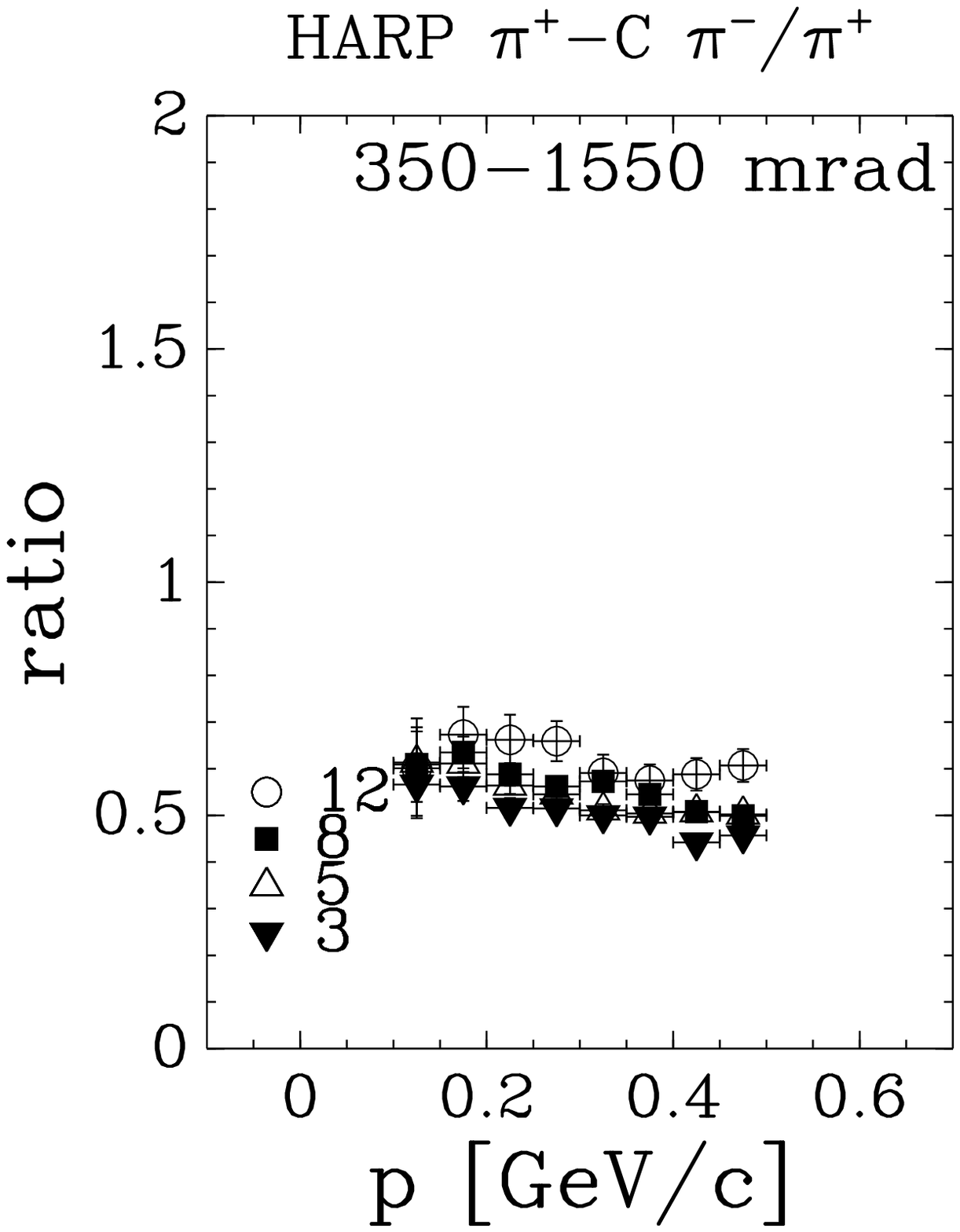}
 \vspace{3mm}
 \includegraphics[width=0.45\textwidth]{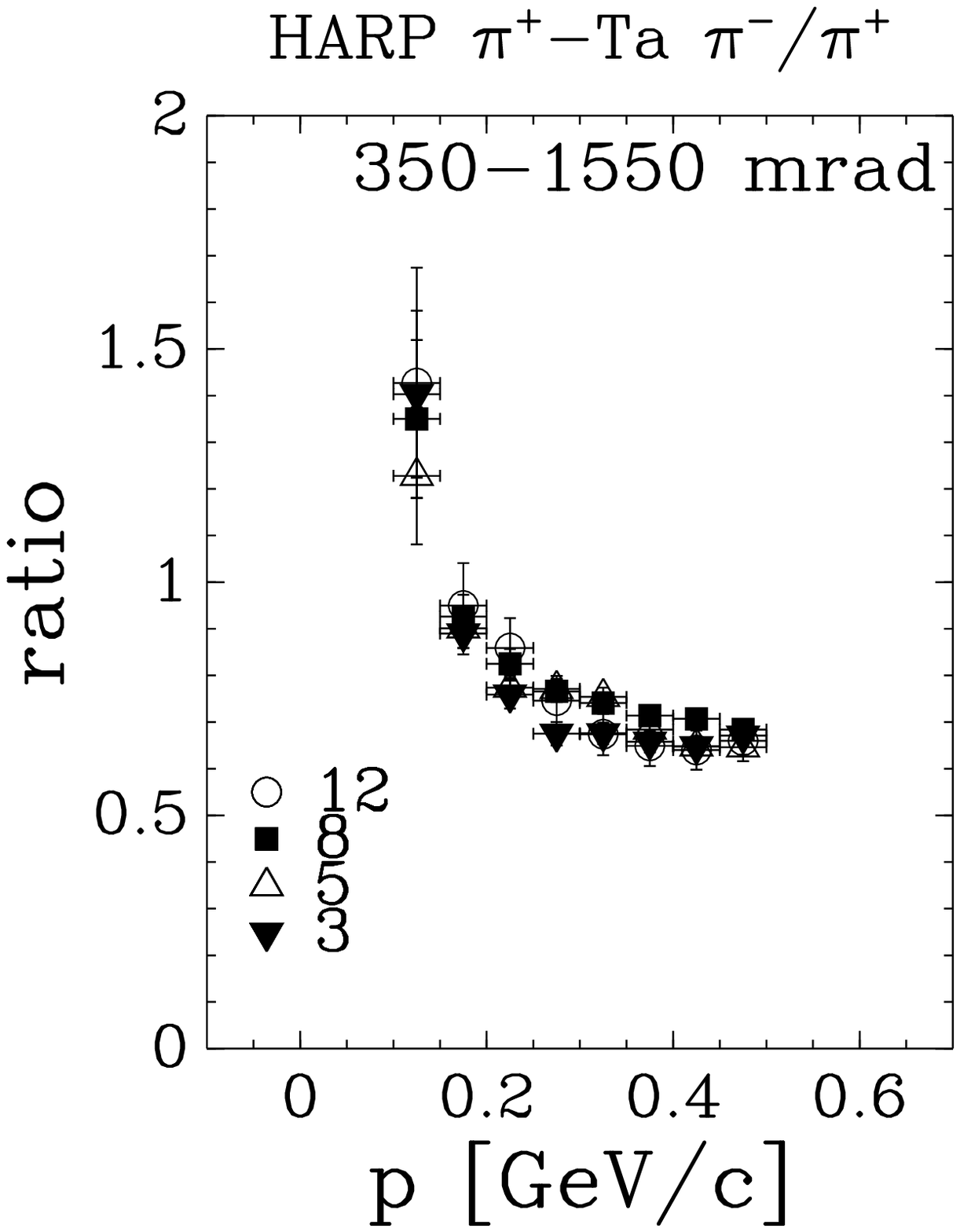}
 ~
 \includegraphics[width=0.45\textwidth]{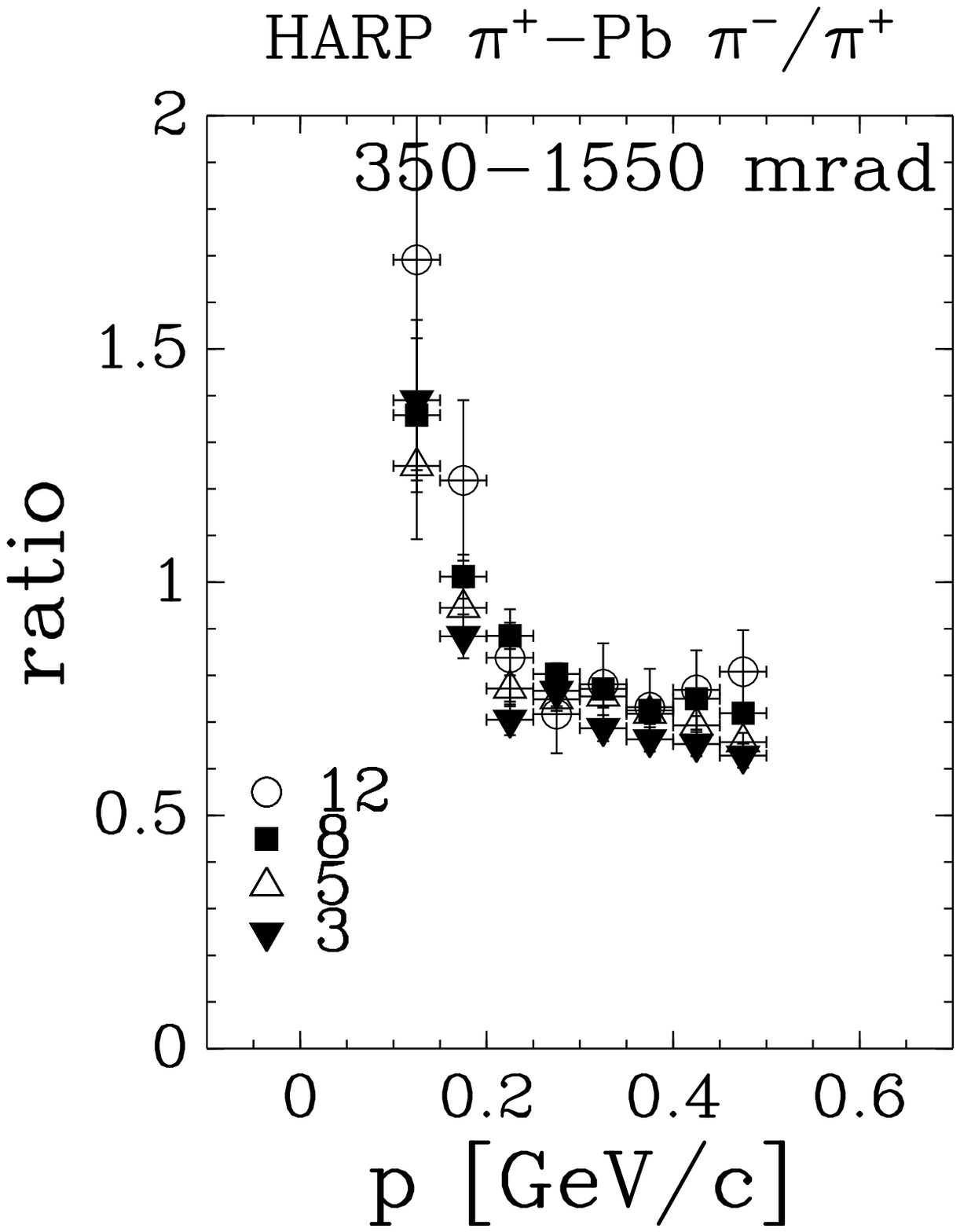}
\end{center}
\caption{
From top-left panel to bottom-right panel, the ratio of the differential cross-sections for \pim and \pip
 production in
\pip--Be (top-left), \pip--C (top-right),  \pip--Ta (bottom-left)   and
 \pip--Pb (bottom-right)   interactions as a function of 
the secondary momentum integrated over the
forward angular region (shown in mrad).
In the figure, the symbol legends 13 and 9 refer to 12.9 and 8.9~\GeVc nominal
beam momentum, respectively.
}
\label{fig:xs-ratio}
\end{figure}

\begin{figure}[tbp!]
\begin{center}
 \includegraphics[width=0.45\textwidth]{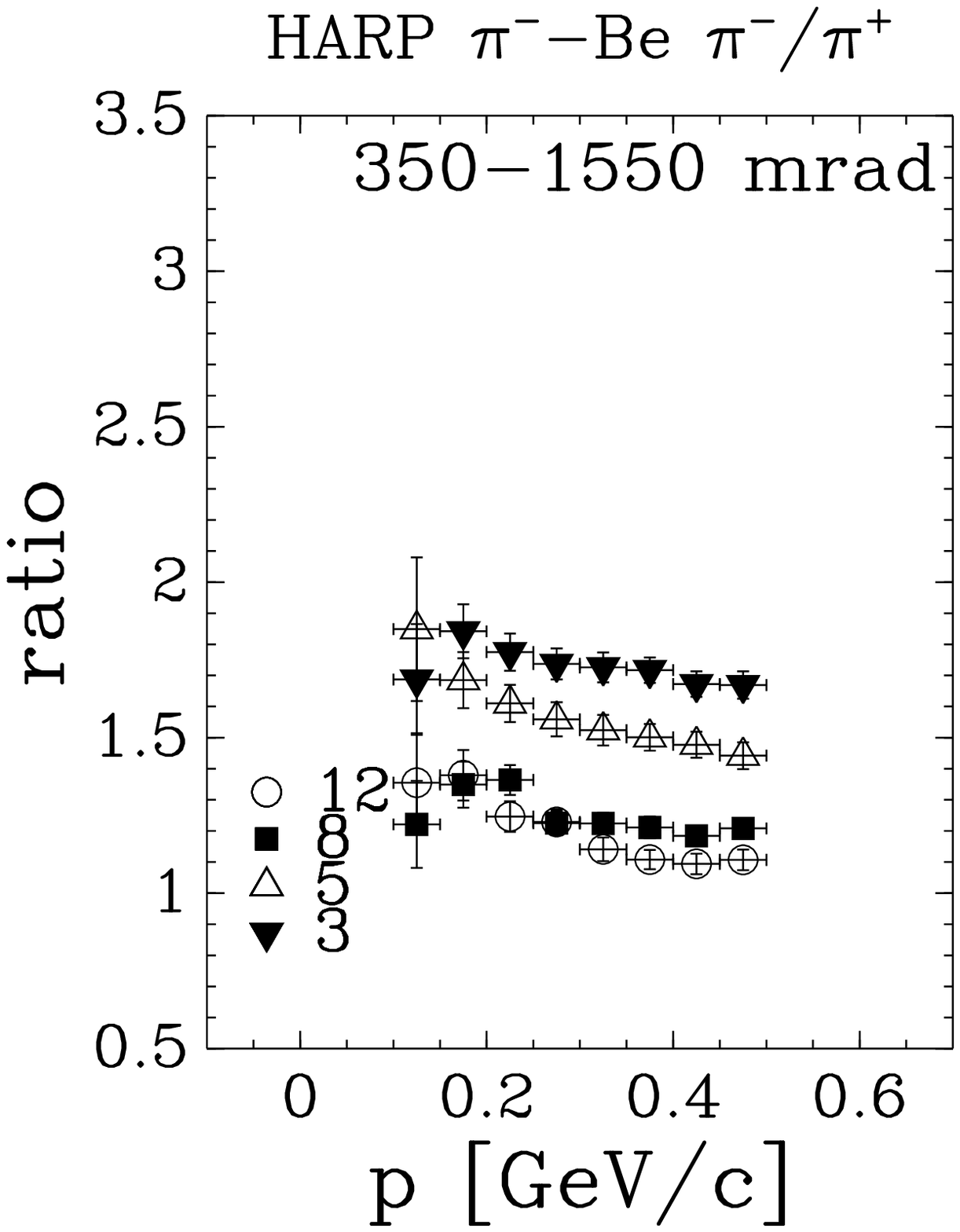}
 ~
 \includegraphics[width=0.45\textwidth]{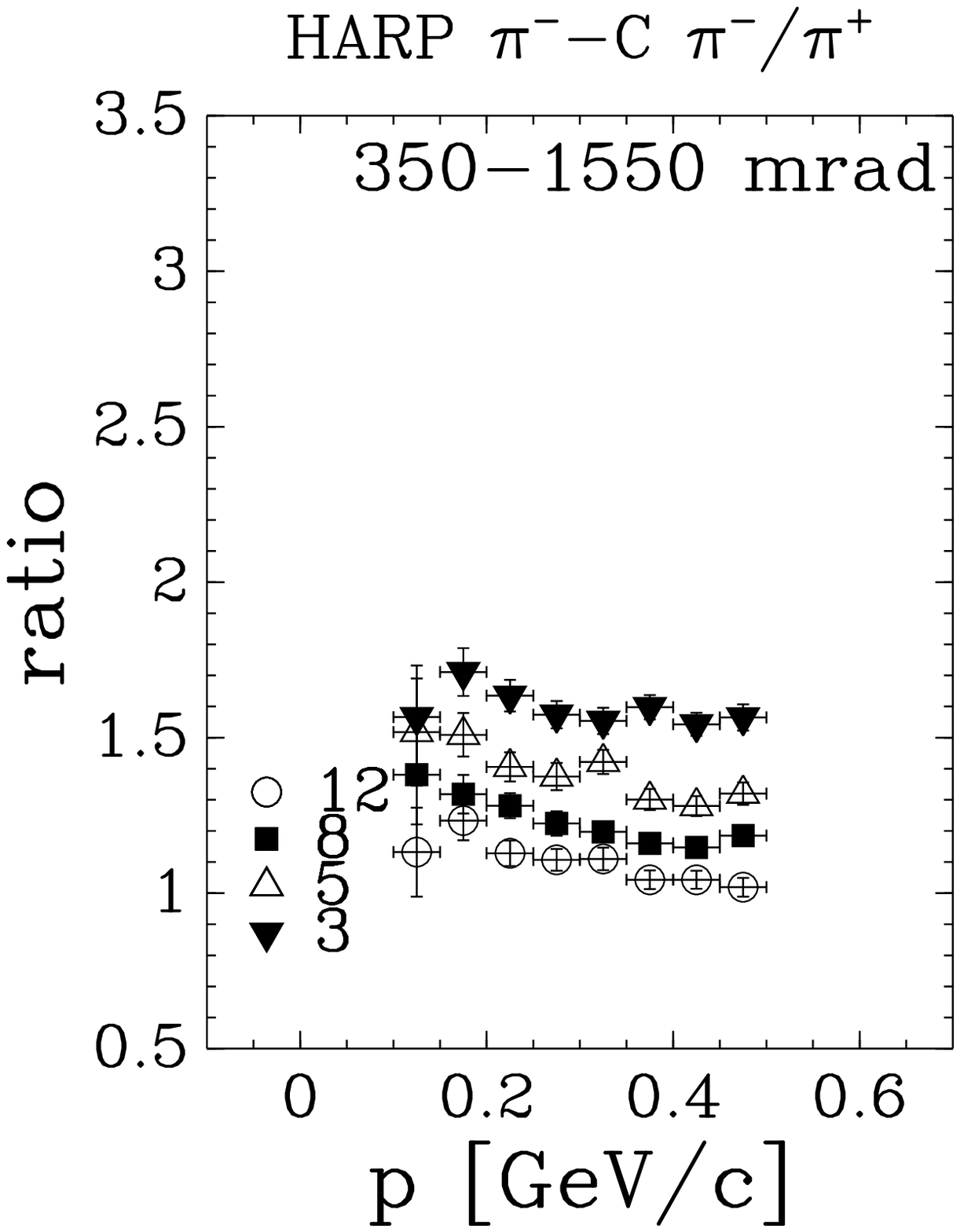}
 \vspace{3mm}
 \includegraphics[width=0.45\textwidth]{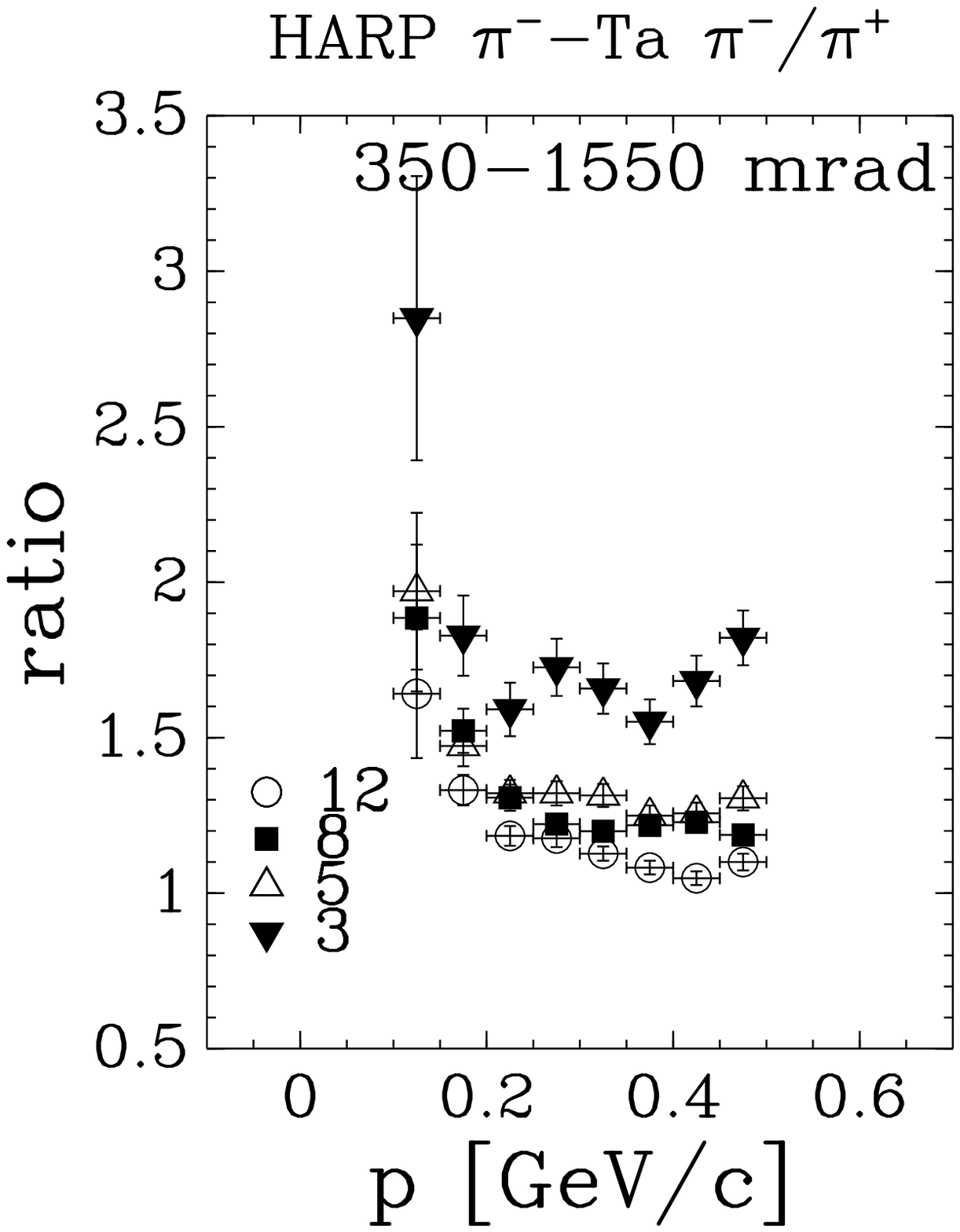}
 \includegraphics[width=0.45\textwidth]{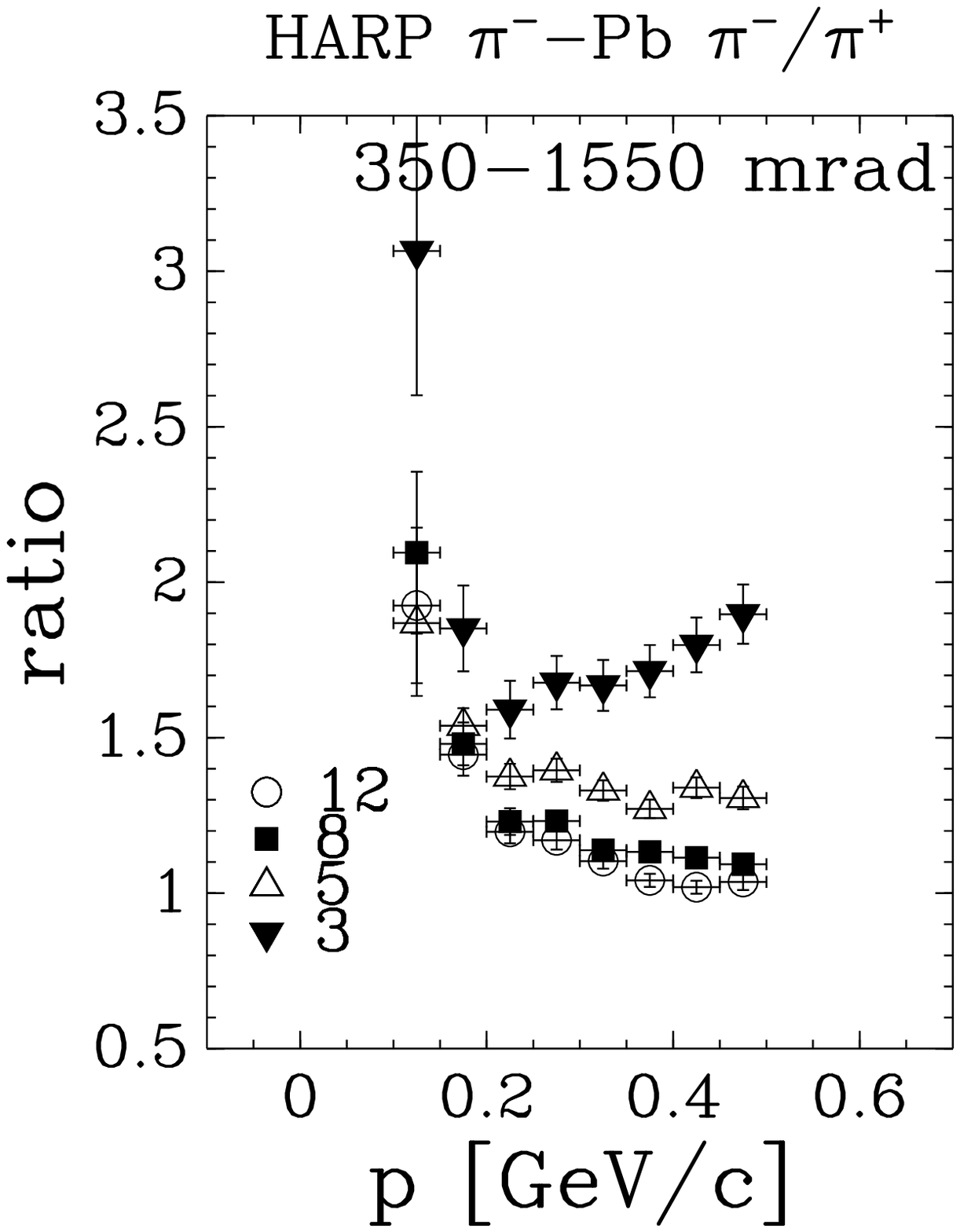}
\end{center}
\caption{
From top-left panel to bottom-right panel, the ratio of the differential cross-sections for \pim and \pip
 production in
\pim--Be (top-left), \pim--C (top-right),  \pim--Ta (bottom-left)   and
 \pim--Pb (bottom-right)   interactions as a function of 
the secondary momentum integrated over the
forward angular region (shown in mrad).
}
\label{fig:xs-pim-ratio}
\end{figure}

The integrated \pim/\pip ratio in the forward direction is displayed in
Figs.~\ref{fig:xs-ratio} and \ref{fig:xs-pim-ratio} as a function of the
secondary momentum for \pip and \pim beams, respectively. 

For the \pip beams, in the momentum range covered in most
bins more \pip's are  produced than \pim's.
In the \pip--Ta and \pip--Pb data the ratio is closer to unity than 
for the \pip--Be and \pip--C data.
The \pim/\pip ratio is larger for higher incoming beam momenta than for
lower momenta and drops with increasing secondary momentum.
Comparing the ratios in the data taken with the \pip beam with the
ratios previously published for proton beam data~\cite{ref:harp:la} one
observes a large similarity. 
The ratios in the p--$A$ data display a somewhat larger beam momentum
dependence than the \pip data.
Especially the 3~\GeVc data are closer to the higher beam momenta in the
\pip data.

Turning now our attention to the \pim/\pip ratios in the \pim beam data,
one observes a much larger beam momentum dependence with a ratio which
is larger than unity everywhere.
In this case, the 3~\GeVc displays, not surprisingly, the highest
\pim/\pip ratio. 
Again, the ratio falls with increasing secondary momentum, except for
the lowest beam momentum for Pb, a trend which is already visible for
Ta. 
Turning back to the double-differential spectra, one can conclude that
this feature is not due to an increase in the production of \pim, but
rather to a suppression of \pip production.

A striking feature is the large \pim/\pip ratio in the lowest bin of
secondary momentum (100~\MeVc--150~\MeVc) for the heavy nuclear targets
(Pb and Ta) in the beams with 8~\GeVc and 12~\GeVc momentum.
This effect had already been observed in the p--Ta
data~\cite{ref:harp:tantalum} and is visible in the p, \pip and \pim
beams.    
For the positive beams these are the only bins where more \pim's are
produced than \pip's.
The E910 Collaboration had made a similar observation for their lowest
momentum bin (100 \MeVc -- 140 \MeVc) in p--Au collisions at 12.3~\GeVc
and 17.5~\GeVc incoming beam momentum~\cite{ref:E910}. 
They offered as explanation $\Lambda^0$ production at rest which would
enhance the \pim yield at low secondary momenta.
Perhaps a more plausible explanation has been put forward in
Ref.~\cite{ref:gallmeister}, where it was shown that this effect was
predicted in the model described in that paper.
These authors invoke an asymmetry in the production of $\Delta$
resonances due to the large neutron excess in these heavy nuclei. 
They tested this hypothesis by simulating data on hypothetical n--p
symmetric heavy nuclei and found that in that case the effect was not
predicted. 
Perhaps the same explanation holds for the marked difference between the
\pip--C data and the other targets, carbon being the only n--p symmetric
target reported here.   
The \pip--C data show a significantly smaller increase of the \pim/\pip
ratio towards small momenta, a feature also visible in the p--C data of
Ref.~\cite{ref:harp:la}, while not present in the \pim--C data.

\begin{figure}[tbp!]
\begin{center}
 \includegraphics[width=0.405\textwidth]{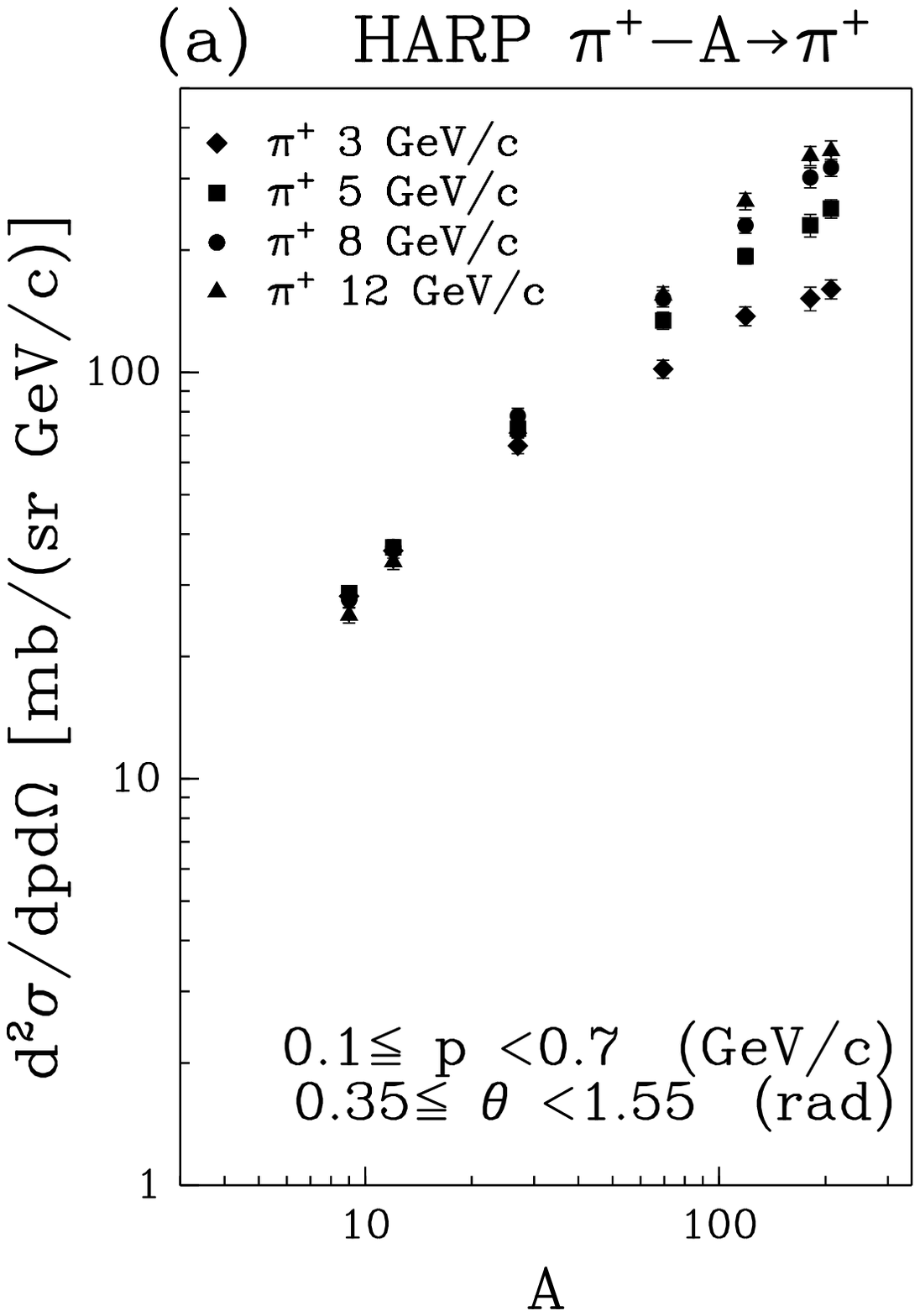}
 ~
 \includegraphics[width=0.405\textwidth]{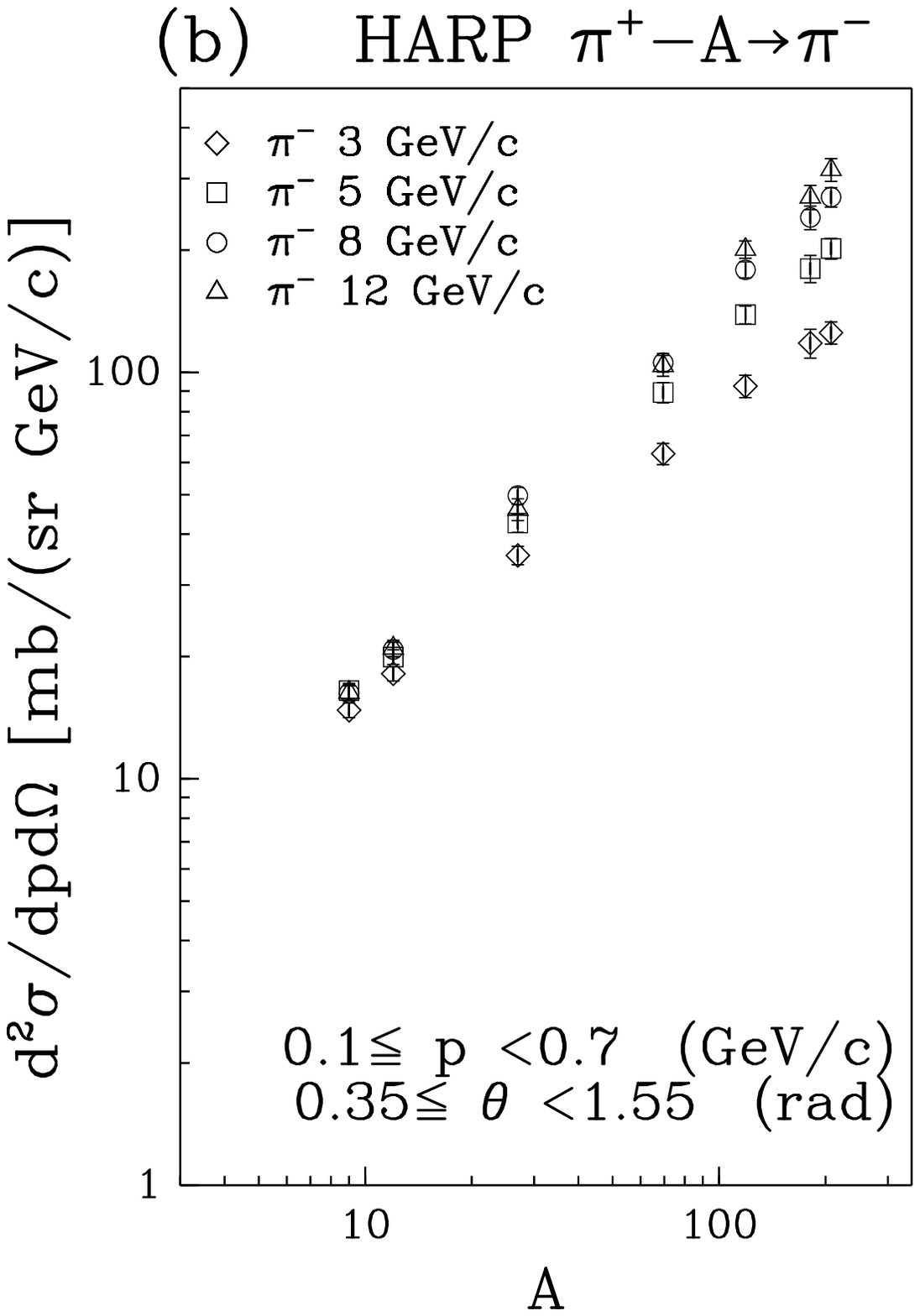}
 ~
 \includegraphics[width=0.405\textwidth]{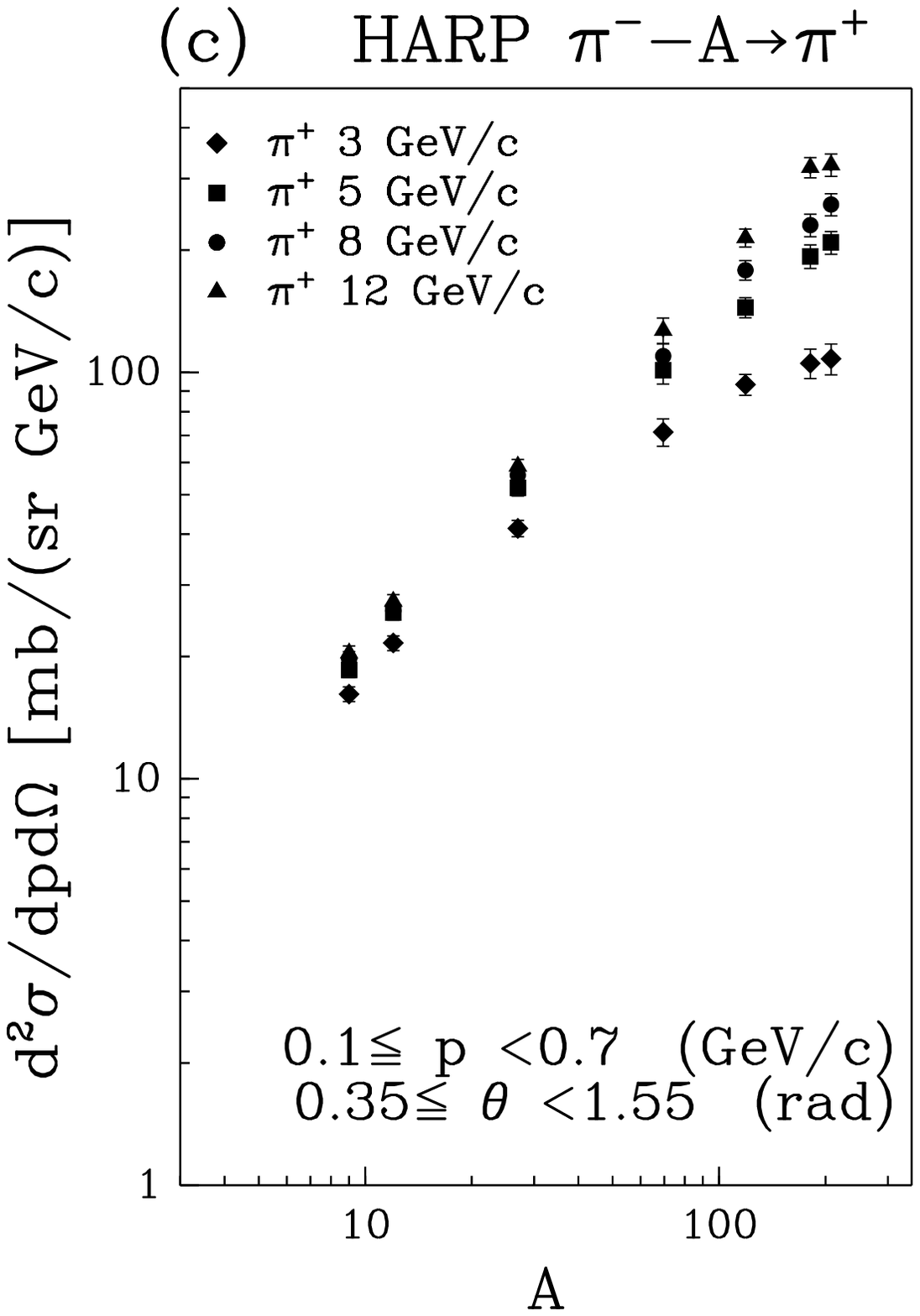}
 ~
 \includegraphics[width=0.405\textwidth]{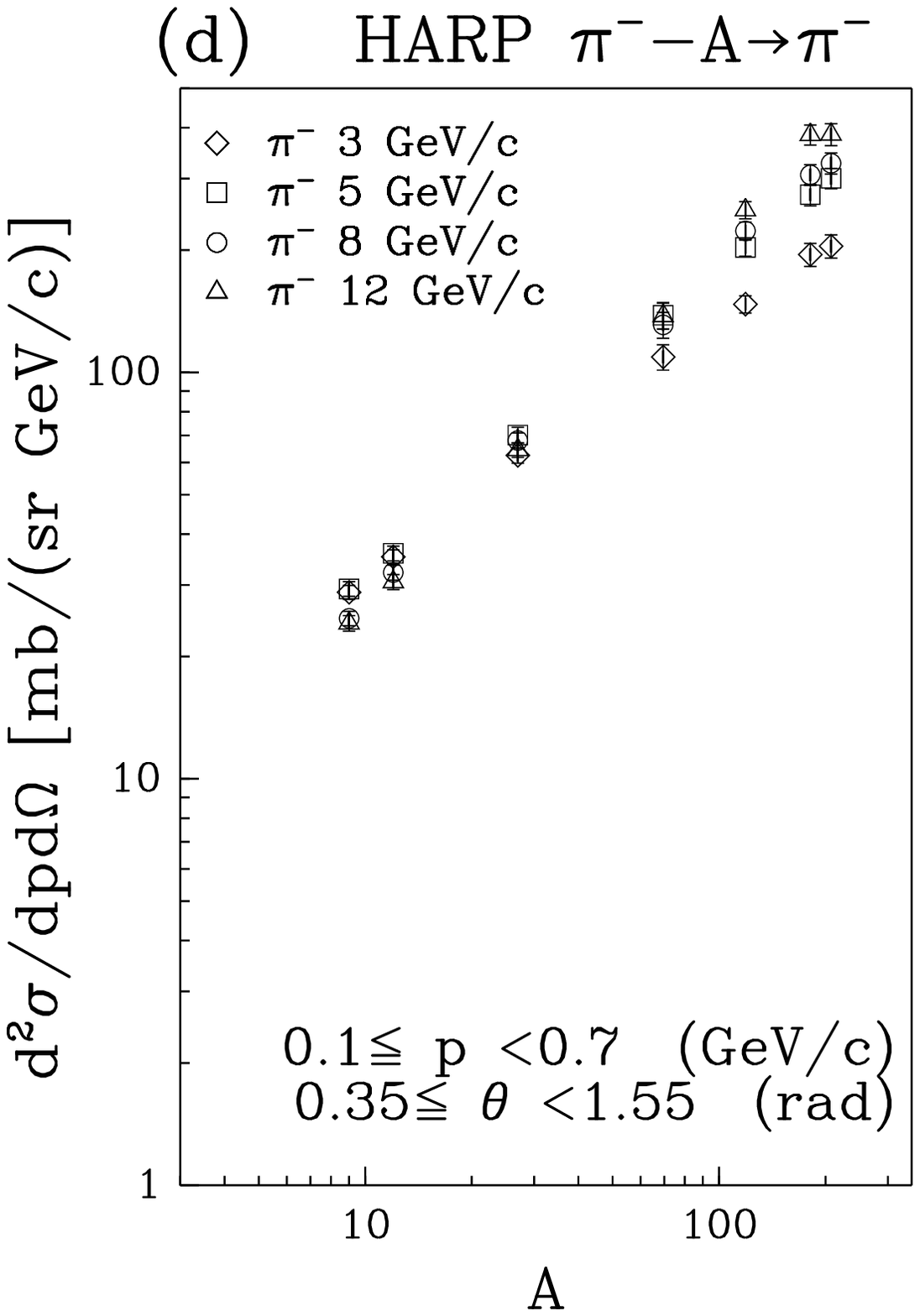}
\end{center}
\caption{
 Dependence on the atomic number $A$ of the pion production yields
 in \pipm--Be, \pipm--Al, \pipm--C, \pipm--Cu, \pipm--Sn, \pipm--Ta, \pipm--Pb
 interactions averaged over the forward angular region 
 ($0.350~\rad \leq \theta < 1.550~\rad$) 
 and momentum region $100~\MeVc \leq p < 700~\MeVc$.
 The top-left panel (a) refers to \pip production in \pip beams, the
 top-right panel (b) to \pim production in \pip beams, while to
 bottom-left (c) and bottom-right (d) panels refer respectively to \pip
 and \pim production in \pim beams.
}
\label{fig:xs-a-dep}
\end{figure}

The dependence of the averaged pion yields on the atomic number $A$ is
shown in Fig.~\ref{fig:xs-a-dep}.  
The \pip yields averaged over the region 
$0.350~\rad \leq \theta < 1.550~\rad$ and $100~\MeVc \leq p < 700~\MeVc$ are
shown in the left panel and the \pim data averaged over the same region
in the right panel for four different beam momenta.
One observes a smooth behaviour of the averaged yields.
The $A$ dependence is slightly different for \pim and \pip production.
The production of opposite-sign pions displays a steeper $A$ dependence
than same-sign pion production.

The experimental uncertainties are summarized for \pip  in
Tables~\ref{tab:errors-3-pip} and \ref{tab:errors-4-pip}, and for \pim in
Tables~\ref{tab:errors-3-pim} and \ref{tab:errors-4-pim} for all used
targets. 
The relative sizes of the different systematic error sources are very
similar for \pim and \pip and for the different beam energies. 
Going from lighter (Be, C) to heavier targets (Ta, Pb)
the corrections for \piz (conversion, concentrated at low secondary
momentum) and absorption/tertiaries are bigger.   
Since the production cross-sections are not very different in the pion
beams from the proton beam data, the discussion and figures shown in
Ref.~\cite{ref:harp:la} give a reliable indication of the momentum and
angular dependence of the systematic error components and need not be
repeated here.

\ifall
\begin{sidewaysfigure}[tbp!]
\begin{center}
 \includegraphics[width=0.49\textwidth]{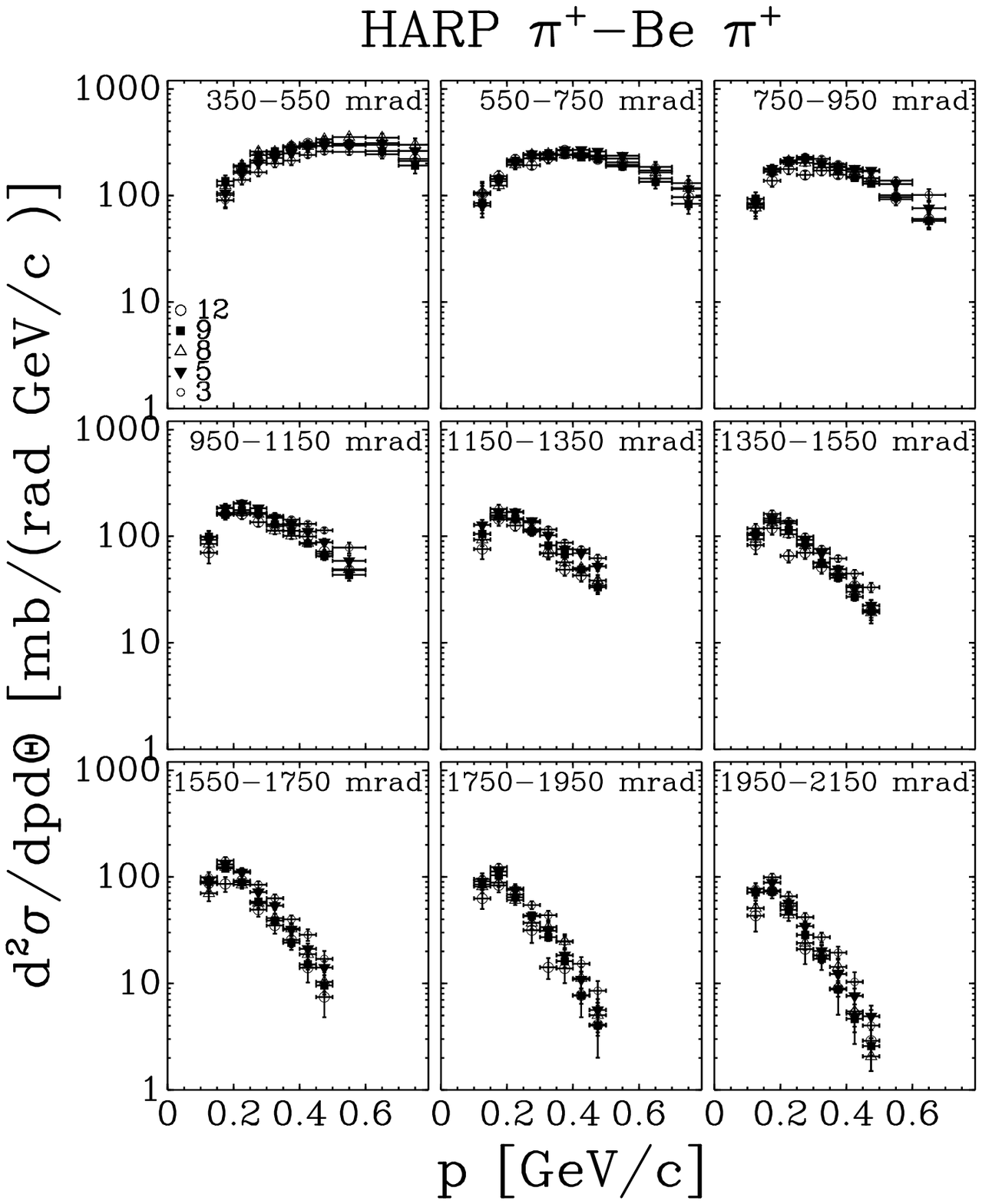}
 \includegraphics[width=0.49\textwidth]{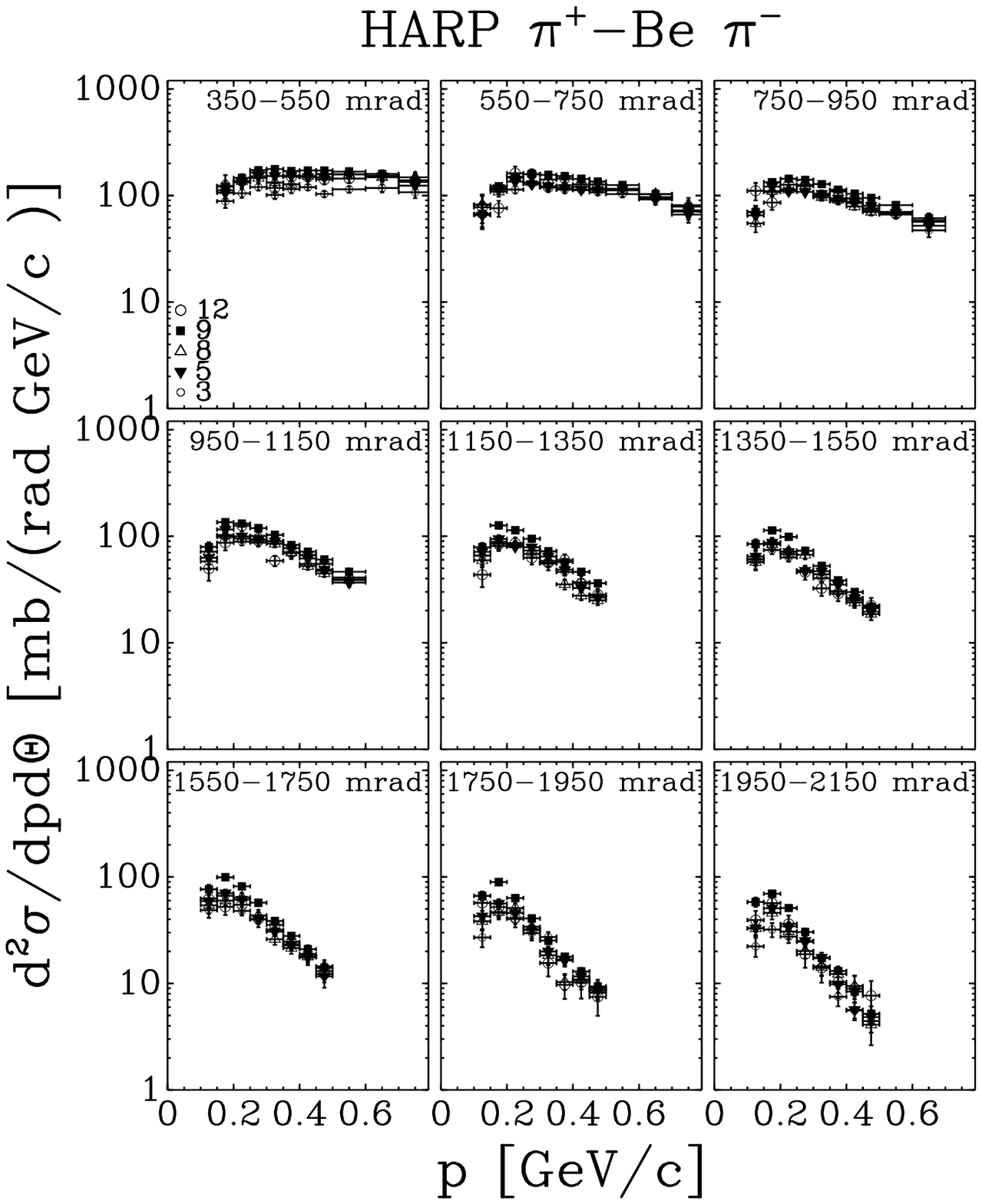}
\caption{
Double-differential cross-sections for \pip production (left) and  \pim
 production (right) in
\pip--Be interactions as a function of momentum displayed in different
angular bins (shown in \mrad in the panels).
In the figure, the symbol legend 9 refers to 8.9~\GeVc nominal
beam momentum.
The error bars represent the combination of statistical and systematic
 uncertainties. 
}
\label{fig:xs-p-th-pbeam-be}
\end{center}
\end{sidewaysfigure}
\begin{sidewaysfigure}[tbp!]
\begin{center}
 \includegraphics[width=0.49\textwidth]{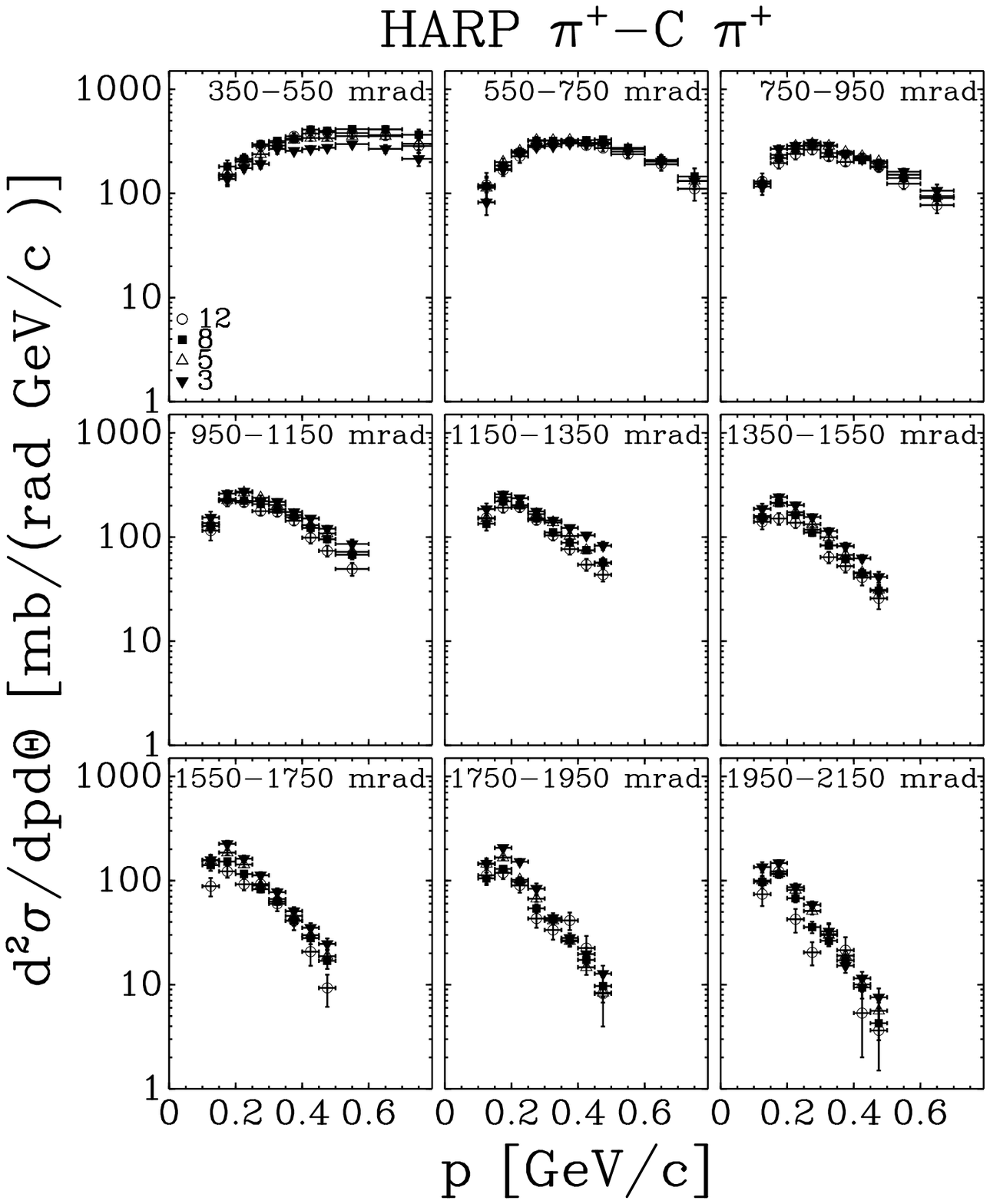}
 \includegraphics[width=0.49\textwidth]{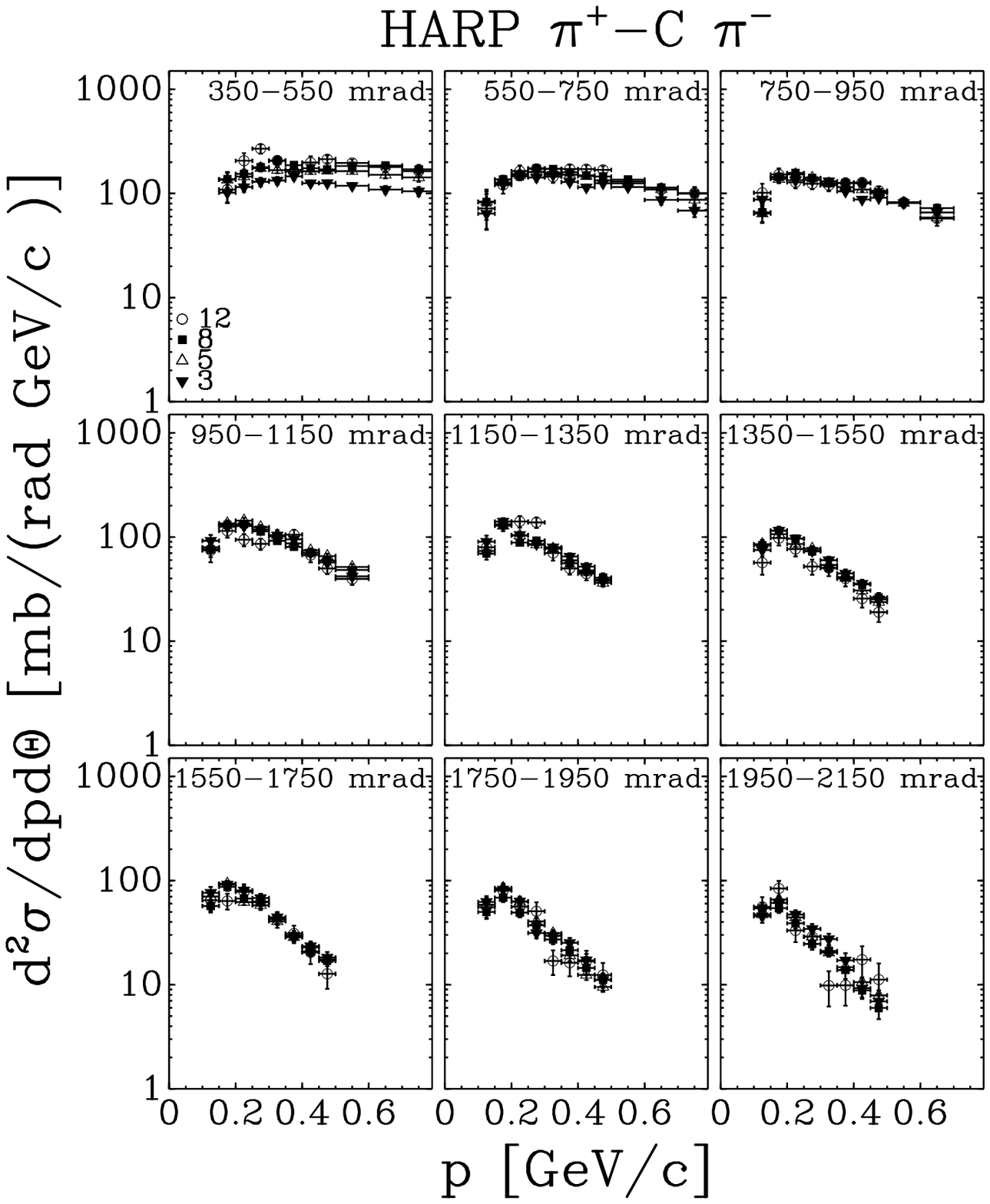}
\caption{
Double-differential cross-sections for \pip production (left) and  \pim
 production (right) in
\pip--C interactions as a function of momentum displayed in different
angular bins (shown in \mrad in the panels).
The error bars represent the combination of statistical and systematic
 uncertainties. 
}
\label{fig:xs-p-th-pbeam-c}
\end{center}
\end{sidewaysfigure}
\begin{sidewaysfigure}[tbp!]
\begin{center}
 \includegraphics[width=0.49\textwidth]{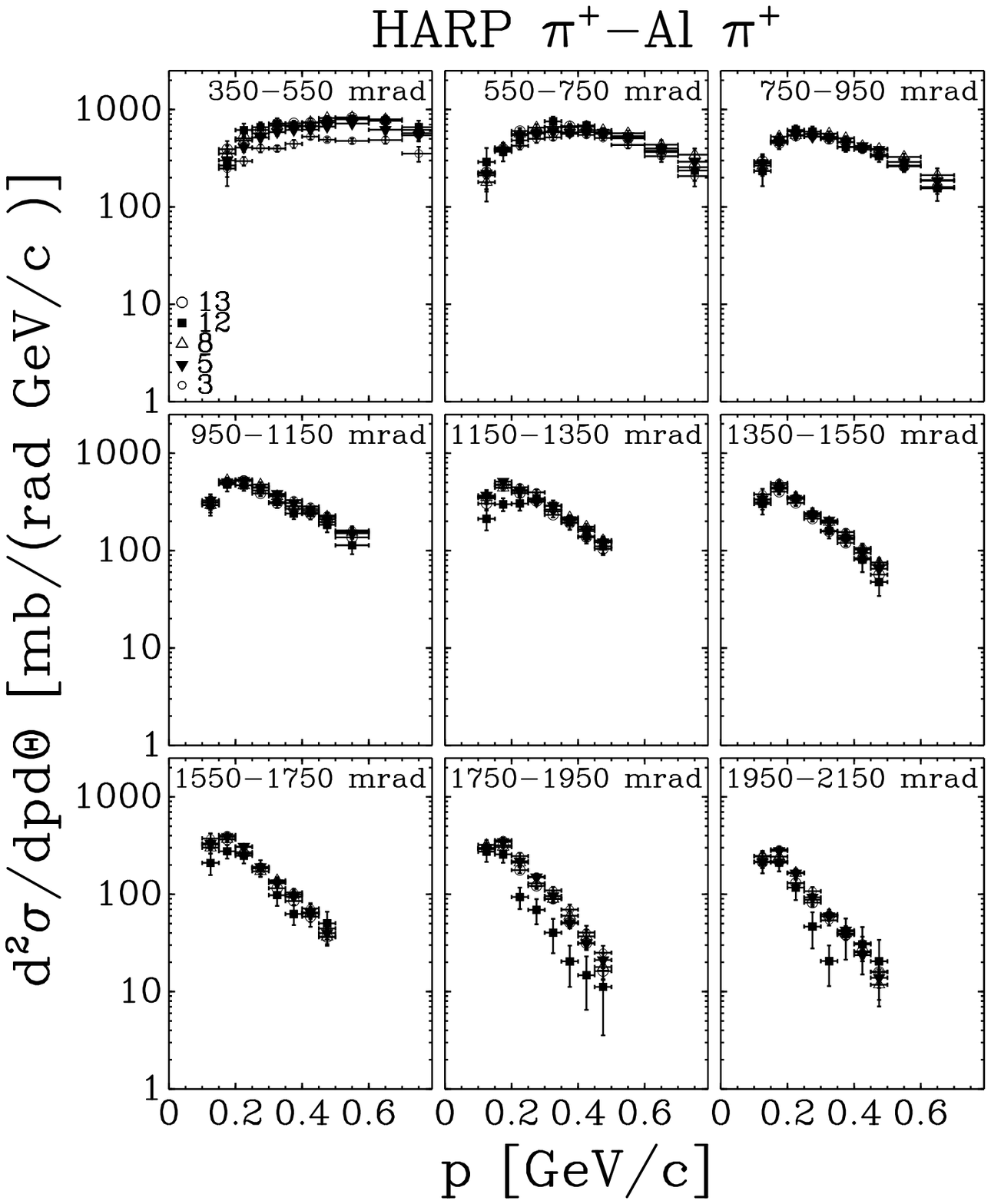}
 \includegraphics[width=0.49\textwidth]{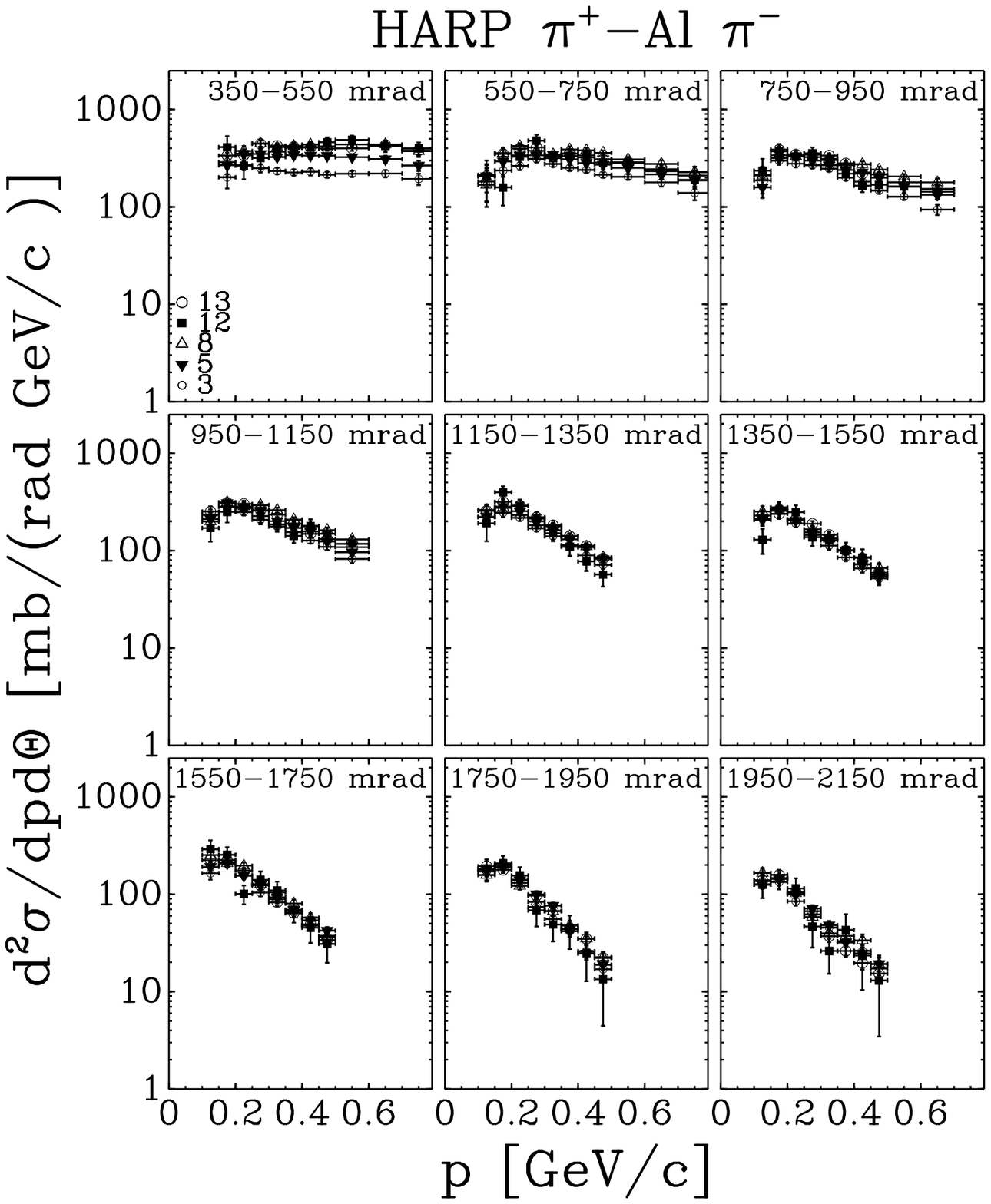}
\caption{
Double-differential cross-sections for \pip production (left) and  \pim
 production (right) in
\pip--Al interactions as a function of momentum displayed in different
angular bins (shown in \mrad in the panels).
In the figure, the symbol legend 13 refers to 12.9~\GeVc nominal
beam momentum.
The error bars represent the combination of statistical and systematic
 uncertainties. 
}
\label{fig:xs-p-th-pbeam-al}
\end{center}
\end{sidewaysfigure}
\begin{sidewaysfigure}[tbp!]
\begin{center}
 \includegraphics[width=0.49\textwidth]{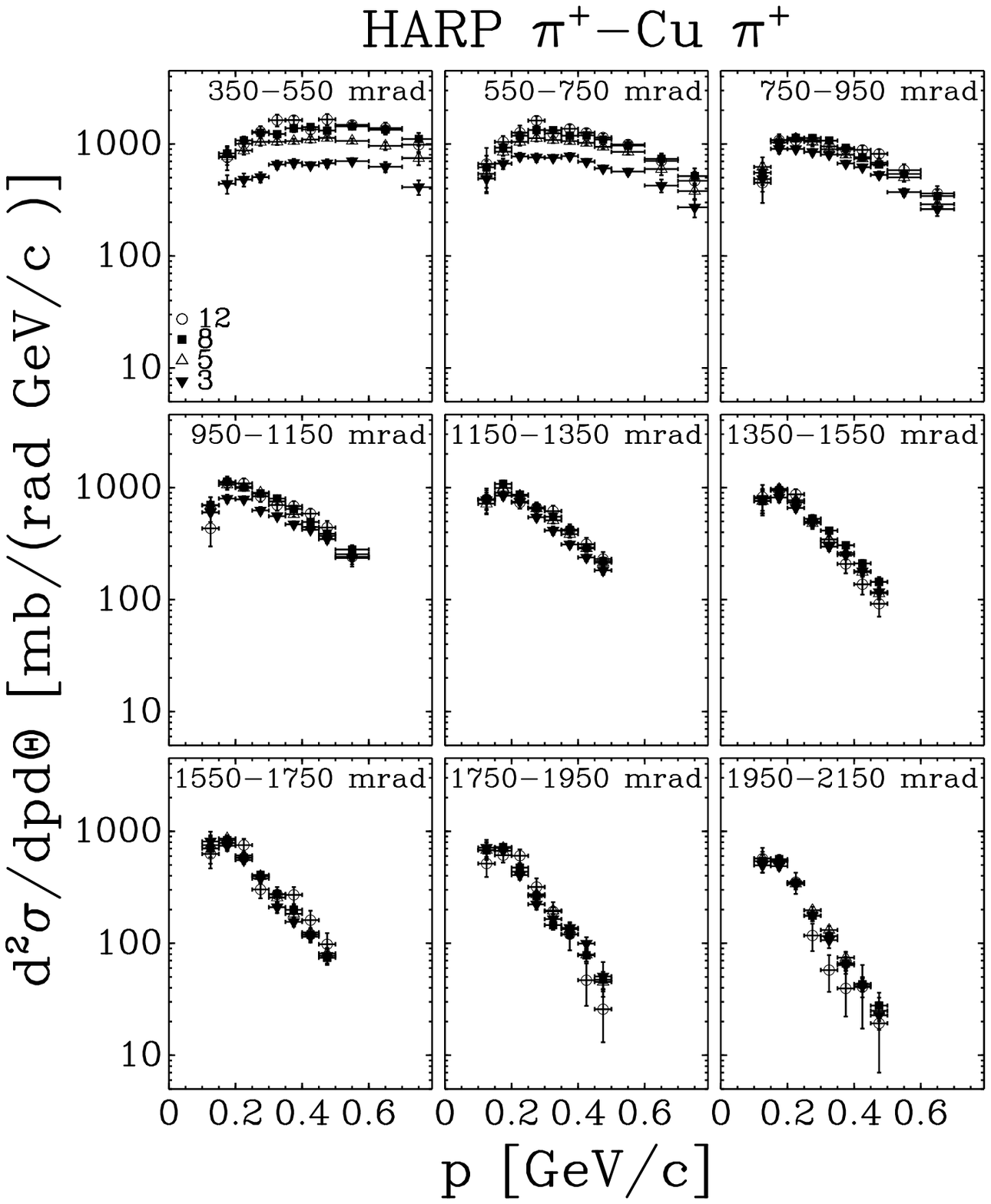}
 \includegraphics[width=0.49\textwidth]{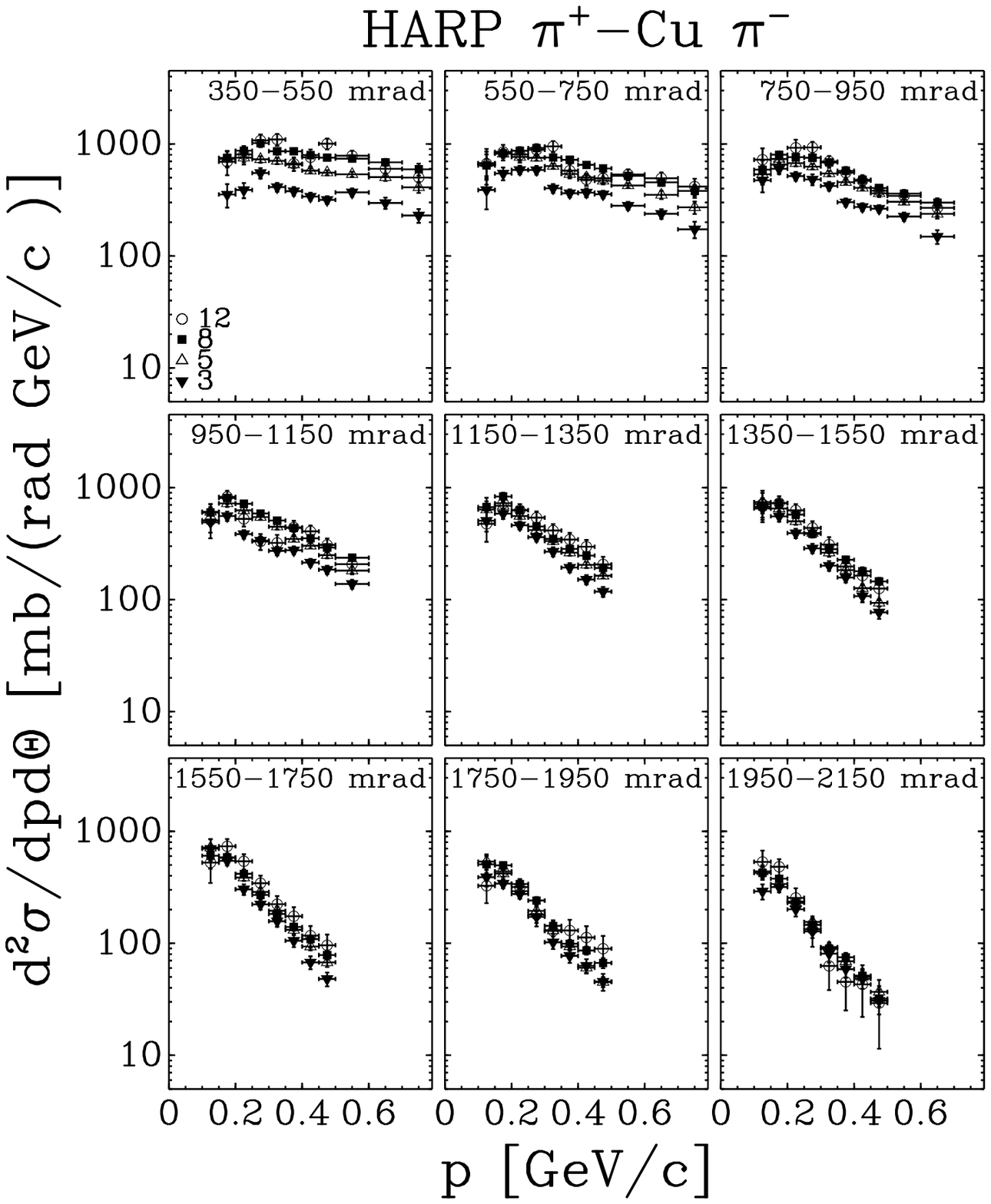}
\caption{
Double-differential cross-sections for \pip production (left) and  \pim
 production (right) in
\pip--Cu interactions as a function of momentum displayed in different
angular bins (shown in \mrad in the panels).
The error bars represent the combination of statistical and systematic
 uncertainties. 
}
\label{fig:xs-p-th-pbeam-cu}
\end{center}
\end{sidewaysfigure}
\begin{sidewaysfigure}[tbp!]
\begin{center}
 \includegraphics[width=0.49\textwidth]{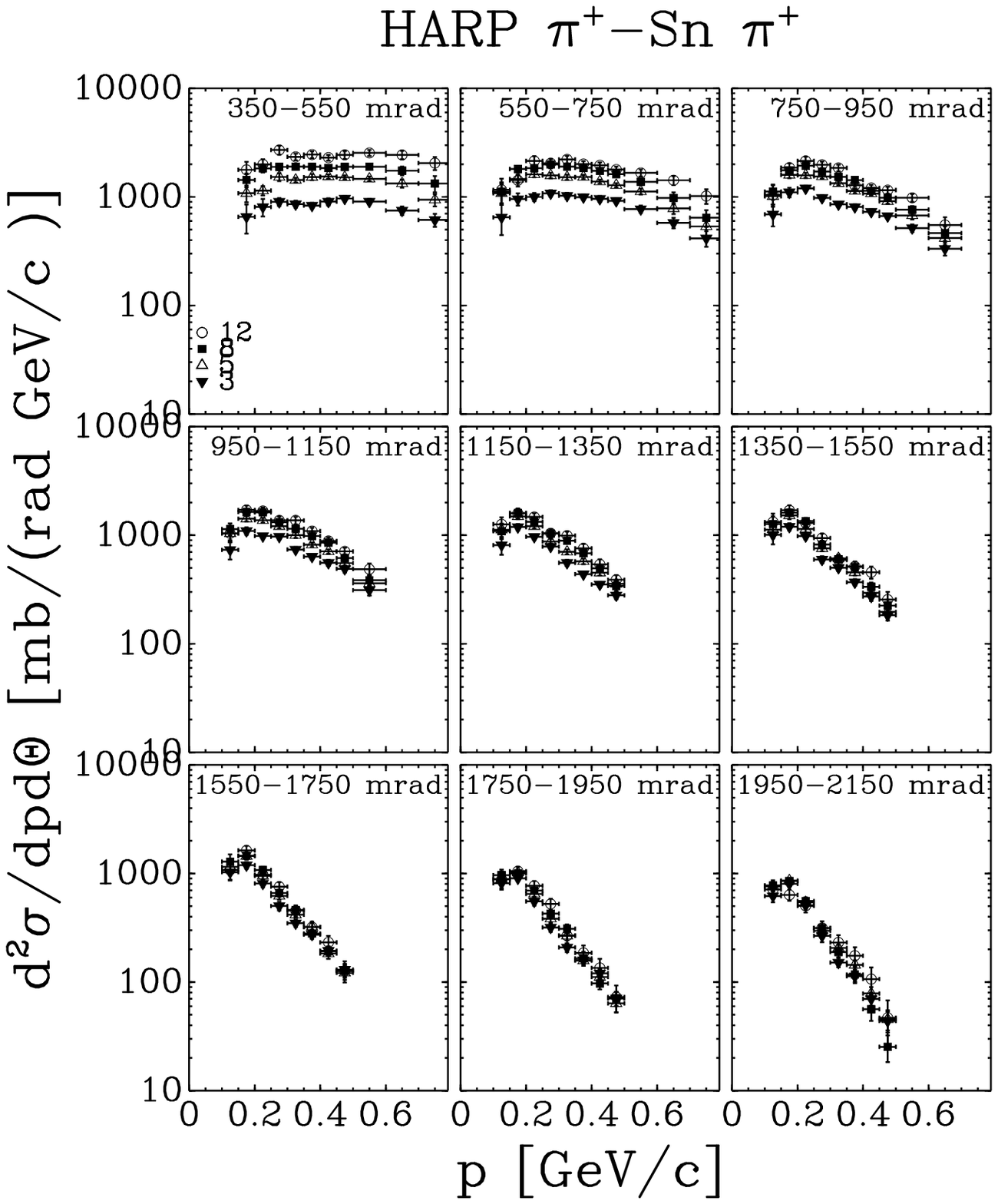}
 \includegraphics[width=0.49\textwidth]{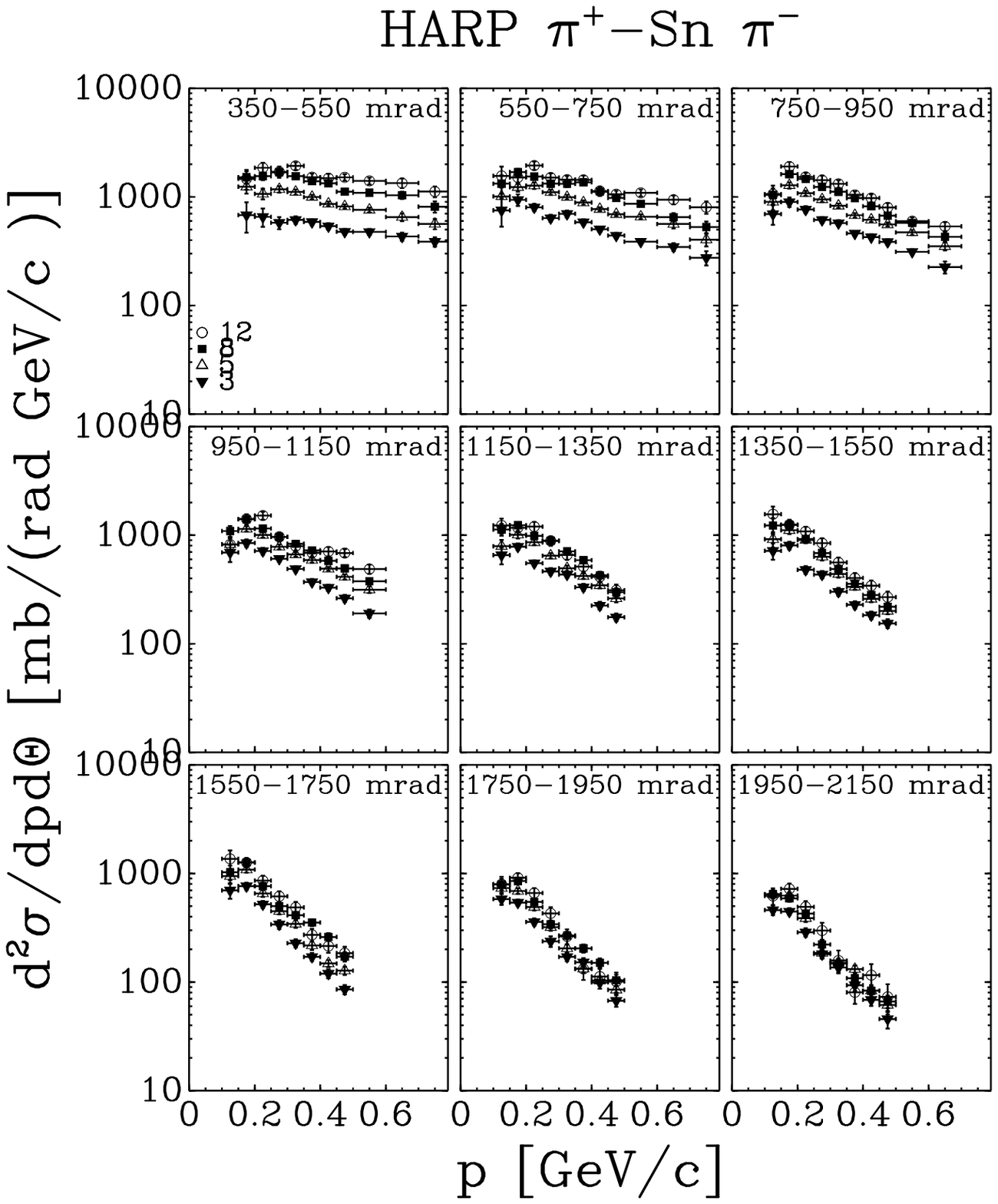}
\caption{
Double-differential cross-sections for \pip production (left) and  \pim
 production (right) in
\pip--Sn interactions as a function of momentum displayed in different
angular bins (shown in \mrad in the panels).
The error bars represent the combination of statistical and systematic
 uncertainties. 
}
\label{fig:xs-p-th-pbeam-sn}
\end{center}
\end{sidewaysfigure}
\begin{sidewaysfigure}[tbp!]
\begin{center}
 \includegraphics[width=0.49\textwidth]{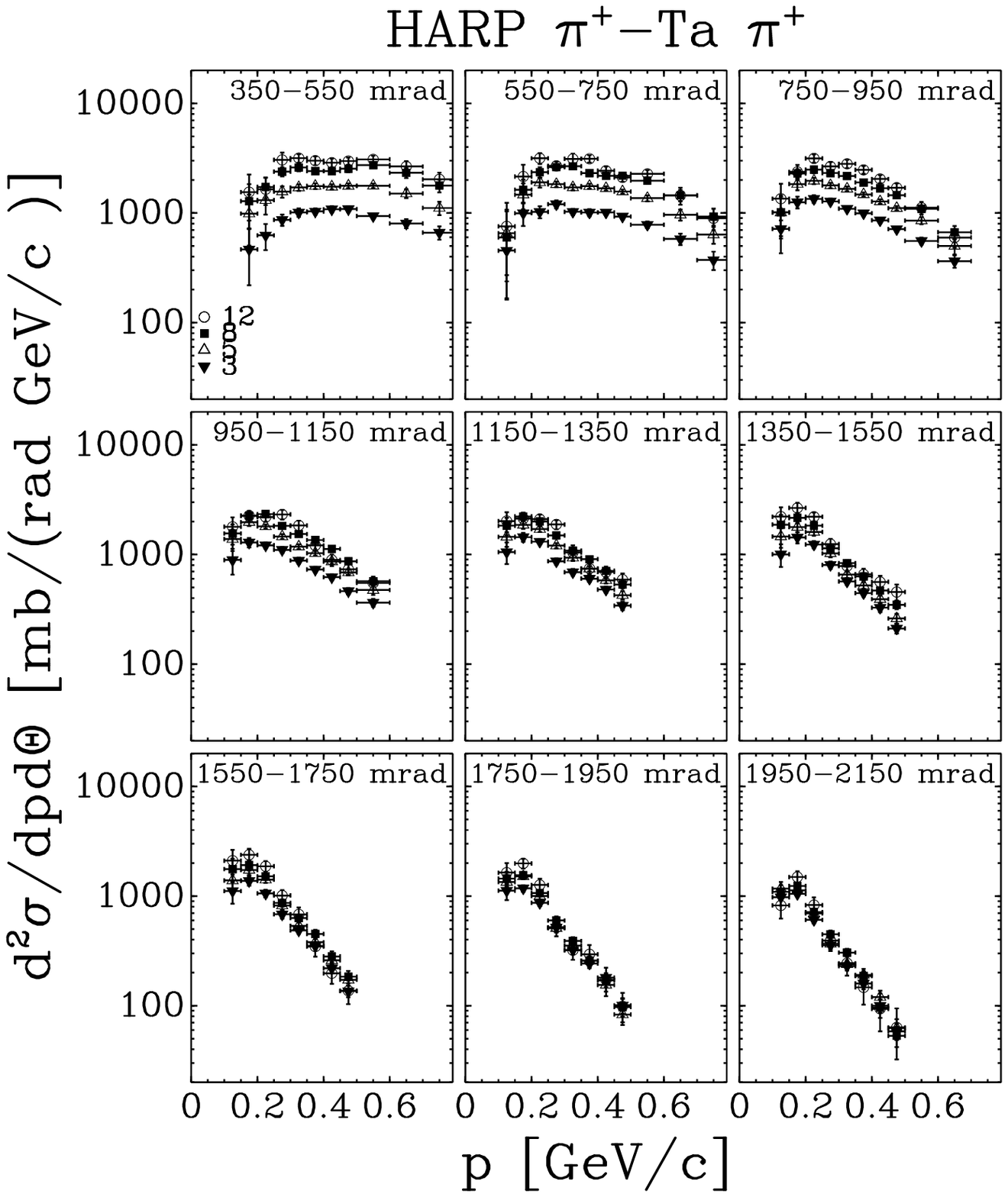}
 \includegraphics[width=0.49\textwidth]{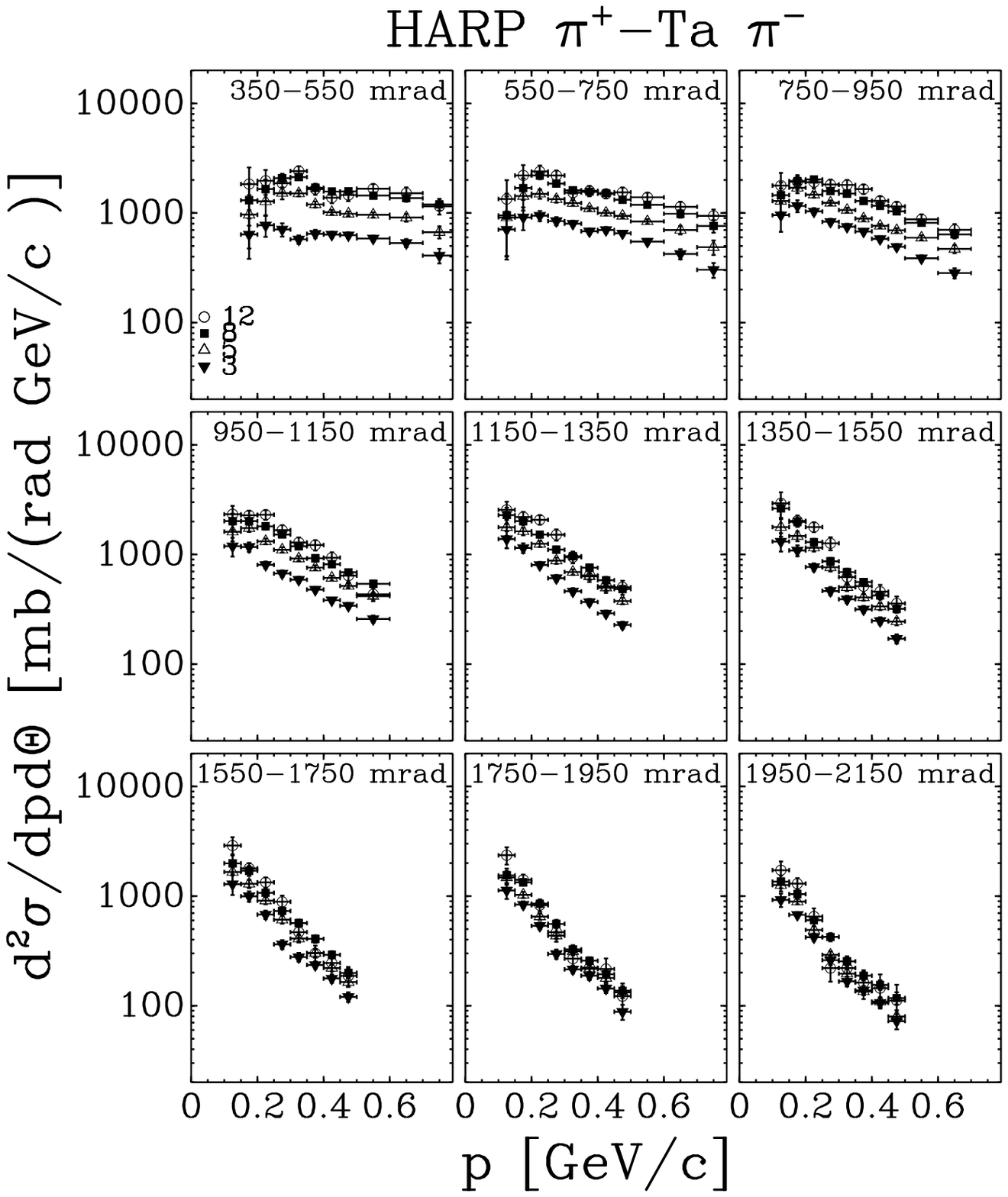}
\caption{
Double-differential cross-sections for \pip production (left) and  \pim
 production (right) in
\pip--Ta interactions as a function of momentum displayed in different
angular bins (shown in \mrad in the panels).
The error bars represent the combination of statistical and systematic
 uncertainties. 
}
\label{fig:xs-p-th-pbeam-ta}
\end{center}
\end{sidewaysfigure}
\begin{sidewaysfigure}[tbp!]
\begin{center}
 \includegraphics[width=0.49\textwidth]{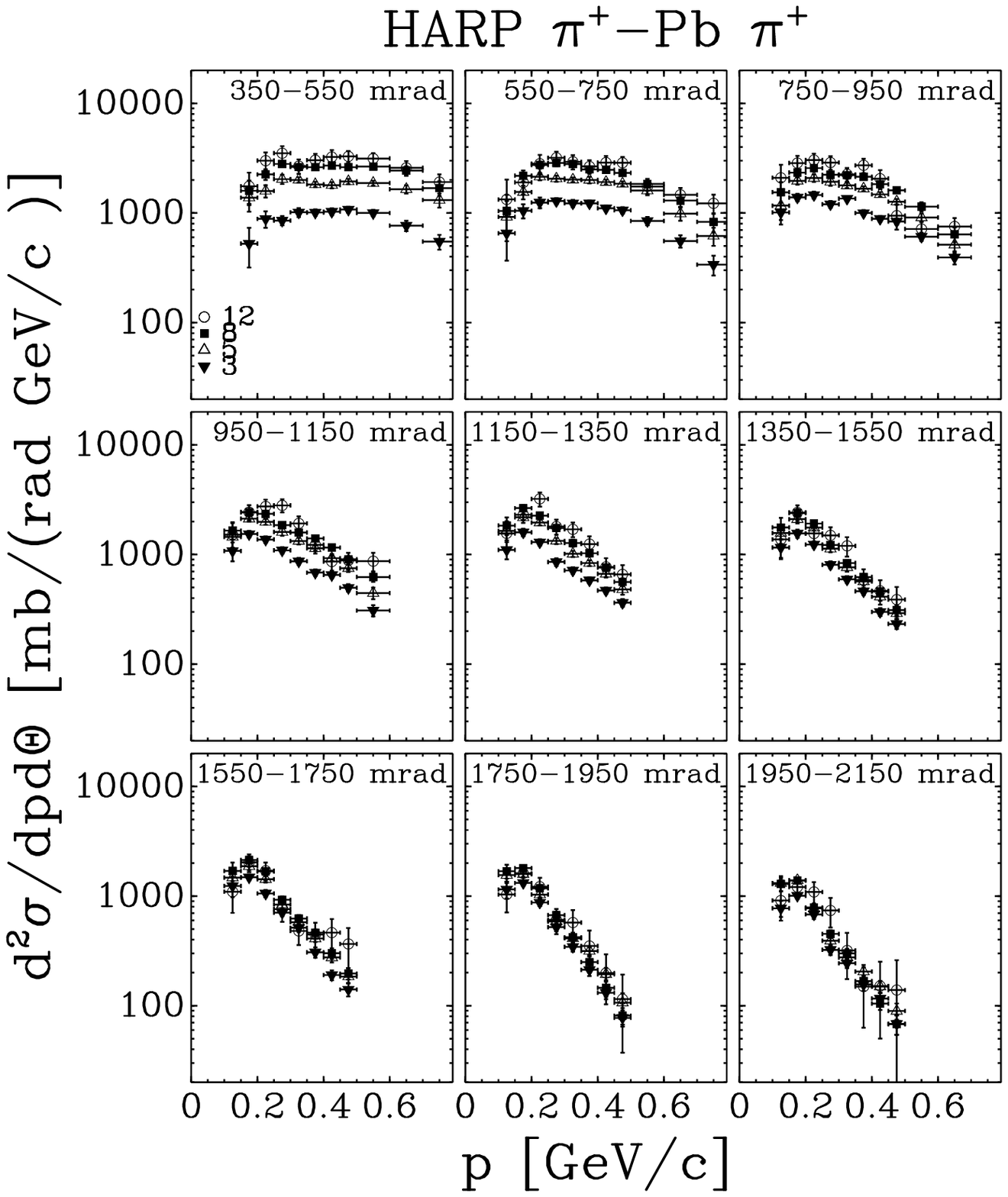}
 \includegraphics[width=0.49\textwidth]{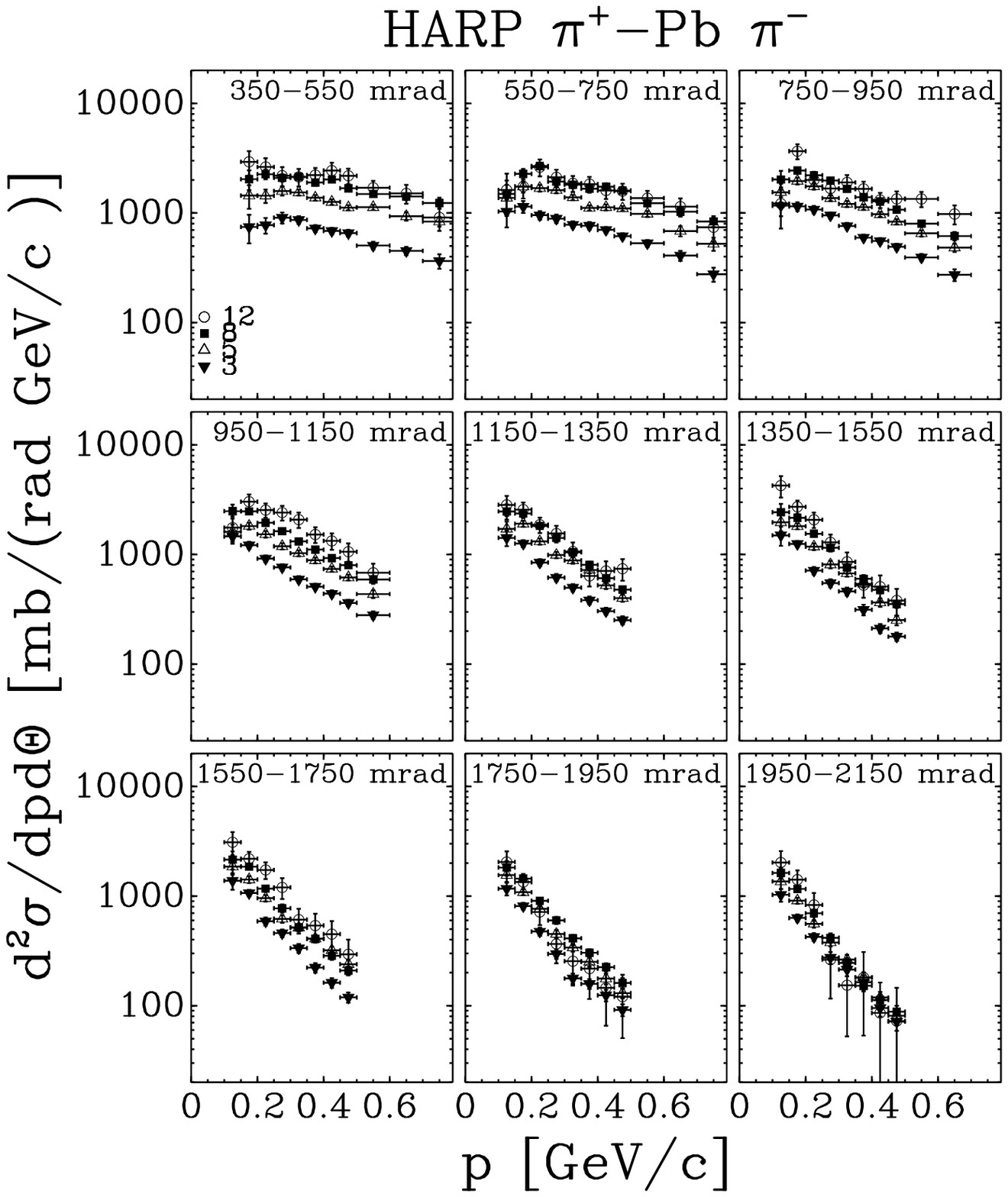}
\caption{
Double-differential cross-sections for \pip production (left) and  \pim
 production (right) in
\pip--Pb interactions as a function of momentum displayed in different
angular bins (shown in \mrad in the panels).
The error bars represent the combination of statistical and systematic
 uncertainties. 
}
\label{fig:xs-p-th-pbeam-pb}
\end{center}
\end{sidewaysfigure}
\begin{sidewaysfigure}[tbp!]
\begin{center}
 \includegraphics[width=0.49\textwidth]{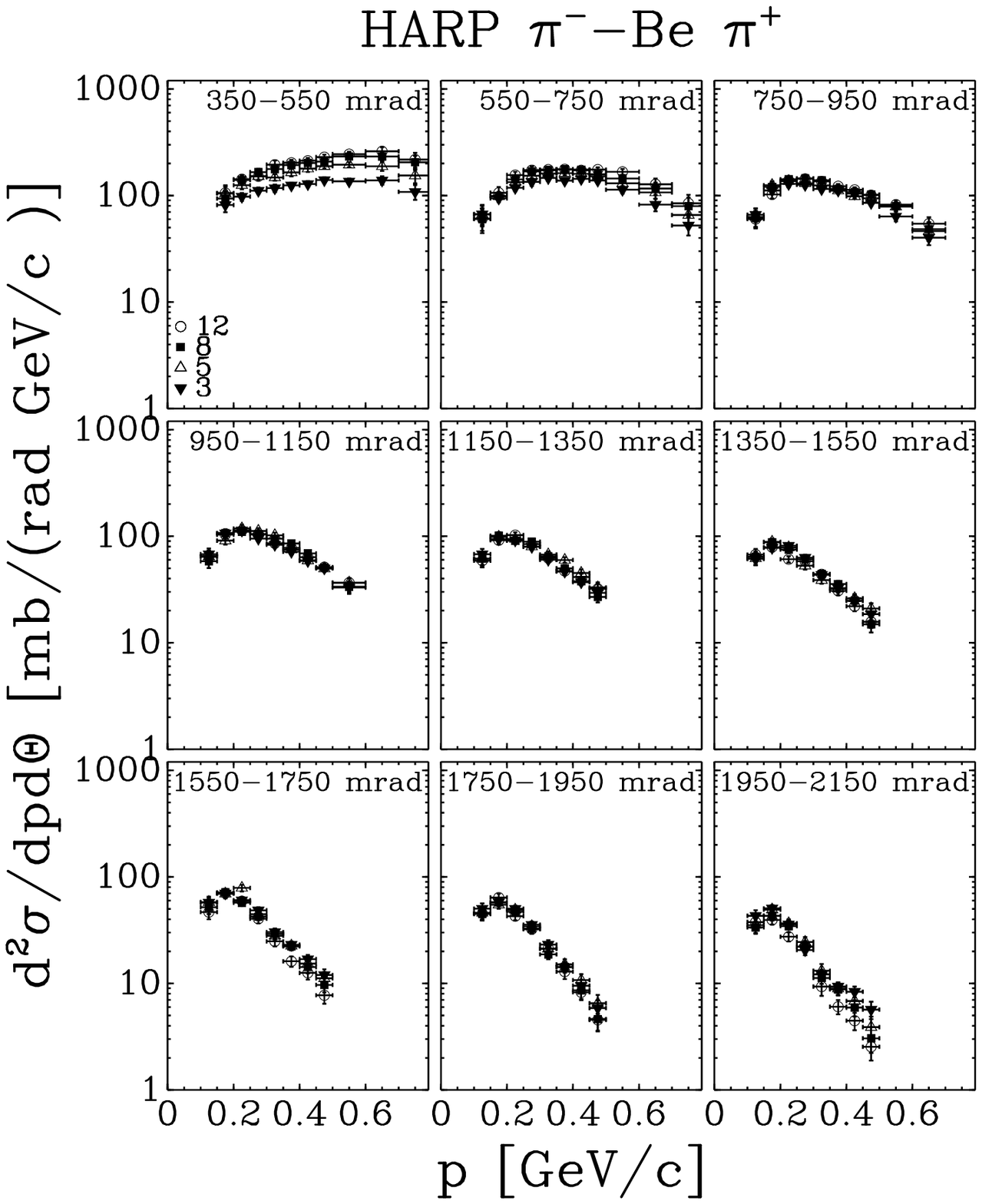}
 \includegraphics[width=0.49\textwidth]{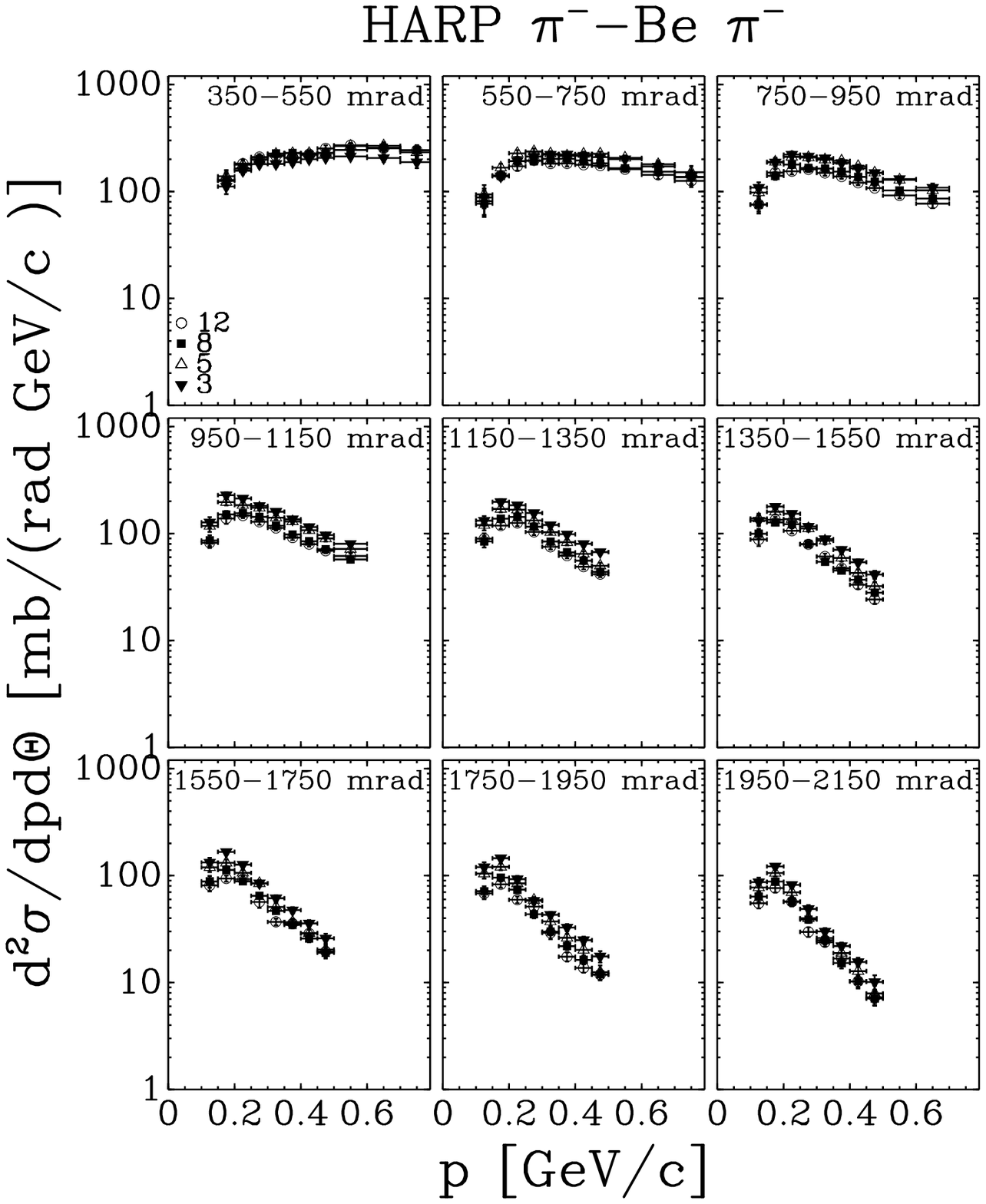}
\caption{
Double-differential cross-sections for \pip production (left) and  \pim
 production (right) in
\pim--Be interactions as a function of momentum displayed in different
angular bins (shown in \mrad in the panels).
The error bars represent the combination of statistical and systematic
 uncertainties. 
}
\label{fig:xs-pim-th-pbeam-be}
\end{center}
\end{sidewaysfigure}
\begin{sidewaysfigure}[tbp!]
\begin{center}
 \includegraphics[width=0.49\textwidth]{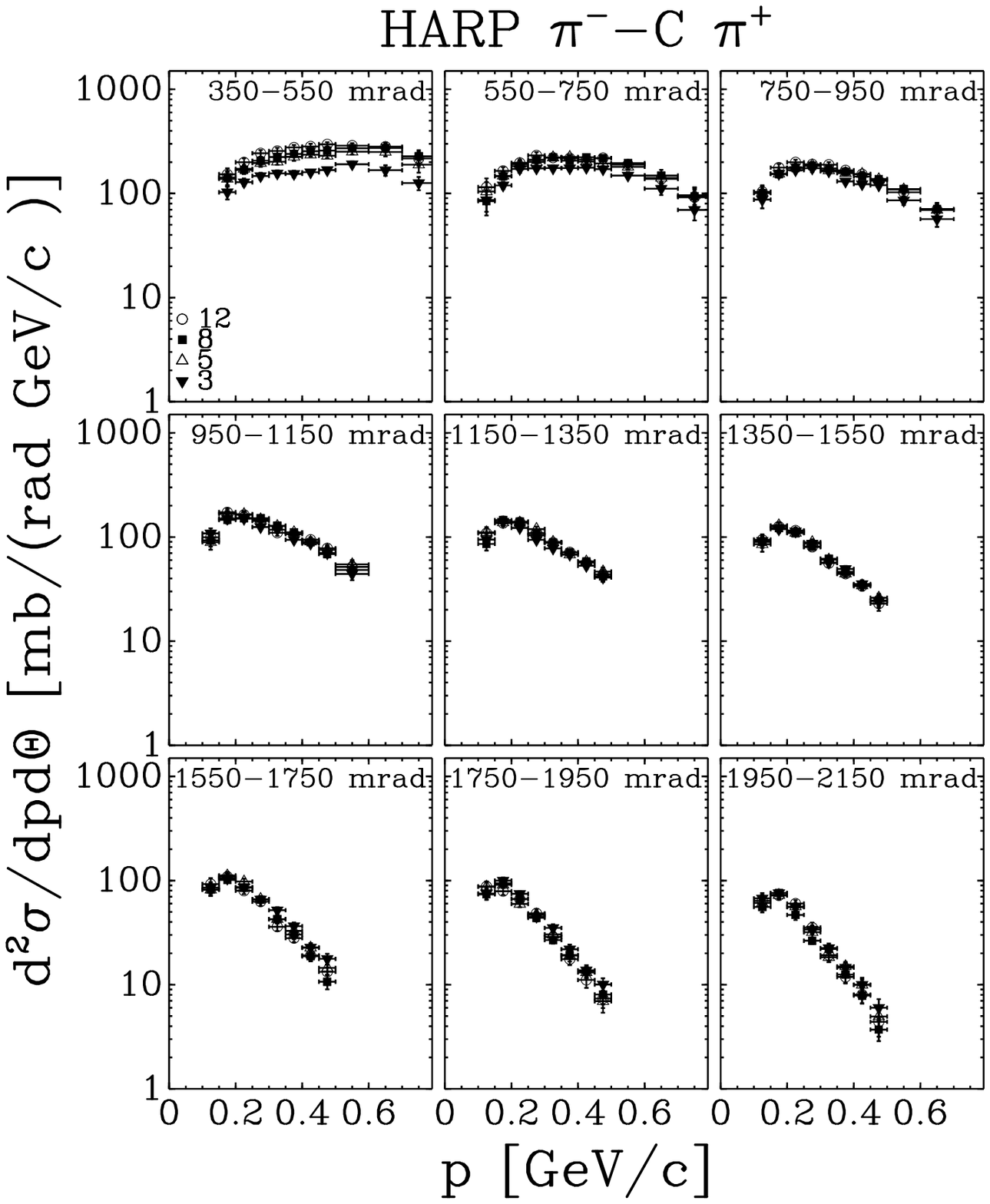}
 \includegraphics[width=0.49\textwidth]{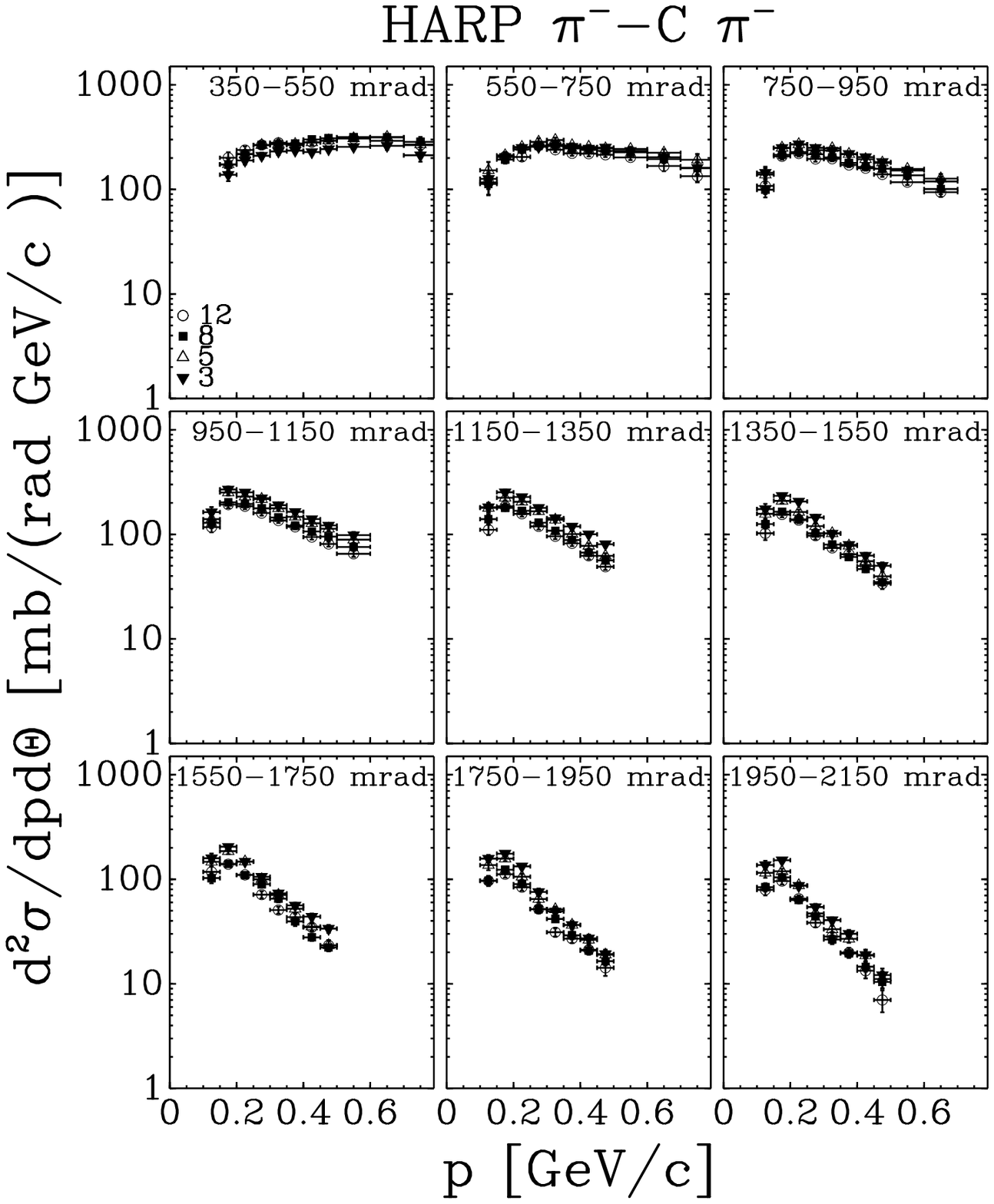}
\caption{
Double-differential cross-sections for \pip production (left) and  \pim
 production (right) in
\pim--C interactions as a function of momentum displayed in different
angular bins (shown in \mrad in the panels).
The error bars represent the combination of statistical and systematic
 uncertainties. 
}
\label{fig:xs-pim-th-pbeam-c}
\end{center}
\end{sidewaysfigure}
\begin{sidewaysfigure}[tbp!]
\begin{center}
 \includegraphics[width=0.49\textwidth]{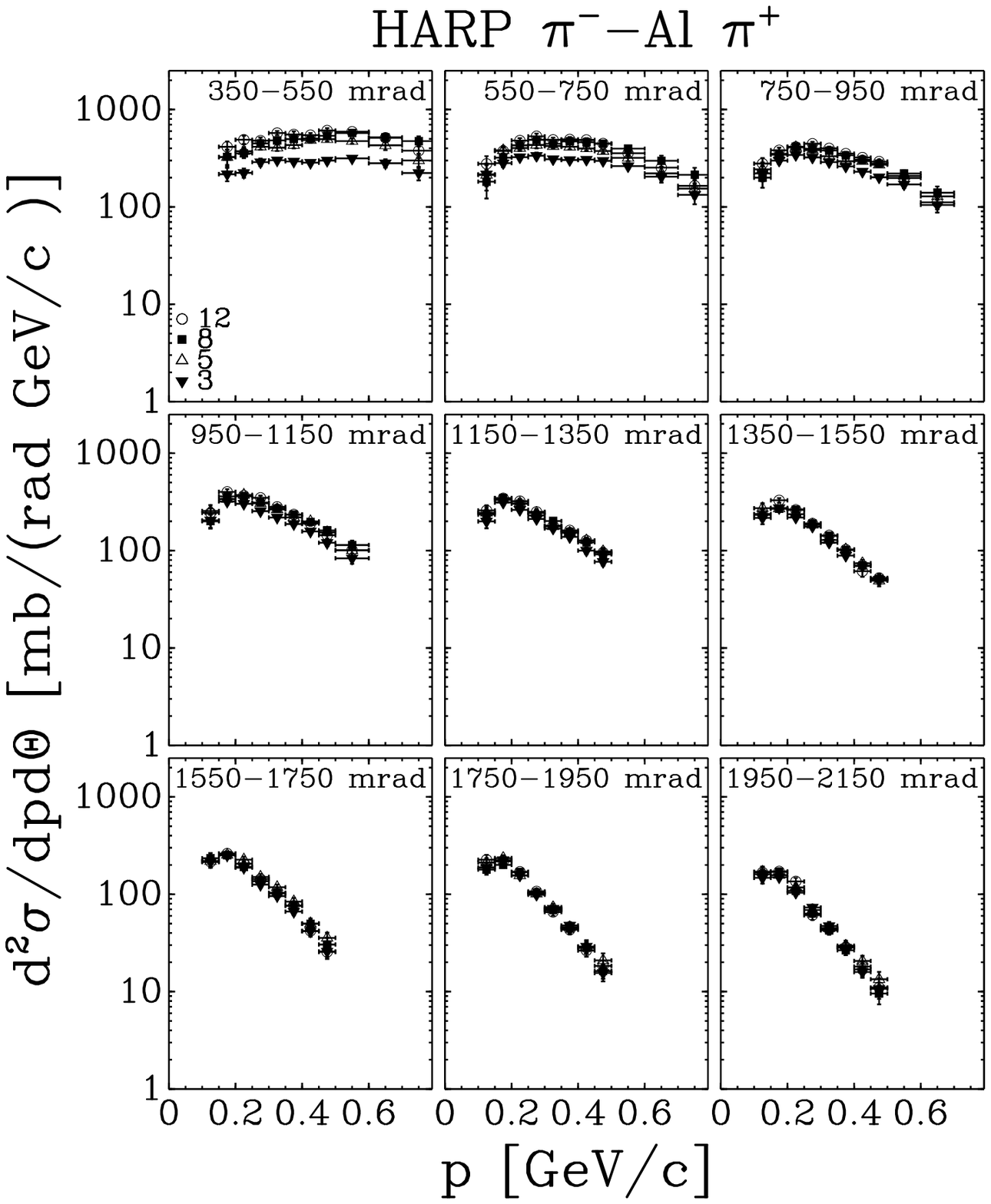}
 \includegraphics[width=0.49\textwidth]{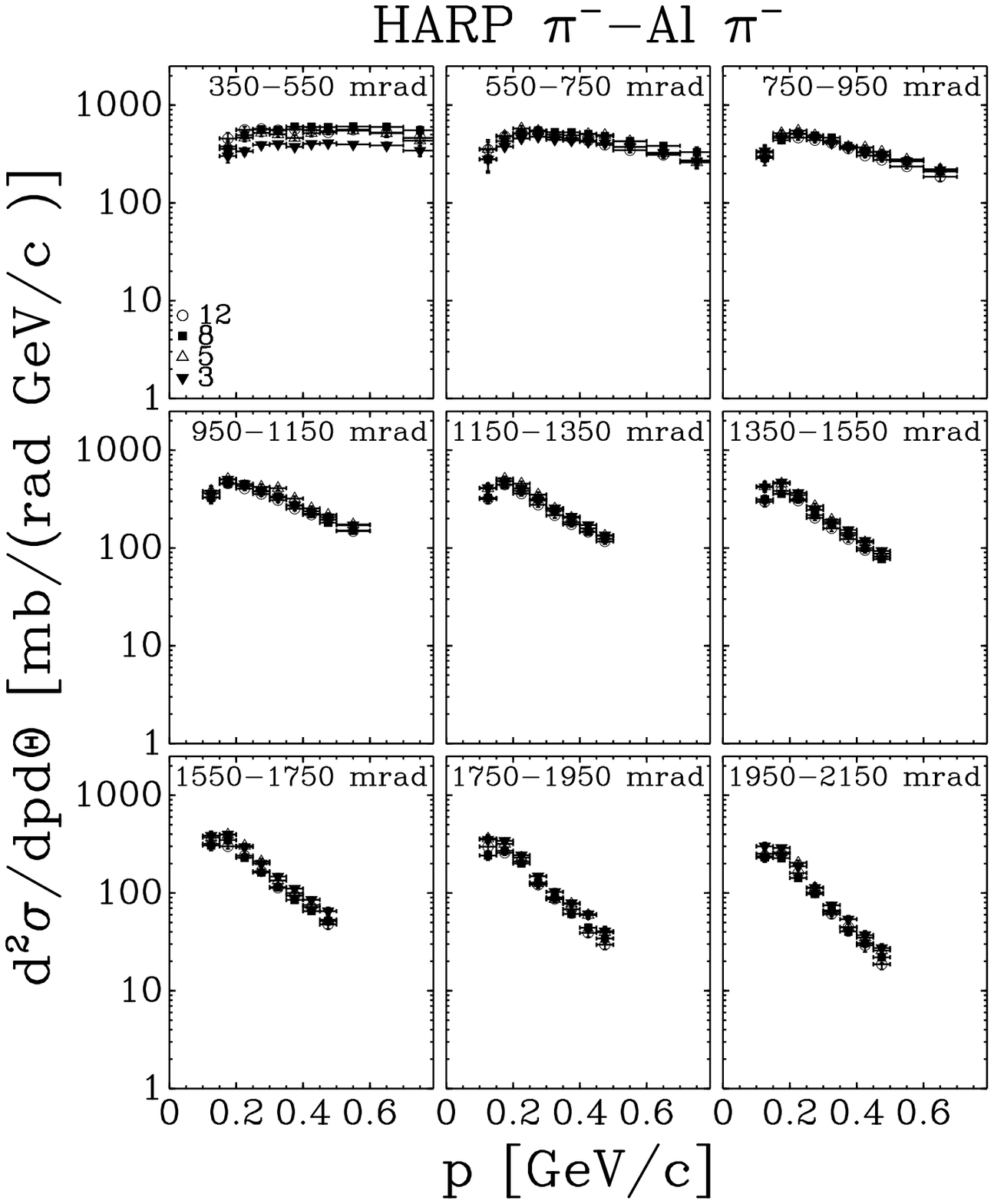}
\caption{
Double-differential cross-sections for \pip production (left) and  \pim
 production (right) in
\pim--Al interactions as a function of momentum displayed in different
angular bins (shown in \mrad in the panels).
The error bars represent the combination of statistical and systematic
 uncertainties. 
}
\label{fig:xs-pim-th-pbeam-al}
\end{center}
\end{sidewaysfigure}
\begin{sidewaysfigure}[tbp!]
\begin{center}
 \includegraphics[width=0.49\textwidth]{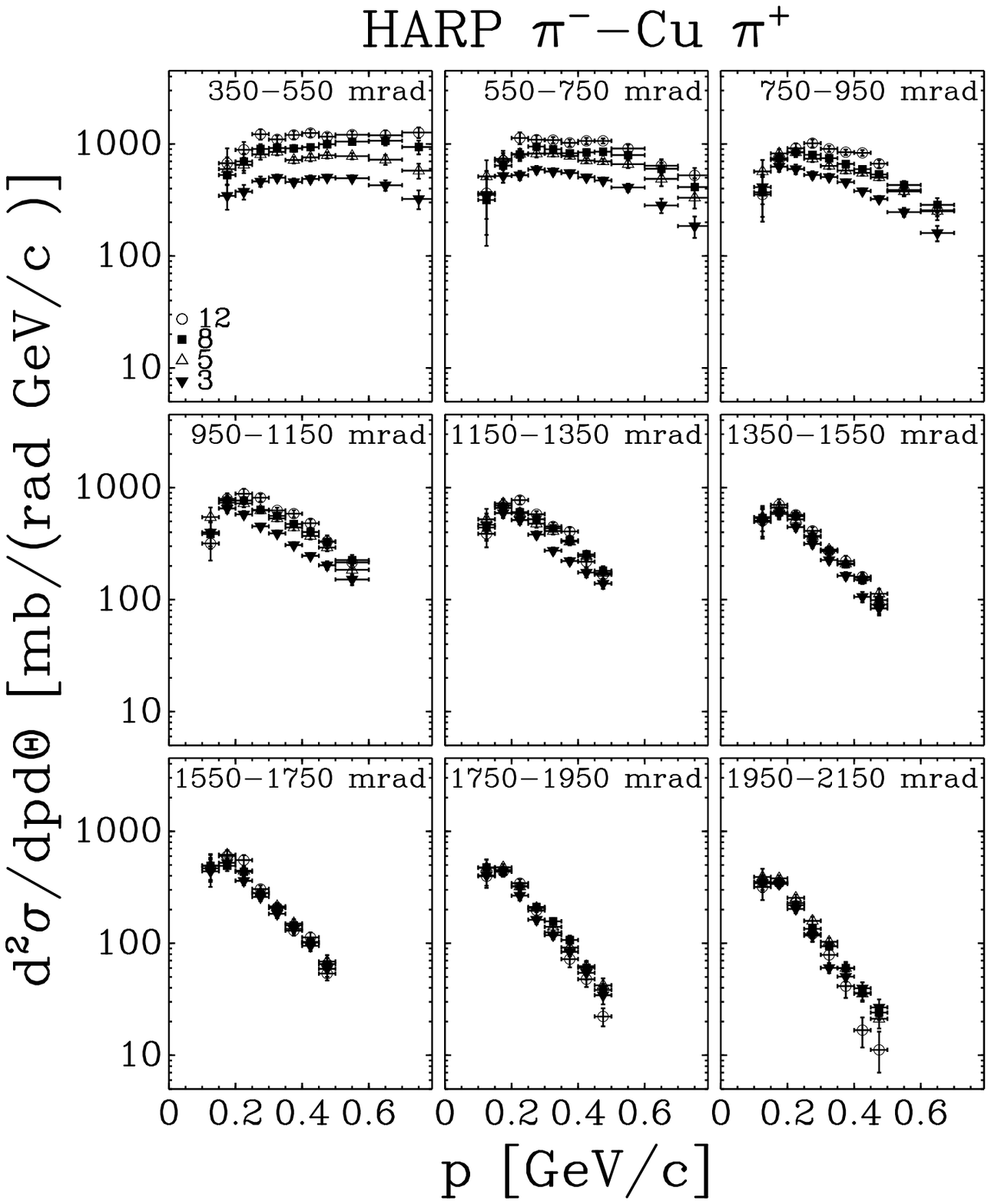}
 \includegraphics[width=0.49\textwidth]{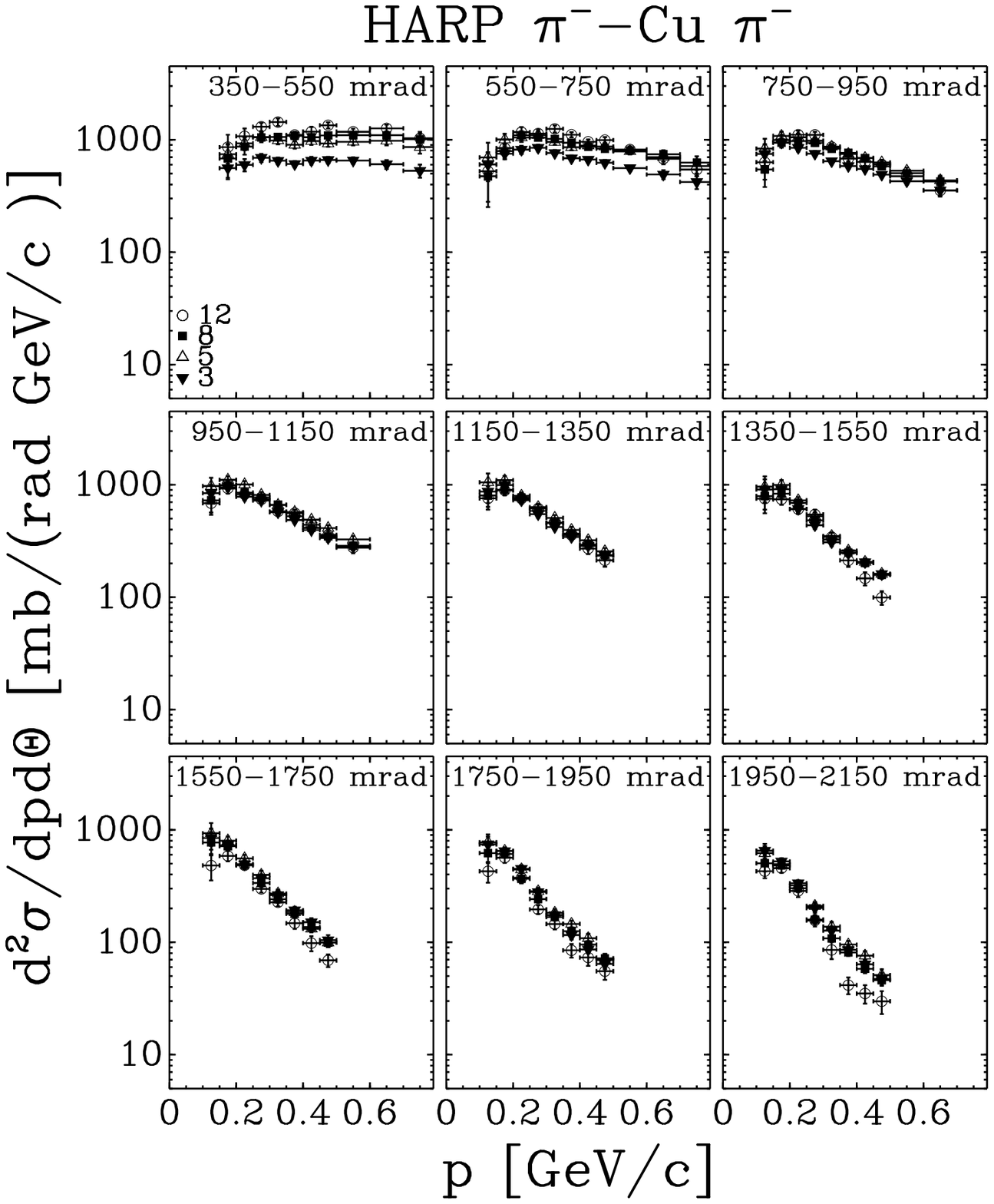}
\caption{
Double-differential cross-sections for \pip production (left) and  \pim
 production (right) in
\pim--Cu interactions as a function of momentum displayed in different
angular bins (shown in \mrad in the panels).
The error bars represent the combination of statistical and systematic
 uncertainties. 
}
\label{fig:xs-pim-th-pbeam-cu}
\end{center}
\end{sidewaysfigure}
\begin{sidewaysfigure}[tbp!]
\begin{center}
 \includegraphics[width=0.49\textwidth]{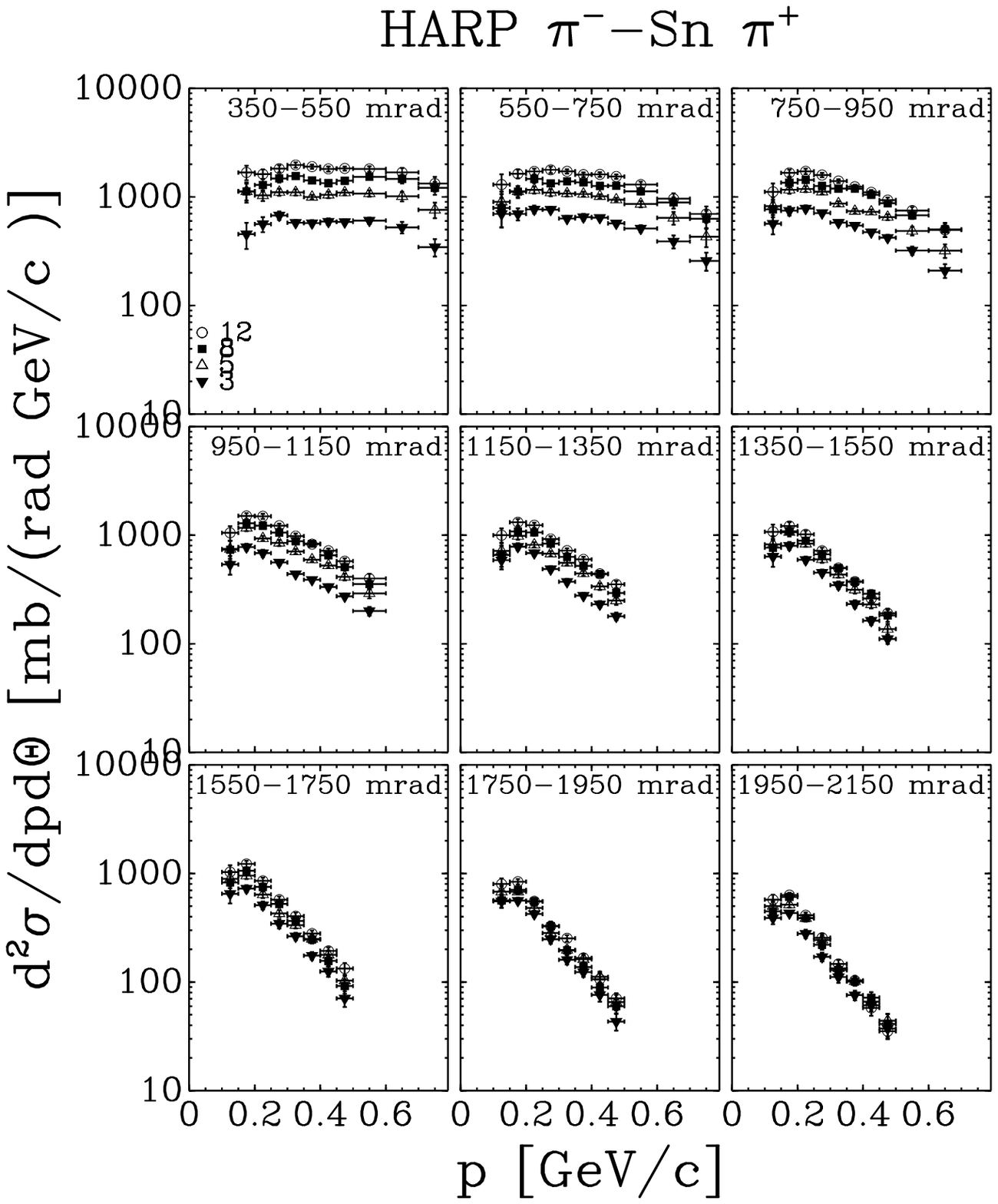}
 \includegraphics[width=0.49\textwidth]{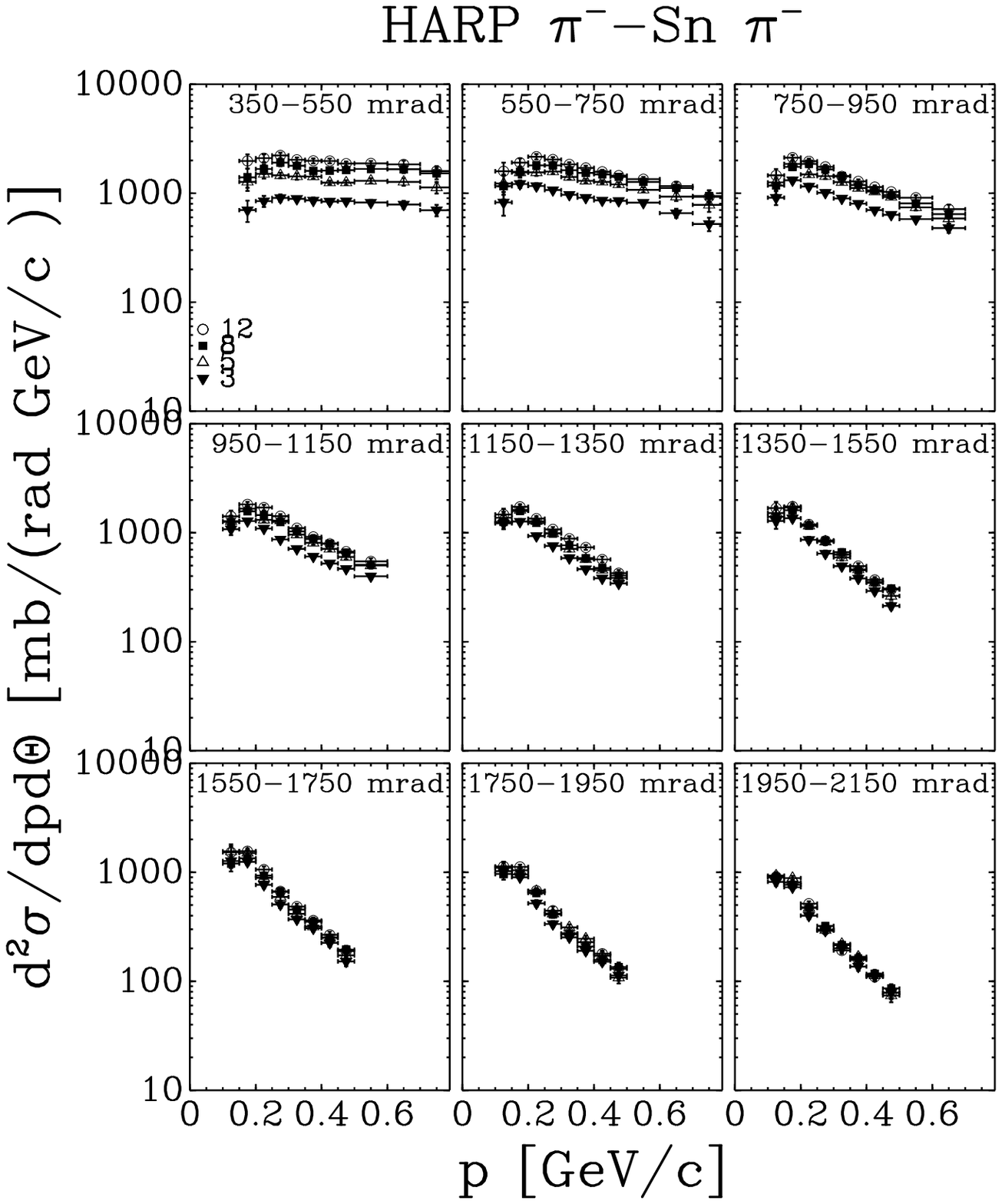}
\caption{
Double-differential cross-sections for \pip production (left) and  \pim
 production (right) in
\pim--Sn interactions as a function of momentum displayed in different
angular bins (shown in \mrad in the panels).
The error bars represent the combination of statistical and systematic
 uncertainties. 
}
\label{fig:xs-pim-th-pbeam-sn}
\end{center}
\end{sidewaysfigure}
\begin{sidewaysfigure}[tbp!]
\begin{center}
 \includegraphics[width=0.49\textwidth]{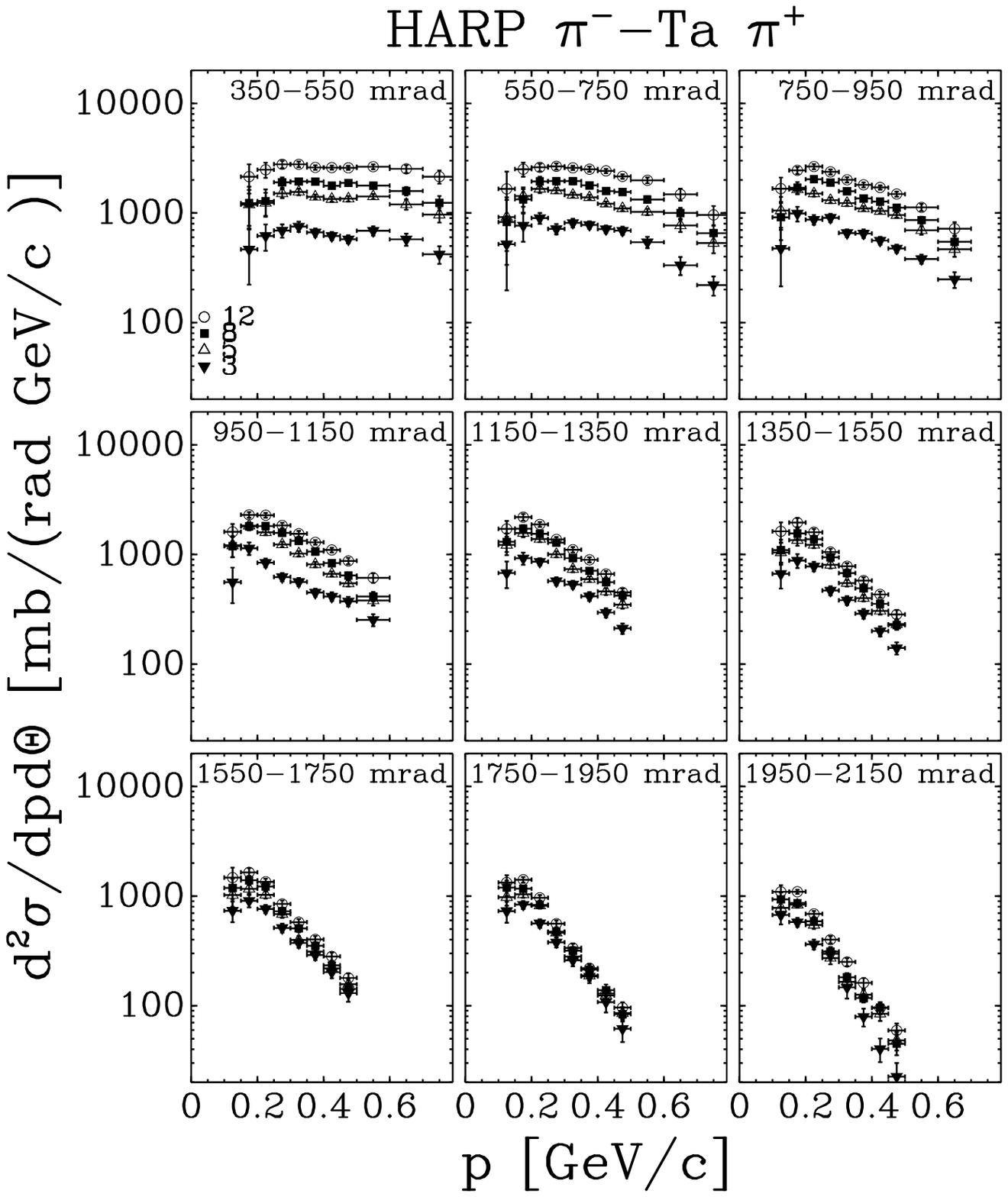}
 \includegraphics[width=0.49\textwidth]{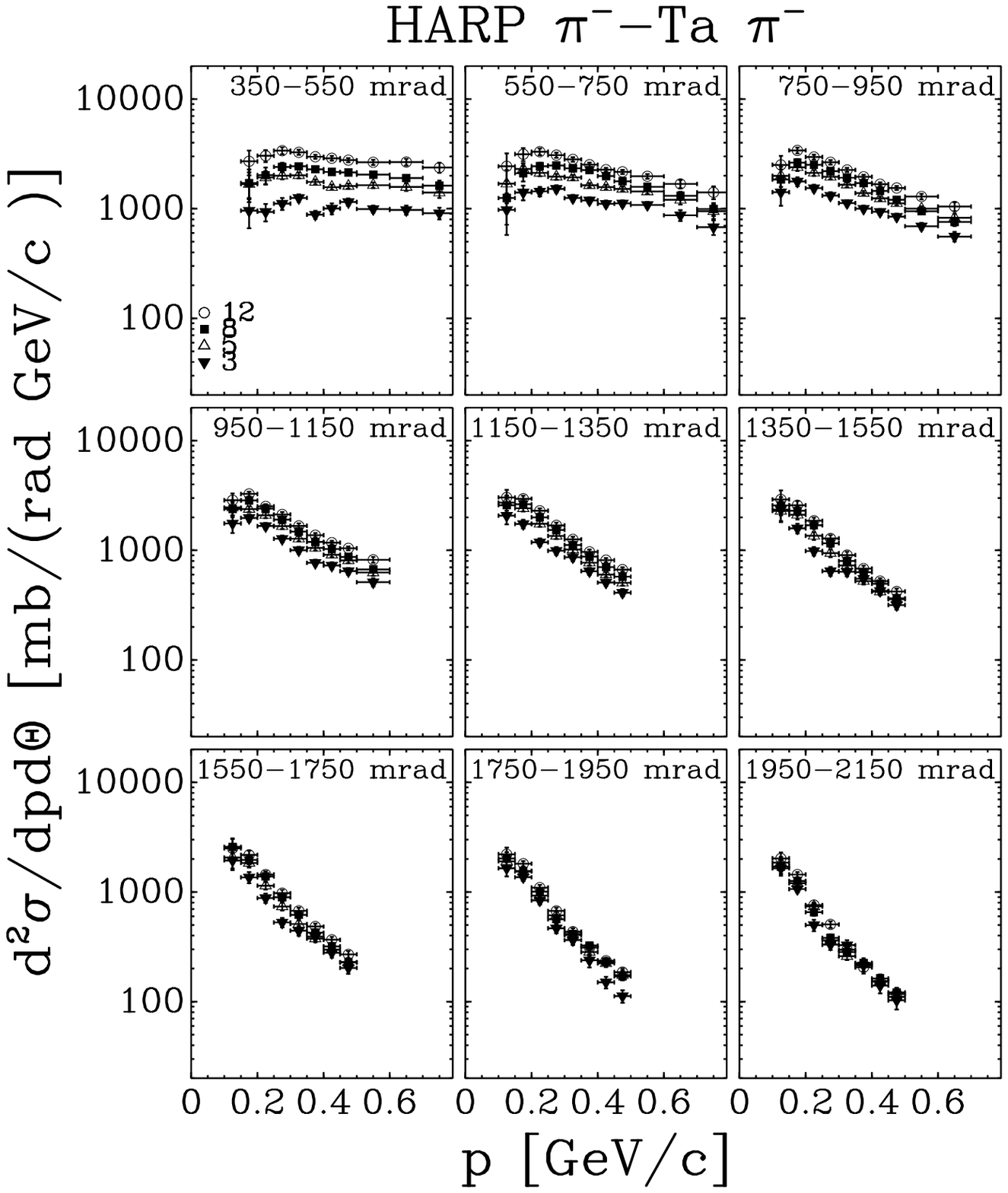}
\caption{
Double-differential cross-sections for \pip production (left) and  \pim
 production (right) in
\pim--Ta interactions as a function of momentum displayed in different
angular bins (shown in \mrad in the panels).
The error bars represent the combination of statistical and systematic
 uncertainties. 
}
\label{fig:xs-pim-th-pbeam-ta}
\end{center}
\end{sidewaysfigure}
\begin{sidewaysfigure}[tbp!]
\begin{center}
 \includegraphics[width=0.49\textwidth]{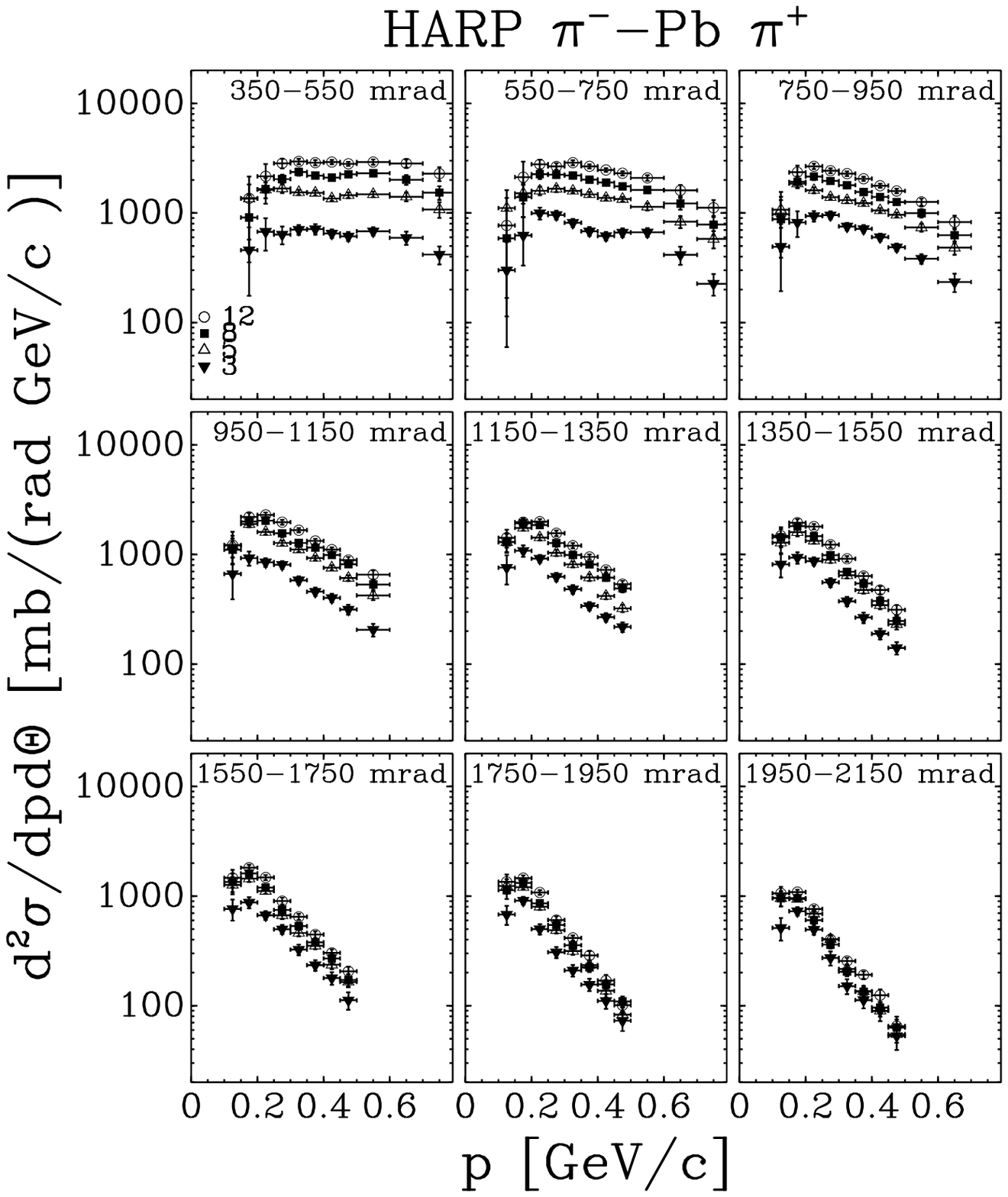}
 \includegraphics[width=0.49\textwidth]{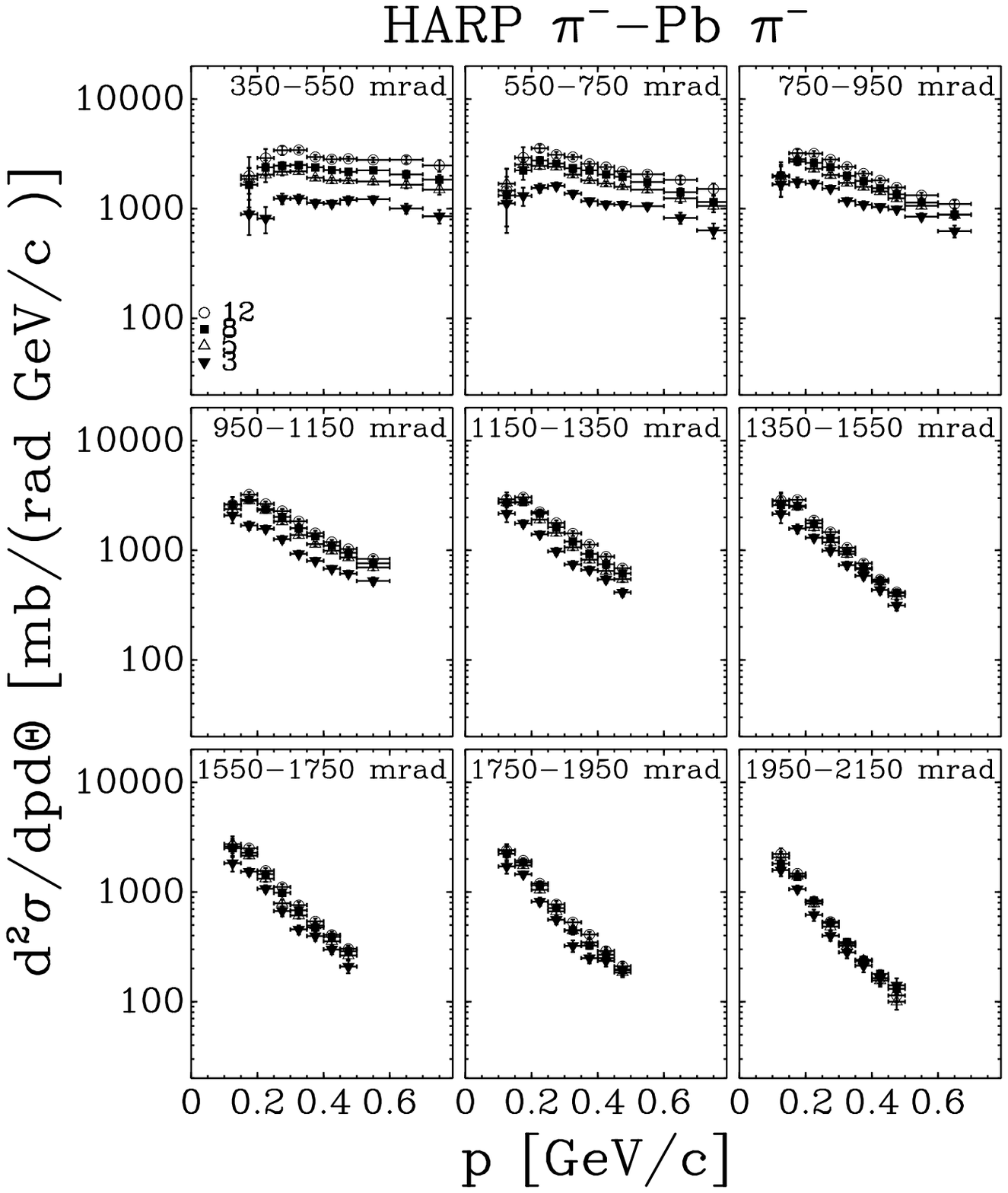}
\caption{
Double-differential cross-sections for \pip production (left) and  \pim
 production (right) in
\pim--Pb interactions as a function of momentum displayed in different
angular bins (shown in \mrad in the panels).
The error bars represent the combination of statistical and systematic
 uncertainties. 
}
\label{fig:xs-pim-th-pbeam-pb}
\end{center}
\end{sidewaysfigure}
\fi
%

\subsection{Comparisons with MC predictions}
\label{sec:compare}

\begin{sidewaysfigure}[tbp!]
\begin{center}
 \includegraphics[width=0.49\textwidth]{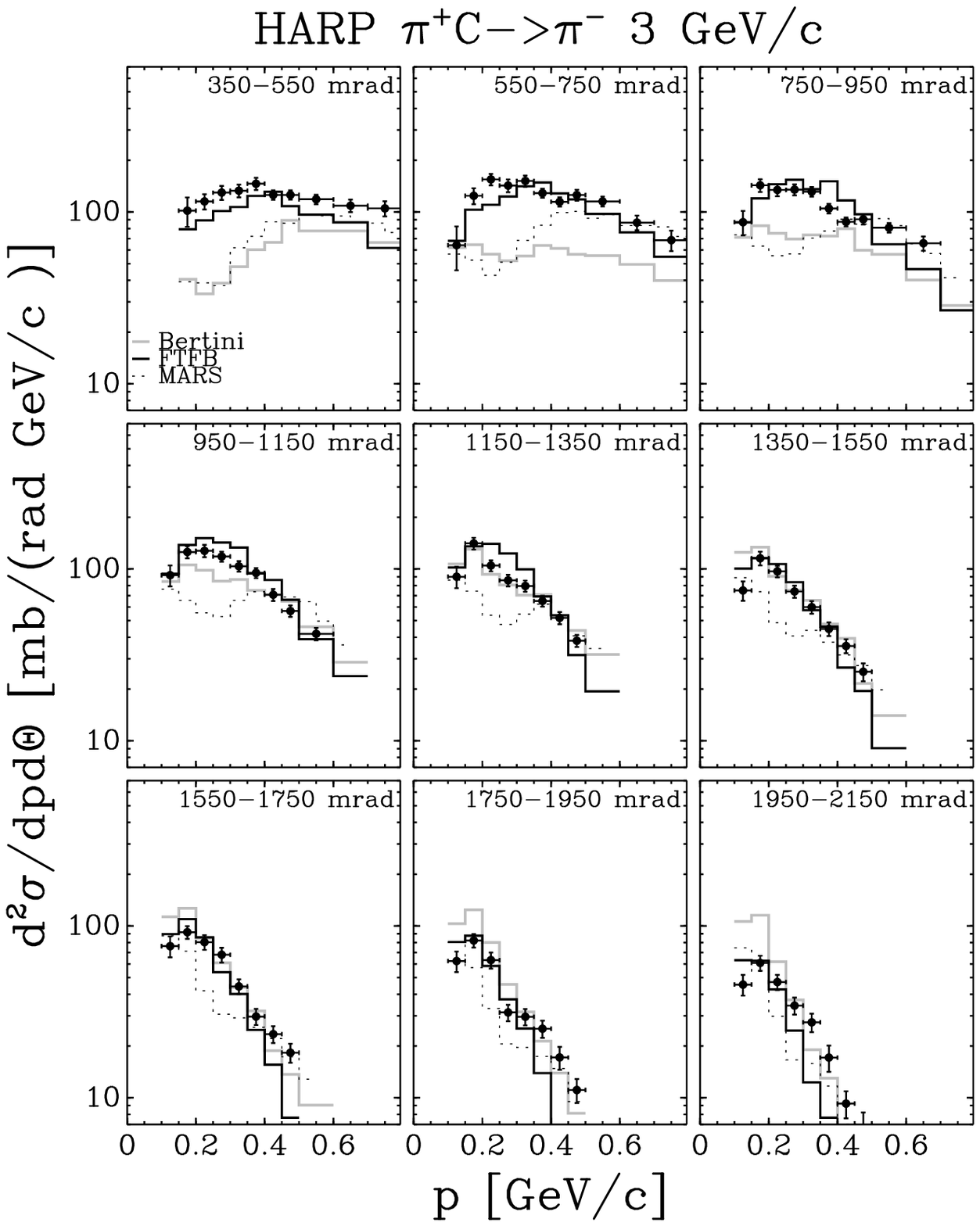}
 \includegraphics[width=0.49\textwidth]{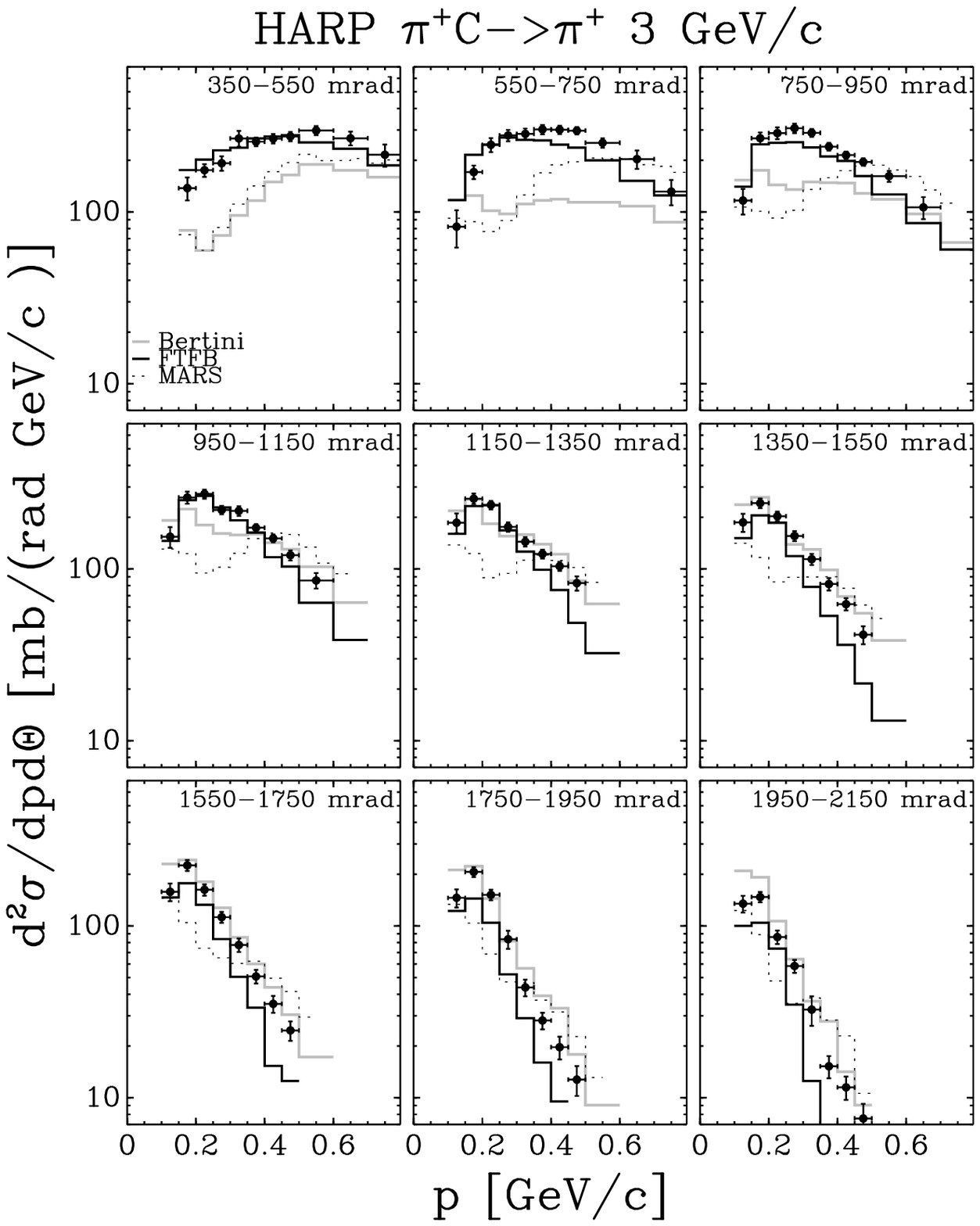}
\end{center}
\caption{
 Comparison of HARP double-differential \pipm production cross sections for \pip--C at 3~\GeVc with
 GEANT4 and MARS MC predictions, using several generator models (see
 text for details).
The left (right) panel shows \pim (\pip) production.
The gray line shows the Bertini model, the black solid line FTFB and
 the dashed line MARS.  
}
\label{fig:G4c3}
\end{sidewaysfigure}

\begin{sidewaysfigure}[tbp!]
\begin{center}
 \includegraphics[width=0.49\textwidth]{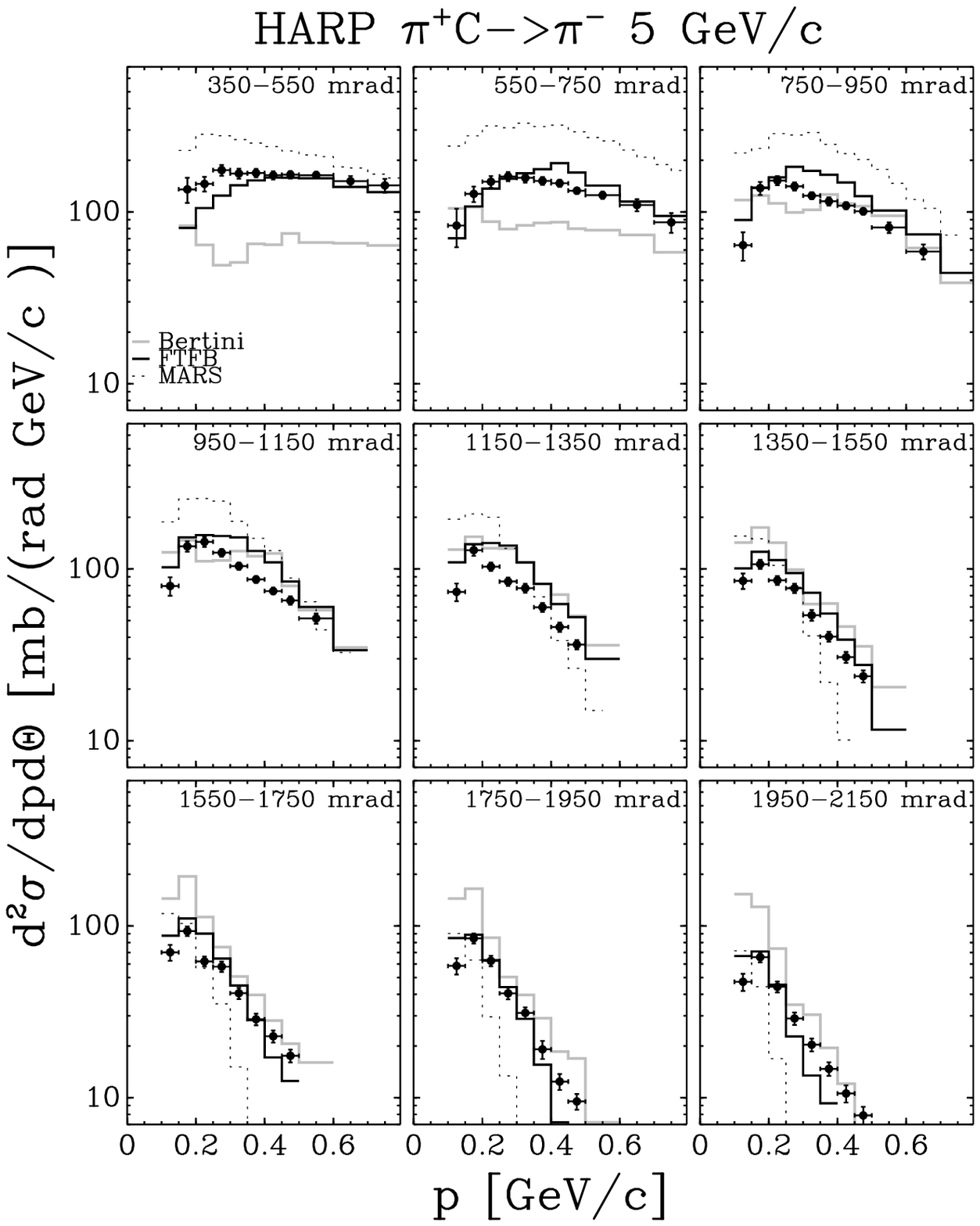}
 \includegraphics[width=0.49\textwidth]{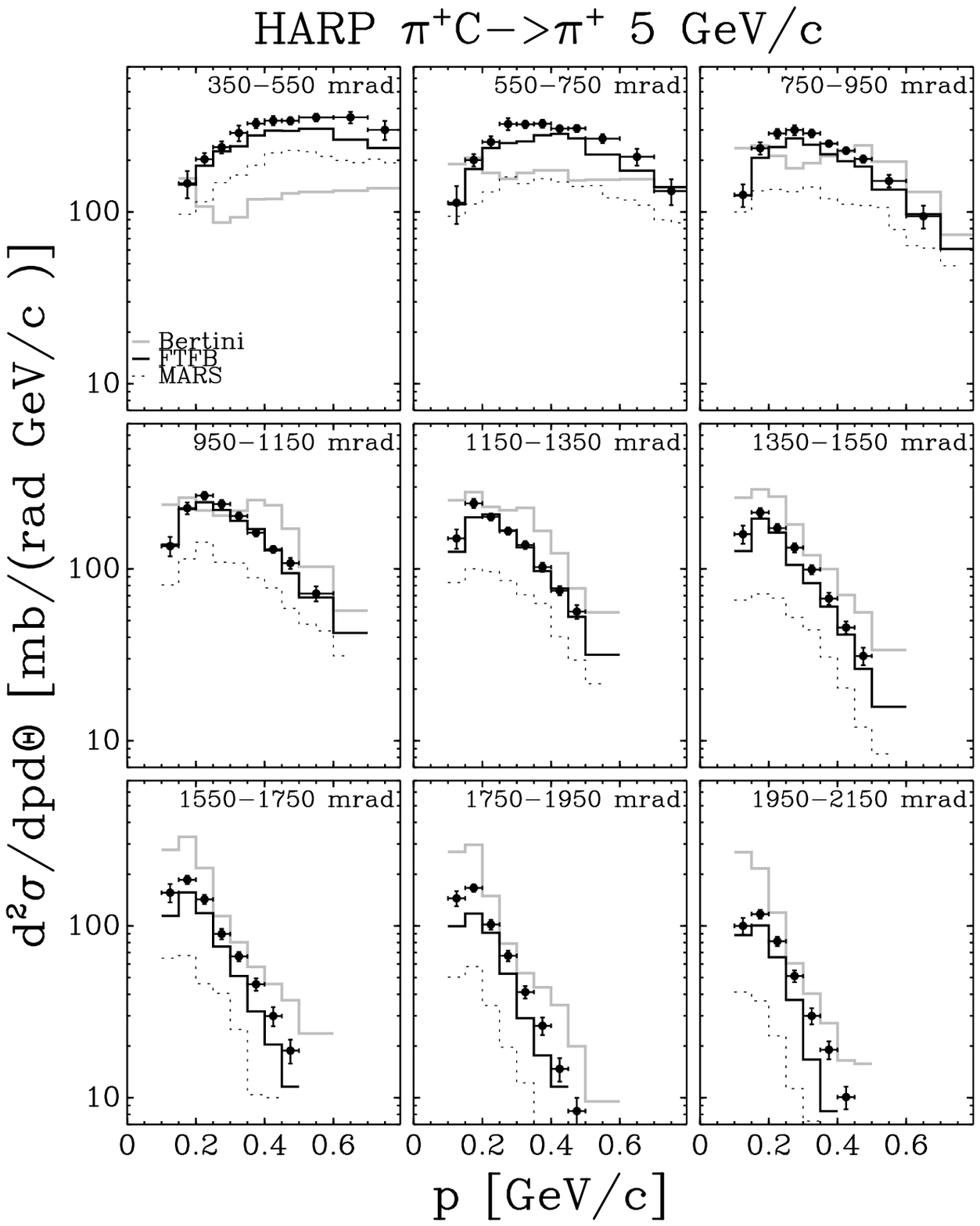}
\end{center}
\caption{
 Comparison of HARP double-differential \pipm production cross sections for \pip--C at 5~\GeVc with
 GEANT4 and MARS MC predictions, using several generator models (see
 text for details).
The left (right) panel shows \pim (\pip) production.
The gray line shows the Bertini model, the black solid line FTFB and
 the dashed line MARS.  
}
\label{fig:G4c5}
\end{sidewaysfigure}

\begin{sidewaysfigure}[tbp!]
\begin{center}
 \includegraphics[width=0.49\textwidth]{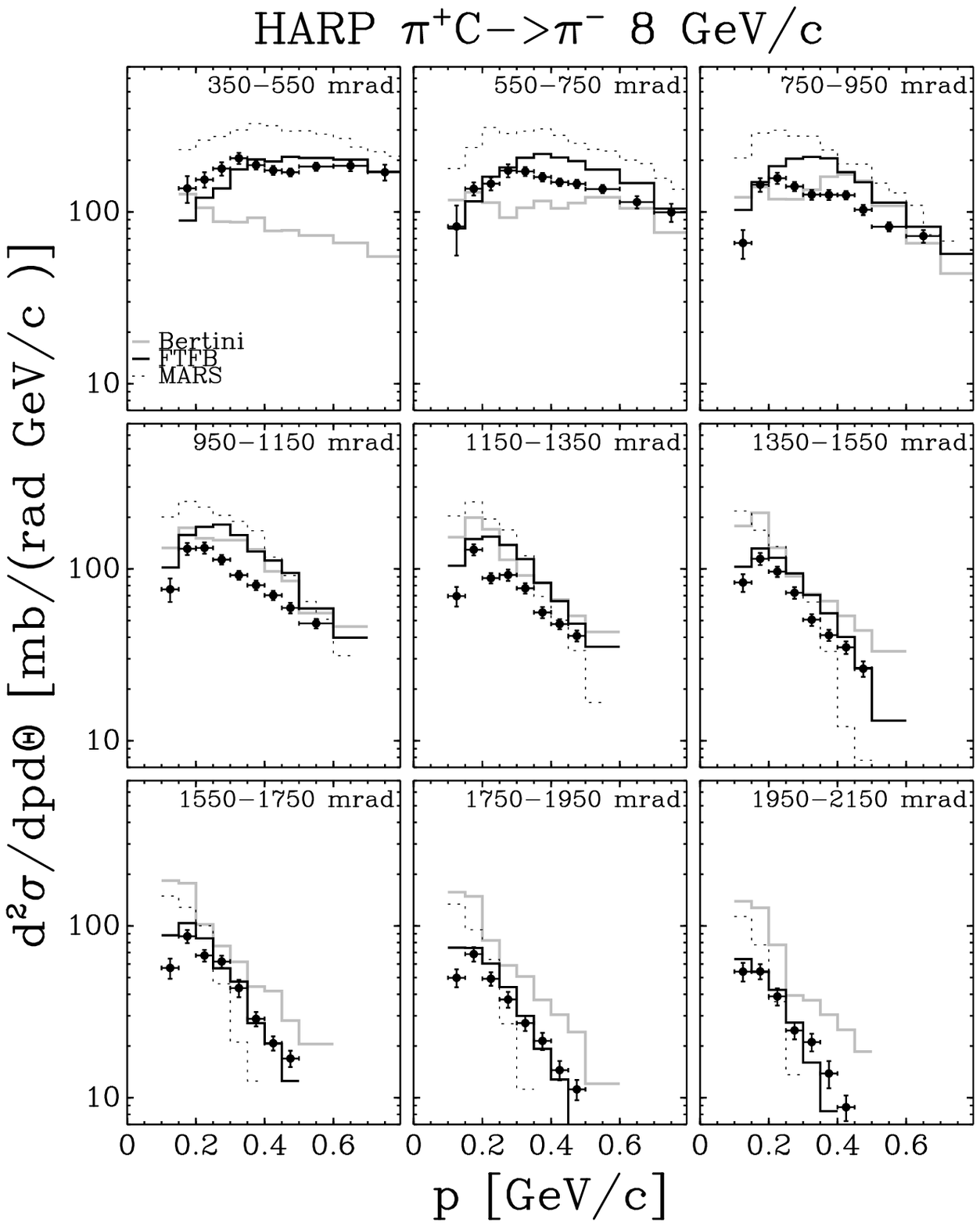}
 \includegraphics[width=0.49\textwidth]{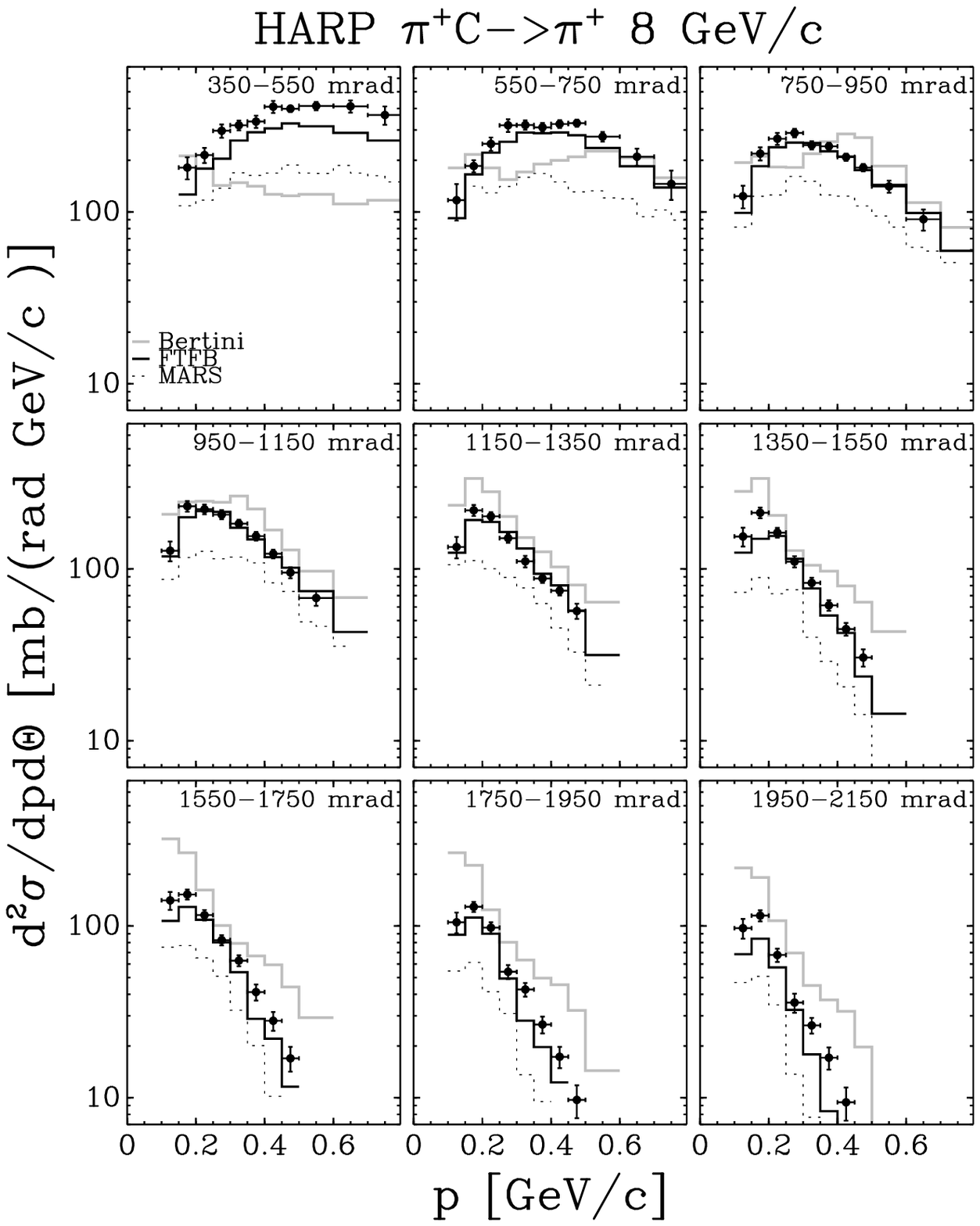}
\end{center}
\caption{
 Comparison of HARP double-differential \pipm production cross sections for \pip--C at 8~\GeVc with
 GEANT4 and MARS MC predictions, using several generator models (see text for details).
The left (right) panel shows \pim (\pip) production.
The gray line shows the Bertini model, the black solid line FTFB and
 the dashed line MARS.  
}
\label{fig:G4c8}
\end{sidewaysfigure}

\begin{sidewaysfigure}[tbp!]
\begin{center}
 \includegraphics[width=0.49\textwidth]{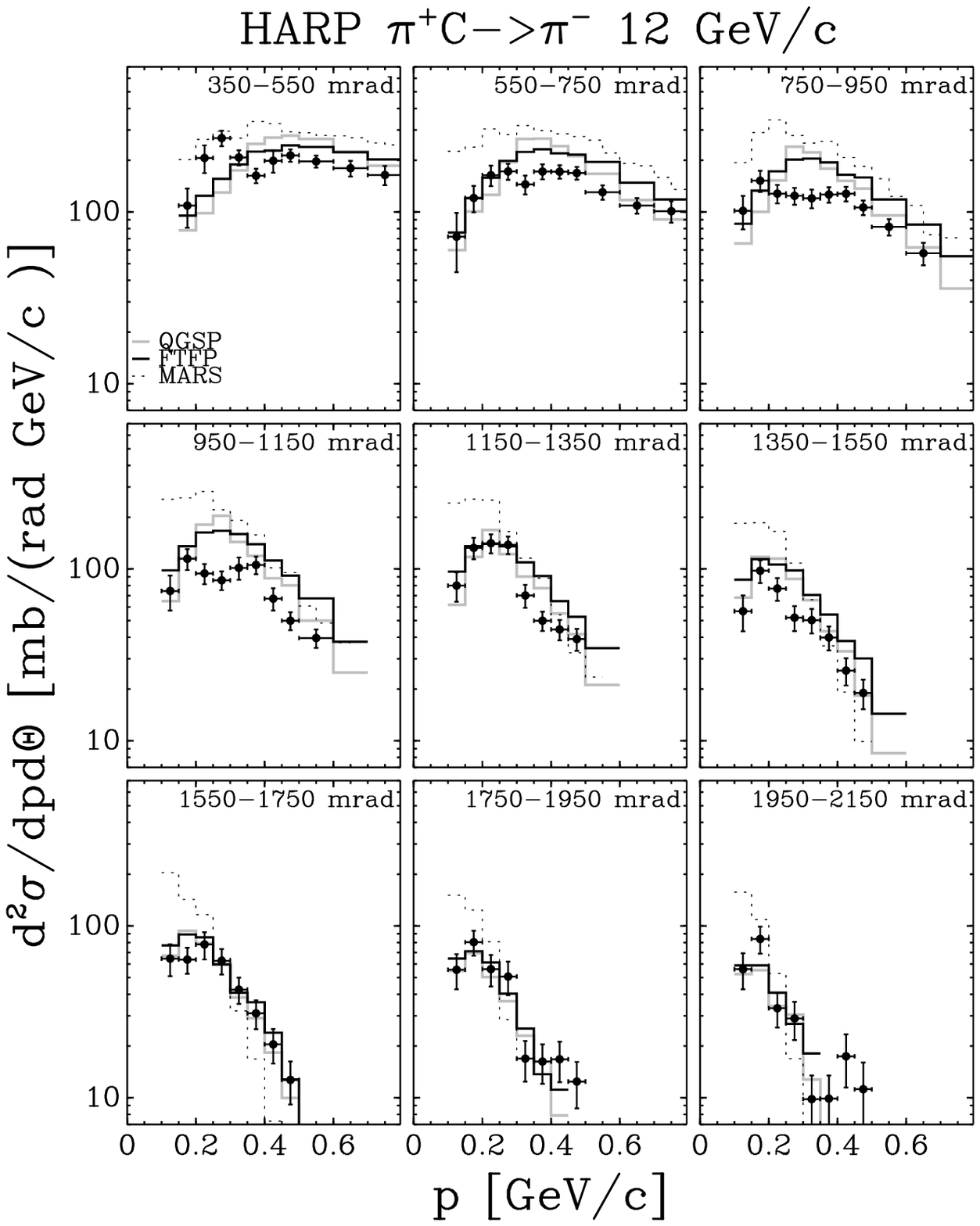}
 \includegraphics[width=0.49\textwidth]{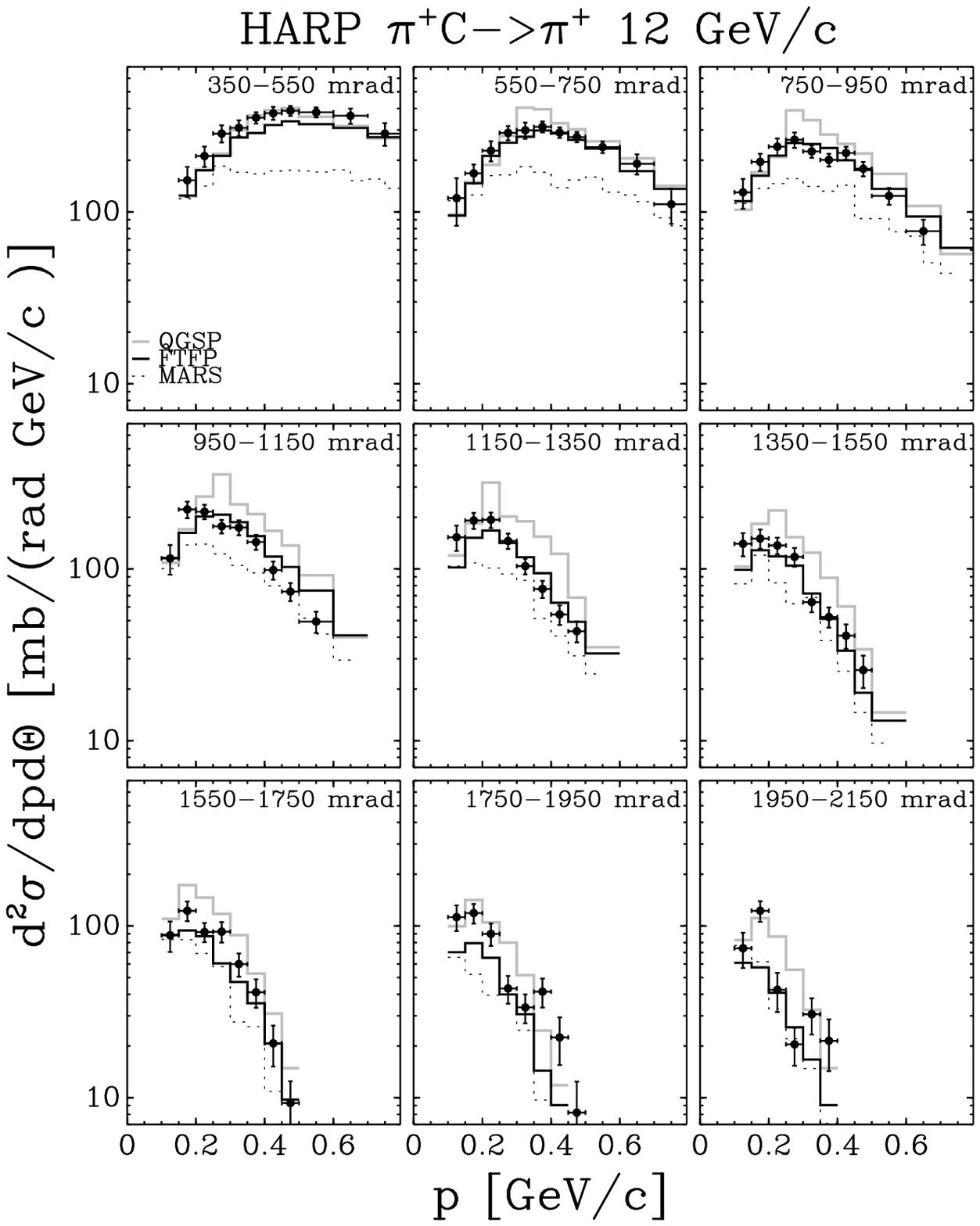}
\end{center}
\caption{
 Comparison of HARP double-differential \pipm production cross sections for \pip--C at 12~\GeVc with
 GEANT4 and MARS MC predictions, using several generator models (see text for details).
The left (right) panel shows \pim (\pip) production.
The gray line shows the QGSP model, the black solid line FTFP and
 the dashed line MARS.  
}
\label{fig:G4c12}
\end{sidewaysfigure}

\begin{sidewaysfigure}[tbp!]
\begin{center}
 \includegraphics[width=0.49\textwidth]{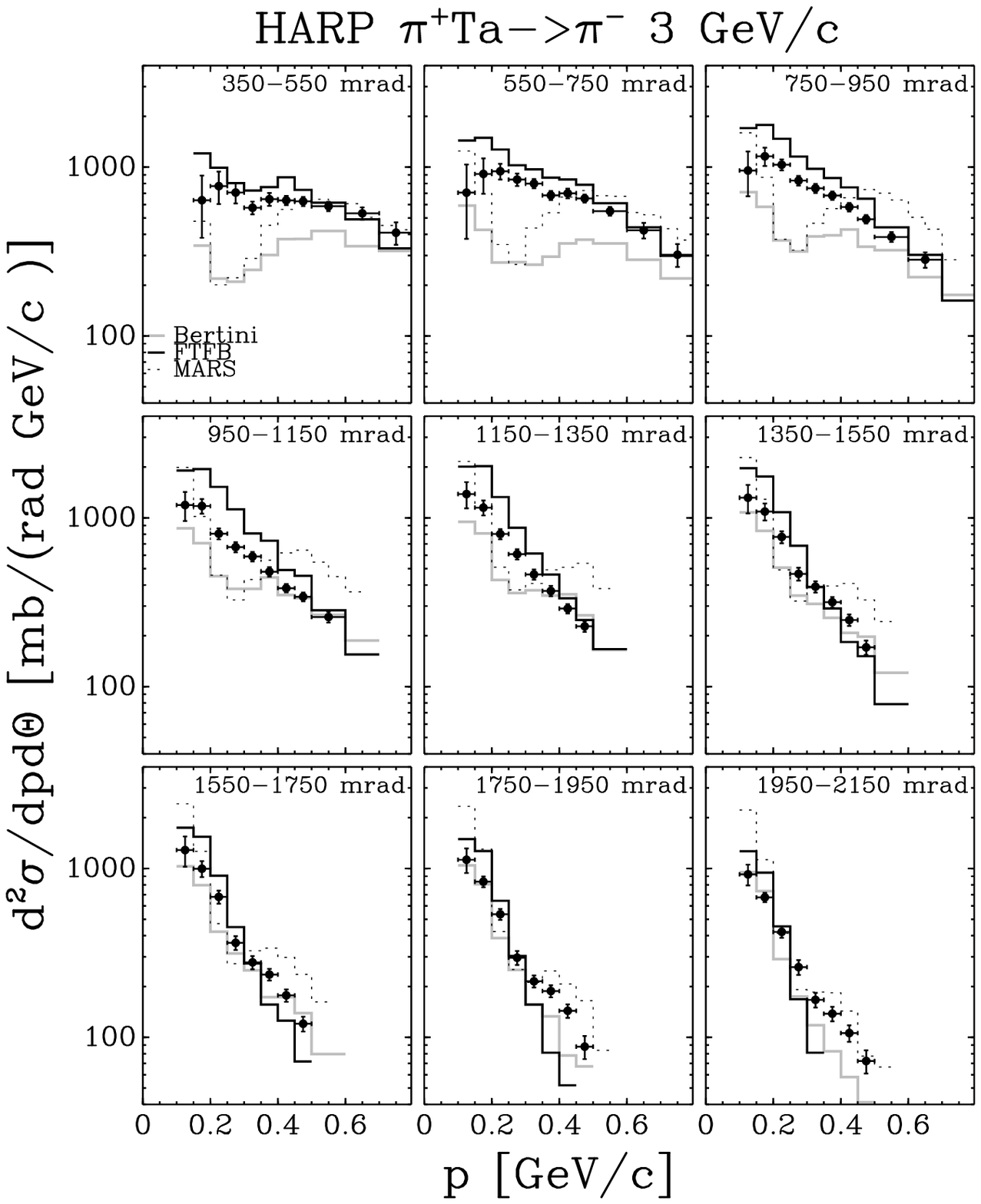}
 \includegraphics[width=0.49\textwidth]{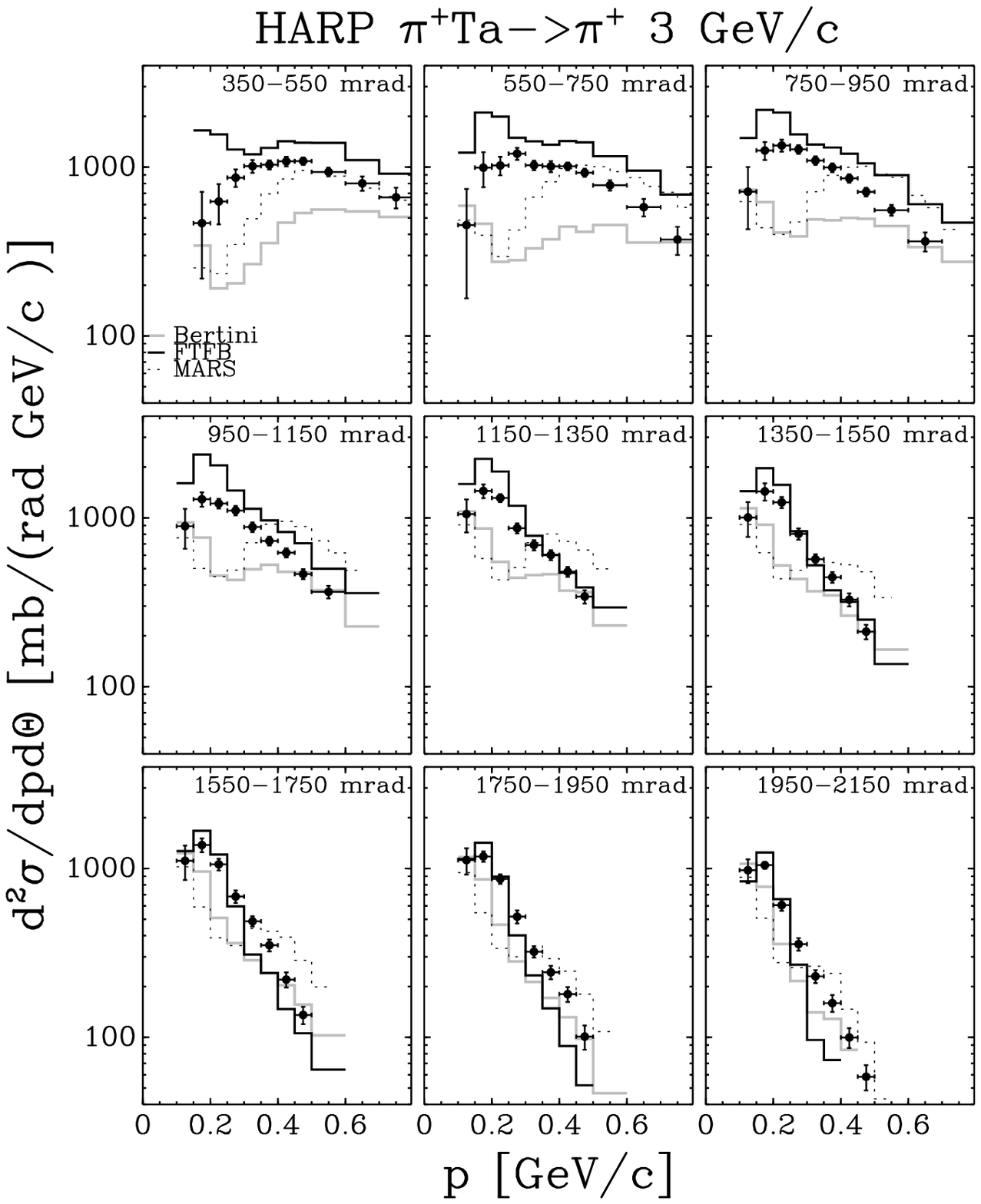}
\end{center}
\caption{
 Comparison of HARP double-differential \pipm production cross sections for \pip--Ta at 3~\GeVc with
 GEANT4 and MARS MC predictions, using several generator models (see text for details).
The left (right) panel shows \pim (\pip) production.
The gray line shows the Bertini model, the black solid line FTFB and
 the dashed line MARS.  
}
\label{fig:G4ta3}
\end{sidewaysfigure}

\begin{sidewaysfigure}[tbp!]
\begin{center}
 \includegraphics[width=0.49\textwidth]{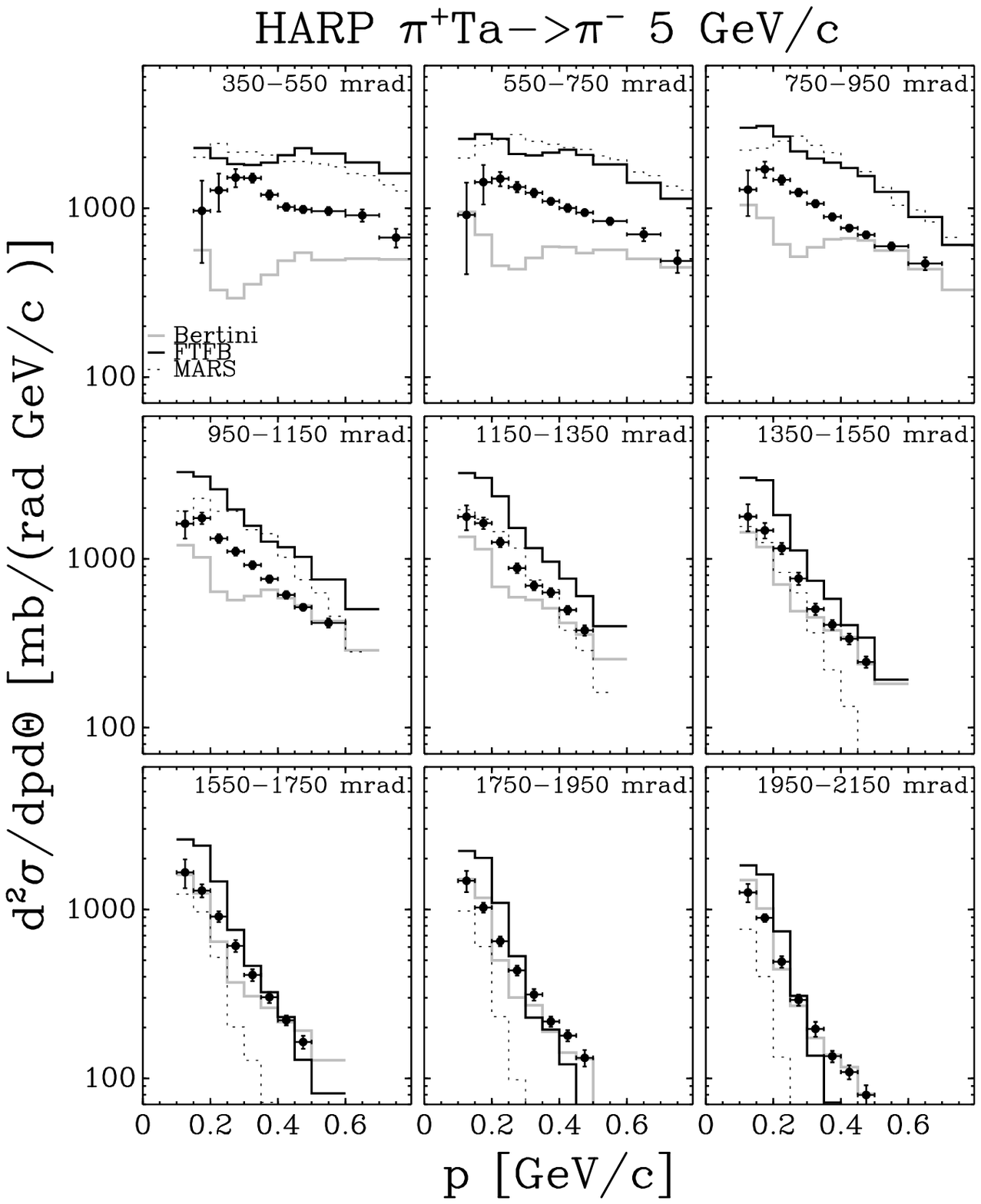}
 \includegraphics[width=0.49\textwidth]{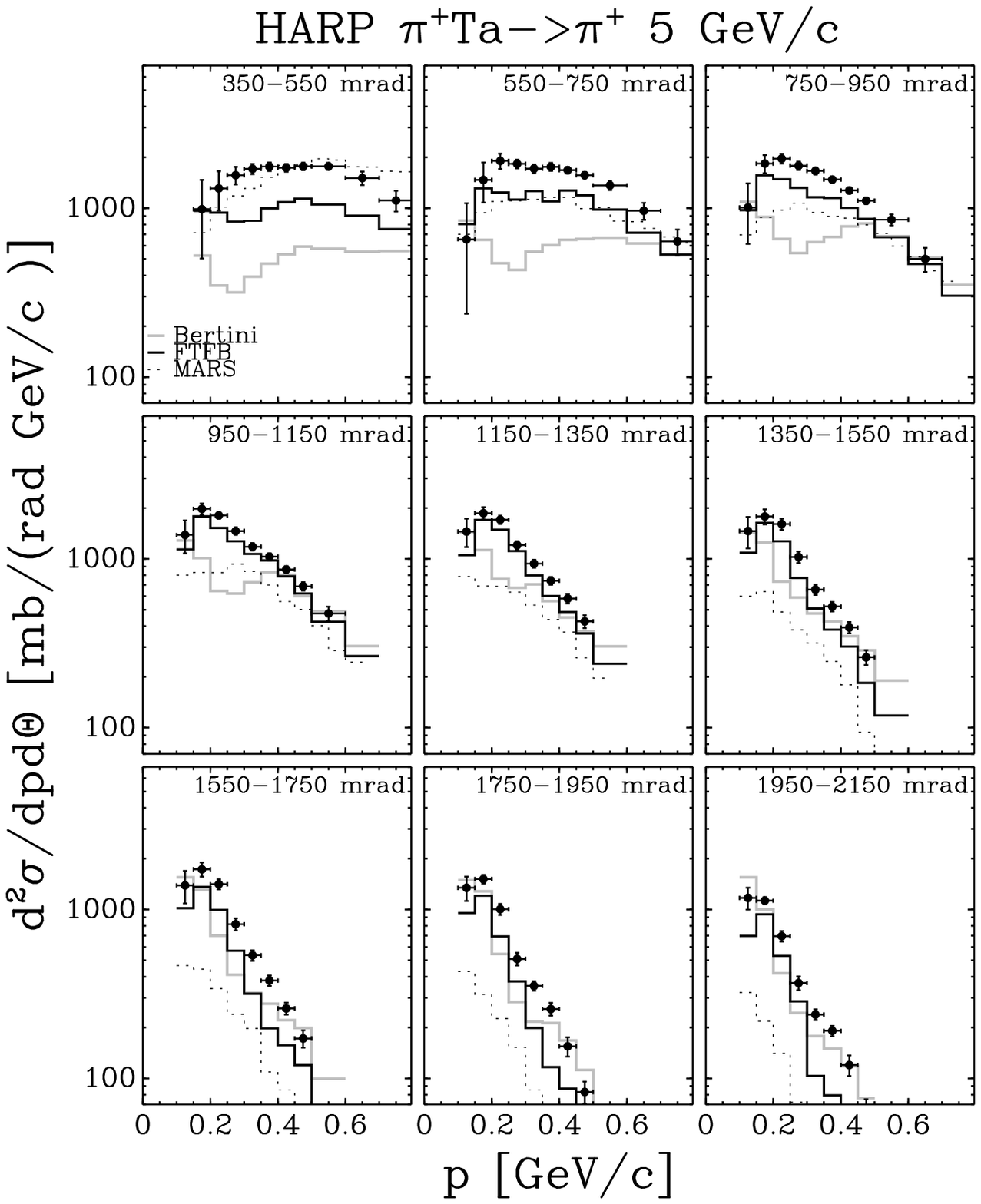}
\end{center}
\caption{
 Comparison of HARP double-differential \pipm production cross sections for \pip--Ta at 5~\GeVc with
 GEANT4 and MARS MC predictions, using several generator models (see text for details).
The left (right) panel shows \pim (\pip) production.
The gray line shows the Bertini model, the black solid line FTFB and
 the dashed line MARS.  
}
\label{fig:G4ta5}
\end{sidewaysfigure}

\begin{sidewaysfigure}[tbp!]
\begin{center}
 \includegraphics[width=0.49\textwidth]{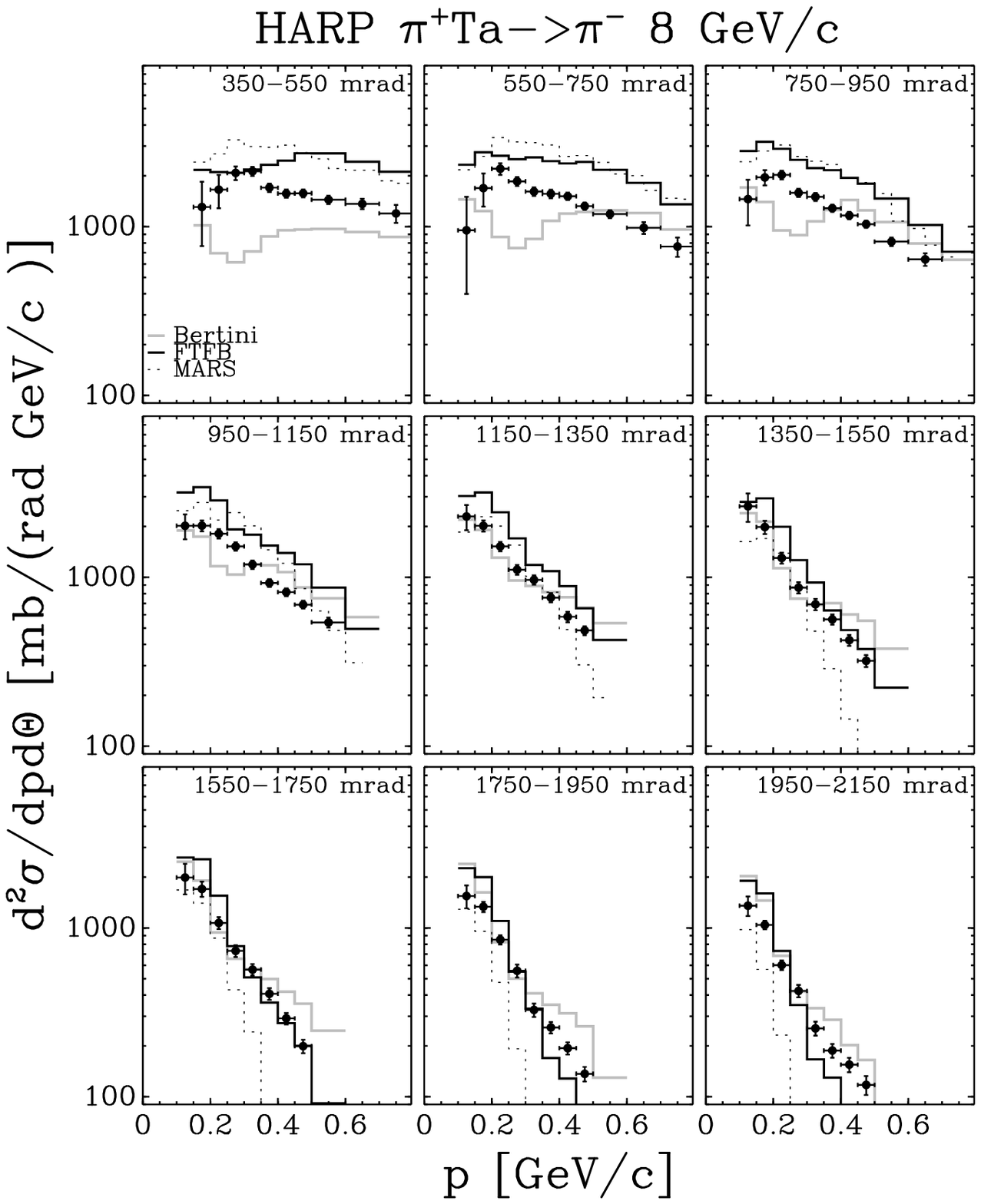}
 \includegraphics[width=0.49\textwidth]{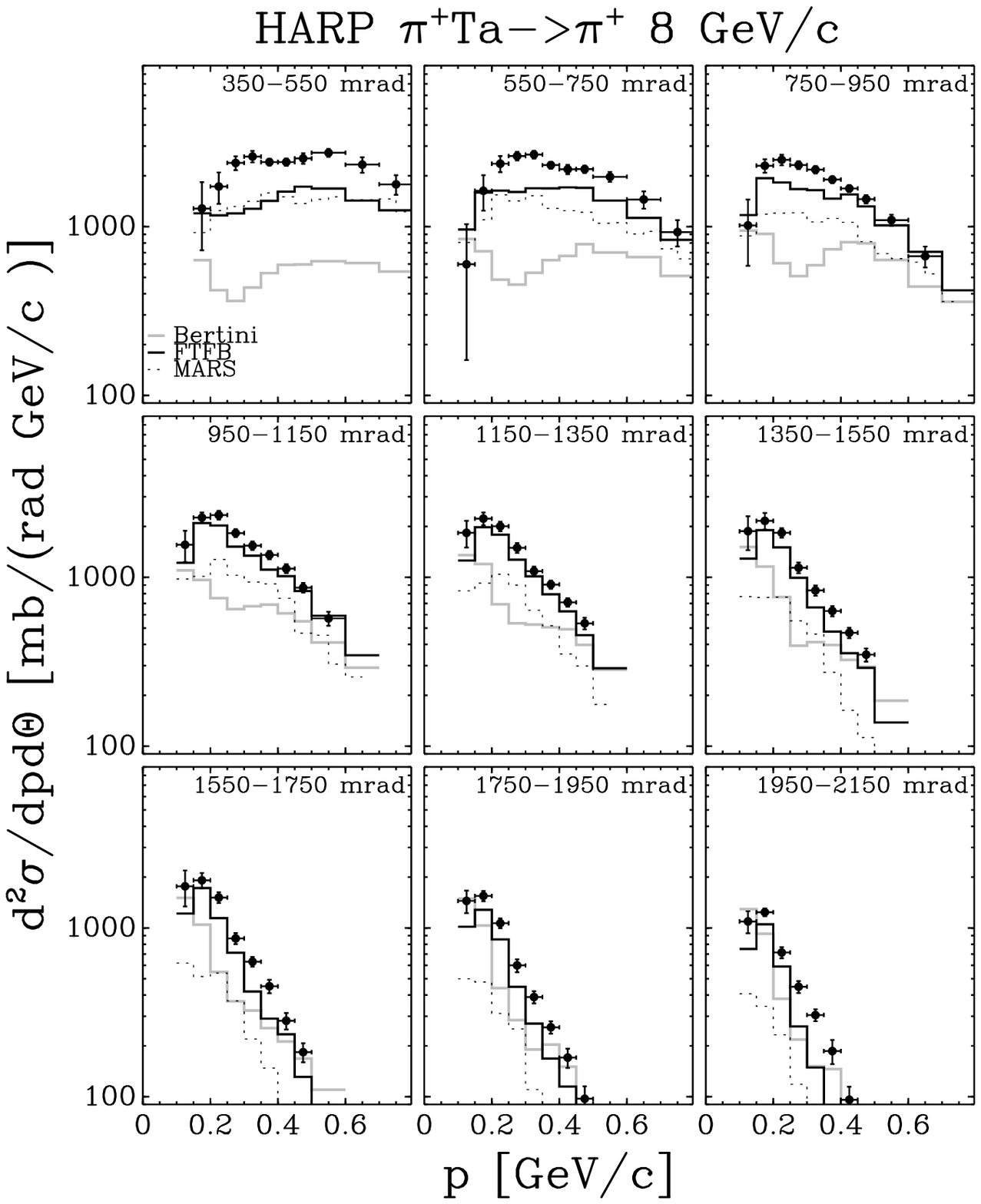}
\end{center}
\caption{
 Comparison of HARP double-differential \pipm production cross sections for \pip--Ta at 8~\GeVc with
 GEANT4 and MARS MC predictions, using several generator models (see text for details).
The left (right) panel shows \pim (\pip) production.
The gray line shows the Bertini model, the black solid line FTFB and
 the dashed line MARS.  
}
\label{fig:G4ta8}
\end{sidewaysfigure}

\begin{sidewaysfigure}[tbp!]
\begin{center}
 \includegraphics[width=0.49\textwidth]{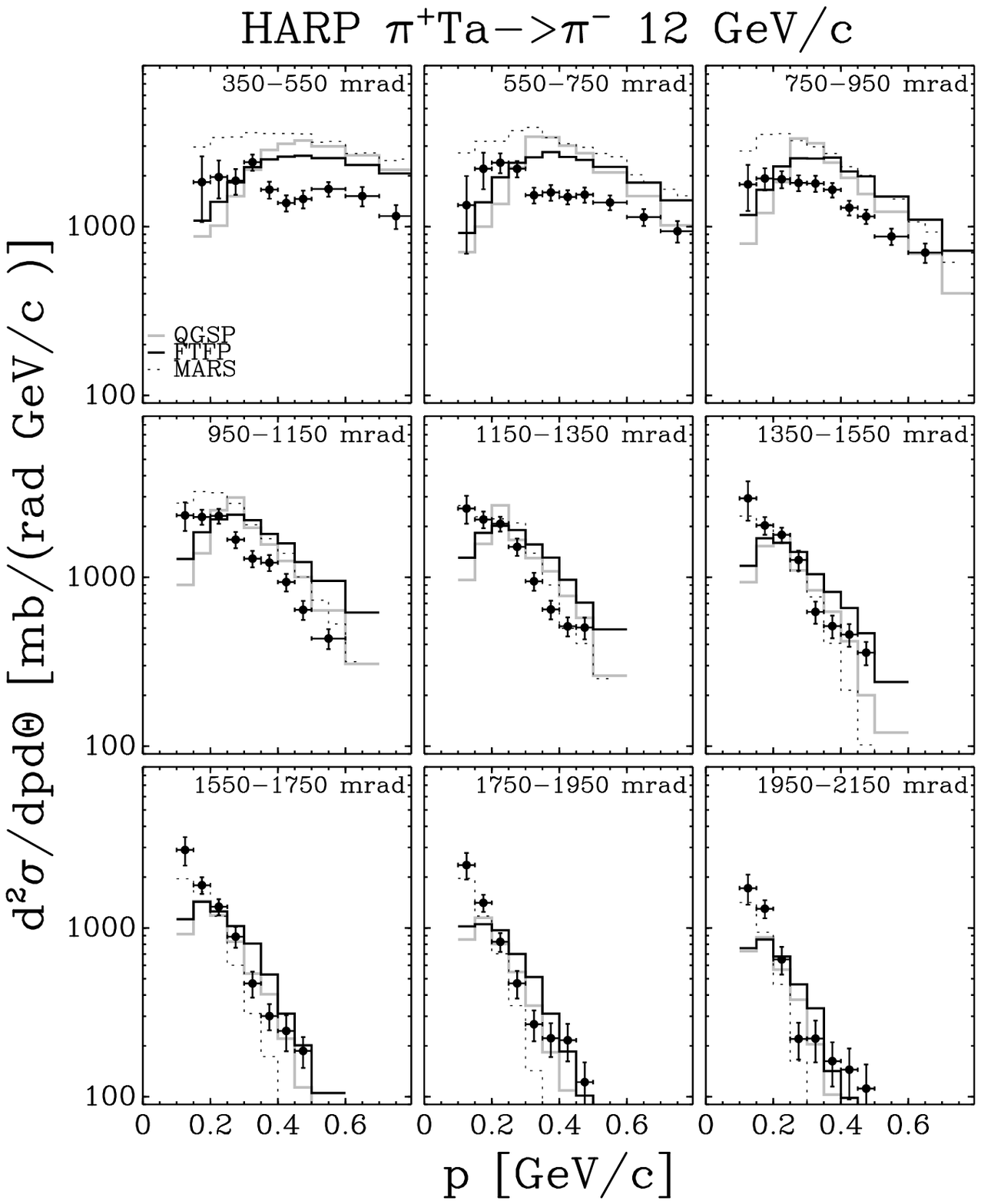}
 \includegraphics[width=0.49\textwidth]{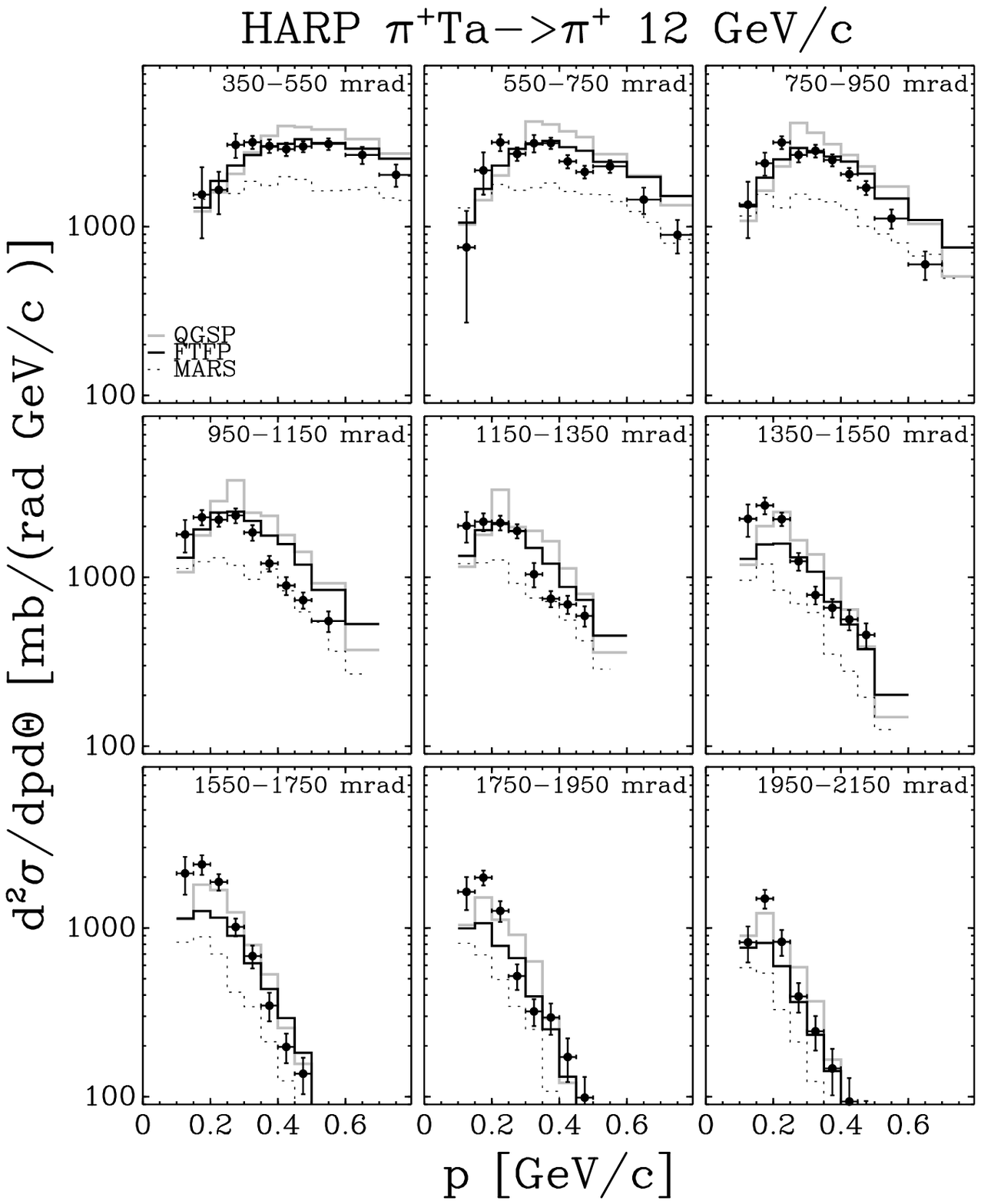}
\end{center}
\caption{
 Comparison of HARP double-differential \pipm production cross sections for \pip--Ta at 12~\GeVc with
 GEANT4 and MARS MC predictions, using several generator models (see text for details).
The left (right) panel shows \pim (\pip) production.
The gray line shows the QGSP model, the black solid line FTFP and
 the dashed line MARS.  
}
\label{fig:G4ta12}
\end{sidewaysfigure}

\begin{sidewaysfigure}[tbp!]
\begin{center}
 \includegraphics[width=0.49\textwidth]{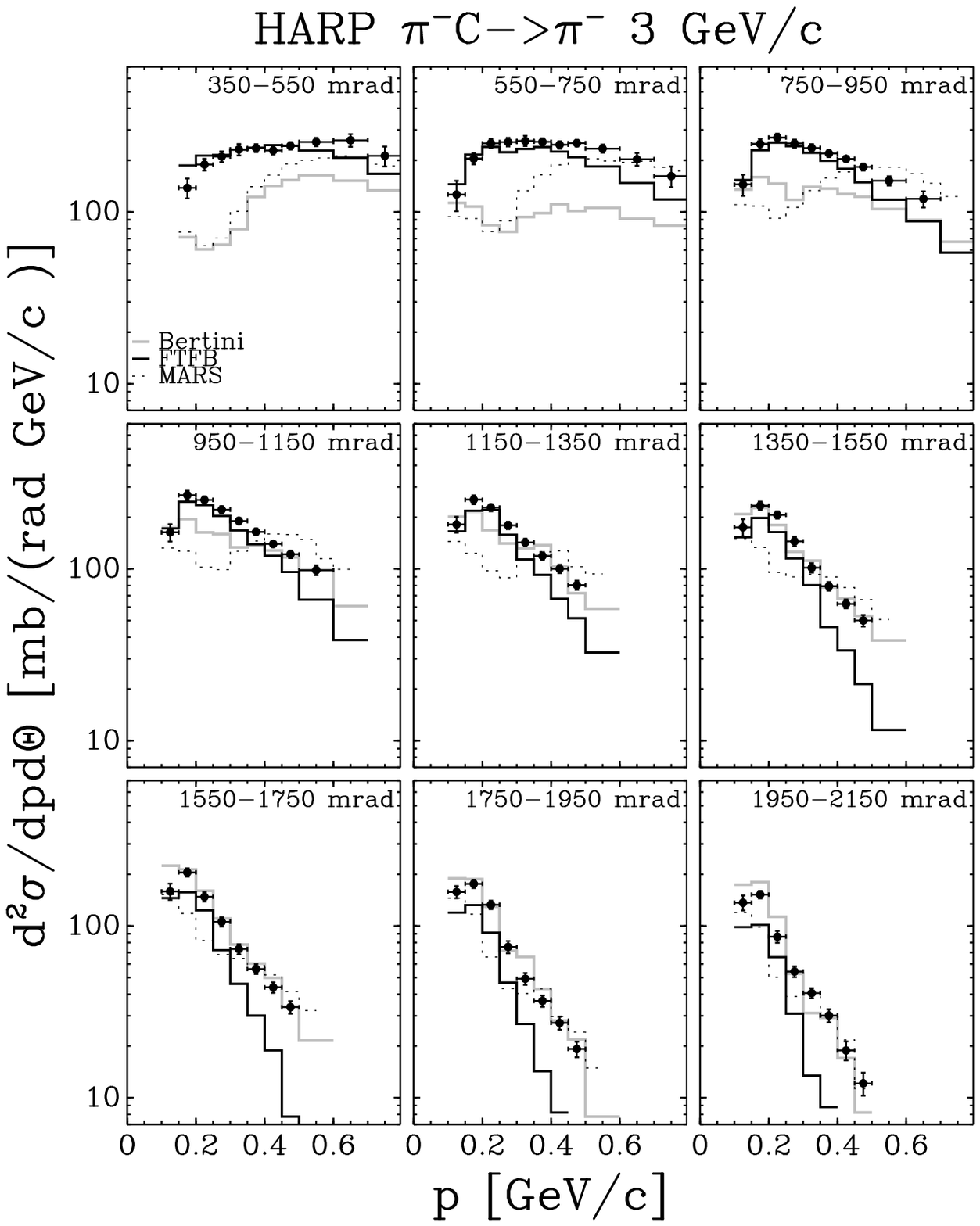}
 \includegraphics[width=0.49\textwidth]{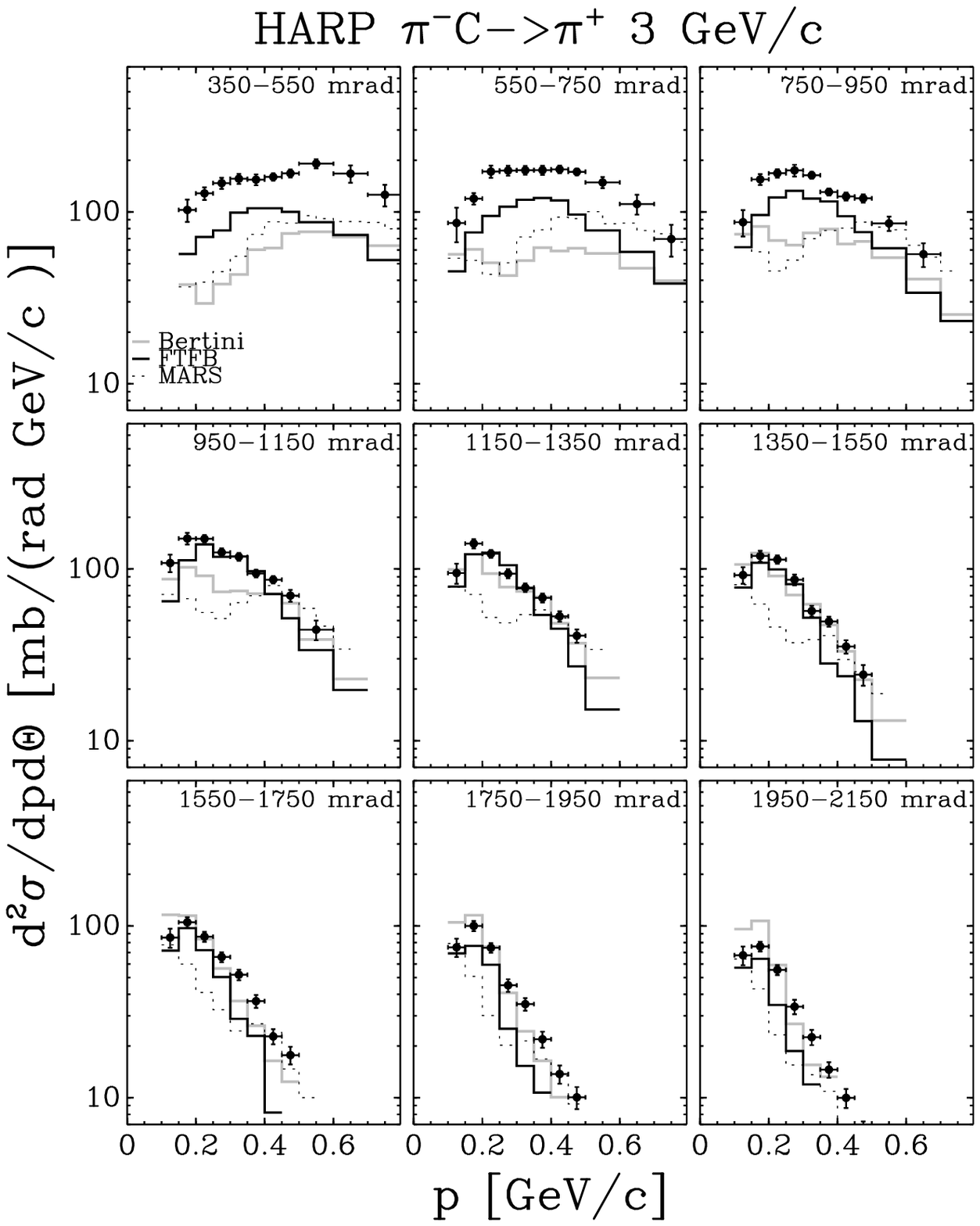}
\end{center}
\caption{
 Comparison of HARP double-differential \pipm production cross sections for \pim--C at 3~\GeVc with
 GEANT4 and MARS MC predictions, using several generator models (see text for details).
The left (right) panel shows \pim (\pip) production.
The gray line shows the Bertini model, the black solid line FTFB and
 the dashed line MARS.  
}
\label{fig:G4c3n}
\end{sidewaysfigure}

\begin{sidewaysfigure}[tbp!]
\begin{center}
 \includegraphics[width=0.49\textwidth]{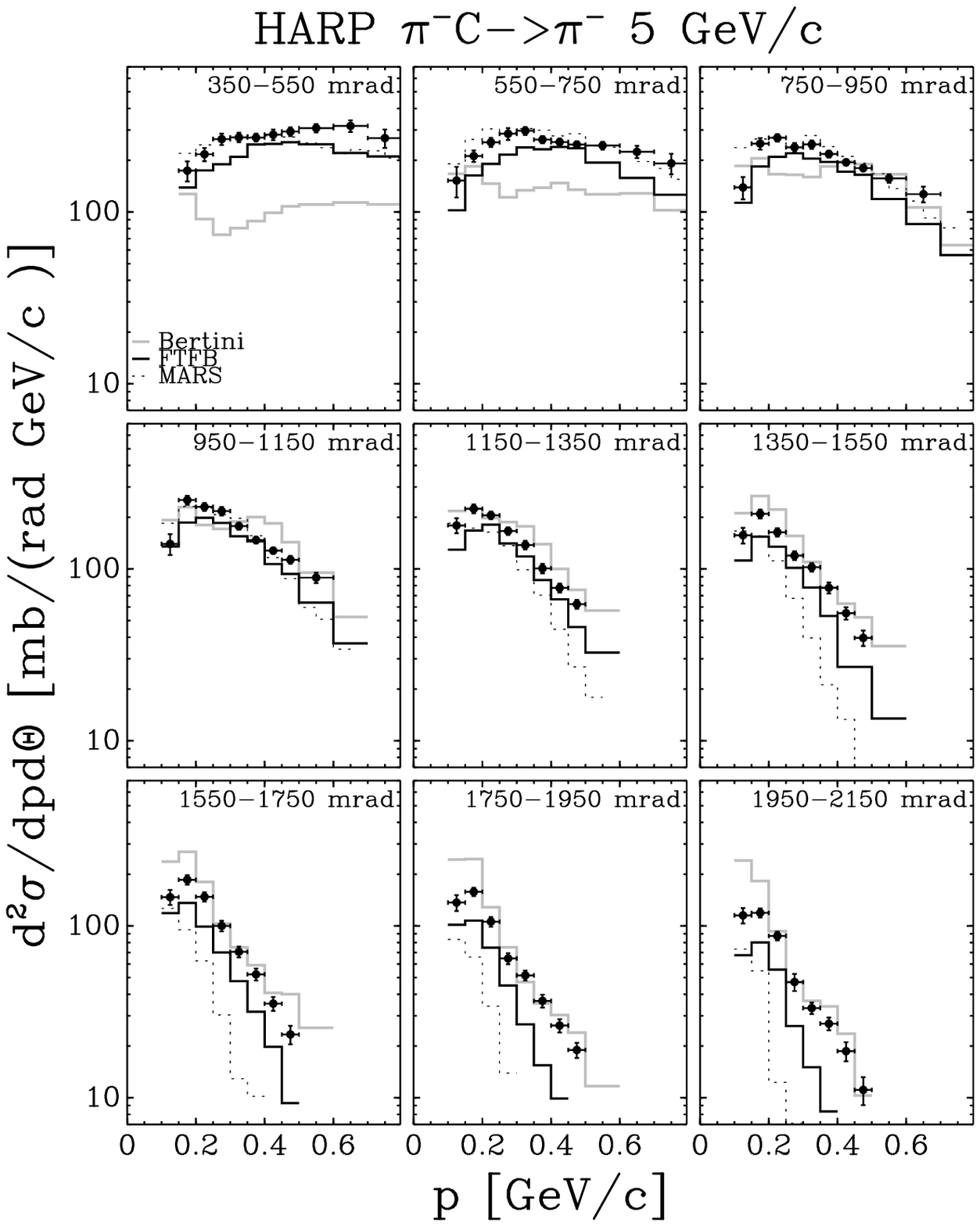}
 \includegraphics[width=0.49\textwidth]{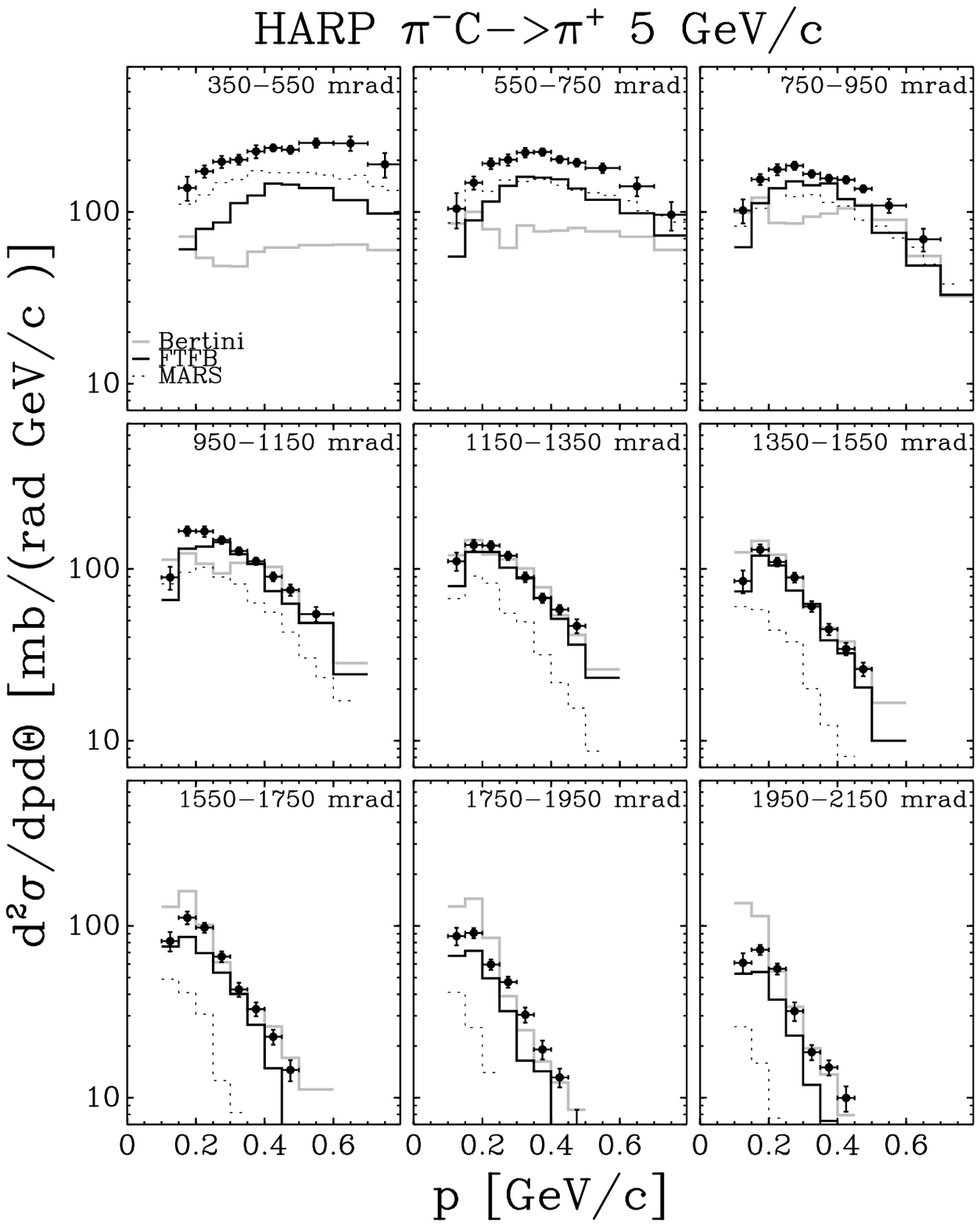}
\end{center}
\caption{
 Comparison of HARP double-differential \pipm production cross sections for \pim--C at 5~\GeVc with
 GEANT4 and MARS MC predictions, using several generator models (see text for details).
The left (right) panel shows \pim (\pip) production.
The gray line shows the Bertini model, the black solid line FTFB and
 the dashed line MARS.  
}
\label{fig:G4c5n}
\end{sidewaysfigure}

\begin{sidewaysfigure}[tbp!]
\begin{center}
 \includegraphics[width=0.49\textwidth]{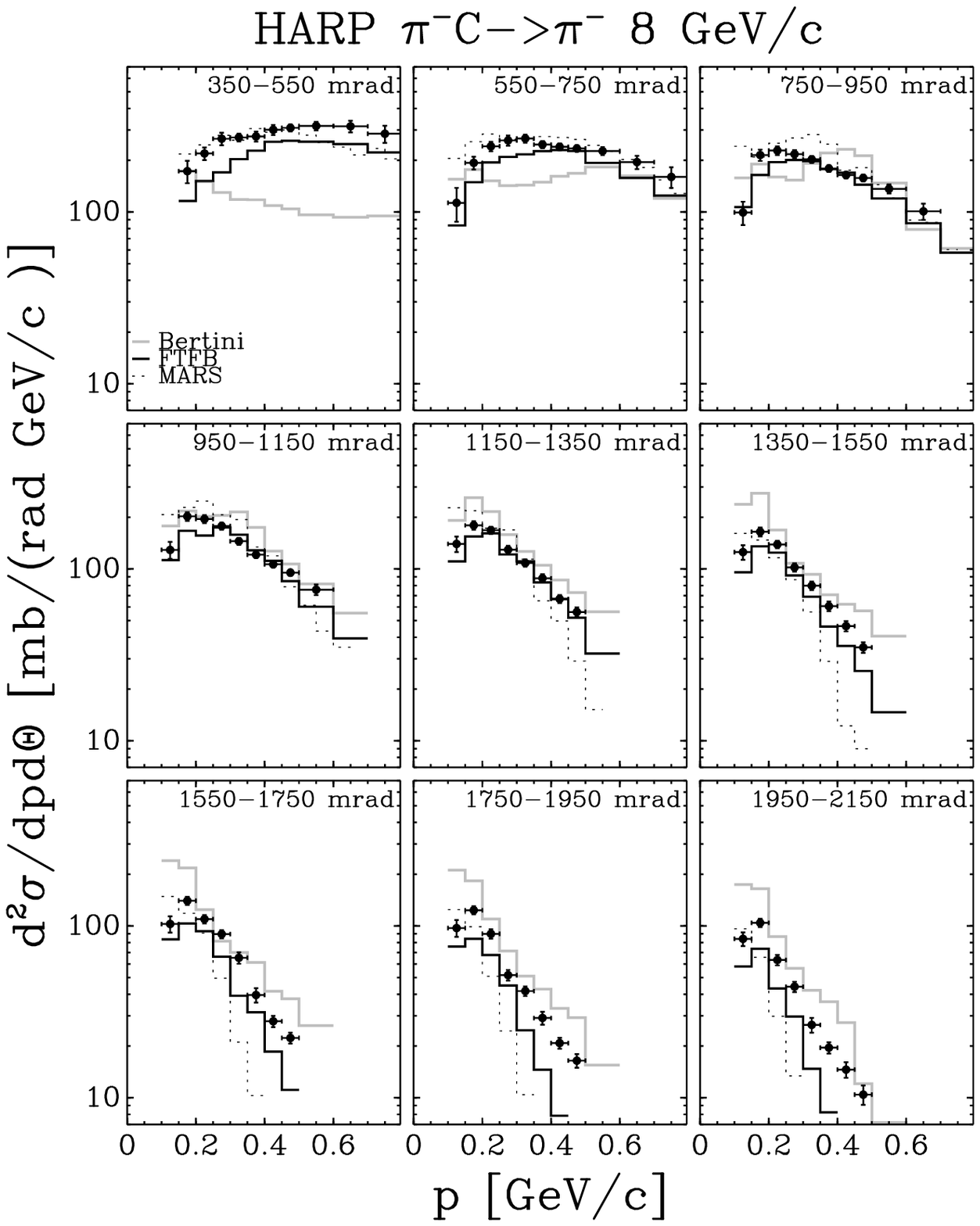}
 \includegraphics[width=0.49\textwidth]{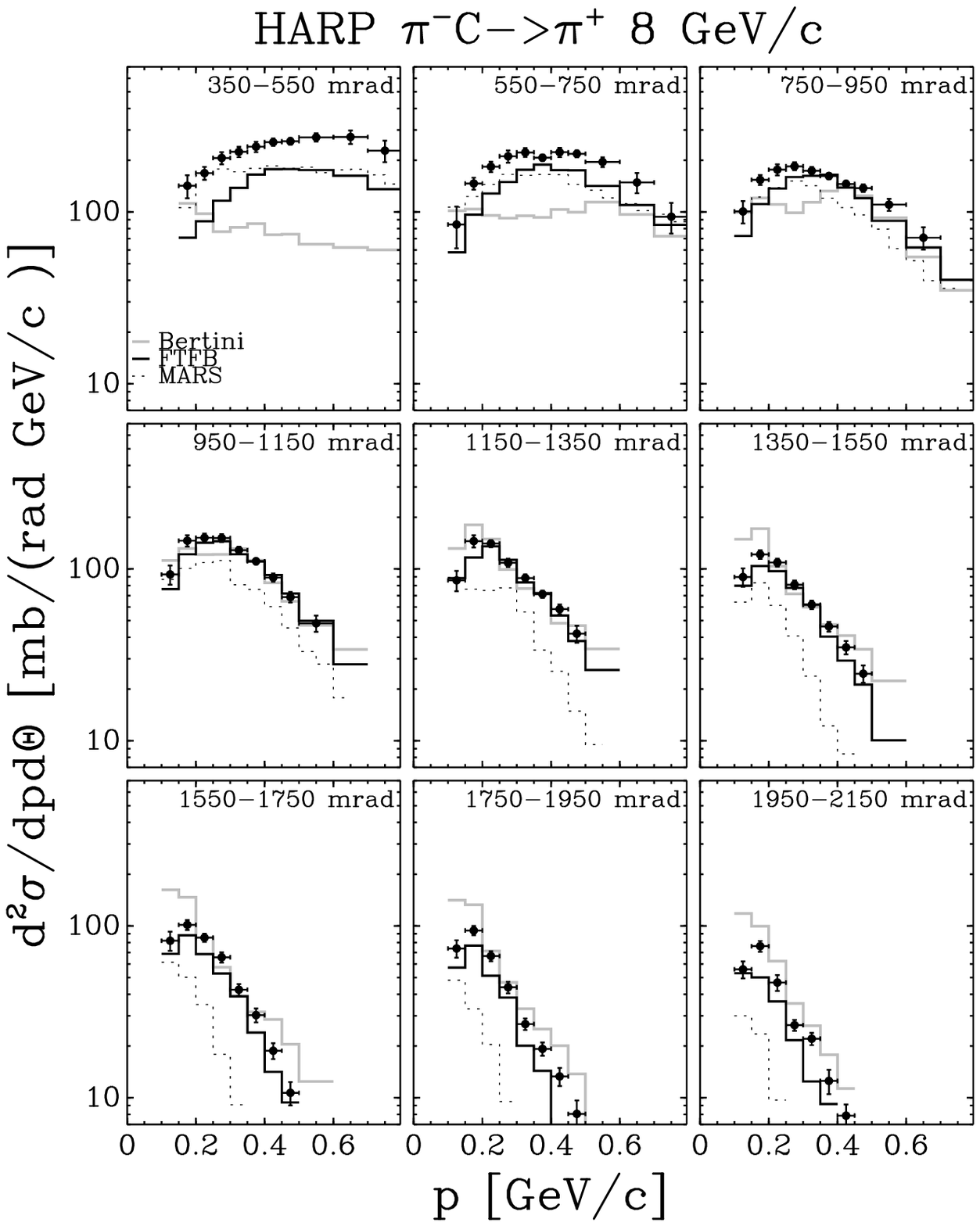}
\end{center}
\caption{
 Comparison of HARP double-differential \pipm production cross sections for \pim--C at 8~\GeVc with
 GEANT4 and MARS MC predictions, using several generator models (see text for details).
The left (right) panel shows \pim (\pip) production.
The gray line shows the Bertini model, the black solid line FTFB and
 the dashed line MARS.  
}
\label{fig:G4c8n}
\end{sidewaysfigure}

\begin{sidewaysfigure}[tbp!]
\begin{center}
 \includegraphics[width=0.49\textwidth]{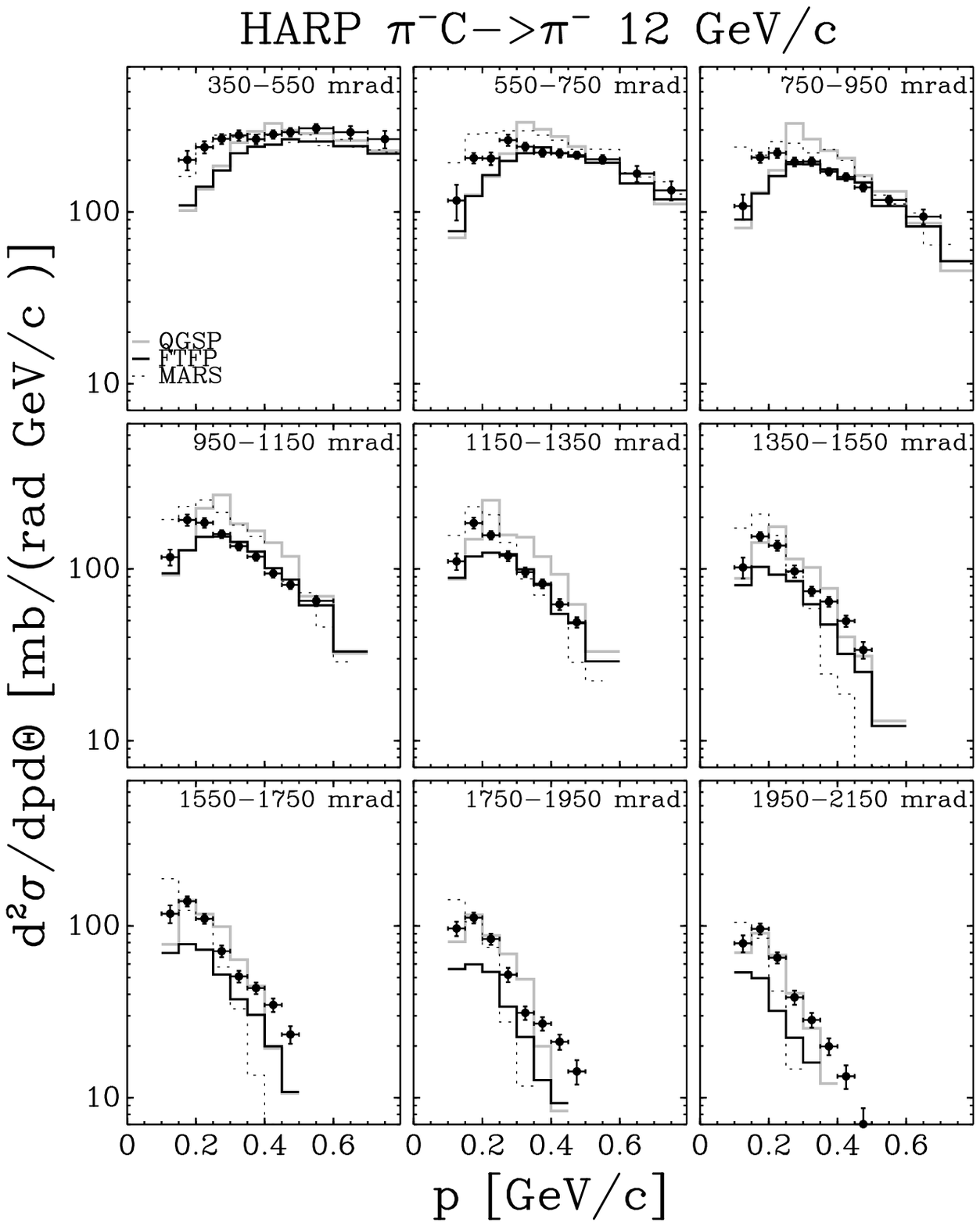}
 \includegraphics[width=0.49\textwidth]{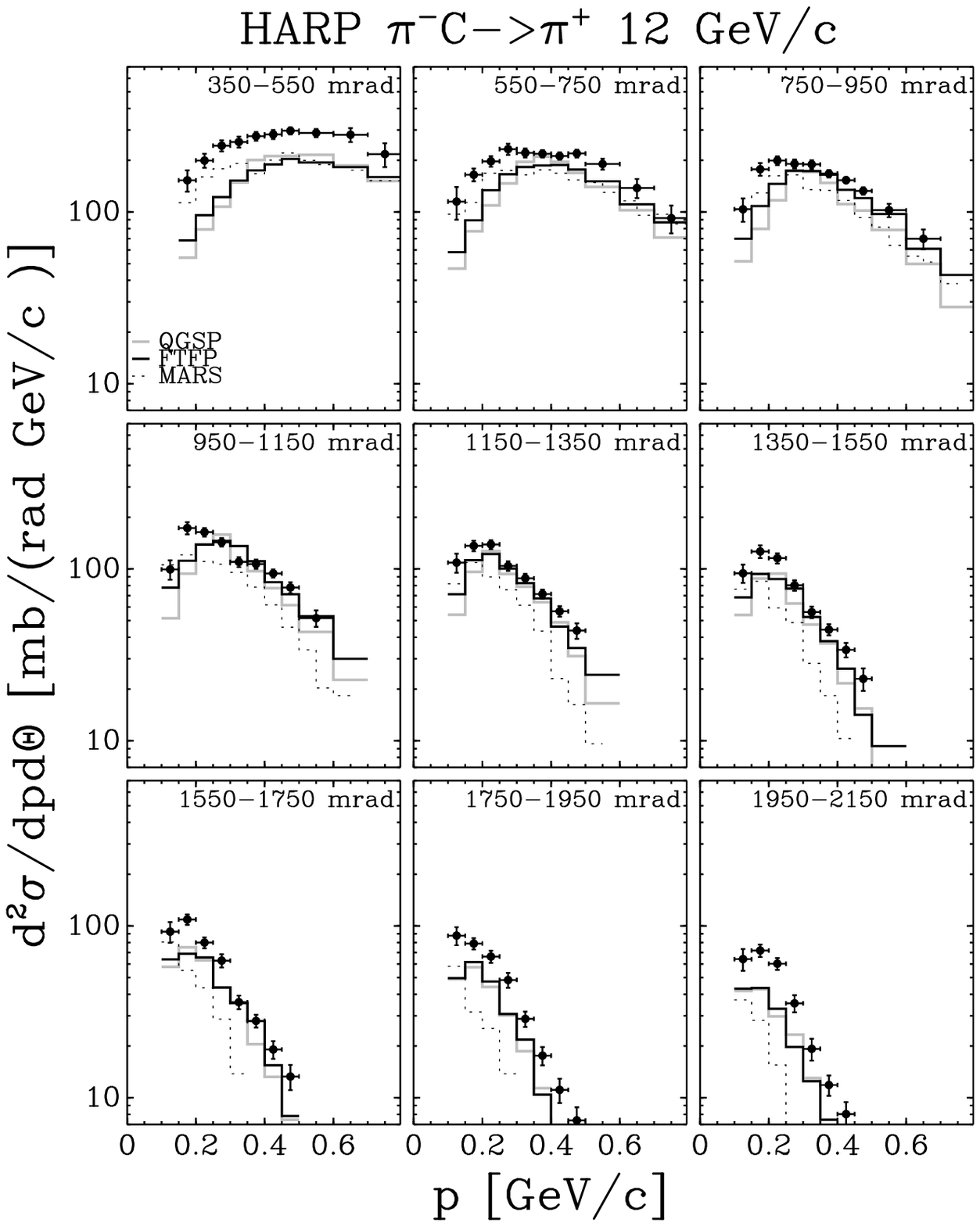}
\end{center}
\caption{
 Comparison of HARP double-differential \pipm production cross sections for \pim--C at 12~\GeVc with
 GEANT4 and MARS MC predictions, using several generator models (see text for details).
The left (right) panel shows \pim (\pip) production.
The gray line shows the Bertini model, the black solid line FTFP and
 the dashed line MARS.  
}
\label{fig:G4c12n}
\end{sidewaysfigure}

\begin{sidewaysfigure}[tbp!]
\begin{center}
 \includegraphics[width=0.49\textwidth]{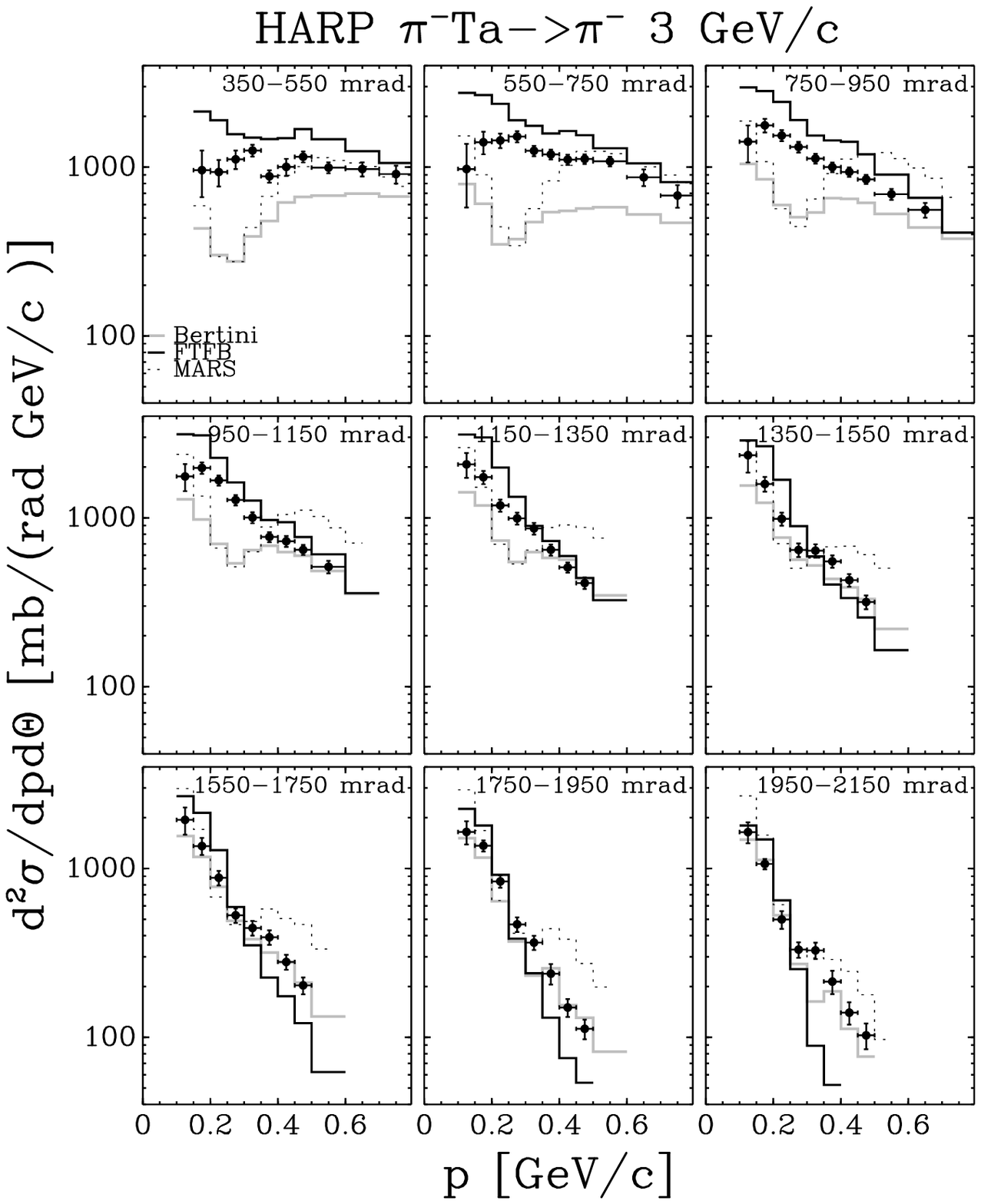}
 \includegraphics[width=0.49\textwidth]{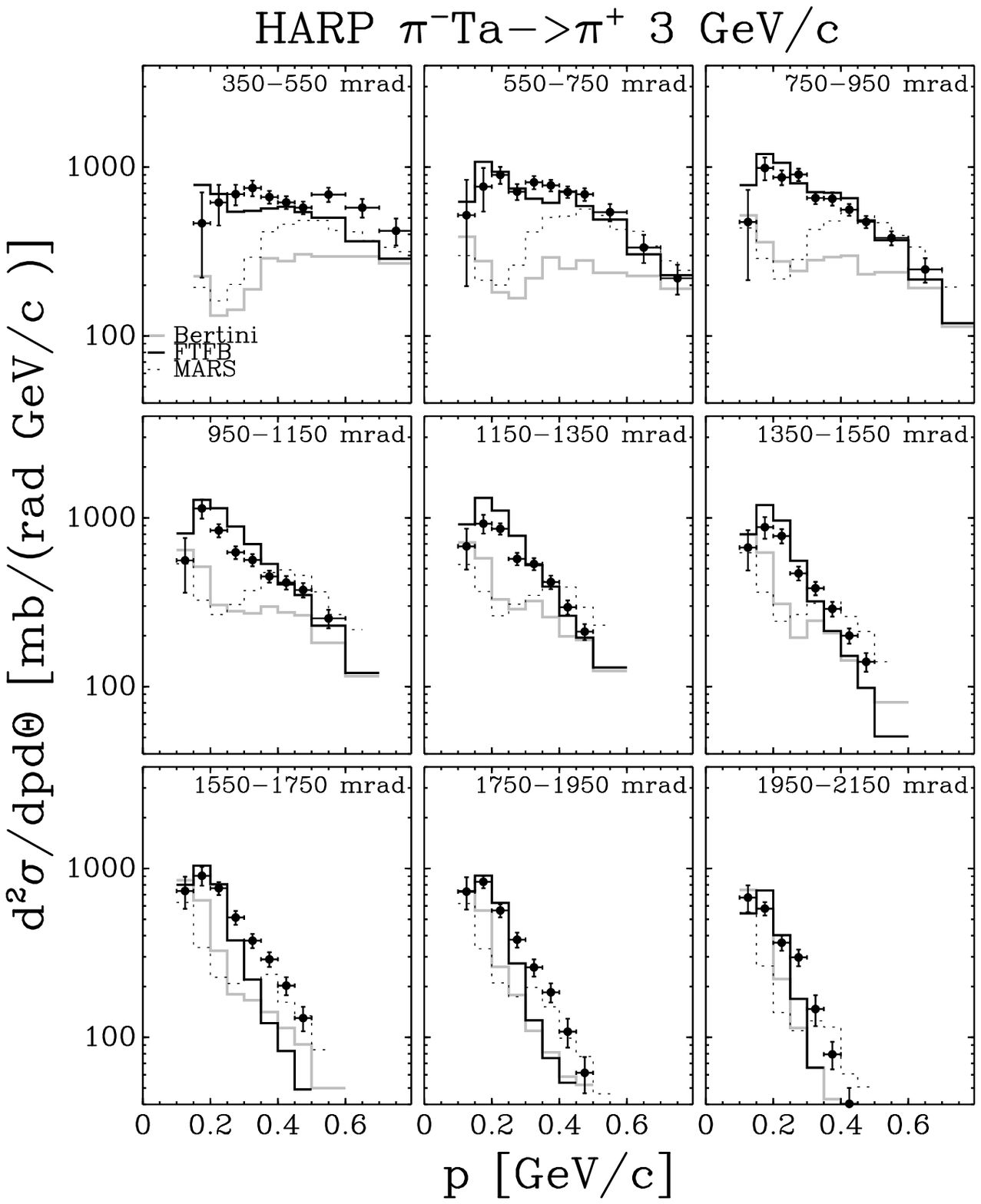}
\end{center}
\caption{
 Comparison of HARP double-differential \pipm production cross sections for \pim--Ta at 3~\GeVc with
 GEANT4 and MARS MC predictions, using several generator models (see text for details).
The left (right) panel shows \pim (\pip) production.
The gray line shows the Bertini model, the black solid line FTFB and
 the dashed line MARS.  
}
\label{fig:G4ta3n}
\end{sidewaysfigure}

\begin{sidewaysfigure}[tbp!]
\begin{center}
 \includegraphics[width=0.49\textwidth]{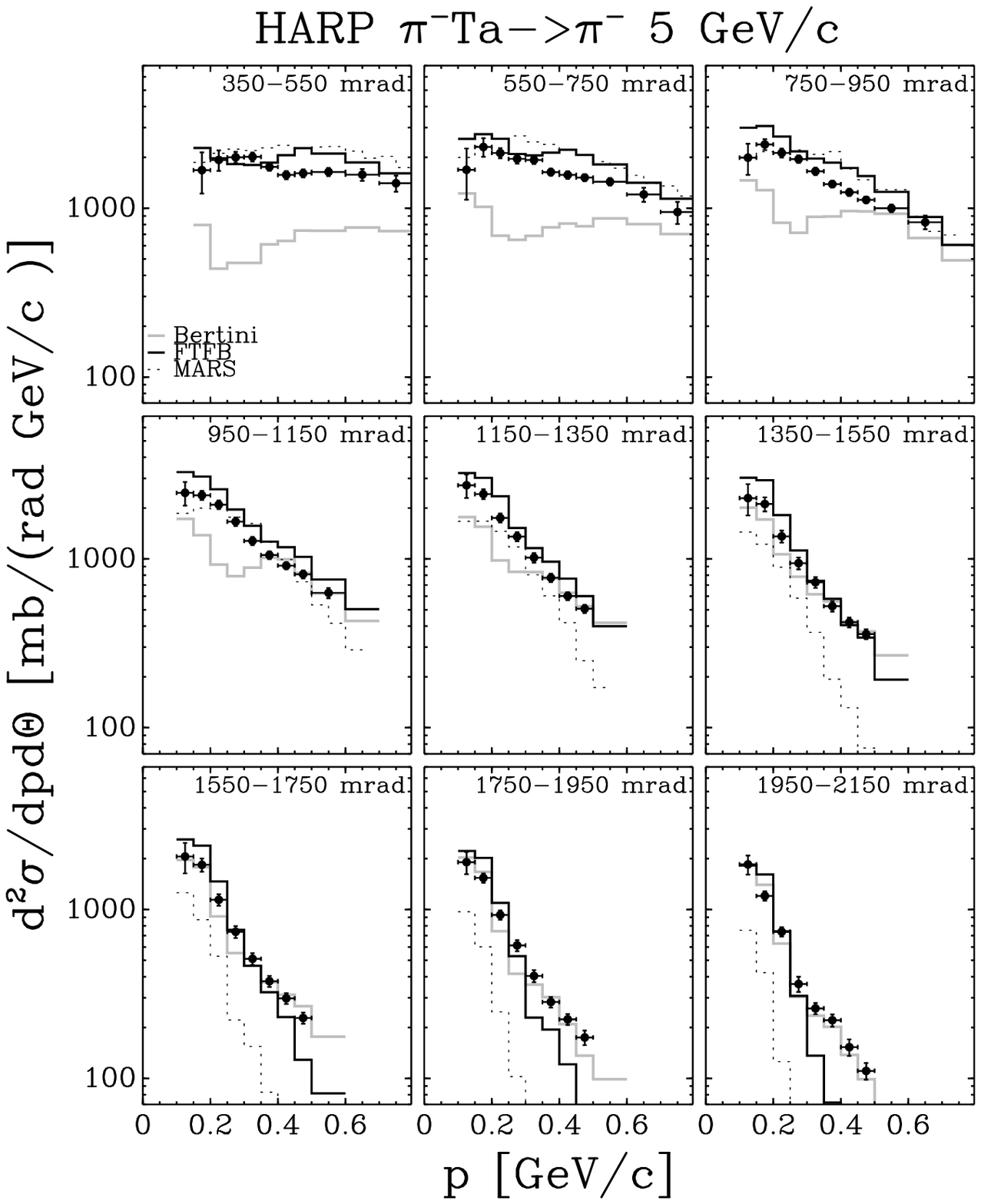}
 \includegraphics[width=0.49\textwidth]{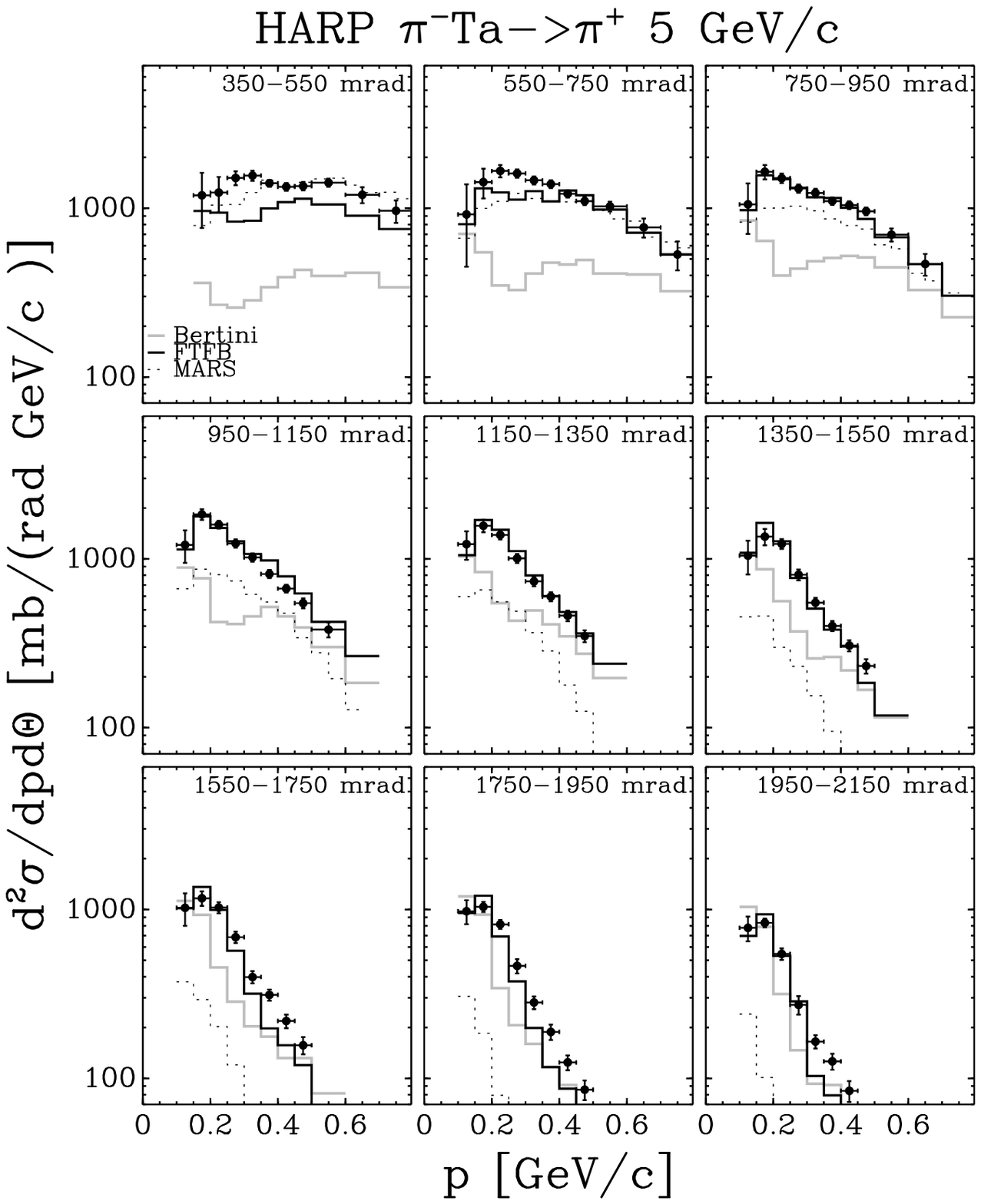}
\end{center}
\caption{
 Comparison of HARP double-differential \pipm production cross sections for \pim--Ta at 5~\GeVc with
 GEANT4 and MARS MC predictions, using several generator models (see text for details).
The left (right) panel shows \pim (\pip) production.
The gray line shows the Bertini model, the black solid line FTFB and
 the dashed line MARS.  
}
\label{fig:G4ta5n}
\end{sidewaysfigure}

\begin{sidewaysfigure}[tbp!]
\begin{center}
 \includegraphics[width=0.49\textwidth]{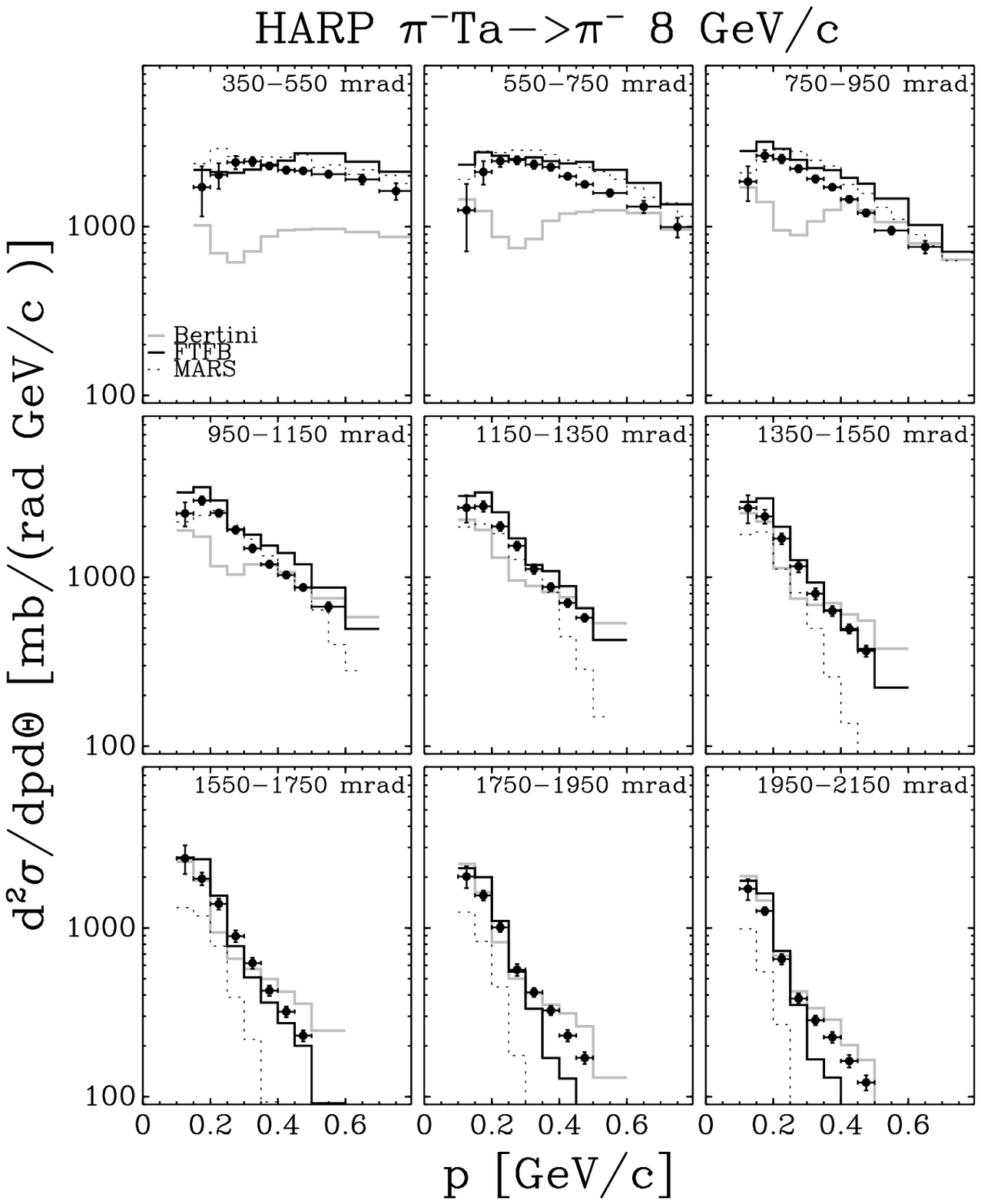}
 \includegraphics[width=0.49\textwidth]{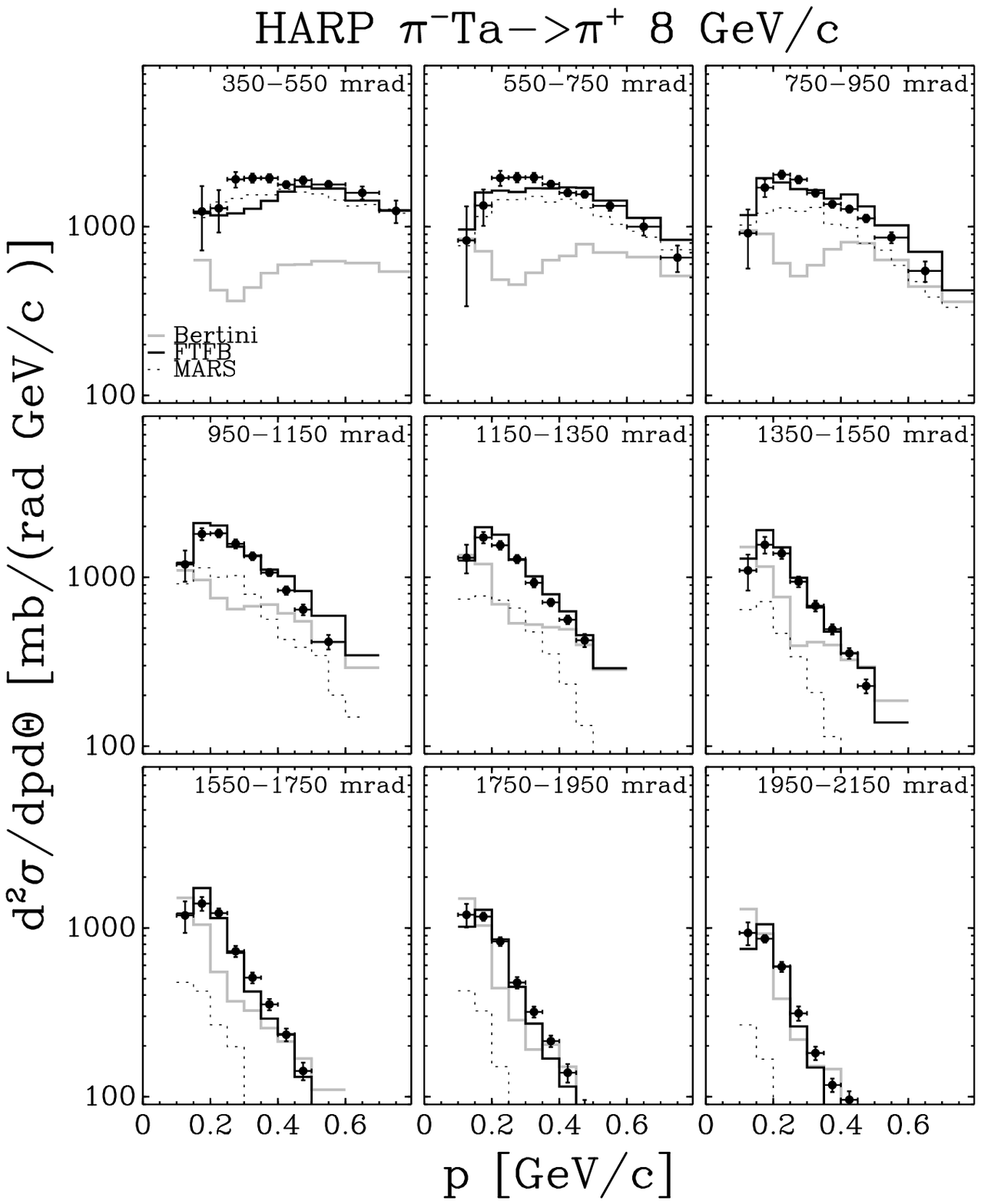}
\end{center}
\caption{
 Comparison of HARP double-differential \pipm production cross sections for \pim--Ta at 8~\GeVc with
 GEANT4 and MARS MC predictions, using several generator models (see text for details).
The left (right) panel shows \pim (\pip) production.
The gray line shows the Bertini model, the black solid line FTFB and
 the dashed line MARS.  
}
\label{fig:G4ta8n}
\end{sidewaysfigure}

\begin{sidewaysfigure}[tbp!]
\begin{center}
 \includegraphics[width=0.49\textwidth]{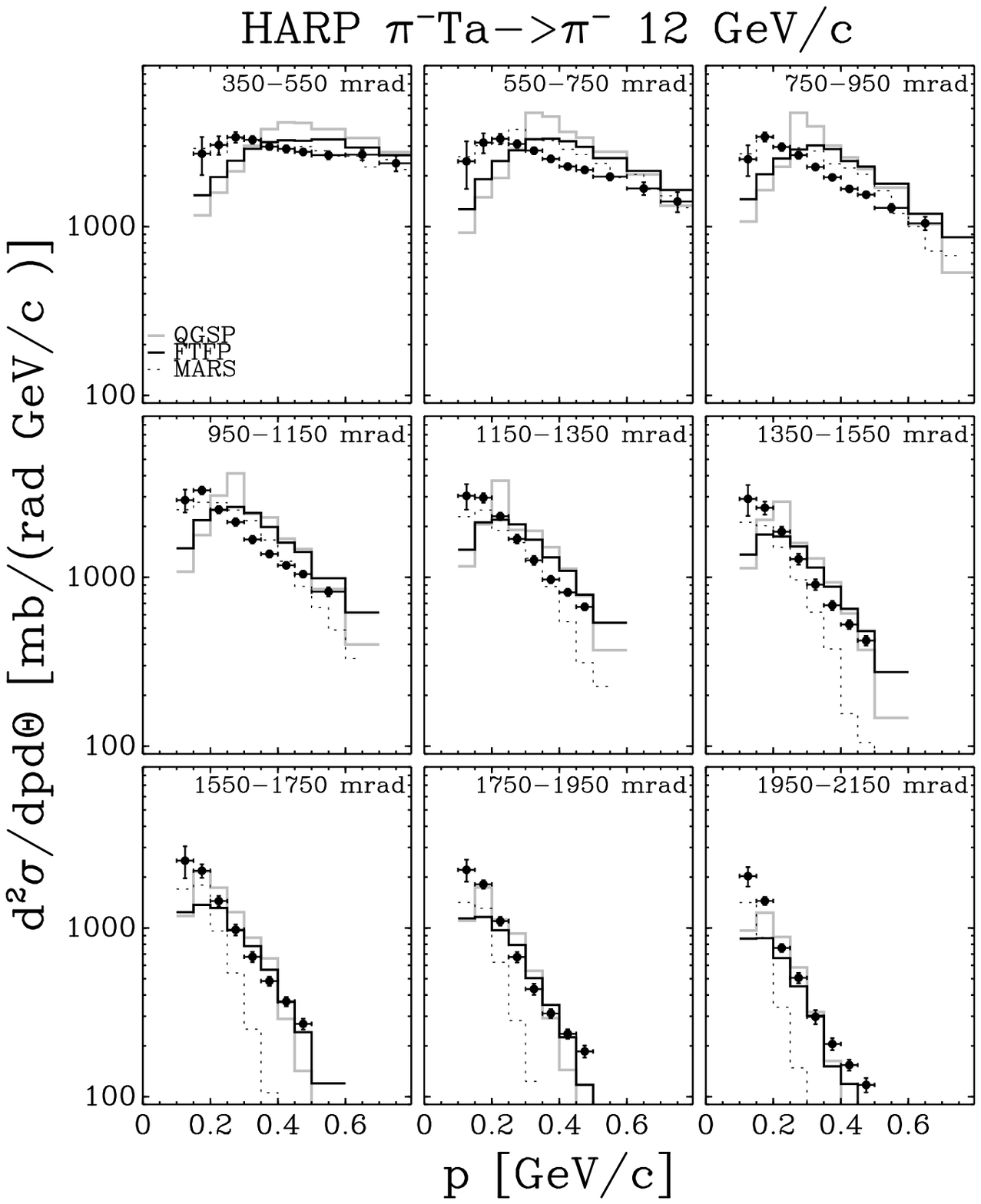}
 \includegraphics[width=0.49\textwidth]{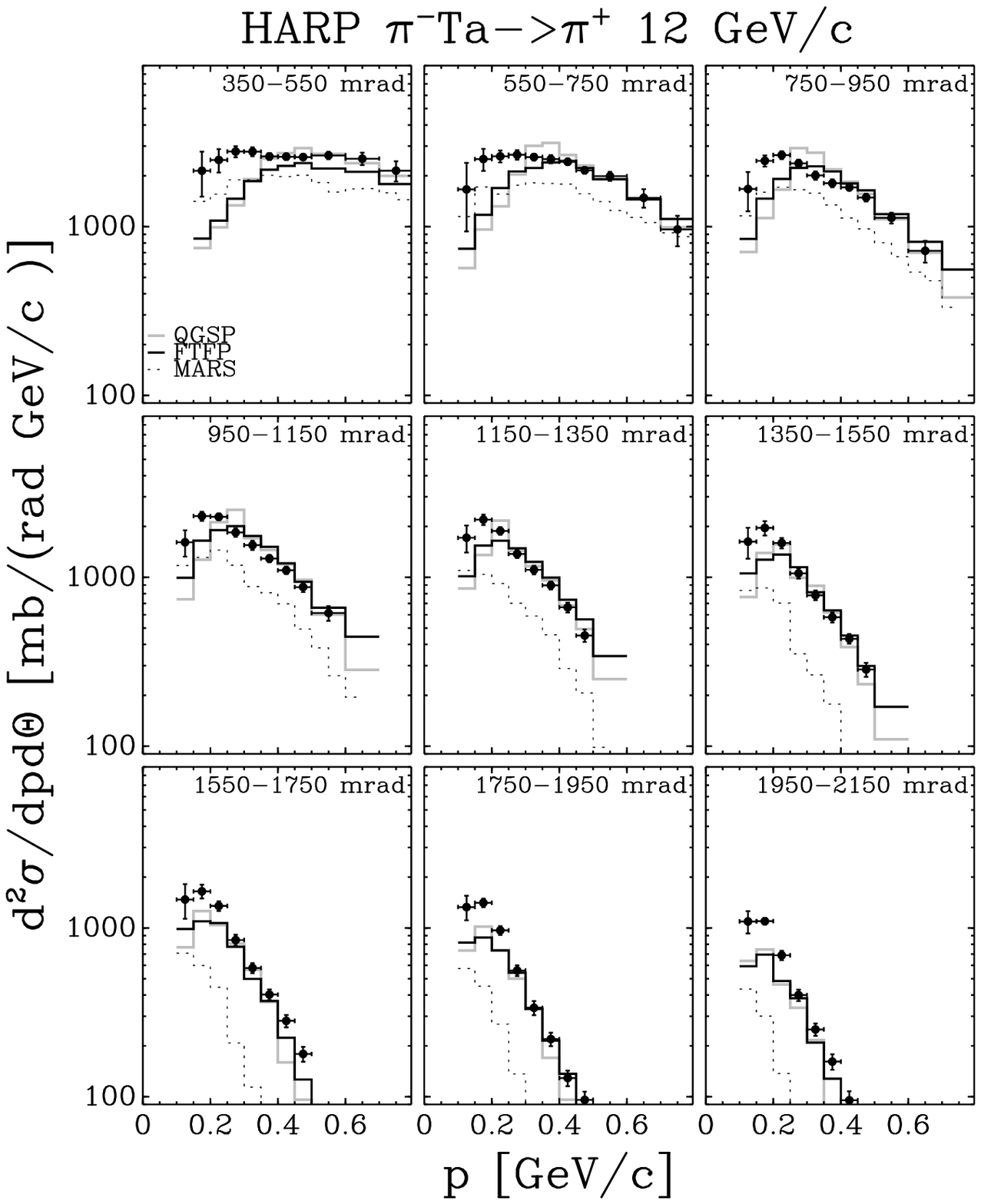}
\end{center}
\caption{
 Comparison of HARP double-differential \pipm production cross sections for \pim--Ta at 12~\GeVc with
 GEANT4 and MARS MC predictions, using several generator models (see text for details).
The left (right) panel shows \pim (\pip) production.
The gray line shows the Bertini model, the black solid line FTFP and
 the dashed line MARS.  
}
\label{fig:G4ta12n}
\end{sidewaysfigure}

In Figures~\ref{fig:G4c3}--\ref{fig:G4ta12n}, comparisons with 
a selected set of Monte Carlo generators of GEANT4~\cite{ref:geant4} 
and MARS~\cite{ref:mars} codes are shown.
We stress that no tuning to our data has been done by the 
GEANT4 or MARS teams. 
The comparisons are shown for the C and Ta targets as examples of a
light and a heavy target. 
 
At intermediate energies (up to 5~\GeVc--10 \GeVc), 
GEANT4 uses two types of intranuclear cascade models: the Bertini 
model~\cite{ref:bert,ref:bert1} (valid up to $\sim 10$ GeV) and the Binary
model~\cite{ref:bin} (valid up to $\sim 3$ GeV). Both models treat the target
nucleus in detail, taking into account density variations and tracking in the
nuclear field. 
The Binary model is based on hadron collisions with nucleons, giving 
resonances that decay according to their quantum numbers. The Bertini
model is based on the cascade code reported in \cite{ref:bert2}
and hadron collisions are assumed to proceed according to free-space partial
cross sections corrected for nuclear field effects and final-state
distributions measured for the incident-particle types.
For the figures the Bertini model was chosen as representative of the
cascade models since it was found to perform better than the Binary
model. 
 
At higher energies, instead, two parton string models, 
the quark-gluon string (QGS)  model~\cite{ref:bert,ref:QGSP} and the Fritiof
(FTF) model~\cite{ref:FTF} are used, in addition to a high-energy 
parametrized model (HEP)
derived from the high-energy part of the GHEISHA code used inside 
GEANT3~\cite{ref:gheisha}.
The parametrized models of GEANT4 (HEP and LEP) are intended to be fast,
but conserve energy and momentum on average and not event by 
event. 
In the figures a low-energy version of the FTF model (``FTFB'') is shown for the
data up to 8~\GeVc, and the higher energy version (``FTFP'') to compare
with the 12~\GeVc data.

The MARS code system~\cite{ref:mars} uses as basic model an inclusive
approach to multiparticle production.  Above 5~GeV
phenomenological particle production models are used: while below 5~GeV
a cascade-exciton model~\cite{ref:casca} combined with the Fermi
breakup model, the coalescence 
model, an evaporation model and a multifragmentation extension are used
instead.  

The comparison between data and models reveals sizable differences. 
Discrepancies up to a factor of three are seen.
One should note that the models have more difficulties with the incoming
pion data than the incoming proton data presented previously~\cite{ref:harp:la}.

Let us first examine the comparison for the carbon target.
The FTFB model provides a fair description of the \pip--C data.
There are some detailed differences, e.g. the absolute level of the
first angular bin in the \pim production in the \pip--C data at 3~\GeVc.
This model also provides a good description of the \pim production in
the \pim--C data, although it has difficulties with the backward
direction, especially for the low-energy data.
It has much more difficulty with the \pip production in
the \pim--C data.
The other models (MARS and Bertini) are both based on a cascade
description at this energy and predict features in the low-energy data
which are not visible in the data.
At 5~\GeVc the MARS prediction is very different from its 3~\GeVc
prediction. 
This is not justified by the data, which look very similar for all beam
energies. 
The Bertini model is not very successful for any of the $\pi$--C data
sets, while MARS gives a reasonable description in the mid-energy
range and for some of the highest energy data sets. 
Especially in the more forward bins the Bertini model underestimates the
production of pions.
At 12~\GeVc the FTFP and QGSP models give a better description for the
opposite-charge pion production, while MARS does better for same-charge
pions. 

In the tantalum data one observes again the structure of the cascade
based models at 3~\GeVc, which is not representative of the data.
The FTFB model does not have this feature and gives a better description
at this energy.
All models have considerable difficulties with the Ta data from 5~\GeVc
onward, with large differences among the models.
The \pim production in the \pim--Ta data is better described from
5~\GeVc onward by FTFB and MARS.

In general, the Bertini model produces a too isotropic pion production.
This may be due to the lack of an explicit diffractive process.
On the contrary, the Fritiof based models do contain these processes and
describe the forward production better.
At low energies MARS uses a cascade model and has similar problems as
the Bertini model.
One notes that with the Ta target for 5~\GeVc and 8~\GeVc, the
predictions for the \pip and \pim beams are identical.  
A similar problem was found with the Pb target for 3~\GeVc and 5~\GeVc. 
This appears to be a technical problem with the model code and the
authors have been informed~\footnote{We used GEANT4 9.2patch01.}.

The overall impression is that all investigated models need to be
considerably improved before they can give a reliable description.
Especially, the choice to change description between 3~\GeVc and 5~\GeVc
does not find support in the data.







\section{Summary and Conclusions}
\label{sec:summary}

An analysis of the production of pions at
large angles with respect to the beam direction for incoming charged pions of
3~\GeVc, 5~\GeVc, 8~\GeVc, 8.9~\GeVc (Be only), 12~\GeVc  and 
12.9~\GeVc (Al only) beam momentum impinging on  thin
(5\% interaction length) beryllium, carbon, aluminium, copper, tin,
tantalum and lead  targets is described.  
The secondary pion yield is measured in a large angular and momentum
range and double-differential cross-sections are obtained.
Results on the dependence of pion production
on the atomic number $A$ and incoming beam
momentum are also presented.
The comparisons of the \pim/\pip production ratios by pion beams of
opposite polarity and proton beams show interesting features as a
function of incoming beam momentum, target nucleus, and momentum of
the secondary pions.

The use of a single detector for a range of beam momenta makes it
possible to measure the dependence of the pion yield on the secondary
particle momentum and emission angle $\theta$ with high precision.
The $A$ dependence of the cross-section can be studied,
using data from a single experiment. 
Some hadronic production models (from GEANT4 and MARS) 
describing this energy range have been
compared with our new results. 
The description is not yet satisfactory for heavy nuclei, while in some
cases the carbon data are reasonably well reproduced.
These new HARP data together with the incoming proton results published
earlier~\cite{ref:harp:la} provide a unique possibility for validating
and tuning hadron production models. 

\section{Acknowledgements}

We gratefully acknowledge the help and support of the PS beam staff
and of the numerous technical collaborators who contributed to the
detector design, construction, commissioning and operation.  
In particular, we would like to thank
G.~Barichello,
R.~Brocard,
K.~Burin,
V.~Carassiti,
F.~Chignoli,
D.~Conventi,
G.~Decreuse,
M.~Delattre,
C.~Detraz,  
A.~Domeniconi,
M.~Dwuznik,   
F.~Evangelisti,
B.~Friend,
A.~Iaciofano,
I.~Krasin, 
D.~Lacroix,
J.-C.~Legrand,
M.~Lobello, 
M.~Lollo,
J.~Loquet,
F.~Marinilli,
J.~Mulon,
L.~Musa,
R.~Nicholson,
A.~Pepato,
P.~Petev, 
X.~Pons,
I.~Rusinov,
M.~Scandurra,
E.~Usenko,
and
R.~van der Vlugt,
for their support in the construction of the detector.
The collaboration acknowledges the major contributions and advice of
M.~Baldo-Ceolin, 
L.~Linssen, 
M.T.~Muciaccia and A. Pullia
during the construction of the experiment.
The collaboration is indebted to 
V.~Ableev,
P.~Arce,   
F.~Bergsma,
P.~Binko,
E.~Boter,
C.~Buttar, 
M.~Calvi, 
M.~Campanelli, 
C.~Cavion, 
A.~Chukanov, 
A.~De~Min,    
M.~Doucet,
D.~D\"{u}llmann,
R.~Engel,   
V.~Ermilova, 
W.~Flegel,
P.~Gruber,  
Y.~Hayato,
P.~Hodgson, 
A.~Ichikawa,
I.~Kato,  
O.~Klimov,
T.~Kobayashi,
D.~Kustov,
M.~Laveder,  
M.~Mass,
H.~Meinhard,
T.~Nakaya,
K.~Nishikawa,
M.~Paganoni,   
F.~Paleari,  
M.~Pasquali,
J.~Pasternak, 
C.~Pattison,  
M.~Placentino,
S.~Robbins,   
G.~Santin,  
V.~Serdiouk,
S.~Simone,
A.~Tornero,  
S.~Troquereau,
S.~Ueda, 
A.~Valassi,
F.~Vannucci  
and
K.~Zuber  
for their contributions to the experiment
and to P. Dini for help in MC production. 

We acknowledge the contributions of 
V.~Ammosov,
G.~Chelkov,
D.~Dedovich,
F.~Dydak,
M.~Gostkin,
A.~Guskov, 
D.~Khartchenko, 
V.~Koreshev,
Z.~Kroumchtein,
I.~Nefedov,
A.~Semak, 
J.~Wotschack,
V.~Zaets and
A.~Zhemchugov
to the work described in this paper.

 The experiment was made possible by grants from
the Institut Interuniversitaire des Sciences Nucl\'eair\-es and the
Interuniversitair Instituut voor Kernwetenschappen (Belgium), 
Ministerio de Educacion y Ciencia, Grant FPA2003-06921-c02-02 and
Generalitat Valenciana, grant GV00-054-1,
CERN (Geneva, Switzerland), 
the German Bundesministerium f\"ur Bildung und Forschung (Germany), 
the Istituto Na\-zio\-na\-le di Fisica Nucleare (Italy), 
INR RAS (Moscow), the Russian Foundation for Basic Research 
(grant 08-02-00018) 
and the Particle Physics and Astronomy Research Council (UK).
We gratefully acknowledge their support.
This work was supported in part by the Swiss National Science Foundation
and the Swiss Agency for Development and Cooperation in the framework of
the programme SCOPES - Scientific co-operation between Eastern Europe
and Switzerland.



\ifall
\clearpage

\begin{appendix}

\section{Cross-section data}\label{app:data}
\begin{table}[hp!]
\begin{center}
  \caption{\label{tab:xsec-pipp-be}
    HARP results for the double-differential $\pi^+$ production
    cross-section in the laboratory system,
    $d^2\sigma^{\pi^+}/(dpd\theta)$ for $\pi^+$--Be interactions. Each row refers to a
    different $(p_{\hbox{\small min}} \le p<p_{\hbox{\small max}},
    \theta_{\hbox{\small min}} \le \theta<\theta_{\hbox{\small max}})$ bin,
    where $p$ and $\theta$ are the pion momentum and polar angle, respectively.
    The central value as well as the square-root of the diagonal elements
    of the covariance matrix are given.}
\vspace{2mm}
\begin{tabular}{rrrr|r@{$\pm$}lr@{$\pm$}lr@{$\pm$}lr@{$\pm$}lr@{$\pm$}l}
\hline
$\theta_{\hbox{\small min}}$ &
$\theta_{\hbox{\small max}}$ &
$p_{\hbox{\small min}}$ &
$p_{\hbox{\small max}}$ &
\multicolumn{10}{c}{$d^2\sigma^{\pi^+}/(dpd\theta)$}
\\
(rad) & (rad) & (\GeVc) & (\GeVc) &
\multicolumn{10}{c}{($\barn/(\GeVc \cdot \rad)$)}
\\
  &  &  &
&\multicolumn{2}{c}{$ \bf{3 \ \GeVc}$}
&\multicolumn{2}{c}{$ \bf{5 \ \GeVc}$}
&\multicolumn{2}{c}{$ \bf{8 \ \GeVc}$}
&\multicolumn{2}{c}{$ \bf{8.9 \ \GeVc}$}
&\multicolumn{2}{c}{$ \bf{12 \ \GeVc}$}
\\
\hline  
 0.35 & 0.55 & 0.15 & 0.20& 0.090 &  0.013& 0.105 &  0.017& 0.124 &  0.020& 0.136 &  0.018& 0.101 &  0.025\\ 
      &      & 0.20 & 0.25& 0.140 &  0.014& 0.163 &  0.017& 0.191 &  0.020& 0.186 &  0.015& 0.168 &  0.022\\ 
      &      & 0.25 & 0.30& 0.165 &  0.014& 0.202 &  0.017& 0.259 &  0.022& 0.240 &  0.018& 0.213 &  0.025\\ 
      &      & 0.30 & 0.35& 0.199 &  0.015& 0.226 &  0.019& 0.257 &  0.021& 0.259 &  0.021& 0.244 &  0.024\\ 
      &      & 0.35 & 0.40& 0.212 &  0.020& 0.249 &  0.019& 0.290 &  0.024& 0.281 &  0.017& 0.287 &  0.032\\ 
      &      & 0.40 & 0.45& 0.240 &  0.015& 0.285 &  0.019& 0.290 &  0.019& 0.298 &  0.013& 0.309 &  0.022\\ 
      &      & 0.45 & 0.50& 0.258 &  0.015& 0.292 &  0.015& 0.338 &  0.024& 0.315 &  0.017& 0.314 &  0.021\\ 
      &      & 0.50 & 0.60& 0.257 &  0.016& 0.295 &  0.015& 0.354 &  0.021& 0.305 &  0.017& 0.303 &  0.021\\ 
      &      & 0.60 & 0.70& 0.244 &  0.021& 0.310 &  0.026& 0.349 &  0.033& 0.262 &  0.026& 0.299 &  0.030\\ 
      &      & 0.70 & 0.80& 0.211 &  0.033& 0.262 &  0.037& 0.300 &  0.044& 0.191 &  0.030& 0.220 &  0.041\\ 
\hline  
 0.55 & 0.75 & 0.10 & 0.15& 0.104 &  0.019& 0.080 &  0.018& 0.107 &  0.023& 0.085 &  0.017& 0.104 &  0.031\\ 
      &      & 0.15 & 0.20& 0.136 &  0.011& 0.143 &  0.014& 0.124 &  0.012& 0.143 &  0.011& 0.152 &  0.018\\ 
      &      & 0.20 & 0.25& 0.193 &  0.016& 0.205 &  0.018& 0.195 &  0.018& 0.212 &  0.015& 0.218 &  0.023\\ 
      &      & 0.25 & 0.30& 0.224 &  0.016& 0.246 &  0.018& 0.241 &  0.021& 0.238 &  0.016& 0.193 &  0.017\\ 
      &      & 0.30 & 0.35& 0.218 &  0.016& 0.241 &  0.018& 0.245 &  0.018& 0.249 &  0.013& 0.226 &  0.026\\ 
      &      & 0.35 & 0.40& 0.240 &  0.016& 0.268 &  0.015& 0.243 &  0.013& 0.241 &  0.011& 0.265 &  0.021\\ 
      &      & 0.40 & 0.45& 0.250 &  0.013& 0.268 &  0.014& 0.241 &  0.014& 0.228 &  0.009& 0.236 &  0.017\\ 
      &      & 0.45 & 0.50& 0.231 &  0.012& 0.259 &  0.012& 0.238 &  0.014& 0.221 &  0.009& 0.219 &  0.016\\ 
      &      & 0.50 & 0.60& 0.204 &  0.014& 0.237 &  0.016& 0.223 &  0.017& 0.187 &  0.014& 0.194 &  0.017\\ 
      &      & 0.60 & 0.70& 0.185 &  0.021& 0.172 &  0.022& 0.165 &  0.021& 0.135 &  0.018& 0.144 &  0.020\\ 
      &      & 0.70 & 0.80& 0.131 &  0.022& 0.115 &  0.022& 0.117 &  0.023& 0.084 &  0.016& 0.097 &  0.022\\ 
\hline  
 0.75 & 0.95 & 0.10 & 0.15& 0.082 &  0.013& 0.085 &  0.015& 0.077 &  0.013& 0.093 &  0.014& 0.079 &  0.018\\ 
      &      & 0.15 & 0.20& 0.175 &  0.017& 0.179 &  0.016& 0.167 &  0.017& 0.168 &  0.012& 0.138 &  0.017\\ 
      &      & 0.20 & 0.25& 0.210 &  0.014& 0.215 &  0.014& 0.209 &  0.015& 0.202 &  0.013& 0.177 &  0.019\\ 
      &      & 0.25 & 0.30& 0.230 &  0.015& 0.217 &  0.013& 0.203 &  0.014& 0.224 &  0.012& 0.156 &  0.014\\ 
      &      & 0.30 & 0.35& 0.221 &  0.013& 0.199 &  0.013& 0.183 &  0.011& 0.198 &  0.009& 0.171 &  0.017\\ 
      &      & 0.35 & 0.40& 0.197 &  0.010& 0.189 &  0.010& 0.184 &  0.012& 0.173 &  0.006& 0.157 &  0.012\\ 
      &      & 0.40 & 0.45& 0.175 &  0.009& 0.179 &  0.008& 0.167 &  0.010& 0.147 &  0.006& 0.150 &  0.012\\ 
      &      & 0.45 & 0.50& 0.169 &  0.009& 0.172 &  0.009& 0.145 &  0.010& 0.129 &  0.006& 0.135 &  0.013\\ 
      &      & 0.50 & 0.60& 0.138 &  0.011& 0.128 &  0.013& 0.101 &  0.012& 0.096 &  0.009& 0.092 &  0.011\\ 
      &      & 0.60 & 0.70& 0.101 &  0.014& 0.076 &  0.014& 0.060 &  0.010& 0.058 &  0.010& 0.060 &  0.010\\ 
\hline  
 0.95 & 1.15 & 0.10 & 0.15& 0.099 &  0.013& 0.092 &  0.014& 0.084 &  0.013& 0.099 &  0.013& 0.070 &  0.014\\ 
      &      & 0.15 & 0.20& 0.185 &  0.017& 0.184 &  0.015& 0.166 &  0.016& 0.159 &  0.009& 0.162 &  0.019\\ 
      &      & 0.20 & 0.25& 0.205 &  0.011& 0.202 &  0.012& 0.167 &  0.011& 0.175 &  0.009& 0.161 &  0.016\\ 
      &      & 0.25 & 0.30& 0.165 &  0.011& 0.183 &  0.010& 0.162 &  0.011& 0.166 &  0.008& 0.135 &  0.012\\ 
      &      & 0.30 & 0.35& 0.155 &  0.009& 0.149 &  0.008& 0.113 &  0.008& 0.129 &  0.006& 0.125 &  0.012\\ 
      &      & 0.35 & 0.40& 0.143 &  0.008& 0.130 &  0.008& 0.101 &  0.007& 0.112 &  0.005& 0.126 &  0.012\\ 
      &      & 0.40 & 0.45& 0.130 &  0.006& 0.110 &  0.006& 0.086 &  0.007& 0.085 &  0.005& 0.099 &  0.010\\ 
      &      & 0.45 & 0.50& 0.113 &  0.006& 0.088 &  0.007& 0.068 &  0.006& 0.065 &  0.006& 0.072 &  0.009\\ 
      &      & 0.50 & 0.60& 0.078 &  0.009& 0.059 &  0.007& 0.049 &  0.006& 0.043 &  0.005& 0.048 &  0.008\\ 
\hline
\end{tabular}
\end{center}
\end{table}

\begin{table}[hp!]
\begin{center}
\begin{tabular}{rrrr|r@{$\pm$}lr@{$\pm$}lr@{$\pm$}lr@{$\pm$}lr@{$\pm$}l}
\hline
$\theta_{\hbox{\small min}}$ &
$\theta_{\hbox{\small max}}$ &
$p_{\hbox{\small min}}$ &
$p_{\hbox{\small max}}$ &
\multicolumn{10}{c}{$d^2\sigma^{\pi^+}/(dpd\theta)$}
\\
(rad) & (rad) & (\GeVc) & (\GeVc) &
\multicolumn{10}{c}{($\barn/(\GeVc \cdot \rad)$)}
\\
  &  &  &
&\multicolumn{2}{c}{$ \bf{3 \ \GeVc}$}
&\multicolumn{2}{c}{$ \bf{5 \ \GeVc}$}
&\multicolumn{2}{c}{$ \bf{8 \ \GeVc}$}
&\multicolumn{2}{c}{$ \bf{8.9 \ \GeVc}$}
&\multicolumn{2}{c}{$ \bf{12 \ \GeVc}$}
\\
\hline
 1.15 & 1.35 & 0.10 & 0.15& 0.105 &  0.012& 0.127 &  0.015& 0.093 &  0.012& 0.105 &  0.013& 0.076 &  0.015\\ 
      &      & 0.15 & 0.20& 0.180 &  0.018& 0.166 &  0.013& 0.151 &  0.015& 0.157 &  0.009& 0.144 &  0.019\\ 
      &      & 0.20 & 0.25& 0.169 &  0.011& 0.168 &  0.010& 0.143 &  0.010& 0.147 &  0.006& 0.126 &  0.013\\ 
      &      & 0.25 & 0.30& 0.135 &  0.008& 0.139 &  0.008& 0.116 &  0.009& 0.111 &  0.006& 0.112 &  0.012\\ 
      &      & 0.30 & 0.35& 0.115 &  0.007& 0.102 &  0.007& 0.069 &  0.007& 0.082 &  0.004& 0.069 &  0.009\\ 
      &      & 0.35 & 0.40& 0.087 &  0.006& 0.076 &  0.005& 0.057 &  0.004& 0.068 &  0.003& 0.048 &  0.006\\ 
      &      & 0.40 & 0.45& 0.075 &  0.005& 0.069 &  0.004& 0.050 &  0.004& 0.049 &  0.004& 0.043 &  0.005\\ 
      &      & 0.45 & 0.50& 0.062 &  0.005& 0.052 &  0.005& 0.038 &  0.005& 0.033 &  0.004& 0.035 &  0.006\\ 
\hline  
 1.35 & 1.55 & 0.10 & 0.15& 0.118 &  0.013& 0.102 &  0.014& 0.094 &  0.013& 0.106 &  0.011& 0.083 &  0.015\\ 
      &      & 0.15 & 0.20& 0.162 &  0.013& 0.146 &  0.010& 0.134 &  0.012& 0.140 &  0.008& 0.119 &  0.015\\ 
      &      & 0.20 & 0.25& 0.138 &  0.010& 0.132 &  0.008& 0.104 &  0.008& 0.114 &  0.006& 0.065 &  0.009\\ 
      &      & 0.25 & 0.30& 0.099 &  0.007& 0.092 &  0.006& 0.082 &  0.008& 0.085 &  0.005& 0.070 &  0.009\\ 
      &      & 0.30 & 0.35& 0.076 &  0.005& 0.069 &  0.005& 0.057 &  0.005& 0.056 &  0.003& 0.051 &  0.007\\ 
      &      & 0.35 & 0.40& 0.062 &  0.004& 0.049 &  0.004& 0.044 &  0.004& 0.041 &  0.003& 0.043 &  0.006\\ 
      &      & 0.40 & 0.45& 0.045 &  0.004& 0.033 &  0.004& 0.030 &  0.004& 0.027 &  0.002& 0.034 &  0.006\\ 
      &      & 0.45 & 0.50& 0.033 &  0.003& 0.022 &  0.003& 0.019 &  0.003& 0.020 &  0.003& 0.020 &  0.005\\ 
\hline  
 1.55 & 1.75 & 0.10 & 0.15& 0.099 &  0.012& 0.090 &  0.011& 0.070 &  0.011& 0.090 &  0.011& 0.087 &  0.014\\ 
      &      & 0.15 & 0.20& 0.142 &  0.011& 0.131 &  0.010& 0.123 &  0.010& 0.120 &  0.006& 0.086 &  0.014\\ 
      &      & 0.20 & 0.25& 0.113 &  0.009& 0.110 &  0.008& 0.085 &  0.007& 0.089 &  0.005& 0.092 &  0.012\\ 
      &      & 0.25 & 0.30& 0.085 &  0.006& 0.072 &  0.005& 0.057 &  0.006& 0.059 &  0.003& 0.049 &  0.007\\ 
      &      & 0.30 & 0.35& 0.063 &  0.005& 0.054 &  0.004& 0.041 &  0.004& 0.039 &  0.003& 0.035 &  0.006\\ 
      &      & 0.35 & 0.40& 0.040 &  0.003& 0.032 &  0.004& 0.033 &  0.004& 0.024 &  0.002& 0.026 &  0.005\\ 
      &      & 0.40 & 0.45& 0.029 &  0.004& 0.021 &  0.003& 0.019 &  0.003& 0.015 &  0.002& 0.014 &  0.004\\ 
      &      & 0.45 & 0.50& 0.017 &  0.003& 0.014 &  0.002& 0.010 &  0.002& 0.010 &  0.001& 0.007 &  0.003\\ 
\hline  
 1.75 & 1.95 & 0.10 & 0.15& 0.095 &  0.013& 0.091 &  0.011& 0.081 &  0.011& 0.085 &  0.009& 0.063 &  0.013\\ 
      &      & 0.15 & 0.20& 0.123 &  0.009& 0.113 &  0.008& 0.088 &  0.008& 0.103 &  0.005& 0.083 &  0.012\\ 
      &      & 0.20 & 0.25& 0.078 &  0.008& 0.075 &  0.007& 0.061 &  0.006& 0.064 &  0.004& 0.068 &  0.010\\ 
      &      & 0.25 & 0.30& 0.054 &  0.005& 0.043 &  0.004& 0.037 &  0.005& 0.043 &  0.003& 0.032 &  0.008\\ 
      &      & 0.30 & 0.35& 0.044 &  0.004& 0.031 &  0.003& 0.034 &  0.004& 0.027 &  0.002& 0.014 &  0.003\\ 
      &      & 0.35 & 0.40& 0.025 &  0.003& 0.018 &  0.002& 0.025 &  0.004& 0.016 &  0.002& 0.014 &  0.004\\ 
      &      & 0.40 & 0.45& 0.015 &  0.002& 0.011 &  0.002& 0.011 &  0.003& 0.008 &  0.001& 0.008 &  0.003\\ 
      &      & 0.45 & 0.50& 0.009 &  0.002& 0.006 &  0.001& 0.005 &  0.002& 0.004 &  0.001& 0.004 &  0.002\\ 
\hline  
 1.95 & 2.15 & 0.10 & 0.15& 0.078 &  0.010& 0.072 &  0.009& 0.051 &  0.009& 0.071 &  0.008& 0.043 &  0.013\\ 
      &      & 0.15 & 0.20& 0.099 &  0.008& 0.088 &  0.006& 0.075 &  0.008& 0.073 &  0.004& 0.074 &  0.011\\ 
      &      & 0.20 & 0.25& 0.066 &  0.007& 0.056 &  0.005& 0.044 &  0.006& 0.049 &  0.003& 0.053 &  0.009\\ 
      &      & 0.25 & 0.30& 0.042 &  0.004& 0.034 &  0.003& 0.024 &  0.003& 0.029 &  0.002& 0.021 &  0.006\\ 
      &      & 0.30 & 0.35& 0.027 &  0.003& 0.020 &  0.002& 0.018 &  0.003& 0.017 &  0.002& 0.018 &  0.005\\ 
      &      & 0.35 & 0.40& 0.019 &  0.003& 0.012 &  0.002& 0.014 &  0.003& 0.009 &  0.001& 0.009 &  0.004\\ 
      &      & 0.40 & 0.45& 0.010 &  0.002& 0.008 &  0.001& 0.005 &  0.002& 0.005 &  0.001& 0.005 &  0.002\\ 
      &      & 0.45 & 0.50& 0.004 &  0.002& 0.005 &  0.001& 0.002 &  0.001& 0.003 &  0.001& 0.003 &  0.002\\ 

\end{tabular}
\end{center}
\end{table}

\begin{table}[hp!]
\begin{center}
  \caption{\label{tab:xsec-pipm-be}
    HARP results for the double-differential $\pi^-$ production
    cross-section in the laboratory system,
    $d^2\sigma^{\pi^-}/(dpd\theta)$ for $\pi^+$--Be interactions. Each row refers to a
    different $(p_{\hbox{\small min}} \le p<p_{\hbox{\small max}},
    \theta_{\hbox{\small min}} \le \theta<\theta_{\hbox{\small max}})$ bin,
    where $p$ and $\theta$ are the pion momentum and polar angle, respectively.
    The central value as well as the square-root of the diagonal elements
    of the covariance matrix are given.}
\vspace{2mm}
\begin{tabular}{rrrr|r@{$\pm$}lr@{$\pm$}lr@{$\pm$}lr@{$\pm$}lr@{$\pm$}l}
\hline
$\theta_{\hbox{\small min}}$ &
$\theta_{\hbox{\small max}}$ &
$p_{\hbox{\small min}}$ &
$p_{\hbox{\small max}}$ &
\multicolumn{10}{c}{$d^2\sigma^{\pi^-}/(dpd\theta)$}
\\
(rad) & (rad) & (\GeVc) & (\GeVc) &
\multicolumn{10}{c}{($\barn/(\GeVc \cdot \rad)$)}
\\
  &  &  &
&\multicolumn{2}{c}{$ \bf{3 \ \GeVc}$}
&\multicolumn{2}{c}{$ \bf{5 \ \GeVc}$}
&\multicolumn{2}{c}{$ \bf{8 \ \GeVc}$}
&\multicolumn{2}{c}{$ \bf{8.9 \ \GeVc}$}
&\multicolumn{2}{c}{$ \bf{12 \ \GeVc}$}
\\
\hline  
 0.35 & 0.55 & 0.15 & 0.20& 0.088 &  0.012& 0.106 &  0.017& 0.112 &  0.018& 0.121 &  0.017& 0.123 &  0.032\\ 
      &      & 0.20 & 0.25& 0.105 &  0.010& 0.133 &  0.013& 0.136 &  0.013& 0.147 &  0.011& 0.136 &  0.020\\ 
      &      & 0.25 & 0.30& 0.120 &  0.009& 0.152 &  0.011& 0.146 &  0.015& 0.173 &  0.012& 0.164 &  0.020\\ 
      &      & 0.30 & 0.35& 0.101 &  0.008& 0.152 &  0.012& 0.132 &  0.010& 0.177 &  0.010& 0.118 &  0.013\\ 
      &      & 0.35 & 0.40& 0.115 &  0.010& 0.153 &  0.009& 0.128 &  0.011& 0.170 &  0.008& 0.150 &  0.022\\ 
      &      & 0.40 & 0.45& 0.120 &  0.008& 0.150 &  0.009& 0.153 &  0.017& 0.172 &  0.010& 0.150 &  0.014\\ 
      &      & 0.45 & 0.50& 0.103 &  0.007& 0.148 &  0.008& 0.160 &  0.010& 0.172 &  0.008& 0.141 &  0.014\\ 
      &      & 0.50 & 0.60& 0.114 &  0.008& 0.157 &  0.010& 0.159 &  0.011& 0.168 &  0.009& 0.144 &  0.013\\ 
      &      & 0.60 & 0.70& 0.117 &  0.010& 0.152 &  0.012& 0.148 &  0.014& 0.159 &  0.013& 0.153 &  0.015\\ 
      &      & 0.70 & 0.80& 0.107 &  0.013& 0.124 &  0.017& 0.148 &  0.017& 0.134 &  0.015& 0.139 &  0.019\\ 
\hline  
 0.55 & 0.75 & 0.10 & 0.15& 0.081 &  0.018& 0.065 &  0.017& 0.082 &  0.020& 0.067 &  0.017& 0.077 &  0.022\\ 
      &      & 0.15 & 0.20& 0.113 &  0.008& 0.118 &  0.011& 0.108 &  0.011& 0.122 &  0.008& 0.076 &  0.013\\ 
      &      & 0.20 & 0.25& 0.114 &  0.010& 0.135 &  0.009& 0.148 &  0.014& 0.150 &  0.011& 0.162 &  0.026\\ 
      &      & 0.25 & 0.30& 0.128 &  0.010& 0.128 &  0.009& 0.133 &  0.010& 0.159 &  0.009& 0.159 &  0.016\\ 
      &      & 0.30 & 0.35& 0.120 &  0.008& 0.122 &  0.009& 0.141 &  0.011& 0.156 &  0.009& 0.120 &  0.012\\ 
      &      & 0.35 & 0.40& 0.113 &  0.008& 0.118 &  0.007& 0.124 &  0.008& 0.152 &  0.008& 0.145 &  0.016\\ 
      &      & 0.40 & 0.45& 0.112 &  0.007& 0.114 &  0.006& 0.120 &  0.009& 0.143 &  0.006& 0.127 &  0.011\\ 
      &      & 0.45 & 0.50& 0.107 &  0.006& 0.113 &  0.007& 0.116 &  0.008& 0.136 &  0.005& 0.127 &  0.012\\ 
      &      & 0.50 & 0.60& 0.103 &  0.007& 0.116 &  0.007& 0.112 &  0.008& 0.125 &  0.006& 0.113 &  0.010\\ 
      &      & 0.60 & 0.70& 0.094 &  0.010& 0.092 &  0.010& 0.103 &  0.010& 0.103 &  0.010& 0.097 &  0.014\\ 
      &      & 0.70 & 0.80& 0.070 &  0.011& 0.066 &  0.011& 0.081 &  0.014& 0.079 &  0.011& 0.073 &  0.012\\ 
\hline  
 0.75 & 0.95 & 0.10 & 0.15& 0.070 &  0.010& 0.065 &  0.012& 0.055 &  0.010& 0.070 &  0.010& 0.111 &  0.021\\ 
      &      & 0.15 & 0.20& 0.108 &  0.009& 0.121 &  0.010& 0.110 &  0.011& 0.133 &  0.009& 0.086 &  0.012\\ 
      &      & 0.20 & 0.25& 0.110 &  0.009& 0.110 &  0.007& 0.114 &  0.011& 0.144 &  0.008& 0.126 &  0.019\\ 
      &      & 0.25 & 0.30& 0.126 &  0.009& 0.108 &  0.007& 0.123 &  0.010& 0.140 &  0.008& 0.124 &  0.013\\ 
      &      & 0.30 & 0.35& 0.100 &  0.007& 0.104 &  0.007& 0.096 &  0.007& 0.128 &  0.005& 0.101 &  0.011\\ 
      &      & 0.35 & 0.40& 0.088 &  0.006& 0.094 &  0.005& 0.091 &  0.007& 0.113 &  0.004& 0.110 &  0.011\\ 
      &      & 0.40 & 0.45& 0.085 &  0.005& 0.087 &  0.005& 0.079 &  0.005& 0.104 &  0.004& 0.094 &  0.009\\ 
      &      & 0.45 & 0.50& 0.076 &  0.004& 0.080 &  0.004& 0.070 &  0.005& 0.095 &  0.003& 0.073 &  0.008\\ 
      &      & 0.50 & 0.60& 0.066 &  0.005& 0.069 &  0.005& 0.070 &  0.005& 0.081 &  0.005& 0.067 &  0.007\\ 
      &      & 0.60 & 0.70& 0.047 &  0.007& 0.052 &  0.006& 0.057 &  0.006& 0.061 &  0.007& 0.059 &  0.007\\ 
\hline  
 0.95 & 1.15 & 0.10 & 0.15& 0.058 &  0.009& 0.063 &  0.009& 0.072 &  0.011& 0.079 &  0.008& 0.050 &  0.011\\ 
      &      & 0.15 & 0.20& 0.099 &  0.008& 0.115 &  0.010& 0.102 &  0.009& 0.135 &  0.009& 0.087 &  0.013\\ 
      &      & 0.20 & 0.25& 0.089 &  0.007& 0.100 &  0.006& 0.095 &  0.008& 0.131 &  0.006& 0.126 &  0.015\\ 
      &      & 0.25 & 0.30& 0.087 &  0.007& 0.094 &  0.007& 0.101 &  0.008& 0.119 &  0.005& 0.091 &  0.010\\ 
      &      & 0.30 & 0.35& 0.089 &  0.006& 0.090 &  0.005& 0.085 &  0.006& 0.103 &  0.004& 0.059 &  0.007\\ 
      &      & 0.35 & 0.40& 0.070 &  0.005& 0.077 &  0.004& 0.070 &  0.005& 0.083 &  0.003& 0.070 &  0.009\\ 
      &      & 0.40 & 0.45& 0.055 &  0.004& 0.062 &  0.004& 0.052 &  0.004& 0.072 &  0.003& 0.066 &  0.007\\ 
      &      & 0.45 & 0.50& 0.047 &  0.003& 0.049 &  0.004& 0.045 &  0.003& 0.060 &  0.003& 0.055 &  0.006\\ 
      &      & 0.50 & 0.60& 0.041 &  0.003& 0.037 &  0.003& 0.039 &  0.003& 0.046 &  0.004& 0.039 &  0.006\\ 
\hline
\end{tabular}
\end{center}
\end{table}

\begin{table}[hp!]
\begin{center}
\begin{tabular}{rrrr|r@{$\pm$}lr@{$\pm$}lr@{$\pm$}lr@{$\pm$}lr@{$\pm$}l}
\hline
$\theta_{\hbox{\small min}}$ &
$\theta_{\hbox{\small max}}$ &
$p_{\hbox{\small min}}$ &
$p_{\hbox{\small max}}$ &
\multicolumn{10}{c}{$d^2\sigma^{\pi^-}/(dpd\theta)$}
\\
(rad) & (rad) & (\GeVc) & (\GeVc) &
\multicolumn{10}{c}{($\barn/(\GeVc \cdot \rad)$)}
\\
  &  &  &
&\multicolumn{2}{c}{$ \bf{3 \ \GeVc}$}
&\multicolumn{2}{c}{$ \bf{5 \ \GeVc}$}
&\multicolumn{2}{c}{$ \bf{8 \ \GeVc}$}
&\multicolumn{2}{c}{$ \bf{8.9 \ \GeVc}$}
&\multicolumn{2}{c}{$ \bf{12 \ \GeVc}$}
\\
\hline
 1.15 & 1.35 & 0.10 & 0.15& 0.066 &  0.008& 0.071 &  0.008& 0.059 &  0.008& 0.079 &  0.007& 0.043 &  0.010\\ 
      &      & 0.15 & 0.20& 0.080 &  0.007& 0.094 &  0.008& 0.088 &  0.010& 0.126 &  0.008& 0.085 &  0.012\\ 
      &      & 0.20 & 0.25& 0.082 &  0.008& 0.080 &  0.005& 0.083 &  0.007& 0.114 &  0.004& 0.086 &  0.011\\ 
      &      & 0.25 & 0.30& 0.068 &  0.006& 0.079 &  0.005& 0.073 &  0.007& 0.095 &  0.004& 0.063 &  0.009\\ 
      &      & 0.30 & 0.35& 0.056 &  0.004& 0.065 &  0.004& 0.059 &  0.007& 0.072 &  0.003& 0.056 &  0.008\\ 
      &      & 0.35 & 0.40& 0.046 &  0.004& 0.048 &  0.004& 0.035 &  0.004& 0.057 &  0.002& 0.060 &  0.008\\ 
      &      & 0.40 & 0.45& 0.037 &  0.003& 0.033 &  0.003& 0.028 &  0.003& 0.046 &  0.002& 0.036 &  0.006\\ 
      &      & 0.45 & 0.50& 0.028 &  0.003& 0.027 &  0.002& 0.025 &  0.003& 0.036 &  0.002& 0.028 &  0.004\\ 
\hline  
 1.35 & 1.55 & 0.10 & 0.15& 0.061 &  0.008& 0.065 &  0.009& 0.057 &  0.008& 0.085 &  0.009& 0.060 &  0.013\\ 
      &      & 0.15 & 0.20& 0.074 &  0.007& 0.087 &  0.007& 0.084 &  0.009& 0.114 &  0.006& 0.084 &  0.012\\ 
      &      & 0.20 & 0.25& 0.063 &  0.005& 0.071 &  0.005& 0.072 &  0.006& 0.099 &  0.004& 0.068 &  0.010\\ 
      &      & 0.25 & 0.30& 0.048 &  0.004& 0.047 &  0.003& 0.067 &  0.006& 0.073 &  0.004& 0.046 &  0.007\\ 
      &      & 0.30 & 0.35& 0.040 &  0.003& 0.047 &  0.004& 0.044 &  0.004& 0.053 &  0.003& 0.032 &  0.005\\ 
      &      & 0.35 & 0.40& 0.030 &  0.003& 0.035 &  0.003& 0.031 &  0.003& 0.039 &  0.002& 0.029 &  0.004\\ 
      &      & 0.40 & 0.45& 0.025 &  0.002& 0.026 &  0.002& 0.024 &  0.003& 0.030 &  0.002& 0.026 &  0.004\\ 
      &      & 0.45 & 0.50& 0.021 &  0.002& 0.020 &  0.002& 0.019 &  0.002& 0.022 &  0.002& 0.022 &  0.004\\ 
\hline  
 1.55 & 1.75 & 0.10 & 0.15& 0.049 &  0.007& 0.059 &  0.008& 0.054 &  0.009& 0.076 &  0.008& 0.062 &  0.012\\ 
      &      & 0.15 & 0.20& 0.066 &  0.007& 0.070 &  0.006& 0.060 &  0.006& 0.099 &  0.005& 0.052 &  0.009\\ 
      &      & 0.20 & 0.25& 0.048 &  0.005& 0.061 &  0.005& 0.064 &  0.006& 0.082 &  0.004& 0.055 &  0.009\\ 
      &      & 0.25 & 0.30& 0.040 &  0.004& 0.040 &  0.003& 0.043 &  0.006& 0.057 &  0.004& 0.040 &  0.006\\ 
      &      & 0.30 & 0.35& 0.032 &  0.003& 0.031 &  0.003& 0.026 &  0.003& 0.039 &  0.002& 0.035 &  0.006\\ 
      &      & 0.35 & 0.40& 0.023 &  0.002& 0.023 &  0.002& 0.021 &  0.003& 0.028 &  0.002& 0.025 &  0.004\\ 
      &      & 0.40 & 0.45& 0.018 &  0.002& 0.018 &  0.002& 0.018 &  0.002& 0.021 &  0.001& 0.019 &  0.004\\ 
      &      & 0.45 & 0.50& 0.013 &  0.002& 0.012 &  0.001& 0.014 &  0.002& 0.014 &  0.001& 0.012 &  0.003\\ 
\hline  
 1.75 & 1.95 & 0.10 & 0.15& 0.027 &  0.005& 0.043 &  0.005& 0.038 &  0.007& 0.066 &  0.006& 0.057 &  0.012\\ 
      &      & 0.15 & 0.20& 0.046 &  0.006& 0.056 &  0.005& 0.047 &  0.006& 0.089 &  0.005& 0.052 &  0.009\\ 
      &      & 0.20 & 0.25& 0.040 &  0.004& 0.046 &  0.004& 0.051 &  0.005& 0.063 &  0.003& 0.041 &  0.007\\ 
      &      & 0.25 & 0.30& 0.030 &  0.003& 0.032 &  0.003& 0.034 &  0.004& 0.041 &  0.003& 0.032 &  0.007\\ 
      &      & 0.30 & 0.35& 0.027 &  0.003& 0.020 &  0.002& 0.018 &  0.004& 0.025 &  0.002& 0.016 &  0.004\\ 
      &      & 0.35 & 0.40& 0.017 &  0.002& 0.016 &  0.002& 0.010 &  0.002& 0.018 &  0.001& 0.010 &  0.003\\ 
      &      & 0.40 & 0.45& 0.011 &  0.002& 0.012 &  0.001& 0.011 &  0.002& 0.013 &  0.001& 0.010 &  0.003\\ 
      &      & 0.45 & 0.50& 0.008 &  0.001& 0.009 &  0.001& 0.009 &  0.002& 0.009 &  0.001& 0.007 &  0.002\\ 
\hline  
 1.95 & 2.15 & 0.10 & 0.15& 0.022 &  0.005& 0.033 &  0.005& 0.033 &  0.005& 0.058 &  0.006& 0.039 &  0.009\\ 
      &      & 0.15 & 0.20& 0.032 &  0.005& 0.051 &  0.005& 0.046 &  0.006& 0.069 &  0.003& 0.057 &  0.010\\ 
      &      & 0.20 & 0.25& 0.027 &  0.004& 0.034 &  0.003& 0.031 &  0.004& 0.051 &  0.002& 0.036 &  0.007\\ 
      &      & 0.25 & 0.30& 0.025 &  0.003& 0.025 &  0.003& 0.020 &  0.003& 0.030 &  0.002& 0.019 &  0.005\\ 
      &      & 0.30 & 0.35& 0.014 &  0.003& 0.017 &  0.002& 0.014 &  0.002& 0.017 &  0.001& 0.014 &  0.004\\ 
      &      & 0.35 & 0.40& 0.007 &  0.001& 0.010 &  0.002& 0.012 &  0.002& 0.013 &  0.001& 0.010 &  0.003\\ 
      &      & 0.40 & 0.45& 0.006 &  0.001& 0.006 &  0.001& 0.009 &  0.002& 0.008 &  0.001& 0.009 &  0.003\\ 
      &      & 0.45 & 0.50& 0.004 &  0.001& 0.005 &  0.001& 0.004 &  0.001& 0.005 &  0.001& 0.008 &  0.003\\ 

\end{tabular}
\end{center}
\end{table}

\clearpage
\begin{table}[hp!]
\begin{center}
  \caption{\label{tab:xsec-pimp-be}
    HARP results for the double-differential $\pi^+$ production
    cross-section in the laboratory system,
    $d^2\sigma^{\pi^+}/(dpd\theta)$ for $\pi^-$--Be interactions. Each row refers to a
    different $(p_{\hbox{\small min}} \le p<p_{\hbox{\small max}},
    \theta_{\hbox{\small min}} \le \theta<\theta_{\hbox{\small max}})$ bin,
    where $p$ and $\theta$ are the pion momentum and polar angle, respectively.
    The central value as well as the square-root of the diagonal elements
    of the covariance matrix are given.}
\vspace{2mm}
\begin{tabular}{rrrr|r@{$\pm$}lr@{$\pm$}lr@{$\pm$}lr@{$\pm$}l}
\hline
$\theta_{\hbox{\small min}}$ &
$\theta_{\hbox{\small max}}$ &
$p_{\hbox{\small min}}$ &
$p_{\hbox{\small max}}$ &
\multicolumn{8}{c}{$d^2\sigma^{\pi^+}/(dpd\theta)$}
\\
(rad) & (rad) & (\GeVc) & (\GeVc) &
\multicolumn{8}{c}{($\barn/(\GeVc \cdot \rad)$)}
\\
  &  &  &
&\multicolumn{2}{c}{$ \bf{3 \ \GeVc}$}
&\multicolumn{2}{c}{$ \bf{5 \ \GeVc}$}
&\multicolumn{2}{c}{$ \bf{8 \ \GeVc}$}
&\multicolumn{2}{c}{$ \bf{12 \ \GeVc}$}
\\
\hline  
 0.35 & 0.55 & 0.15 & 0.20& 0.081 &  0.012& 0.105 &  0.017& 0.094 &  0.016& 0.106 &  0.019\\ 
      &      & 0.20 & 0.25& 0.098 &  0.008& 0.125 &  0.011& 0.140 &  0.013& 0.142 &  0.013\\ 
      &      & 0.25 & 0.30& 0.111 &  0.008& 0.152 &  0.012& 0.166 &  0.013& 0.151 &  0.013\\ 
      &      & 0.30 & 0.35& 0.117 &  0.009& 0.148 &  0.010& 0.177 &  0.013& 0.195 &  0.018\\ 
      &      & 0.35 & 0.40& 0.124 &  0.007& 0.165 &  0.013& 0.193 &  0.013& 0.203 &  0.013\\ 
      &      & 0.40 & 0.45& 0.128 &  0.008& 0.179 &  0.012& 0.202 &  0.010& 0.212 &  0.013\\ 
      &      & 0.45 & 0.50& 0.140 &  0.009& 0.188 &  0.010& 0.208 &  0.014& 0.229 &  0.013\\ 
      &      & 0.50 & 0.60& 0.136 &  0.008& 0.195 &  0.011& 0.233 &  0.013& 0.245 &  0.014\\ 
      &      & 0.60 & 0.70& 0.139 &  0.013& 0.188 &  0.018& 0.232 &  0.021& 0.260 &  0.024\\ 
      &      & 0.70 & 0.80& 0.108 &  0.017& 0.154 &  0.024& 0.205 &  0.031& 0.218 &  0.034\\ 
\hline  
 0.55 & 0.75 & 0.10 & 0.15& 0.067 &  0.014& 0.067 &  0.015& 0.060 &  0.016& 0.061 &  0.014\\ 
      &      & 0.15 & 0.20& 0.094 &  0.008& 0.098 &  0.009& 0.099 &  0.009& 0.106 &  0.014\\ 
      &      & 0.20 & 0.25& 0.118 &  0.008& 0.134 &  0.012& 0.142 &  0.012& 0.156 &  0.011\\ 
      &      & 0.25 & 0.30& 0.134 &  0.011& 0.153 &  0.012& 0.168 &  0.011& 0.173 &  0.014\\ 
      &      & 0.30 & 0.35& 0.146 &  0.010& 0.161 &  0.012& 0.173 &  0.013& 0.175 &  0.012\\ 
      &      & 0.35 & 0.40& 0.138 &  0.007& 0.162 &  0.008& 0.176 &  0.012& 0.176 &  0.009\\ 
      &      & 0.40 & 0.45& 0.143 &  0.008& 0.160 &  0.008& 0.175 &  0.007& 0.173 &  0.010\\ 
      &      & 0.45 & 0.50& 0.138 &  0.006& 0.152 &  0.008& 0.158 &  0.007& 0.175 &  0.008\\ 
      &      & 0.50 & 0.60& 0.113 &  0.009& 0.130 &  0.009& 0.145 &  0.010& 0.167 &  0.011\\ 
      &      & 0.60 & 0.70& 0.082 &  0.011& 0.107 &  0.013& 0.116 &  0.015& 0.128 &  0.017\\ 
      &      & 0.70 & 0.80& 0.052 &  0.010& 0.066 &  0.014& 0.079 &  0.016& 0.084 &  0.017\\ 
\hline  
 0.75 & 0.95 & 0.10 & 0.15& 0.066 &  0.010& 0.063 &  0.013& 0.061 &  0.010& 0.060 &  0.011\\ 
      &      & 0.15 & 0.20& 0.112 &  0.009& 0.125 &  0.011& 0.120 &  0.010& 0.102 &  0.009\\ 
      &      & 0.20 & 0.25& 0.129 &  0.009& 0.136 &  0.008& 0.142 &  0.008& 0.139 &  0.011\\ 
      &      & 0.25 & 0.30& 0.126 &  0.007& 0.130 &  0.010& 0.145 &  0.012& 0.143 &  0.009\\ 
      &      & 0.30 & 0.35& 0.115 &  0.007& 0.124 &  0.007& 0.139 &  0.006& 0.136 &  0.007\\ 
      &      & 0.35 & 0.40& 0.111 &  0.006& 0.111 &  0.006& 0.116 &  0.006& 0.122 &  0.006\\ 
      &      & 0.40 & 0.45& 0.104 &  0.005& 0.098 &  0.005& 0.108 &  0.005& 0.113 &  0.006\\ 
      &      & 0.45 & 0.50& 0.085 &  0.005& 0.094 &  0.005& 0.102 &  0.005& 0.101 &  0.006\\ 
      &      & 0.50 & 0.60& 0.064 &  0.006& 0.078 &  0.007& 0.079 &  0.007& 0.082 &  0.008\\ 
      &      & 0.60 & 0.70& 0.040 &  0.006& 0.046 &  0.009& 0.048 &  0.008& 0.054 &  0.008\\ 
\hline  
 0.95 & 1.15 & 0.10 & 0.15& 0.068 &  0.009& 0.064 &  0.009& 0.058 &  0.008& 0.065 &  0.008\\ 
      &      & 0.15 & 0.20& 0.107 &  0.007& 0.091 &  0.007& 0.107 &  0.009& 0.103 &  0.008\\ 
      &      & 0.20 & 0.25& 0.113 &  0.006& 0.118 &  0.010& 0.110 &  0.006& 0.113 &  0.008\\ 
      &      & 0.25 & 0.30& 0.095 &  0.005& 0.112 &  0.007& 0.105 &  0.006& 0.104 &  0.006\\ 
      &      & 0.30 & 0.35& 0.083 &  0.005& 0.102 &  0.006& 0.095 &  0.005& 0.086 &  0.005\\ 
      &      & 0.35 & 0.40& 0.071 &  0.003& 0.075 &  0.006& 0.086 &  0.004& 0.078 &  0.004\\ 
      &      & 0.40 & 0.45& 0.059 &  0.003& 0.059 &  0.004& 0.069 &  0.004& 0.064 &  0.004\\ 
      &      & 0.45 & 0.50& 0.050 &  0.003& 0.050 &  0.003& 0.051 &  0.004& 0.051 &  0.004\\ 
      &      & 0.50 & 0.60& 0.033 &  0.004& 0.037 &  0.004& 0.033 &  0.004& 0.037 &  0.004\\ 
\hline
\end{tabular}
\end{center}
\end{table}

\begin{table}[hp!]
\begin{center}
\begin{tabular}{rrrr|r@{$\pm$}lr@{$\pm$}lr@{$\pm$}lr@{$\pm$}l}
\hline
$\theta_{\hbox{\small min}}$ &
$\theta_{\hbox{\small max}}$ &
$p_{\hbox{\small min}}$ &
$p_{\hbox{\small max}}$ &
\multicolumn{8}{c}{$d^2\sigma^{\pi^+}/(dpd\theta)$}
\\
(rad) & (rad) & (\GeVc) & (\GeVc) &
\multicolumn{8}{c}{(\barn/($\GeVc \cdot \rad$))}
\\
  &  &  &
&\multicolumn{2}{c}{$ \bf{3 \ \GeVc}$}
&\multicolumn{2}{c}{$ \bf{5 \ \GeVc}$}
&\multicolumn{2}{c}{$ \bf{8 \ \GeVc}$}
&\multicolumn{2}{c}{$ \bf{12 \ \GeVc}$}
\\
\hline
 1.15 & 1.35 & 0.10 & 0.15& 0.068 &  0.008& 0.061 &  0.009& 0.062 &  0.008& 0.059 &  0.008\\ 
      &      & 0.15 & 0.20& 0.094 &  0.007& 0.099 &  0.009& 0.101 &  0.008& 0.092 &  0.008\\ 
      &      & 0.20 & 0.25& 0.092 &  0.006& 0.094 &  0.006& 0.090 &  0.006& 0.102 &  0.007\\ 
      &      & 0.25 & 0.30& 0.080 &  0.004& 0.080 &  0.005& 0.089 &  0.005& 0.084 &  0.006\\ 
      &      & 0.30 & 0.35& 0.060 &  0.004& 0.067 &  0.005& 0.065 &  0.004& 0.062 &  0.005\\ 
      &      & 0.35 & 0.40& 0.047 &  0.003& 0.060 &  0.004& 0.050 &  0.003& 0.047 &  0.003\\ 
      &      & 0.40 & 0.45& 0.037 &  0.002& 0.045 &  0.004& 0.038 &  0.003& 0.041 &  0.003\\ 
      &      & 0.45 & 0.50& 0.029 &  0.002& 0.033 &  0.004& 0.027 &  0.003& 0.032 &  0.003\\ 
\hline  
 1.35 & 1.55 & 0.10 & 0.15& 0.061 &  0.007& 0.062 &  0.009& 0.064 &  0.008& 0.067 &  0.009\\ 
      &      & 0.15 & 0.20& 0.078 &  0.006& 0.088 &  0.007& 0.088 &  0.007& 0.079 &  0.006\\ 
      &      & 0.20 & 0.25& 0.081 &  0.005& 0.078 &  0.005& 0.075 &  0.004& 0.061 &  0.004\\ 
      &      & 0.25 & 0.30& 0.063 &  0.004& 0.053 &  0.004& 0.061 &  0.004& 0.058 &  0.004\\ 
      &      & 0.30 & 0.35& 0.043 &  0.003& 0.039 &  0.003& 0.044 &  0.003& 0.044 &  0.004\\ 
      &      & 0.35 & 0.40& 0.033 &  0.002& 0.032 &  0.003& 0.036 &  0.003& 0.031 &  0.003\\ 
      &      & 0.40 & 0.45& 0.025 &  0.002& 0.026 &  0.002& 0.025 &  0.003& 0.022 &  0.002\\ 
      &      & 0.45 & 0.50& 0.019 &  0.002& 0.021 &  0.003& 0.015 &  0.002& 0.016 &  0.002\\ 
\hline  
 1.55 & 1.75 & 0.10 & 0.15& 0.057 &  0.007& 0.058 &  0.007& 0.052 &  0.007& 0.047 &  0.007\\ 
      &      & 0.15 & 0.20& 0.070 &  0.005& 0.069 &  0.007& 0.072 &  0.005& 0.070 &  0.006\\ 
      &      & 0.20 & 0.25& 0.060 &  0.004& 0.079 &  0.006& 0.057 &  0.004& 0.059 &  0.004\\ 
      &      & 0.25 & 0.30& 0.049 &  0.003& 0.045 &  0.005& 0.042 &  0.003& 0.040 &  0.004\\ 
      &      & 0.30 & 0.35& 0.030 &  0.002& 0.028 &  0.002& 0.029 &  0.002& 0.025 &  0.003\\ 
      &      & 0.35 & 0.40& 0.023 &  0.002& 0.023 &  0.002& 0.022 &  0.002& 0.016 &  0.002\\ 
      &      & 0.40 & 0.45& 0.017 &  0.002& 0.015 &  0.002& 0.014 &  0.002& 0.013 &  0.002\\ 
      &      & 0.45 & 0.50& 0.012 &  0.002& 0.011 &  0.002& 0.010 &  0.001& 0.008 &  0.001\\ 
\hline  
 1.75 & 1.95 & 0.10 & 0.15& 0.050 &  0.006& 0.046 &  0.007& 0.045 &  0.005& 0.045 &  0.006\\ 
      &      & 0.15 & 0.20& 0.058 &  0.004& 0.055 &  0.005& 0.057 &  0.004& 0.064 &  0.005\\ 
      &      & 0.20 & 0.25& 0.050 &  0.004& 0.049 &  0.004& 0.047 &  0.003& 0.043 &  0.004\\ 
      &      & 0.25 & 0.30& 0.036 &  0.002& 0.034 &  0.004& 0.033 &  0.003& 0.032 &  0.003\\ 
      &      & 0.30 & 0.35& 0.023 &  0.002& 0.021 &  0.002& 0.019 &  0.002& 0.021 &  0.002\\ 
      &      & 0.35 & 0.40& 0.014 &  0.002& 0.015 &  0.002& 0.014 &  0.001& 0.013 &  0.002\\ 
      &      & 0.40 & 0.45& 0.010 &  0.001& 0.011 &  0.001& 0.009 &  0.001& 0.008 &  0.001\\ 
      &      & 0.45 & 0.50& 0.006 &  0.001& 0.007 &  0.001& 0.005 &  0.001& 0.005 &  0.001\\ 
\hline  
 1.95 & 2.15 & 0.10 & 0.15& 0.043 &  0.005& 0.035 &  0.006& 0.034 &  0.004& 0.038 &  0.005\\ 
      &      & 0.15 & 0.20& 0.051 &  0.004& 0.049 &  0.004& 0.043 &  0.004& 0.039 &  0.004\\ 
      &      & 0.20 & 0.25& 0.036 &  0.003& 0.037 &  0.003& 0.035 &  0.003& 0.027 &  0.003\\ 
      &      & 0.25 & 0.30& 0.021 &  0.002& 0.024 &  0.003& 0.022 &  0.003& 0.022 &  0.003\\ 
      &      & 0.30 & 0.35& 0.012 &  0.001& 0.013 &  0.002& 0.011 &  0.001& 0.009 &  0.002\\ 
      &      & 0.35 & 0.40& 0.009 &  0.001& 0.009 &  0.001& 0.009 &  0.001& 0.006 &  0.001\\ 
      &      & 0.40 & 0.45& 0.008 &  0.001& 0.007 &  0.001& 0.006 &  0.001& 0.004 &  0.001\\ 
      &      & 0.45 & 0.50& 0.006 &  0.001& 0.004 &  0.001& 0.003 &  0.001& 0.003 &  0.001\\ 

\end{tabular}
\end{center}
\end{table}

\begin{table}[hp!]
\begin{center}
  \caption{\label{tab:xsec-pimm-be}
    HARP results for the double-differential $\pi^-$ production
    cross-section in the laboratory system,
    $d^2\sigma^{\pi^-}/(dpd\theta)$ for $\pi^-$--Be interactions. Each row refers to a
    different $(p_{\hbox{\small min}} \le p<p_{\hbox{\small max}},
    \theta_{\hbox{\small min}} \le \theta<\theta_{\hbox{\small max}})$ bin,
    where $p$ and $\theta$ are the pion momentum and polar angle, respectively.
    The central value as well as the square-root of the diagonal elements
    of the covariance matrix are given.}
\vspace{2mm}
\begin{tabular}{rrrr|r@{$\pm$}lr@{$\pm$}lr@{$\pm$}lr@{$\pm$}l}
\hline
$\theta_{\hbox{\small min}}$ &
$\theta_{\hbox{\small max}}$ &
$p_{\hbox{\small min}}$ &
$p_{\hbox{\small max}}$ &
\multicolumn{8}{c}{$d^2\sigma^{\pi^-}/(dpd\theta)$}
\\
(rad) & (rad) & (\GeVc) & (\GeVc) &
\multicolumn{8}{c}{($\barn/(\GeVc \cdot \rad)$)}
\\
  &  &  &
&\multicolumn{2}{c}{$ \bf{3 \ \GeVc}$}
&\multicolumn{2}{c}{$ \bf{5 \ \GeVc}$}
&\multicolumn{2}{c}{$ \bf{8 \ \GeVc}$}
&\multicolumn{2}{c}{$ \bf{12 \ \GeVc}$}
\\
\hline  
 0.35 & 0.55 & 0.15 & 0.20& 0.111 &  0.017& 0.139 &  0.019& 0.126 &  0.021& 0.128 &  0.024\\ 
      &      & 0.20 & 0.25& 0.156 &  0.012& 0.178 &  0.016& 0.163 &  0.014& 0.182 &  0.016\\ 
      &      & 0.25 & 0.30& 0.180 &  0.013& 0.198 &  0.014& 0.201 &  0.018& 0.210 &  0.014\\ 
      &      & 0.30 & 0.35& 0.181 &  0.012& 0.224 &  0.018& 0.226 &  0.014& 0.216 &  0.018\\ 
      &      & 0.35 & 0.40& 0.188 &  0.012& 0.227 &  0.013& 0.229 &  0.015& 0.207 &  0.011\\ 
      &      & 0.40 & 0.45& 0.200 &  0.015& 0.228 &  0.015& 0.220 &  0.011& 0.223 &  0.017\\ 
      &      & 0.45 & 0.50& 0.206 &  0.010& 0.229 &  0.014& 0.228 &  0.012& 0.253 &  0.018\\ 
      &      & 0.50 & 0.60& 0.213 &  0.012& 0.266 &  0.016& 0.245 &  0.014& 0.270 &  0.014\\ 
      &      & 0.60 & 0.70& 0.206 &  0.016& 0.268 &  0.021& 0.257 &  0.020& 0.253 &  0.020\\ 
      &      & 0.70 & 0.80& 0.188 &  0.023& 0.233 &  0.030& 0.243 &  0.028& 0.243 &  0.028\\ 
\hline  
 0.55 & 0.75 & 0.10 & 0.15& 0.088 &  0.018& 0.094 &  0.021& 0.077 &  0.019& 0.081 &  0.021\\ 
      &      & 0.15 & 0.20& 0.139 &  0.010& 0.167 &  0.015& 0.140 &  0.013& 0.141 &  0.013\\ 
      &      & 0.20 & 0.25& 0.193 &  0.015& 0.227 &  0.016& 0.192 &  0.013& 0.172 &  0.015\\ 
      &      & 0.25 & 0.30& 0.212 &  0.014& 0.236 &  0.016& 0.193 &  0.014& 0.195 &  0.012\\ 
      &      & 0.30 & 0.35& 0.221 &  0.015& 0.226 &  0.013& 0.202 &  0.011& 0.182 &  0.011\\ 
      &      & 0.35 & 0.40& 0.222 &  0.012& 0.216 &  0.013& 0.202 &  0.012& 0.182 &  0.011\\ 
      &      & 0.40 & 0.45& 0.216 &  0.010& 0.224 &  0.012& 0.195 &  0.009& 0.178 &  0.009\\ 
      &      & 0.45 & 0.50& 0.209 &  0.009& 0.225 &  0.011& 0.183 &  0.008& 0.174 &  0.008\\ 
      &      & 0.50 & 0.60& 0.201 &  0.010& 0.206 &  0.011& 0.165 &  0.009& 0.162 &  0.009\\ 
      &      & 0.60 & 0.70& 0.171 &  0.016& 0.179 &  0.015& 0.153 &  0.013& 0.143 &  0.012\\ 
      &      & 0.70 & 0.80& 0.137 &  0.021& 0.151 &  0.022& 0.136 &  0.018& 0.126 &  0.016\\ 
\hline  
 0.75 & 0.95 & 0.10 & 0.15& 0.108 &  0.013& 0.098 &  0.014& 0.075 &  0.010& 0.076 &  0.013\\ 
      &      & 0.15 & 0.20& 0.187 &  0.013& 0.189 &  0.017& 0.140 &  0.012& 0.150 &  0.011\\ 
      &      & 0.20 & 0.25& 0.221 &  0.015& 0.210 &  0.012& 0.178 &  0.011& 0.155 &  0.011\\ 
      &      & 0.25 & 0.30& 0.212 &  0.012& 0.208 &  0.015& 0.165 &  0.008& 0.161 &  0.009\\ 
      &      & 0.30 & 0.35& 0.205 &  0.009& 0.198 &  0.012& 0.164 &  0.009& 0.149 &  0.009\\ 
      &      & 0.35 & 0.40& 0.185 &  0.008& 0.193 &  0.009& 0.153 &  0.008& 0.138 &  0.007\\ 
      &      & 0.40 & 0.45& 0.164 &  0.007& 0.172 &  0.007& 0.137 &  0.005& 0.121 &  0.006\\ 
      &      & 0.45 & 0.50& 0.149 &  0.006& 0.150 &  0.007& 0.123 &  0.005& 0.108 &  0.005\\ 
      &      & 0.50 & 0.60& 0.131 &  0.007& 0.128 &  0.008& 0.102 &  0.006& 0.092 &  0.005\\ 
      &      & 0.60 & 0.70& 0.109 &  0.010& 0.103 &  0.011& 0.086 &  0.007& 0.077 &  0.007\\ 
\hline  
 0.95 & 1.15 & 0.10 & 0.15& 0.127 &  0.015& 0.118 &  0.014& 0.086 &  0.010& 0.082 &  0.009\\ 
      &      & 0.15 & 0.20& 0.228 &  0.014& 0.197 &  0.013& 0.151 &  0.012& 0.138 &  0.014\\ 
      &      & 0.20 & 0.25& 0.212 &  0.010& 0.184 &  0.011& 0.155 &  0.009& 0.147 &  0.008\\ 
      &      & 0.25 & 0.30& 0.181 &  0.008& 0.174 &  0.009& 0.143 &  0.007& 0.129 &  0.008\\ 
      &      & 0.30 & 0.35& 0.160 &  0.007& 0.141 &  0.007& 0.118 &  0.005& 0.113 &  0.006\\ 
      &      & 0.35 & 0.40& 0.136 &  0.006& 0.130 &  0.007& 0.098 &  0.004& 0.092 &  0.005\\ 
      &      & 0.40 & 0.45& 0.115 &  0.005& 0.107 &  0.006& 0.085 &  0.004& 0.079 &  0.004\\ 
      &      & 0.45 & 0.50& 0.097 &  0.005& 0.090 &  0.006& 0.071 &  0.003& 0.069 &  0.004\\ 
      &      & 0.50 & 0.60& 0.080 &  0.005& 0.072 &  0.006& 0.057 &  0.004& 0.062 &  0.004\\ 
\hline
\end{tabular}
\end{center}
\end{table}

\begin{table}[hp!]
\begin{center}
\begin{tabular}{rrrr|r@{$\pm$}lr@{$\pm$}lr@{$\pm$}lr@{$\pm$}l}
\hline
$\theta_{\hbox{\small min}}$ &
$\theta_{\hbox{\small max}}$ &
$p_{\hbox{\small min}}$ &
$p_{\hbox{\small max}}$ &
\multicolumn{8}{c}{$d^2\sigma^{\pi^-}/(dpd\theta)$}
\\
(rad) & (rad) & (\GeVc) & (\GeVc) &
\multicolumn{8}{c}{(\barn/($\GeVc \cdot \rad$))}
\\
  &  &  &
&\multicolumn{2}{c}{$ \bf{3 \ \GeVc}$}
&\multicolumn{2}{c}{$ \bf{5 \ \GeVc}$}
&\multicolumn{2}{c}{$ \bf{8 \ \GeVc}$}
&\multicolumn{2}{c}{$ \bf{12 \ \GeVc}$}
\\
\hline
 1.15 & 1.35 & 0.10 & 0.15& 0.132 &  0.013& 0.120 &  0.013& 0.084 &  0.010& 0.090 &  0.011\\ 
      &      & 0.15 & 0.20& 0.197 &  0.012& 0.170 &  0.012& 0.138 &  0.010& 0.118 &  0.008\\ 
      &      & 0.20 & 0.25& 0.184 &  0.009& 0.167 &  0.009& 0.143 &  0.008& 0.125 &  0.008\\ 
      &      & 0.25 & 0.30& 0.155 &  0.007& 0.132 &  0.007& 0.116 &  0.006& 0.104 &  0.007\\ 
      &      & 0.30 & 0.35& 0.120 &  0.006& 0.103 &  0.005& 0.084 &  0.005& 0.075 &  0.004\\ 
      &      & 0.35 & 0.40& 0.098 &  0.005& 0.083 &  0.005& 0.067 &  0.003& 0.062 &  0.004\\ 
      &      & 0.40 & 0.45& 0.080 &  0.004& 0.064 &  0.004& 0.056 &  0.003& 0.049 &  0.003\\ 
      &      & 0.45 & 0.50& 0.067 &  0.004& 0.050 &  0.003& 0.044 &  0.003& 0.042 &  0.003\\ 
\hline  
 1.35 & 1.55 & 0.10 & 0.15& 0.132 &  0.015& 0.138 &  0.015& 0.100 &  0.011& 0.088 &  0.012\\ 
      &      & 0.15 & 0.20& 0.179 &  0.009& 0.160 &  0.010& 0.127 &  0.008& 0.134 &  0.009\\ 
      &      & 0.20 & 0.25& 0.153 &  0.008& 0.136 &  0.008& 0.122 &  0.006& 0.106 &  0.006\\ 
      &      & 0.25 & 0.30& 0.116 &  0.007& 0.112 &  0.007& 0.080 &  0.006& 0.079 &  0.005\\ 
      &      & 0.30 & 0.35& 0.089 &  0.005& 0.086 &  0.006& 0.054 &  0.003& 0.061 &  0.004\\ 
      &      & 0.35 & 0.40& 0.071 &  0.004& 0.059 &  0.004& 0.045 &  0.003& 0.047 &  0.003\\ 
      &      & 0.40 & 0.45& 0.054 &  0.003& 0.043 &  0.003& 0.037 &  0.002& 0.033 &  0.003\\ 
      &      & 0.45 & 0.50& 0.041 &  0.004& 0.032 &  0.003& 0.028 &  0.002& 0.024 &  0.002\\ 
\hline  
 1.55 & 1.75 & 0.10 & 0.15& 0.133 &  0.014& 0.119 &  0.012& 0.088 &  0.011& 0.081 &  0.010\\ 
      &      & 0.15 & 0.20& 0.166 &  0.009& 0.132 &  0.009& 0.113 &  0.007& 0.094 &  0.006\\ 
      &      & 0.20 & 0.25& 0.127 &  0.007& 0.106 &  0.006& 0.088 &  0.005& 0.093 &  0.006\\ 
      &      & 0.25 & 0.30& 0.085 &  0.006& 0.085 &  0.006& 0.065 &  0.004& 0.057 &  0.007\\ 
      &      & 0.30 & 0.35& 0.061 &  0.004& 0.051 &  0.005& 0.047 &  0.003& 0.037 &  0.003\\ 
      &      & 0.35 & 0.40& 0.047 &  0.003& 0.037 &  0.003& 0.035 &  0.002& 0.036 &  0.003\\ 
      &      & 0.40 & 0.45& 0.036 &  0.003& 0.029 &  0.002& 0.026 &  0.002& 0.029 &  0.002\\ 
      &      & 0.45 & 0.50& 0.026 &  0.003& 0.020 &  0.002& 0.019 &  0.002& 0.019 &  0.002\\ 
\hline  
 1.75 & 1.95 & 0.10 & 0.15& 0.120 &  0.013& 0.104 &  0.011& 0.071 &  0.008& 0.068 &  0.008\\ 
      &      & 0.15 & 0.20& 0.145 &  0.007& 0.121 &  0.008& 0.096 &  0.007& 0.083 &  0.006\\ 
      &      & 0.20 & 0.25& 0.093 &  0.007& 0.084 &  0.007& 0.074 &  0.005& 0.059 &  0.004\\ 
      &      & 0.25 & 0.30& 0.058 &  0.004& 0.059 &  0.004& 0.044 &  0.004& 0.051 &  0.004\\ 
      &      & 0.30 & 0.35& 0.043 &  0.003& 0.037 &  0.004& 0.030 &  0.002& 0.029 &  0.004\\ 
      &      & 0.35 & 0.40& 0.033 &  0.002& 0.026 &  0.002& 0.022 &  0.002& 0.017 &  0.002\\ 
      &      & 0.40 & 0.45& 0.025 &  0.002& 0.020 &  0.002& 0.016 &  0.001& 0.014 &  0.001\\ 
      &      & 0.45 & 0.50& 0.018 &  0.002& 0.013 &  0.002& 0.012 &  0.001& 0.012 &  0.001\\ 
\hline  
 1.95 & 2.15 & 0.10 & 0.15& 0.087 &  0.009& 0.077 &  0.007& 0.064 &  0.007& 0.055 &  0.006\\ 
      &      & 0.15 & 0.20& 0.122 &  0.006& 0.106 &  0.008& 0.088 &  0.006& 0.076 &  0.007\\ 
      &      & 0.20 & 0.25& 0.082 &  0.004& 0.070 &  0.006& 0.057 &  0.005& 0.057 &  0.005\\ 
      &      & 0.25 & 0.30& 0.049 &  0.004& 0.040 &  0.003& 0.039 &  0.003& 0.030 &  0.003\\ 
      &      & 0.30 & 0.35& 0.030 &  0.002& 0.026 &  0.003& 0.025 &  0.002& 0.024 &  0.002\\ 
      &      & 0.35 & 0.40& 0.022 &  0.002& 0.019 &  0.002& 0.015 &  0.002& 0.017 &  0.002\\ 
      &      & 0.40 & 0.45& 0.016 &  0.002& 0.013 &  0.002& 0.010 &  0.001& 0.010 &  0.002\\ 
      &      & 0.45 & 0.50& 0.010 &  0.002& 0.008 &  0.001& 0.007 &  0.001& 0.007 &  0.001\\ 

\end{tabular}
\end{center}
\end{table}
\clearpage
\begin{table}[hp!]
\begin{center}
  \caption{\label{tab:xsec-pipp-c}
    HARP results for the double-differential $\pi^+$ production
    cross-section in the laboratory system,
    $d^2\sigma^{\pi^+}/(dpd\theta)$ for $\pi^+$--C interactions. Each row refers to a
    different $(p_{\hbox{\small min}} \le p<p_{\hbox{\small max}},
    \theta_{\hbox{\small min}} \le \theta<\theta_{\hbox{\small max}})$ bin,
    where $p$ and $\theta$ are the pion momentum and polar angle, respectively.
    The central value as well as the square-root of the diagonal elements
    of the covariance matrix are given.}
\vspace{2mm}
\begin{tabular}{rrrr|r@{$\pm$}lr@{$\pm$}lr@{$\pm$}lr@{$\pm$}l}
\hline
$\theta_{\hbox{\small min}}$ &
$\theta_{\hbox{\small max}}$ &
$p_{\hbox{\small min}}$ &
$p_{\hbox{\small max}}$ &
\multicolumn{8}{c}{$d^2\sigma^{\pi^+}/(dpd\theta)$}
\\
(rad) & (rad) & (\GeVc) & (\GeVc) &
\multicolumn{8}{c}{($\barn/(\GeVc \cdot \rad)$)}
\\
  &  &  &
&\multicolumn{2}{c}{$ \bf{3 \ \GeVc}$}
&\multicolumn{2}{c}{$ \bf{5 \ \GeVc}$}
&\multicolumn{2}{c}{$ \bf{8 \ \GeVc}$}
&\multicolumn{2}{c}{$ \bf{12 \ \GeVc}$}
\\
\hline  
 0.35 & 0.55 & 0.15 & 0.20& 0.138 &  0.021& 0.147 &  0.027& 0.181 &  0.027& 0.153 &  0.030\\ 
      &      & 0.20 & 0.25& 0.175 &  0.015& 0.203 &  0.017& 0.214 &  0.022& 0.211 &  0.029\\ 
      &      & 0.25 & 0.30& 0.192 &  0.019& 0.238 &  0.019& 0.297 &  0.027& 0.286 &  0.033\\ 
      &      & 0.30 & 0.35& 0.268 &  0.028& 0.288 &  0.030& 0.319 &  0.021& 0.309 &  0.032\\ 
      &      & 0.35 & 0.40& 0.256 &  0.015& 0.328 &  0.020& 0.336 &  0.028& 0.354 &  0.027\\ 
      &      & 0.40 & 0.45& 0.267 &  0.017& 0.340 &  0.021& 0.410 &  0.033& 0.376 &  0.032\\ 
      &      & 0.45 & 0.50& 0.275 &  0.017& 0.339 &  0.017& 0.399 &  0.017& 0.388 &  0.026\\ 
      &      & 0.50 & 0.60& 0.298 &  0.018& 0.355 &  0.018& 0.414 &  0.024& 0.380 &  0.026\\ 
      &      & 0.60 & 0.70& 0.269 &  0.024& 0.355 &  0.027& 0.412 &  0.033& 0.363 &  0.037\\ 
      &      & 0.70 & 0.80& 0.215 &  0.032& 0.300 &  0.038& 0.367 &  0.045& 0.286 &  0.043\\ 
\hline  
 0.55 & 0.75 & 0.10 & 0.15& 0.082 &  0.020& 0.113 &  0.028& 0.117 &  0.028& 0.120 &  0.037\\ 
      &      & 0.15 & 0.20& 0.171 &  0.015& 0.201 &  0.016& 0.185 &  0.015& 0.168 &  0.021\\ 
      &      & 0.20 & 0.25& 0.246 &  0.023& 0.256 &  0.020& 0.249 &  0.022& 0.227 &  0.031\\ 
      &      & 0.25 & 0.30& 0.279 &  0.021& 0.325 &  0.026& 0.319 &  0.027& 0.289 &  0.027\\ 
      &      & 0.30 & 0.35& 0.285 &  0.021& 0.324 &  0.016& 0.321 &  0.020& 0.299 &  0.033\\ 
      &      & 0.35 & 0.40& 0.302 &  0.019& 0.326 &  0.017& 0.311 &  0.019& 0.313 &  0.022\\ 
      &      & 0.40 & 0.45& 0.301 &  0.016& 0.306 &  0.013& 0.325 &  0.017& 0.290 &  0.020\\ 
      &      & 0.45 & 0.50& 0.297 &  0.015& 0.306 &  0.015& 0.329 &  0.016& 0.273 &  0.019\\ 
      &      & 0.50 & 0.60& 0.252 &  0.016& 0.267 &  0.016& 0.274 &  0.018& 0.239 &  0.019\\ 
      &      & 0.60 & 0.70& 0.203 &  0.025& 0.209 &  0.023& 0.210 &  0.024& 0.191 &  0.026\\ 
      &      & 0.70 & 0.80& 0.131 &  0.022& 0.132 &  0.023& 0.146 &  0.028& 0.111 &  0.026\\ 
\hline  
 0.75 & 0.95 & 0.10 & 0.15& 0.116 &  0.020& 0.126 &  0.019& 0.124 &  0.019& 0.130 &  0.026\\ 
      &      & 0.15 & 0.20& 0.269 &  0.021& 0.235 &  0.019& 0.219 &  0.019& 0.196 &  0.022\\ 
      &      & 0.20 & 0.25& 0.288 &  0.023& 0.286 &  0.018& 0.267 &  0.021& 0.240 &  0.028\\ 
      &      & 0.25 & 0.30& 0.307 &  0.019& 0.300 &  0.020& 0.288 &  0.016& 0.263 &  0.027\\ 
      &      & 0.30 & 0.35& 0.289 &  0.015& 0.287 &  0.015& 0.245 &  0.014& 0.225 &  0.018\\ 
      &      & 0.35 & 0.40& 0.240 &  0.013& 0.250 &  0.009& 0.241 &  0.012& 0.201 &  0.017\\ 
      &      & 0.40 & 0.45& 0.214 &  0.011& 0.228 &  0.009& 0.209 &  0.010& 0.220 &  0.019\\ 
      &      & 0.45 & 0.50& 0.196 &  0.010& 0.203 &  0.009& 0.181 &  0.009& 0.179 &  0.017\\ 
      &      & 0.50 & 0.60& 0.162 &  0.012& 0.152 &  0.013& 0.141 &  0.011& 0.124 &  0.014\\ 
      &      & 0.60 & 0.70& 0.106 &  0.016& 0.094 &  0.014& 0.091 &  0.013& 0.077 &  0.013\\ 
\hline  
 0.95 & 1.15 & 0.10 & 0.15& 0.154 &  0.021& 0.136 &  0.017& 0.128 &  0.017& 0.115 &  0.023\\ 
      &      & 0.15 & 0.20& 0.260 &  0.021& 0.226 &  0.018& 0.232 &  0.016& 0.222 &  0.025\\ 
      &      & 0.20 & 0.25& 0.273 &  0.016& 0.267 &  0.014& 0.223 &  0.014& 0.215 &  0.021\\ 
      &      & 0.25 & 0.30& 0.220 &  0.013& 0.239 &  0.013& 0.207 &  0.013& 0.177 &  0.016\\ 
      &      & 0.30 & 0.35& 0.218 &  0.014& 0.203 &  0.010& 0.183 &  0.010& 0.174 &  0.018\\ 
      &      & 0.35 & 0.40& 0.174 &  0.009& 0.163 &  0.008& 0.155 &  0.009& 0.143 &  0.014\\ 
      &      & 0.40 & 0.45& 0.151 &  0.008& 0.130 &  0.006& 0.122 &  0.007& 0.098 &  0.012\\ 
      &      & 0.45 & 0.50& 0.120 &  0.008& 0.108 &  0.008& 0.095 &  0.007& 0.074 &  0.009\\ 
      &      & 0.50 & 0.60& 0.086 &  0.009& 0.072 &  0.007& 0.068 &  0.007& 0.049 &  0.007\\ 
\hline
\end{tabular}
\end{center}
\end{table}

\begin{table}[hp!]
\begin{center}
\begin{tabular}{rrrr|r@{$\pm$}lr@{$\pm$}lr@{$\pm$}lr@{$\pm$}l}
\hline
$\theta_{\hbox{\small min}}$ &
$\theta_{\hbox{\small max}}$ &
$p_{\hbox{\small min}}$ &
$p_{\hbox{\small max}}$ &
\multicolumn{8}{c}{$d^2\sigma^{\pi^+}/(dpd\theta)$}
\\
(rad) & (rad) & (\GeVc) & (\GeVc) &
\multicolumn{8}{c}{(\barn/($\GeVc \cdot \rad$))}
\\
  &  &  &
&\multicolumn{2}{c}{$ \bf{3 \ \GeVc}$}
&\multicolumn{2}{c}{$ \bf{5 \ \GeVc}$}
&\multicolumn{2}{c}{$ \bf{8 \ \GeVc}$}
&\multicolumn{2}{c}{$ \bf{12 \ \GeVc}$}
\\
\hline
 1.15 & 1.35 & 0.10 & 0.15& 0.186 &  0.024& 0.150 &  0.019& 0.134 &  0.019& 0.153 &  0.026\\ 
      &      & 0.15 & 0.20& 0.257 &  0.018& 0.241 &  0.015& 0.219 &  0.016& 0.191 &  0.020\\ 
      &      & 0.20 & 0.25& 0.236 &  0.014& 0.201 &  0.010& 0.202 &  0.012& 0.193 &  0.020\\ 
      &      & 0.25 & 0.30& 0.176 &  0.011& 0.166 &  0.008& 0.151 &  0.009& 0.146 &  0.015\\ 
      &      & 0.30 & 0.35& 0.144 &  0.009& 0.138 &  0.007& 0.111 &  0.008& 0.104 &  0.011\\ 
      &      & 0.35 & 0.40& 0.122 &  0.007& 0.102 &  0.006& 0.088 &  0.005& 0.077 &  0.009\\ 
      &      & 0.40 & 0.45& 0.104 &  0.006& 0.075 &  0.005& 0.075 &  0.005& 0.054 &  0.007\\ 
      &      & 0.45 & 0.50& 0.083 &  0.008& 0.056 &  0.005& 0.057 &  0.006& 0.043 &  0.006\\ 
\hline  
 1.35 & 1.55 & 0.10 & 0.15& 0.186 &  0.022& 0.159 &  0.019& 0.155 &  0.019& 0.140 &  0.022\\ 
      &      & 0.15 & 0.20& 0.241 &  0.016& 0.213 &  0.013& 0.212 &  0.015& 0.150 &  0.019\\ 
      &      & 0.20 & 0.25& 0.203 &  0.013& 0.173 &  0.010& 0.163 &  0.011& 0.137 &  0.015\\ 
      &      & 0.25 & 0.30& 0.155 &  0.011& 0.133 &  0.008& 0.110 &  0.008& 0.118 &  0.015\\ 
      &      & 0.30 & 0.35& 0.114 &  0.008& 0.099 &  0.006& 0.083 &  0.006& 0.064 &  0.008\\ 
      &      & 0.35 & 0.40& 0.082 &  0.007& 0.067 &  0.005& 0.061 &  0.004& 0.053 &  0.007\\ 
      &      & 0.40 & 0.45& 0.062 &  0.005& 0.046 &  0.004& 0.045 &  0.004& 0.041 &  0.007\\ 
      &      & 0.45 & 0.50& 0.041 &  0.005& 0.031 &  0.004& 0.031 &  0.004& 0.026 &  0.006\\ 
\hline  
 1.55 & 1.75 & 0.10 & 0.15& 0.158 &  0.019& 0.156 &  0.019& 0.141 &  0.017& 0.088 &  0.018\\ 
      &      & 0.15 & 0.20& 0.225 &  0.016& 0.186 &  0.011& 0.152 &  0.010& 0.122 &  0.016\\ 
      &      & 0.20 & 0.25& 0.162 &  0.012& 0.143 &  0.009& 0.116 &  0.008& 0.092 &  0.012\\ 
      &      & 0.25 & 0.30& 0.112 &  0.008& 0.090 &  0.006& 0.083 &  0.006& 0.093 &  0.013\\ 
      &      & 0.30 & 0.35& 0.077 &  0.007& 0.066 &  0.004& 0.063 &  0.005& 0.060 &  0.009\\ 
      &      & 0.35 & 0.40& 0.051 &  0.005& 0.046 &  0.004& 0.041 &  0.004& 0.041 &  0.008\\ 
      &      & 0.40 & 0.45& 0.035 &  0.004& 0.030 &  0.004& 0.028 &  0.004& 0.021 &  0.006\\ 
      &      & 0.45 & 0.50& 0.025 &  0.003& 0.019 &  0.003& 0.017 &  0.003& 0.009 &  0.003\\ 
\hline  
 1.75 & 1.95 & 0.10 & 0.15& 0.146 &  0.018& 0.145 &  0.015& 0.105 &  0.015& 0.112 &  0.019\\ 
      &      & 0.15 & 0.20& 0.206 &  0.014& 0.166 &  0.008& 0.129 &  0.009& 0.119 &  0.015\\ 
      &      & 0.20 & 0.25& 0.152 &  0.011& 0.102 &  0.007& 0.098 &  0.007& 0.090 &  0.013\\ 
      &      & 0.25 & 0.30& 0.084 &  0.010& 0.067 &  0.005& 0.054 &  0.005& 0.043 &  0.008\\ 
      &      & 0.30 & 0.35& 0.044 &  0.005& 0.041 &  0.003& 0.043 &  0.004& 0.034 &  0.006\\ 
      &      & 0.35 & 0.40& 0.028 &  0.003& 0.026 &  0.003& 0.027 &  0.003& 0.041 &  0.008\\ 
      &      & 0.40 & 0.45& 0.020 &  0.003& 0.015 &  0.002& 0.017 &  0.002& 0.022 &  0.007\\ 
      &      & 0.45 & 0.50& 0.013 &  0.003& 0.008 &  0.002& 0.010 &  0.002& 0.008 &  0.004\\ 
\hline  
 1.95 & 2.15 & 0.10 & 0.15& 0.135 &  0.015& 0.100 &  0.011& 0.097 &  0.013& 0.074 &  0.017\\ 
      &      & 0.15 & 0.20& 0.147 &  0.010& 0.117 &  0.007& 0.115 &  0.008& 0.123 &  0.017\\ 
      &      & 0.20 & 0.25& 0.086 &  0.008& 0.081 &  0.005& 0.068 &  0.006& 0.042 &  0.011\\ 
      &      & 0.25 & 0.30& 0.058 &  0.005& 0.051 &  0.004& 0.036 &  0.005& 0.020 &  0.005\\ 
      &      & 0.30 & 0.35& 0.033 &  0.006& 0.030 &  0.003& 0.026 &  0.003& 0.031 &  0.007\\ 
      &      & 0.35 & 0.40& 0.015 &  0.002& 0.019 &  0.002& 0.017 &  0.003& 0.021 &  0.007\\ 
      &      & 0.40 & 0.45& 0.012 &  0.002& 0.010 &  0.002& 0.009 &  0.002& 0.005 &  0.003\\ 
      &      & 0.45 & 0.50& 0.008 &  0.002& 0.006 &  0.001& 0.004 &  0.001& 0.004 &  0.002\\

\end{tabular}
\end{center}
\end{table}

\begin{table}[hp!]
\begin{center}
  \caption{\label{tab:xsec-pipm-c}
    HARP results for the double-differential $\pi^-$ production
    cross-section in the laboratory system,
    $d^2\sigma^{\pi^-}/(dpd\theta)$ for $\pi^+$--C interactions. Each row refers to a
    different $(p_{\hbox{\small min}} \le p<p_{\hbox{\small max}},
    \theta_{\hbox{\small min}} \le \theta<\theta_{\hbox{\small max}})$ bin,
    where $p$ and $\theta$ are the pion momentum and polar angle, respectively.
    The central value as well as the square-root of the diagonal elements
    of the covariance matrix are given.}
\vspace{2mm}
\begin{tabular}{rrrr|r@{$\pm$}lr@{$\pm$}lr@{$\pm$}lr@{$\pm$}l}
\hline
$\theta_{\hbox{\small min}}$ &
$\theta_{\hbox{\small max}}$ &
$p_{\hbox{\small min}}$ &
$p_{\hbox{\small max}}$ &
\multicolumn{8}{c}{$d^2\sigma^{\pi^-}/(dpd\theta)$}
\\
(rad) & (rad) & (\GeVc) & (\GeVc) &
\multicolumn{8}{c}{($\barn/(\GeVc \cdot \rad)$)}
\\
  &  &  &
&\multicolumn{2}{c}{$ \bf{3 \ \GeVc}$}
&\multicolumn{2}{c}{$ \bf{5 \ \GeVc}$}
&\multicolumn{2}{c}{$ \bf{8 \ \GeVc}$}
&\multicolumn{2}{c}{$ \bf{12 \ \GeVc}$}
\\
\hline  
 0.35 & 0.55 & 0.15 & 0.20& 0.102 &  0.020& 0.135 &  0.022& 0.137 &  0.024& 0.109 &  0.028\\ 
      &      & 0.20 & 0.25& 0.115 &  0.012& 0.146 &  0.014& 0.154 &  0.016& 0.206 &  0.038\\ 
      &      & 0.25 & 0.30& 0.130 &  0.012& 0.175 &  0.013& 0.179 &  0.016& 0.269 &  0.028\\ 
      &      & 0.30 & 0.35& 0.133 &  0.011& 0.167 &  0.011& 0.205 &  0.015& 0.208 &  0.021\\ 
      &      & 0.35 & 0.40& 0.146 &  0.012& 0.169 &  0.009& 0.187 &  0.010& 0.163 &  0.016\\ 
      &      & 0.40 & 0.45& 0.126 &  0.008& 0.164 &  0.009& 0.175 &  0.010& 0.198 &  0.029\\ 
      &      & 0.45 & 0.50& 0.126 &  0.008& 0.165 &  0.008& 0.170 &  0.009& 0.214 &  0.018\\ 
      &      & 0.50 & 0.60& 0.118 &  0.007& 0.164 &  0.007& 0.184 &  0.010& 0.197 &  0.016\\ 
      &      & 0.60 & 0.70& 0.109 &  0.009& 0.151 &  0.011& 0.186 &  0.013& 0.180 &  0.019\\ 
      &      & 0.70 & 0.80& 0.105 &  0.011& 0.143 &  0.013& 0.170 &  0.018& 0.164 &  0.021\\ 
\hline  
 0.55 & 0.75 & 0.10 & 0.15& 0.064 &  0.018& 0.083 &  0.021& 0.082 &  0.027& 0.072 &  0.027\\ 
      &      & 0.15 & 0.20& 0.124 &  0.013& 0.127 &  0.013& 0.136 &  0.012& 0.121 &  0.021\\ 
      &      & 0.20 & 0.25& 0.155 &  0.012& 0.150 &  0.012& 0.146 &  0.012& 0.164 &  0.022\\ 
      &      & 0.25 & 0.30& 0.143 &  0.012& 0.161 &  0.010& 0.174 &  0.015& 0.172 &  0.019\\ 
      &      & 0.30 & 0.35& 0.151 &  0.011& 0.158 &  0.010& 0.172 &  0.010& 0.145 &  0.018\\ 
      &      & 0.35 & 0.40& 0.129 &  0.008& 0.152 &  0.008& 0.160 &  0.009& 0.172 &  0.017\\ 
      &      & 0.40 & 0.45& 0.115 &  0.007& 0.147 &  0.007& 0.149 &  0.008& 0.171 &  0.015\\ 
      &      & 0.45 & 0.50& 0.125 &  0.009& 0.133 &  0.005& 0.146 &  0.008& 0.169 &  0.014\\ 
      &      & 0.50 & 0.60& 0.115 &  0.007& 0.125 &  0.006& 0.136 &  0.008& 0.130 &  0.013\\ 
      &      & 0.60 & 0.70& 0.087 &  0.009& 0.110 &  0.009& 0.114 &  0.009& 0.109 &  0.011\\ 
      &      & 0.70 & 0.80& 0.069 &  0.009& 0.087 &  0.011& 0.100 &  0.012& 0.101 &  0.014\\ 
\hline  
 0.75 & 0.95 & 0.10 & 0.15& 0.087 &  0.014& 0.064 &  0.012& 0.066 &  0.013& 0.102 &  0.022\\ 
      &      & 0.15 & 0.20& 0.143 &  0.012& 0.138 &  0.012& 0.144 &  0.013& 0.152 &  0.021\\ 
      &      & 0.20 & 0.25& 0.134 &  0.011& 0.153 &  0.009& 0.157 &  0.012& 0.128 &  0.016\\ 
      &      & 0.25 & 0.30& 0.135 &  0.010& 0.141 &  0.008& 0.141 &  0.008& 0.124 &  0.014\\ 
      &      & 0.30 & 0.35& 0.131 &  0.009& 0.124 &  0.006& 0.126 &  0.008& 0.120 &  0.015\\ 
      &      & 0.35 & 0.40& 0.105 &  0.007& 0.115 &  0.006& 0.126 &  0.009& 0.126 &  0.013\\ 
      &      & 0.40 & 0.45& 0.088 &  0.005& 0.109 &  0.005& 0.125 &  0.007& 0.128 &  0.012\\ 
      &      & 0.45 & 0.50& 0.091 &  0.006& 0.101 &  0.004& 0.103 &  0.007& 0.106 &  0.010\\ 
      &      & 0.50 & 0.60& 0.081 &  0.005& 0.081 &  0.006& 0.082 &  0.005& 0.082 &  0.009\\ 
      &      & 0.60 & 0.70& 0.066 &  0.006& 0.059 &  0.006& 0.072 &  0.006& 0.057 &  0.009\\ 
\hline  
 0.95 & 1.15 & 0.10 & 0.15& 0.092 &  0.013& 0.080 &  0.010& 0.076 &  0.012& 0.074 &  0.017\\ 
      &      & 0.15 & 0.20& 0.126 &  0.010& 0.135 &  0.010& 0.131 &  0.010& 0.115 &  0.016\\ 
      &      & 0.20 & 0.25& 0.127 &  0.011& 0.144 &  0.009& 0.133 &  0.010& 0.094 &  0.013\\ 
      &      & 0.25 & 0.30& 0.118 &  0.008& 0.124 &  0.006& 0.113 &  0.007& 0.086 &  0.011\\ 
      &      & 0.30 & 0.35& 0.104 &  0.007& 0.104 &  0.005& 0.092 &  0.006& 0.101 &  0.015\\ 
      &      & 0.35 & 0.40& 0.095 &  0.006& 0.087 &  0.004& 0.080 &  0.005& 0.105 &  0.012\\ 
      &      & 0.40 & 0.45& 0.071 &  0.006& 0.074 &  0.003& 0.070 &  0.004& 0.067 &  0.010\\ 
      &      & 0.45 & 0.50& 0.057 &  0.004& 0.066 &  0.004& 0.059 &  0.004& 0.050 &  0.006\\ 
      &      & 0.50 & 0.60& 0.042 &  0.003& 0.052 &  0.004& 0.048 &  0.003& 0.040 &  0.005\\ 
\hline
\end{tabular}
\end{center}
\end{table}

\begin{table}[hp!]
\begin{center}
\begin{tabular}{rrrr|r@{$\pm$}lr@{$\pm$}lr@{$\pm$}lr@{$\pm$}l}
\hline
$\theta_{\hbox{\small min}}$ &
$\theta_{\hbox{\small max}}$ &
$p_{\hbox{\small min}}$ &
$p_{\hbox{\small max}}$ &
\multicolumn{8}{c}{$d^2\sigma^{\pi^-}/(dpd\theta)$}
\\
(rad) & (rad) & (\GeVc) & (\GeVc) &
\multicolumn{8}{c}{(\barn/($\GeVc \cdot \rad$))}
\\
  &  &  &
&\multicolumn{2}{c}{$ \bf{3 \ \GeVc}$}
&\multicolumn{2}{c}{$ \bf{5 \ \GeVc}$}
&\multicolumn{2}{c}{$ \bf{8 \ \GeVc}$}
&\multicolumn{2}{c}{$ \bf{12 \ \GeVc}$}
\\
\hline
 1.15 & 1.35 & 0.10 & 0.15& 0.090 &  0.013& 0.074 &  0.009& 0.069 &  0.009& 0.080 &  0.016\\ 
      &      & 0.15 & 0.20& 0.141 &  0.011& 0.128 &  0.009& 0.129 &  0.010& 0.133 &  0.019\\ 
      &      & 0.20 & 0.25& 0.105 &  0.008& 0.103 &  0.006& 0.088 &  0.006& 0.141 &  0.018\\ 
      &      & 0.25 & 0.30& 0.086 &  0.007& 0.084 &  0.005& 0.092 &  0.007& 0.138 &  0.016\\ 
      &      & 0.30 & 0.35& 0.080 &  0.006& 0.077 &  0.005& 0.077 &  0.005& 0.070 &  0.011\\ 
      &      & 0.35 & 0.40& 0.065 &  0.005& 0.060 &  0.004& 0.056 &  0.004& 0.050 &  0.006\\ 
      &      & 0.40 & 0.45& 0.052 &  0.004& 0.046 &  0.003& 0.048 &  0.003& 0.044 &  0.006\\ 
      &      & 0.45 & 0.50& 0.038 &  0.003& 0.036 &  0.002& 0.041 &  0.003& 0.039 &  0.006\\ 
\hline  
 1.35 & 1.55 & 0.10 & 0.15& 0.075 &  0.010& 0.085 &  0.009& 0.083 &  0.010& 0.057 &  0.013\\ 
      &      & 0.15 & 0.20& 0.115 &  0.011& 0.107 &  0.007& 0.114 &  0.009& 0.098 &  0.015\\ 
      &      & 0.20 & 0.25& 0.097 &  0.008& 0.086 &  0.005& 0.096 &  0.007& 0.077 &  0.012\\ 
      &      & 0.25 & 0.30& 0.074 &  0.006& 0.077 &  0.005& 0.073 &  0.006& 0.052 &  0.009\\ 
      &      & 0.30 & 0.35& 0.060 &  0.005& 0.054 &  0.004& 0.051 &  0.004& 0.050 &  0.008\\ 
      &      & 0.35 & 0.40& 0.045 &  0.004& 0.040 &  0.003& 0.041 &  0.003& 0.040 &  0.006\\ 
      &      & 0.40 & 0.45& 0.036 &  0.003& 0.031 &  0.002& 0.035 &  0.003& 0.026 &  0.005\\ 
      &      & 0.45 & 0.50& 0.025 &  0.003& 0.024 &  0.002& 0.026 &  0.003& 0.019 &  0.004\\ 
\hline  
 1.55 & 1.75 & 0.10 & 0.15& 0.076 &  0.011& 0.070 &  0.007& 0.057 &  0.008& 0.064 &  0.014\\ 
      &      & 0.15 & 0.20& 0.092 &  0.008& 0.093 &  0.006& 0.087 &  0.008& 0.064 &  0.011\\ 
      &      & 0.20 & 0.25& 0.080 &  0.008& 0.062 &  0.004& 0.067 &  0.005& 0.078 &  0.014\\ 
      &      & 0.25 & 0.30& 0.068 &  0.007& 0.058 &  0.004& 0.062 &  0.005& 0.063 &  0.011\\ 
      &      & 0.30 & 0.35& 0.044 &  0.004& 0.041 &  0.003& 0.043 &  0.005& 0.043 &  0.007\\ 
      &      & 0.35 & 0.40& 0.030 &  0.003& 0.029 &  0.002& 0.029 &  0.003& 0.031 &  0.006\\ 
      &      & 0.40 & 0.45& 0.023 &  0.003& 0.023 &  0.002& 0.021 &  0.002& 0.020 &  0.005\\ 
      &      & 0.45 & 0.50& 0.018 &  0.002& 0.018 &  0.002& 0.017 &  0.002& 0.013 &  0.004\\ 
\hline  
 1.75 & 1.95 & 0.10 & 0.15& 0.062 &  0.009& 0.058 &  0.006& 0.050 &  0.006& 0.056 &  0.013\\ 
      &      & 0.15 & 0.20& 0.082 &  0.007& 0.085 &  0.006& 0.069 &  0.007& 0.080 &  0.013\\ 
      &      & 0.20 & 0.25& 0.063 &  0.007& 0.063 &  0.004& 0.049 &  0.004& 0.056 &  0.012\\ 
      &      & 0.25 & 0.30& 0.031 &  0.003& 0.041 &  0.003& 0.037 &  0.004& 0.051 &  0.011\\ 
      &      & 0.30 & 0.35& 0.030 &  0.003& 0.031 &  0.002& 0.027 &  0.003& 0.017 &  0.004\\ 
      &      & 0.35 & 0.40& 0.025 &  0.003& 0.019 &  0.002& 0.021 &  0.002& 0.016 &  0.004\\ 
      &      & 0.40 & 0.45& 0.017 &  0.003& 0.012 &  0.001& 0.014 &  0.002& 0.017 &  0.004\\ 
      &      & 0.45 & 0.50& 0.011 &  0.002& 0.010 &  0.001& 0.011 &  0.002& 0.012 &  0.004\\ 
\hline  
 1.95 & 2.15 & 0.10 & 0.15& 0.046 &  0.006& 0.047 &  0.005& 0.054 &  0.007& 0.056 &  0.013\\ 
      &      & 0.15 & 0.20& 0.061 &  0.006& 0.066 &  0.004& 0.054 &  0.005& 0.084 &  0.015\\ 
      &      & 0.20 & 0.25& 0.047 &  0.005& 0.044 &  0.003& 0.039 &  0.004& 0.033 &  0.008\\ 
      &      & 0.25 & 0.30& 0.034 &  0.004& 0.029 &  0.002& 0.025 &  0.003& 0.029 &  0.007\\ 
      &      & 0.30 & 0.35& 0.027 &  0.003& 0.020 &  0.002& 0.021 &  0.003& 0.010 &  0.004\\ 
      &      & 0.35 & 0.40& 0.017 &  0.003& 0.015 &  0.001& 0.014 &  0.002& 0.010 &  0.004\\ 
      &      & 0.40 & 0.45& 0.009 &  0.002& 0.011 &  0.001& 0.009 &  0.001& 0.017 &  0.006\\ 
      &      & 0.45 & 0.50& 0.007 &  0.001& 0.008 &  0.001& 0.006 &  0.001& 0.011 &  0.005\\

\end{tabular}
\end{center}
\end{table}
\clearpage
\begin{table}[hp!]
\begin{center}
  \caption{\label{tab:xsec-pimp-c}
    HARP results for the double-differential $\pi^+$ production
    cross-section in the laboratory system,
    $d^2\sigma^{\pi^+}/(dpd\theta)$ for $\pi^-$--C interactions. Each row refers to a
    different $(p_{\hbox{\small min}} \le p<p_{\hbox{\small max}},
    \theta_{\hbox{\small min}} \le \theta<\theta_{\hbox{\small max}})$ bin,
    where $p$ and $\theta$ are the pion momentum and polar angle, respectively.
    The central value as well as the square-root of the diagonal elements
    of the covariance matrix are given.}
\vspace{2mm}
\begin{tabular}{rrrr|r@{$\pm$}lr@{$\pm$}lr@{$\pm$}lr@{$\pm$}l}
\hline
$\theta_{\hbox{\small min}}$ &
$\theta_{\hbox{\small max}}$ &
$p_{\hbox{\small min}}$ &
$p_{\hbox{\small max}}$ &
\multicolumn{8}{c}{$d^2\sigma^{\pi^+}/(dpd\theta)$}
\\
(rad) & (rad) & (\GeVc) & (\GeVc) &
\multicolumn{8}{c}{($\barn/(\GeVc \cdot \rad)$)}
\\
  &  &  &
&\multicolumn{2}{c}{$ \bf{3 \ \GeVc}$}
&\multicolumn{2}{c}{$ \bf{5 \ \GeVc}$}
&\multicolumn{2}{c}{$ \bf{8 \ \GeVc}$}
&\multicolumn{2}{c}{$ \bf{12 \ \GeVc}$}
\\
\hline  
 0.35 & 0.55 & 0.15 & 0.20& 0.103 &  0.015& 0.138 &  0.022& 0.142 &  0.022& 0.153 &  0.022\\ 
      &      & 0.20 & 0.25& 0.128 &  0.011& 0.172 &  0.014& 0.168 &  0.014& 0.199 &  0.019\\ 
      &      & 0.25 & 0.30& 0.147 &  0.011& 0.196 &  0.016& 0.206 &  0.017& 0.243 &  0.018\\ 
      &      & 0.30 & 0.35& 0.156 &  0.011& 0.202 &  0.013& 0.224 &  0.015& 0.256 &  0.019\\ 
      &      & 0.35 & 0.40& 0.154 &  0.011& 0.225 &  0.019& 0.240 &  0.017& 0.276 &  0.016\\ 
      &      & 0.40 & 0.45& 0.160 &  0.008& 0.236 &  0.011& 0.255 &  0.013& 0.283 &  0.017\\ 
      &      & 0.45 & 0.50& 0.168 &  0.009& 0.230 &  0.012& 0.258 &  0.012& 0.297 &  0.014\\ 
      &      & 0.50 & 0.60& 0.191 &  0.012& 0.252 &  0.015& 0.272 &  0.015& 0.288 &  0.016\\ 
      &      & 0.60 & 0.70& 0.167 &  0.019& 0.251 &  0.024& 0.273 &  0.025& 0.281 &  0.026\\ 
      &      & 0.70 & 0.80& 0.126 &  0.018& 0.189 &  0.031& 0.227 &  0.033& 0.217 &  0.034\\ 
\hline  
 0.55 & 0.75 & 0.10 & 0.15& 0.086 &  0.020& 0.104 &  0.024& 0.084 &  0.023& 0.115 &  0.025\\ 
      &      & 0.15 & 0.20& 0.120 &  0.009& 0.148 &  0.013& 0.147 &  0.012& 0.165 &  0.014\\ 
      &      & 0.20 & 0.25& 0.172 &  0.014& 0.191 &  0.014& 0.183 &  0.013& 0.197 &  0.014\\ 
      &      & 0.25 & 0.30& 0.174 &  0.012& 0.201 &  0.015& 0.211 &  0.017& 0.232 &  0.017\\ 
      &      & 0.30 & 0.35& 0.175 &  0.010& 0.222 &  0.014& 0.222 &  0.013& 0.221 &  0.013\\ 
      &      & 0.35 & 0.40& 0.175 &  0.011& 0.223 &  0.011& 0.207 &  0.009& 0.218 &  0.011\\ 
      &      & 0.40 & 0.45& 0.177 &  0.009& 0.202 &  0.009& 0.223 &  0.014& 0.212 &  0.011\\ 
      &      & 0.45 & 0.50& 0.171 &  0.008& 0.194 &  0.010& 0.218 &  0.010& 0.219 &  0.012\\ 
      &      & 0.50 & 0.60& 0.149 &  0.011& 0.180 &  0.012& 0.196 &  0.013& 0.190 &  0.014\\ 
      &      & 0.60 & 0.70& 0.111 &  0.015& 0.141 &  0.018& 0.149 &  0.020& 0.138 &  0.018\\ 
      &      & 0.70 & 0.80& 0.070 &  0.015& 0.096 &  0.018& 0.094 &  0.019& 0.092 &  0.017\\ 
\hline  
 0.75 & 0.95 & 0.10 & 0.15& 0.087 &  0.015& 0.102 &  0.016& 0.101 &  0.015& 0.104 &  0.016\\ 
      &      & 0.15 & 0.20& 0.155 &  0.011& 0.155 &  0.011& 0.154 &  0.010& 0.177 &  0.015\\ 
      &      & 0.20 & 0.25& 0.168 &  0.009& 0.177 &  0.013& 0.176 &  0.013& 0.199 &  0.011\\ 
      &      & 0.25 & 0.30& 0.175 &  0.014& 0.186 &  0.010& 0.185 &  0.010& 0.190 &  0.012\\ 
      &      & 0.30 & 0.35& 0.164 &  0.008& 0.167 &  0.008& 0.174 &  0.010& 0.189 &  0.011\\ 
      &      & 0.35 & 0.40& 0.131 &  0.006& 0.157 &  0.007& 0.162 &  0.007& 0.167 &  0.008\\ 
      &      & 0.40 & 0.45& 0.123 &  0.007& 0.154 &  0.008& 0.146 &  0.007& 0.153 &  0.006\\ 
      &      & 0.45 & 0.50& 0.120 &  0.006& 0.136 &  0.007& 0.138 &  0.007& 0.133 &  0.006\\ 
      &      & 0.50 & 0.60& 0.086 &  0.008& 0.109 &  0.010& 0.110 &  0.009& 0.102 &  0.009\\ 
      &      & 0.60 & 0.70& 0.057 &  0.009& 0.069 &  0.011& 0.071 &  0.011& 0.070 &  0.009\\ 
\hline  
 0.95 & 1.15 & 0.10 & 0.15& 0.108 &  0.013& 0.089 &  0.014& 0.093 &  0.012& 0.099 &  0.013\\ 
      &      & 0.15 & 0.20& 0.150 &  0.012& 0.166 &  0.011& 0.146 &  0.011& 0.173 &  0.014\\ 
      &      & 0.20 & 0.25& 0.150 &  0.007& 0.165 &  0.012& 0.152 &  0.009& 0.164 &  0.009\\ 
      &      & 0.25 & 0.30& 0.125 &  0.007& 0.148 &  0.008& 0.151 &  0.008& 0.143 &  0.008\\ 
      &      & 0.30 & 0.35& 0.118 &  0.006& 0.127 &  0.007& 0.128 &  0.006& 0.110 &  0.007\\ 
      &      & 0.35 & 0.40& 0.094 &  0.005& 0.111 &  0.006& 0.111 &  0.005& 0.107 &  0.007\\ 
      &      & 0.40 & 0.45& 0.087 &  0.004& 0.090 &  0.005& 0.089 &  0.005& 0.094 &  0.005\\ 
      &      & 0.45 & 0.50& 0.070 &  0.006& 0.075 &  0.006& 0.069 &  0.005& 0.078 &  0.006\\ 
      &      & 0.50 & 0.60& 0.044 &  0.006& 0.055 &  0.005& 0.048 &  0.005& 0.052 &  0.006\\ 
\hline
\end{tabular}
\end{center}
\end{table}

\begin{table}[hp!]
\begin{center}
\begin{tabular}{rrrr|r@{$\pm$}lr@{$\pm$}lr@{$\pm$}lr@{$\pm$}l}
\hline
$\theta_{\hbox{\small min}}$ &
$\theta_{\hbox{\small max}}$ &
$p_{\hbox{\small min}}$ &
$p_{\hbox{\small max}}$ &
\multicolumn{8}{c}{$d^2\sigma^{\pi^+}/(dpd\theta)$}
\\
(rad) & (rad) & (\GeVc) & (\GeVc) &
\multicolumn{8}{c}{(\barn/($\GeVc \cdot \rad$))}
\\
  &  &  &
&\multicolumn{2}{c}{$ \bf{3 \ \GeVc}$}
&\multicolumn{2}{c}{$ \bf{5 \ \GeVc}$}
&\multicolumn{2}{c}{$ \bf{8 \ \GeVc}$}
&\multicolumn{2}{c}{$ \bf{12 \ \GeVc}$}
\\
\hline
 1.15 & 1.35 & 0.10 & 0.15& 0.095 &  0.013& 0.111 &  0.013& 0.086 &  0.012& 0.109 &  0.014\\ 
      &      & 0.15 & 0.20& 0.140 &  0.009& 0.138 &  0.010& 0.145 &  0.012& 0.136 &  0.009\\ 
      &      & 0.20 & 0.25& 0.122 &  0.006& 0.137 &  0.008& 0.140 &  0.007& 0.139 &  0.008\\ 
      &      & 0.25 & 0.30& 0.094 &  0.006& 0.119 &  0.007& 0.109 &  0.006& 0.104 &  0.007\\ 
      &      & 0.30 & 0.35& 0.078 &  0.004& 0.090 &  0.006& 0.088 &  0.005& 0.088 &  0.006\\ 
      &      & 0.35 & 0.40& 0.068 &  0.004& 0.068 &  0.004& 0.071 &  0.004& 0.071 &  0.004\\ 
      &      & 0.40 & 0.45& 0.053 &  0.004& 0.058 &  0.004& 0.058 &  0.004& 0.057 &  0.004\\ 
      &      & 0.45 & 0.50& 0.041 &  0.004& 0.047 &  0.004& 0.042 &  0.005& 0.044 &  0.005\\ 
\hline  
 1.35 & 1.55 & 0.10 & 0.15& 0.092 &  0.010& 0.085 &  0.013& 0.090 &  0.011& 0.094 &  0.011\\ 
      &      & 0.15 & 0.20& 0.119 &  0.009& 0.129 &  0.009& 0.121 &  0.007& 0.126 &  0.011\\ 
      &      & 0.20 & 0.25& 0.113 &  0.006& 0.110 &  0.006& 0.109 &  0.006& 0.115 &  0.008\\ 
      &      & 0.25 & 0.30& 0.087 &  0.006& 0.089 &  0.006& 0.081 &  0.005& 0.080 &  0.006\\ 
      &      & 0.30 & 0.35& 0.057 &  0.004& 0.061 &  0.004& 0.062 &  0.004& 0.056 &  0.004\\ 
      &      & 0.35 & 0.40& 0.049 &  0.003& 0.045 &  0.003& 0.046 &  0.003& 0.044 &  0.003\\ 
      &      & 0.40 & 0.45& 0.035 &  0.003& 0.034 &  0.003& 0.035 &  0.003& 0.034 &  0.003\\ 
      &      & 0.45 & 0.50& 0.024 &  0.003& 0.026 &  0.002& 0.025 &  0.003& 0.023 &  0.003\\ 
\hline  
 1.55 & 1.75 & 0.10 & 0.15& 0.086 &  0.011& 0.082 &  0.011& 0.082 &  0.010& 0.092 &  0.013\\ 
      &      & 0.15 & 0.20& 0.105 &  0.007& 0.112 &  0.010& 0.101 &  0.007& 0.109 &  0.008\\ 
      &      & 0.20 & 0.25& 0.087 &  0.006& 0.098 &  0.007& 0.085 &  0.005& 0.080 &  0.006\\ 
      &      & 0.25 & 0.30& 0.066 &  0.004& 0.066 &  0.005& 0.066 &  0.004& 0.063 &  0.006\\ 
      &      & 0.30 & 0.35& 0.052 &  0.004& 0.043 &  0.004& 0.043 &  0.003& 0.036 &  0.003\\ 
      &      & 0.35 & 0.40& 0.036 &  0.003& 0.033 &  0.003& 0.030 &  0.003& 0.028 &  0.002\\ 
      &      & 0.40 & 0.45& 0.023 &  0.002& 0.023 &  0.002& 0.019 &  0.002& 0.019 &  0.002\\ 
      &      & 0.45 & 0.50& 0.018 &  0.002& 0.015 &  0.002& 0.011 &  0.002& 0.013 &  0.002\\ 
\hline  
 1.75 & 1.95 & 0.10 & 0.15& 0.075 &  0.009& 0.087 &  0.010& 0.074 &  0.009& 0.088 &  0.011\\ 
      &      & 0.15 & 0.20& 0.100 &  0.007& 0.091 &  0.006& 0.094 &  0.006& 0.079 &  0.006\\ 
      &      & 0.20 & 0.25& 0.075 &  0.005& 0.060 &  0.004& 0.067 &  0.005& 0.066 &  0.005\\ 
      &      & 0.25 & 0.30& 0.045 &  0.004& 0.047 &  0.003& 0.044 &  0.003& 0.048 &  0.005\\ 
      &      & 0.30 & 0.35& 0.035 &  0.003& 0.030 &  0.003& 0.027 &  0.002& 0.029 &  0.003\\ 
      &      & 0.35 & 0.40& 0.022 &  0.002& 0.019 &  0.002& 0.019 &  0.002& 0.018 &  0.002\\ 
      &      & 0.40 & 0.45& 0.014 &  0.002& 0.013 &  0.002& 0.013 &  0.002& 0.011 &  0.002\\ 
      &      & 0.45 & 0.50& 0.010 &  0.001& 0.007 &  0.002& 0.008 &  0.002& 0.007 &  0.001\\ 
\hline  
 1.95 & 2.15 & 0.10 & 0.15& 0.067 &  0.008& 0.061 &  0.008& 0.056 &  0.006& 0.064 &  0.009\\ 
      &      & 0.15 & 0.20& 0.076 &  0.005& 0.073 &  0.005& 0.076 &  0.006& 0.072 &  0.006\\ 
      &      & 0.20 & 0.25& 0.055 &  0.004& 0.056 &  0.004& 0.047 &  0.005& 0.060 &  0.005\\ 
      &      & 0.25 & 0.30& 0.034 &  0.003& 0.032 &  0.004& 0.026 &  0.002& 0.035 &  0.004\\ 
      &      & 0.30 & 0.35& 0.023 &  0.002& 0.018 &  0.002& 0.022 &  0.002& 0.019 &  0.003\\ 
      &      & 0.35 & 0.40& 0.015 &  0.002& 0.015 &  0.002& 0.013 &  0.002& 0.012 &  0.002\\ 
      &      & 0.40 & 0.45& 0.010 &  0.001& 0.010 &  0.002& 0.008 &  0.001& 0.008 &  0.001\\ 
      &      & 0.45 & 0.50& 0.006 &  0.001& 0.005 &  0.001& 0.004 &  0.001& 0.004 &  0.001\\ 

\end{tabular}
\end{center}
\end{table}

\begin{table}[hp!]
\begin{center}
  \caption{\label{tab:xsec-pimm-c}
    HARP results for the double-differential $\pi^-$ production
    cross-section in the laboratory system,
    $d^2\sigma^{\pi^-}/(dpd\theta)$ for $\pi^-$--C interactions. Each row refers to a
    different $(p_{\hbox{\small min}} \le p<p_{\hbox{\small max}},
    \theta_{\hbox{\small min}} \le \theta<\theta_{\hbox{\small max}})$ bin,
    where $p$ and $\theta$ are the pion momentum and polar angle, respectively.
    The central value as well as the square-root of the diagonal elements
    of the covariance matrix are given.}
\vspace{2mm}
\begin{tabular}{rrrr|r@{$\pm$}lr@{$\pm$}lr@{$\pm$}lr@{$\pm$}l}
\hline
$\theta_{\hbox{\small min}}$ &
$\theta_{\hbox{\small max}}$ &
$p_{\hbox{\small min}}$ &
$p_{\hbox{\small max}}$ &
\multicolumn{8}{c}{$d^2\sigma^{\pi^-}/(dpd\theta)$}
\\
(rad) & (rad) & (\GeVc) & (\GeVc) &
\multicolumn{8}{c}{($\barn/(\GeVc \cdot \rad)$)}
\\
  &  &  &
&\multicolumn{2}{c}{$ \bf{3 \ \GeVc}$}
&\multicolumn{2}{c}{$ \bf{5 \ \GeVc}$}
&\multicolumn{2}{c}{$ \bf{8 \ \GeVc}$}
&\multicolumn{2}{c}{$ \bf{12 \ \GeVc}$}
\\
\hline  
 0.35 & 0.55 & 0.15 & 0.20& 0.138 &  0.018& 0.174 &  0.023& 0.173 &  0.026& 0.201 &  0.026\\ 
      &      & 0.20 & 0.25& 0.189 &  0.015& 0.217 &  0.019& 0.219 &  0.018& 0.238 &  0.019\\ 
      &      & 0.25 & 0.30& 0.210 &  0.015& 0.265 &  0.020& 0.267 &  0.022& 0.266 &  0.018\\ 
      &      & 0.30 & 0.35& 0.231 &  0.018& 0.272 &  0.017& 0.272 &  0.014& 0.279 &  0.020\\ 
      &      & 0.35 & 0.40& 0.235 &  0.013& 0.271 &  0.015& 0.275 &  0.018& 0.265 &  0.017\\ 
      &      & 0.40 & 0.45& 0.228 &  0.012& 0.282 &  0.020& 0.301 &  0.020& 0.283 &  0.016\\ 
      &      & 0.45 & 0.50& 0.243 &  0.012& 0.294 &  0.017& 0.309 &  0.015& 0.291 &  0.017\\ 
      &      & 0.50 & 0.60& 0.255 &  0.014& 0.308 &  0.017& 0.317 &  0.018& 0.307 &  0.019\\ 
      &      & 0.60 & 0.70& 0.262 &  0.022& 0.317 &  0.025& 0.315 &  0.024& 0.291 &  0.024\\ 
      &      & 0.70 & 0.80& 0.212 &  0.028& 0.269 &  0.033& 0.285 &  0.033& 0.265 &  0.031\\ 
\hline  
 0.55 & 0.75 & 0.10 & 0.15& 0.126 &  0.025& 0.152 &  0.031& 0.113 &  0.025& 0.117 &  0.027\\ 
      &      & 0.15 & 0.20& 0.204 &  0.015& 0.212 &  0.016& 0.193 &  0.016& 0.206 &  0.014\\ 
      &      & 0.20 & 0.25& 0.251 &  0.016& 0.254 &  0.016& 0.241 &  0.015& 0.205 &  0.017\\ 
      &      & 0.25 & 0.30& 0.255 &  0.015& 0.285 &  0.022& 0.261 &  0.018& 0.262 &  0.020\\ 
      &      & 0.30 & 0.35& 0.259 &  0.018& 0.297 &  0.016& 0.268 &  0.015& 0.240 &  0.013\\ 
      &      & 0.35 & 0.40& 0.256 &  0.013& 0.264 &  0.013& 0.247 &  0.012& 0.221 &  0.012\\ 
      &      & 0.40 & 0.45& 0.246 &  0.012& 0.255 &  0.012& 0.240 &  0.011& 0.219 &  0.012\\ 
      &      & 0.45 & 0.50& 0.252 &  0.011& 0.246 &  0.011& 0.234 &  0.010& 0.214 &  0.010\\ 
      &      & 0.50 & 0.60& 0.234 &  0.013& 0.243 &  0.013& 0.226 &  0.012& 0.203 &  0.012\\ 
      &      & 0.60 & 0.70& 0.203 &  0.017& 0.224 &  0.018& 0.195 &  0.017& 0.167 &  0.018\\ 
      &      & 0.70 & 0.80& 0.161 &  0.022& 0.192 &  0.026& 0.160 &  0.022& 0.134 &  0.017\\ 
\hline  
 0.75 & 0.95 & 0.10 & 0.15& 0.144 &  0.020& 0.139 &  0.021& 0.099 &  0.015& 0.108 &  0.018\\ 
      &      & 0.15 & 0.20& 0.249 &  0.016& 0.250 &  0.019& 0.214 &  0.016& 0.208 &  0.014\\ 
      &      & 0.20 & 0.25& 0.271 &  0.014& 0.270 &  0.014& 0.228 &  0.013& 0.221 &  0.014\\ 
      &      & 0.25 & 0.30& 0.250 &  0.014& 0.238 &  0.014& 0.217 &  0.012& 0.195 &  0.011\\ 
      &      & 0.30 & 0.35& 0.236 &  0.013& 0.247 &  0.014& 0.202 &  0.009& 0.196 &  0.011\\ 
      &      & 0.35 & 0.40& 0.219 &  0.010& 0.218 &  0.010& 0.179 &  0.008& 0.172 &  0.008\\ 
      &      & 0.40 & 0.45& 0.204 &  0.008& 0.195 &  0.008& 0.164 &  0.007& 0.160 &  0.008\\ 
      &      & 0.45 & 0.50& 0.183 &  0.009& 0.180 &  0.008& 0.158 &  0.007& 0.139 &  0.007\\ 
      &      & 0.50 & 0.60& 0.152 &  0.009& 0.157 &  0.009& 0.136 &  0.008& 0.117 &  0.007\\ 
      &      & 0.60 & 0.70& 0.119 &  0.013& 0.127 &  0.013& 0.101 &  0.011& 0.094 &  0.009\\ 
\hline  
 0.95 & 1.15 & 0.10 & 0.15& 0.163 &  0.019& 0.140 &  0.019& 0.129 &  0.015& 0.117 &  0.013\\ 
      &      & 0.15 & 0.20& 0.269 &  0.017& 0.251 &  0.015& 0.202 &  0.012& 0.193 &  0.015\\ 
      &      & 0.20 & 0.25& 0.252 &  0.013& 0.230 &  0.012& 0.195 &  0.010& 0.186 &  0.012\\ 
      &      & 0.25 & 0.30& 0.221 &  0.011& 0.217 &  0.012& 0.178 &  0.008& 0.160 &  0.008\\ 
      &      & 0.30 & 0.35& 0.190 &  0.008& 0.177 &  0.008& 0.145 &  0.007& 0.136 &  0.007\\ 
      &      & 0.35 & 0.40& 0.164 &  0.008& 0.147 &  0.007& 0.121 &  0.005& 0.118 &  0.006\\ 
      &      & 0.40 & 0.45& 0.140 &  0.006& 0.128 &  0.006& 0.106 &  0.004& 0.094 &  0.005\\ 
      &      & 0.45 & 0.50& 0.122 &  0.006& 0.113 &  0.006& 0.095 &  0.004& 0.081 &  0.005\\ 
      &      & 0.50 & 0.60& 0.098 &  0.006& 0.089 &  0.006& 0.076 &  0.005& 0.065 &  0.004\\ 
\hline
\end{tabular}
\end{center}
\end{table}

\begin{table}[hp!]
\begin{center}
\begin{tabular}{rrrr|r@{$\pm$}lr@{$\pm$}lr@{$\pm$}lr@{$\pm$}l}
\hline
$\theta_{\hbox{\small min}}$ &
$\theta_{\hbox{\small max}}$ &
$p_{\hbox{\small min}}$ &
$p_{\hbox{\small max}}$ &
\multicolumn{8}{c}{$d^2\sigma^{\pi^-}/(dpd\theta)$}
\\
(rad) & (rad) & (\GeVc) & (\GeVc) &
\multicolumn{8}{c}{(\barn/($\GeVc \cdot \rad$))}
\\
  &  &  &
&\multicolumn{2}{c}{$ \bf{3 \ \GeVc}$}
&\multicolumn{2}{c}{$ \bf{5 \ \GeVc}$}
&\multicolumn{2}{c}{$ \bf{8 \ \GeVc}$}
&\multicolumn{2}{c}{$ \bf{12 \ \GeVc}$}
\\
\hline
 1.15 & 1.35 & 0.10 & 0.15& 0.182 &  0.019& 0.179 &  0.018& 0.140 &  0.014& 0.111 &  0.012\\ 
      &      & 0.15 & 0.20& 0.253 &  0.015& 0.224 &  0.013& 0.179 &  0.010& 0.185 &  0.014\\ 
      &      & 0.20 & 0.25& 0.227 &  0.011& 0.205 &  0.011& 0.168 &  0.008& 0.157 &  0.009\\ 
      &      & 0.25 & 0.30& 0.179 &  0.009& 0.166 &  0.009& 0.129 &  0.007& 0.119 &  0.008\\ 
      &      & 0.30 & 0.35& 0.143 &  0.007& 0.138 &  0.008& 0.108 &  0.005& 0.096 &  0.006\\ 
      &      & 0.35 & 0.40& 0.119 &  0.006& 0.101 &  0.006& 0.088 &  0.005& 0.082 &  0.005\\ 
      &      & 0.40 & 0.45& 0.100 &  0.006& 0.078 &  0.005& 0.067 &  0.004& 0.062 &  0.005\\ 
      &      & 0.45 & 0.50& 0.081 &  0.005& 0.062 &  0.004& 0.056 &  0.004& 0.049 &  0.003\\ 
\hline  
 1.35 & 1.55 & 0.10 & 0.15& 0.175 &  0.020& 0.157 &  0.016& 0.125 &  0.012& 0.102 &  0.014\\ 
      &      & 0.15 & 0.20& 0.233 &  0.013& 0.209 &  0.013& 0.165 &  0.010& 0.155 &  0.009\\ 
      &      & 0.20 & 0.25& 0.206 &  0.010& 0.164 &  0.009& 0.139 &  0.007& 0.137 &  0.009\\ 
      &      & 0.25 & 0.30& 0.145 &  0.009& 0.120 &  0.007& 0.102 &  0.006& 0.097 &  0.008\\ 
      &      & 0.30 & 0.35& 0.102 &  0.006& 0.102 &  0.006& 0.080 &  0.005& 0.074 &  0.005\\ 
      &      & 0.35 & 0.40& 0.079 &  0.005& 0.078 &  0.006& 0.061 &  0.004& 0.065 &  0.004\\ 
      &      & 0.40 & 0.45& 0.063 &  0.004& 0.055 &  0.005& 0.047 &  0.003& 0.050 &  0.004\\ 
      &      & 0.45 & 0.50& 0.050 &  0.004& 0.040 &  0.004& 0.035 &  0.002& 0.034 &  0.004\\ 
\hline  
 1.55 & 1.75 & 0.10 & 0.15& 0.159 &  0.017& 0.147 &  0.015& 0.103 &  0.011& 0.118 &  0.014\\ 
      &      & 0.15 & 0.20& 0.205 &  0.011& 0.185 &  0.011& 0.140 &  0.008& 0.139 &  0.009\\ 
      &      & 0.20 & 0.25& 0.148 &  0.009& 0.148 &  0.009& 0.109 &  0.006& 0.110 &  0.007\\ 
      &      & 0.25 & 0.30& 0.106 &  0.007& 0.100 &  0.007& 0.090 &  0.005& 0.071 &  0.006\\ 
      &      & 0.30 & 0.35& 0.073 &  0.005& 0.071 &  0.005& 0.065 &  0.005& 0.051 &  0.004\\ 
      &      & 0.35 & 0.40& 0.056 &  0.004& 0.052 &  0.004& 0.040 &  0.004& 0.043 &  0.003\\ 
      &      & 0.40 & 0.45& 0.044 &  0.003& 0.035 &  0.003& 0.028 &  0.002& 0.035 &  0.003\\ 
      &      & 0.45 & 0.50& 0.034 &  0.003& 0.023 &  0.003& 0.022 &  0.002& 0.023 &  0.003\\ 
\hline  
 1.75 & 1.95 & 0.10 & 0.15& 0.158 &  0.013& 0.137 &  0.014& 0.097 &  0.011& 0.097 &  0.009\\ 
      &      & 0.15 & 0.20& 0.176 &  0.010& 0.158 &  0.009& 0.123 &  0.007& 0.112 &  0.008\\ 
      &      & 0.20 & 0.25& 0.133 &  0.008& 0.106 &  0.007& 0.090 &  0.006& 0.084 &  0.006\\ 
      &      & 0.25 & 0.30& 0.075 &  0.006& 0.065 &  0.005& 0.052 &  0.004& 0.052 &  0.005\\ 
      &      & 0.30 & 0.35& 0.049 &  0.004& 0.052 &  0.003& 0.042 &  0.003& 0.031 &  0.003\\ 
      &      & 0.35 & 0.40& 0.037 &  0.003& 0.037 &  0.003& 0.029 &  0.003& 0.027 &  0.002\\ 
      &      & 0.40 & 0.45& 0.027 &  0.002& 0.026 &  0.002& 0.021 &  0.002& 0.021 &  0.002\\ 
      &      & 0.45 & 0.50& 0.019 &  0.002& 0.019 &  0.002& 0.016 &  0.002& 0.014 &  0.002\\ 
\hline  
 1.95 & 2.15 & 0.10 & 0.15& 0.137 &  0.014& 0.115 &  0.012& 0.084 &  0.008& 0.079 &  0.009\\ 
      &      & 0.15 & 0.20& 0.152 &  0.008& 0.119 &  0.007& 0.104 &  0.006& 0.096 &  0.007\\ 
      &      & 0.20 & 0.25& 0.087 &  0.007& 0.087 &  0.005& 0.063 &  0.004& 0.065 &  0.005\\ 
      &      & 0.25 & 0.30& 0.054 &  0.004& 0.047 &  0.005& 0.044 &  0.003& 0.038 &  0.004\\ 
      &      & 0.30 & 0.35& 0.041 &  0.003& 0.033 &  0.003& 0.027 &  0.003& 0.028 &  0.003\\ 
      &      & 0.35 & 0.40& 0.030 &  0.003& 0.027 &  0.002& 0.020 &  0.001& 0.020 &  0.002\\ 
      &      & 0.40 & 0.45& 0.019 &  0.002& 0.019 &  0.002& 0.015 &  0.002& 0.013 &  0.002\\ 
      &      & 0.45 & 0.50& 0.012 &  0.002& 0.011 &  0.002& 0.010 &  0.001& 0.007 &  0.002\\

\end{tabular}
\end{center}
\end{table}
\clearpage
\begin{table}[hp!]
\begin{center}
  \caption{\label{tab:xsec-pipp-al}
    HARP results for the double-differential $\pi^+$ production
    cross-section in the laboratory system,
    $d^2\sigma^{\pi^+}/(dpd\theta)$ for $\pi^+$--Al interactions. Each row refers to a
    different $(p_{\hbox{\small min}} \le p<p_{\hbox{\small max}},
    \theta_{\hbox{\small min}} \le \theta<\theta_{\hbox{\small max}})$ bin,
    where $p$ and $\theta$ are the pion momentum and polar angle, respectively.
    The central value as well as the square-root of the diagonal elements
    of the covariance matrix are given.}
\vspace{2mm}
\begin{tabular}{rrrr|r@{$\pm$}lr@{$\pm$}lr@{$\pm$}lr@{$\pm$}lr@{$\pm$}l}
\hline
$\theta_{\hbox{\small min}}$ &
$\theta_{\hbox{\small max}}$ &
$p_{\hbox{\small min}}$ &
$p_{\hbox{\small max}}$ &
\multicolumn{10}{c}{$d^2\sigma^{\pi^+}/(dpd\theta)$}
\\
(rad) & (rad) & (\GeVc) & (\GeVc) &
\multicolumn{10}{c}{($\barn/(\GeVc \cdot \rad)$)}
\\
  &  &  &
&\multicolumn{2}{c}{$ \bf{3 \ \GeVc}$}
&\multicolumn{2}{c}{$ \bf{5 \ \GeVc}$}
&\multicolumn{2}{c}{$ \bf{8 \ \GeVc}$}
&\multicolumn{2}{c}{$ \bf{12 \ \GeVc}$}
&\multicolumn{2}{c}{$ \bf{12.9 \ \GeVc}$}
\\
\hline  
 0.35 & 0.55 & 0.15 & 0.20& 0.247 &  0.045& 0.301 &  0.061& 0.354 &  0.072& 0.267 &  0.103& 0.391 &  0.082\\ 
      &      & 0.20 & 0.25& 0.296 &  0.030& 0.407 &  0.041& 0.480 &  0.049& 0.615 &  0.102& 0.505 &  0.049\\ 
      &      & 0.25 & 0.30& 0.401 &  0.037& 0.514 &  0.043& 0.569 &  0.056& 0.659 &  0.080& 0.616 &  0.059\\ 
      &      & 0.30 & 0.35& 0.398 &  0.028& 0.587 &  0.049& 0.683 &  0.056& 0.716 &  0.098& 0.711 &  0.051\\ 
      &      & 0.35 & 0.40& 0.444 &  0.043& 0.617 &  0.037& 0.667 &  0.037& 0.687 &  0.067& 0.721 &  0.045\\ 
      &      & 0.40 & 0.45& 0.527 &  0.035& 0.622 &  0.034& 0.741 &  0.071& 0.673 &  0.079& 0.728 &  0.031\\ 
      &      & 0.45 & 0.50& 0.491 &  0.026& 0.667 &  0.038& 0.817 &  0.043& 0.770 &  0.072& 0.738 &  0.050\\ 
      &      & 0.50 & 0.60& 0.477 &  0.029& 0.719 &  0.039& 0.828 &  0.047& 0.793 &  0.072& 0.823 &  0.044\\ 
      &      & 0.60 & 0.70& 0.486 &  0.042& 0.620 &  0.056& 0.793 &  0.078& 0.773 &  0.090& 0.769 &  0.085\\ 
      &      & 0.70 & 0.80& 0.353 &  0.063& 0.544 &  0.068& 0.619 &  0.098& 0.659 &  0.109& 0.572 &  0.099\\ 
\hline  
 0.55 & 0.75 & 0.10 & 0.15& 0.211 &  0.067& 0.221 &  0.069& 0.180 &  0.066& 0.289 &  0.113& 0.227 &  0.076\\ 
      &      & 0.15 & 0.20& 0.382 &  0.036& 0.384 &  0.037& 0.406 &  0.049& 0.368 &  0.074& 0.394 &  0.056\\ 
      &      & 0.20 & 0.25& 0.424 &  0.036& 0.543 &  0.059& 0.563 &  0.045& 0.482 &  0.071& 0.602 &  0.042\\ 
      &      & 0.25 & 0.30& 0.501 &  0.046& 0.567 &  0.032& 0.653 &  0.056& 0.575 &  0.071& 0.583 &  0.030\\ 
      &      & 0.30 & 0.35& 0.517 &  0.033& 0.596 &  0.035& 0.663 &  0.034& 0.756 &  0.089& 0.586 &  0.045\\ 
      &      & 0.35 & 0.40& 0.563 &  0.033& 0.580 &  0.037& 0.661 &  0.041& 0.581 &  0.057& 0.674 &  0.040\\ 
      &      & 0.40 & 0.45& 0.549 &  0.031& 0.588 &  0.027& 0.666 &  0.037& 0.685 &  0.067& 0.615 &  0.025\\ 
      &      & 0.45 & 0.50& 0.509 &  0.027& 0.582 &  0.026& 0.613 &  0.034& 0.562 &  0.056& 0.576 &  0.023\\ 
      &      & 0.50 & 0.60& 0.434 &  0.033& 0.534 &  0.031& 0.570 &  0.037& 0.518 &  0.054& 0.506 &  0.034\\ 
      &      & 0.60 & 0.70& 0.331 &  0.042& 0.398 &  0.047& 0.435 &  0.048& 0.377 &  0.066& 0.366 &  0.050\\ 
      &      & 0.70 & 0.80& 0.207 &  0.045& 0.291 &  0.054& 0.344 &  0.055& 0.237 &  0.052& 0.256 &  0.047\\ 
\hline  
 0.75 & 0.95 & 0.10 & 0.15& 0.279 &  0.055& 0.284 &  0.055& 0.265 &  0.057& 0.234 &  0.071& 0.294 &  0.054\\ 
      &      & 0.15 & 0.20& 0.467 &  0.038& 0.482 &  0.040& 0.527 &  0.043& 0.468 &  0.077& 0.451 &  0.035\\ 
      &      & 0.20 & 0.25& 0.596 &  0.041& 0.548 &  0.031& 0.544 &  0.033& 0.617 &  0.069& 0.540 &  0.043\\ 
      &      & 0.25 & 0.30& 0.525 &  0.032& 0.525 &  0.028& 0.562 &  0.042& 0.593 &  0.075& 0.558 &  0.032\\ 
      &      & 0.30 & 0.35& 0.524 &  0.028& 0.517 &  0.028& 0.566 &  0.031& 0.512 &  0.055& 0.498 &  0.021\\ 
      &      & 0.35 & 0.40& 0.414 &  0.021& 0.463 &  0.021& 0.507 &  0.026& 0.410 &  0.042& 0.462 &  0.022\\ 
      &      & 0.40 & 0.45& 0.387 &  0.019& 0.423 &  0.018& 0.422 &  0.021& 0.402 &  0.043& 0.406 &  0.018\\ 
      &      & 0.45 & 0.50& 0.340 &  0.018& 0.386 &  0.019& 0.395 &  0.021& 0.331 &  0.041& 0.345 &  0.023\\ 
      &      & 0.50 & 0.60& 0.266 &  0.022& 0.289 &  0.026& 0.326 &  0.027& 0.262 &  0.034& 0.261 &  0.025\\ 
      &      & 0.60 & 0.70& 0.189 &  0.024& 0.185 &  0.029& 0.212 &  0.037& 0.155 &  0.040& 0.160 &  0.024\\ 
\hline  
 0.95 & 1.15 & 0.10 & 0.15& 0.310 &  0.049& 0.328 &  0.050& 0.292 &  0.046& 0.294 &  0.065& 0.313 &  0.045\\ 
      &      & 0.15 & 0.20& 0.499 &  0.038& 0.505 &  0.033& 0.525 &  0.037& 0.478 &  0.075& 0.495 &  0.032\\ 
      &      & 0.20 & 0.25& 0.537 &  0.032& 0.523 &  0.028& 0.467 &  0.028& 0.469 &  0.058& 0.521 &  0.031\\ 
      &      & 0.25 & 0.30& 0.442 &  0.026& 0.446 &  0.026& 0.475 &  0.027& 0.414 &  0.052& 0.386 &  0.021\\ 
      &      & 0.30 & 0.35& 0.371 &  0.021& 0.383 &  0.020& 0.360 &  0.024& 0.315 &  0.041& 0.307 &  0.014\\ 
      &      & 0.35 & 0.40& 0.329 &  0.019& 0.308 &  0.014& 0.285 &  0.017& 0.240 &  0.030& 0.268 &  0.012\\ 
      &      & 0.40 & 0.45& 0.283 &  0.017& 0.266 &  0.012& 0.255 &  0.014& 0.243 &  0.033& 0.239 &  0.013\\ 
      &      & 0.45 & 0.50& 0.229 &  0.017& 0.211 &  0.014& 0.222 &  0.016& 0.184 &  0.030& 0.197 &  0.016\\ 
      &      & 0.50 & 0.60& 0.161 &  0.017& 0.150 &  0.018& 0.154 &  0.018& 0.113 &  0.021& 0.137 &  0.015\\ 
\hline
\end{tabular}
\end{center}
\end{table}

\begin{table}[hp!]
\begin{center}
\begin{tabular}{rrrr|r@{$\pm$}lr@{$\pm$}lr@{$\pm$}lr@{$\pm$}lr@{$\pm$}l}
\hline
$\theta_{\hbox{\small min}}$ &
$\theta_{\hbox{\small max}}$ &
$p_{\hbox{\small min}}$ &
$p_{\hbox{\small max}}$ &
\multicolumn{10}{c}{$d^2\sigma^{\pi^+}/(dpd\theta)$}
\\
(rad) & (rad) & (\GeVc) & (\GeVc) &
\multicolumn{10}{c}{($\barn/(\GeVc \cdot \rad)$)}
\\
  &  &  &
&\multicolumn{2}{c}{$ \bf{3 \ \GeVc}$}
&\multicolumn{2}{c}{$ \bf{5 \ \GeVc}$}
&\multicolumn{2}{c}{$ \bf{8 \ \GeVc}$}
&\multicolumn{2}{c}{$ \bf{12 \ \GeVc}$}
&\multicolumn{2}{c}{$ \bf{12.9 \ \GeVc}$}
\\
\hline
 1.15 & 1.35 & 0.10 & 0.15& 0.345 &  0.046& 0.368 &  0.051& 0.355 &  0.046& 0.212 &  0.051& 0.304 &  0.046\\ 
      &      & 0.15 & 0.20& 0.471 &  0.034& 0.511 &  0.024& 0.442 &  0.031& 0.299 &  0.045& 0.476 &  0.027\\ 
      &      & 0.20 & 0.25& 0.447 &  0.029& 0.408 &  0.023& 0.418 &  0.023& 0.304 &  0.047& 0.388 &  0.021\\ 
      &      & 0.25 & 0.30& 0.396 &  0.026& 0.318 &  0.018& 0.334 &  0.021& 0.343 &  0.046& 0.322 &  0.018\\ 
      &      & 0.30 & 0.35& 0.264 &  0.018& 0.288 &  0.016& 0.253 &  0.015& 0.289 &  0.038& 0.233 &  0.014\\ 
      &      & 0.35 & 0.40& 0.212 &  0.012& 0.207 &  0.015& 0.219 &  0.013& 0.195 &  0.031& 0.192 &  0.011\\ 
      &      & 0.40 & 0.45& 0.166 &  0.013& 0.158 &  0.011& 0.176 &  0.013& 0.140 &  0.022& 0.137 &  0.011\\ 
      &      & 0.45 & 0.50& 0.122 &  0.013& 0.122 &  0.010& 0.127 &  0.012& 0.110 &  0.020& 0.104 &  0.010\\ 
\hline  
 1.35 & 1.55 & 0.10 & 0.15& 0.379 &  0.051& 0.337 &  0.045& 0.333 &  0.049& 0.301 &  0.066& 0.313 &  0.050\\ 
      &      & 0.15 & 0.20& 0.495 &  0.030& 0.432 &  0.025& 0.474 &  0.030& 0.445 &  0.068& 0.402 &  0.024\\ 
      &      & 0.20 & 0.25& 0.350 &  0.024& 0.342 &  0.017& 0.355 &  0.022& 0.342 &  0.054& 0.308 &  0.017\\ 
      &      & 0.25 & 0.30& 0.226 &  0.017& 0.236 &  0.013& 0.242 &  0.017& 0.237 &  0.036& 0.213 &  0.014\\ 
      &      & 0.30 & 0.35& 0.192 &  0.013& 0.205 &  0.012& 0.197 &  0.014& 0.160 &  0.027& 0.157 &  0.011\\ 
      &      & 0.35 & 0.40& 0.155 &  0.012& 0.131 &  0.010& 0.142 &  0.011& 0.136 &  0.025& 0.120 &  0.008\\ 
      &      & 0.40 & 0.45& 0.106 &  0.011& 0.094 &  0.009& 0.102 &  0.009& 0.081 &  0.021& 0.084 &  0.009\\ 
      &      & 0.45 & 0.50& 0.071 &  0.008& 0.065 &  0.008& 0.075 &  0.008& 0.048 &  0.013& 0.057 &  0.008\\ 
\hline  
 1.55 & 1.75 & 0.10 & 0.15& 0.374 &  0.048& 0.332 &  0.039& 0.301 &  0.039& 0.210 &  0.053& 0.324 &  0.041\\ 
      &      & 0.15 & 0.20& 0.405 &  0.025& 0.390 &  0.021& 0.368 &  0.025& 0.276 &  0.044& 0.365 &  0.022\\ 
      &      & 0.20 & 0.25& 0.305 &  0.020& 0.309 &  0.017& 0.270 &  0.021& 0.248 &  0.041& 0.261 &  0.016\\ 
      &      & 0.25 & 0.30& 0.187 &  0.015& 0.191 &  0.015& 0.172 &  0.013& 0.187 &  0.036& 0.184 &  0.012\\ 
      &      & 0.30 & 0.35& 0.134 &  0.010& 0.130 &  0.008& 0.138 &  0.011& 0.098 &  0.022& 0.115 &  0.010\\ 
      &      & 0.35 & 0.40& 0.104 &  0.009& 0.094 &  0.007& 0.099 &  0.009& 0.063 &  0.014& 0.085 &  0.008\\ 
      &      & 0.40 & 0.45& 0.071 &  0.008& 0.065 &  0.009& 0.064 &  0.008& 0.064 &  0.017& 0.059 &  0.007\\ 
      &      & 0.45 & 0.50& 0.045 &  0.007& 0.040 &  0.006& 0.037 &  0.007& 0.050 &  0.016& 0.036 &  0.006\\ 
\hline  
 1.75 & 1.95 & 0.10 & 0.15& 0.296 &  0.031& 0.299 &  0.037& 0.321 &  0.038& 0.274 &  0.060& 0.291 &  0.036\\ 
      &      & 0.15 & 0.20& 0.359 &  0.024& 0.346 &  0.019& 0.305 &  0.024& 0.256 &  0.046& 0.318 &  0.019\\ 
      &      & 0.20 & 0.25& 0.246 &  0.020& 0.218 &  0.012& 0.212 &  0.017& 0.094 &  0.023& 0.177 &  0.012\\ 
      &      & 0.25 & 0.30& 0.151 &  0.012& 0.150 &  0.009& 0.129 &  0.012& 0.069 &  0.020& 0.122 &  0.009\\ 
      &      & 0.30 & 0.35& 0.109 &  0.009& 0.096 &  0.012& 0.089 &  0.008& 0.040 &  0.016& 0.090 &  0.008\\ 
      &      & 0.35 & 0.40& 0.070 &  0.009& 0.050 &  0.006& 0.060 &  0.006& 0.020 &  0.009& 0.051 &  0.007\\ 
      &      & 0.40 & 0.45& 0.040 &  0.007& 0.031 &  0.004& 0.037 &  0.006& 0.015 &  0.008& 0.031 &  0.005\\ 
      &      & 0.45 & 0.50& 0.025 &  0.004& 0.021 &  0.004& 0.018 &  0.004& 0.011 &  0.008& 0.016 &  0.003\\ 
\hline  
 1.95 & 2.15 & 0.10 & 0.15& 0.247 &  0.030& 0.216 &  0.029& 0.245 &  0.032& 0.215 &  0.051& 0.214 &  0.029\\ 
      &      & 0.15 & 0.20& 0.288 &  0.020& 0.279 &  0.015& 0.244 &  0.017& 0.212 &  0.040& 0.218 &  0.015\\ 
      &      & 0.20 & 0.25& 0.167 &  0.015& 0.164 &  0.011& 0.166 &  0.016& 0.117 &  0.030& 0.130 &  0.011\\ 
      &      & 0.25 & 0.30& 0.108 &  0.011& 0.093 &  0.008& 0.088 &  0.010& 0.047 &  0.019& 0.082 &  0.007\\ 
      &      & 0.30 & 0.35& 0.054 &  0.006& 0.060 &  0.005& 0.062 &  0.007& 0.021 &  0.009& 0.059 &  0.006\\ 
      &      & 0.35 & 0.40& 0.042 &  0.005& 0.043 &  0.004& 0.040 &  0.006& 0.039 &  0.017& 0.038 &  0.004\\ 
      &      & 0.40 & 0.45& 0.031 &  0.005& 0.024 &  0.003& 0.025 &  0.005& 0.031 &  0.016& 0.026 &  0.004\\ 
      &      & 0.45 & 0.50& 0.016 &  0.004& 0.014 &  0.003& 0.012 &  0.004& 0.020 &  0.013& 0.016 &  0.004\\ 
\end{tabular}
\end{center}
\end{table}

\begin{table}[hp!]
\begin{center}
  \caption{\label{tab:xsec-pipm-al}
    HARP results for the double-differential $\pi^-$ production
    cross-section in the laboratory system,
    $d^2\sigma^{\pi^-}/(dpd\theta)$ for $\pi^+$--Al interactions. Each row refers to a
    different $(p_{\hbox{\small min}} \le p<p_{\hbox{\small max}},
    \theta_{\hbox{\small min}} \le \theta<\theta_{\hbox{\small max}})$ bin,
    where $p$ and $\theta$ are the pion momentum and polar angle, respectively.
    The central value as well as the square-root of the diagonal elements
    of the covariance matrix are given.}
\vspace{2mm}
\begin{tabular}{rrrr|r@{$\pm$}lr@{$\pm$}lr@{$\pm$}lr@{$\pm$}lr@{$\pm$}l}
\hline
$\theta_{\hbox{\small min}}$ &
$\theta_{\hbox{\small max}}$ &
$p_{\hbox{\small min}}$ &
$p_{\hbox{\small max}}$ &
\multicolumn{10}{c}{$d^2\sigma^{\pi^-}/(dpd\theta)$}
\\
(rad) & (rad) & (\GeVc) & (\GeVc) &
\multicolumn{10}{c}{($\barn/(\GeVc \cdot \rad)$)}
\\
  &  &  &
&\multicolumn{2}{c}{$ \bf{3 \ \GeVc}$}
&\multicolumn{2}{c}{$ \bf{5 \ \GeVc}$}
&\multicolumn{2}{c}{$ \bf{8 \ \GeVc}$}
&\multicolumn{2}{c}{$ \bf{12 \ \GeVc}$}
&\multicolumn{2}{c}{$ \bf{12.9 \ \GeVc}$}
\\
\hline  
 0.35 & 0.55 & 0.15 & 0.20& 0.201 &  0.047& 0.268 &  0.065& 0.288 &  0.073& 0.410 &  0.122& 0.337 &  0.078\\ 
      &      & 0.20 & 0.25& 0.269 &  0.030& 0.353 &  0.038& 0.375 &  0.047& 0.265 &  0.073& 0.339 &  0.049\\ 
      &      & 0.25 & 0.30& 0.250 &  0.024& 0.350 &  0.026& 0.448 &  0.033& 0.318 &  0.075& 0.445 &  0.041\\ 
      &      & 0.30 & 0.35& 0.234 &  0.018& 0.325 &  0.022& 0.410 &  0.028& 0.369 &  0.055& 0.424 &  0.024\\ 
      &      & 0.35 & 0.40& 0.226 &  0.016& 0.340 &  0.028& 0.420 &  0.034& 0.404 &  0.055& 0.414 &  0.030\\ 
      &      & 0.40 & 0.45& 0.230 &  0.020& 0.337 &  0.017& 0.436 &  0.024& 0.411 &  0.051& 0.407 &  0.020\\ 
      &      & 0.45 & 0.50& 0.214 &  0.014& 0.335 &  0.017& 0.418 &  0.023& 0.461 &  0.055& 0.398 &  0.021\\ 
      &      & 0.50 & 0.60& 0.219 &  0.015& 0.324 &  0.020& 0.438 &  0.029& 0.487 &  0.052& 0.399 &  0.024\\ 
      &      & 0.60 & 0.70& 0.220 &  0.019& 0.310 &  0.026& 0.438 &  0.038& 0.422 &  0.055& 0.433 &  0.034\\ 
      &      & 0.70 & 0.80& 0.194 &  0.026& 0.266 &  0.032& 0.374 &  0.048& 0.399 &  0.058& 0.381 &  0.050\\ 
\hline  
 0.55 & 0.75 & 0.10 & 0.15& 0.214 &  0.055& 0.207 &  0.067& 0.169 &  0.069& 0.206 &  0.094& 0.182 &  0.068\\ 
      &      & 0.15 & 0.20& 0.237 &  0.033& 0.294 &  0.037& 0.357 &  0.046& 0.158 &  0.055& 0.350 &  0.045\\ 
      &      & 0.20 & 0.25& 0.264 &  0.025& 0.338 &  0.025& 0.421 &  0.031& 0.327 &  0.058& 0.398 &  0.038\\ 
      &      & 0.25 & 0.30& 0.311 &  0.024& 0.339 &  0.023& 0.376 &  0.027& 0.479 &  0.071& 0.409 &  0.022\\ 
      &      & 0.30 & 0.35& 0.276 &  0.018& 0.343 &  0.020& 0.345 &  0.025& 0.316 &  0.042& 0.333 &  0.018\\ 
      &      & 0.35 & 0.40& 0.252 &  0.016& 0.309 &  0.015& 0.389 &  0.029& 0.335 &  0.044& 0.354 &  0.022\\ 
      &      & 0.40 & 0.45& 0.240 &  0.014& 0.283 &  0.015& 0.381 &  0.021& 0.339 &  0.042& 0.322 &  0.014\\ 
      &      & 0.45 & 0.50& 0.214 &  0.013& 0.272 &  0.014& 0.359 &  0.021& 0.284 &  0.039& 0.309 &  0.016\\ 
      &      & 0.50 & 0.60& 0.204 &  0.012& 0.251 &  0.014& 0.308 &  0.020& 0.288 &  0.036& 0.282 &  0.016\\ 
      &      & 0.60 & 0.70& 0.179 &  0.017& 0.215 &  0.018& 0.277 &  0.025& 0.233 &  0.042& 0.244 &  0.023\\ 
      &      & 0.70 & 0.80& 0.139 &  0.022& 0.187 &  0.026& 0.228 &  0.032& 0.190 &  0.034& 0.210 &  0.028\\ 
\hline  
 0.75 & 0.95 & 0.10 & 0.15& 0.214 &  0.044& 0.159 &  0.036& 0.210 &  0.047& 0.235 &  0.076& 0.187 &  0.050\\ 
      &      & 0.15 & 0.20& 0.298 &  0.028& 0.317 &  0.031& 0.339 &  0.028& 0.374 &  0.068& 0.387 &  0.032\\ 
      &      & 0.20 & 0.25& 0.277 &  0.023& 0.333 &  0.024& 0.332 &  0.032& 0.328 &  0.051& 0.346 &  0.022\\ 
      &      & 0.25 & 0.30& 0.268 &  0.021& 0.299 &  0.016& 0.324 &  0.022& 0.356 &  0.053& 0.343 &  0.024\\ 
      &      & 0.30 & 0.35& 0.246 &  0.016& 0.268 &  0.016& 0.311 &  0.020& 0.303 &  0.041& 0.337 &  0.019\\ 
      &      & 0.35 & 0.40& 0.200 &  0.012& 0.239 &  0.011& 0.268 &  0.015& 0.220 &  0.032& 0.280 &  0.016\\ 
      &      & 0.40 & 0.45& 0.181 &  0.011& 0.226 &  0.013& 0.274 &  0.019& 0.165 &  0.023& 0.220 &  0.010\\ 
      &      & 0.45 & 0.50& 0.147 &  0.011& 0.200 &  0.011& 0.242 &  0.016& 0.169 &  0.026& 0.216 &  0.012\\ 
      &      & 0.50 & 0.60& 0.128 &  0.010& 0.161 &  0.009& 0.205 &  0.012& 0.163 &  0.023& 0.179 &  0.012\\ 
      &      & 0.60 & 0.70& 0.094 &  0.011& 0.133 &  0.013& 0.179 &  0.017& 0.143 &  0.024& 0.153 &  0.014\\ 
\hline  
 0.95 & 1.15 & 0.10 & 0.15& 0.232 &  0.033& 0.211 &  0.034& 0.199 &  0.034& 0.171 &  0.047& 0.252 &  0.034\\ 
      &      & 0.15 & 0.20& 0.282 &  0.024& 0.307 &  0.023& 0.312 &  0.025& 0.250 &  0.056& 0.310 &  0.023\\ 
      &      & 0.20 & 0.25& 0.248 &  0.018& 0.272 &  0.018& 0.283 &  0.024& 0.278 &  0.046& 0.301 &  0.019\\ 
      &      & 0.25 & 0.30& 0.207 &  0.014& 0.260 &  0.014& 0.293 &  0.019& 0.226 &  0.040& 0.259 &  0.014\\ 
      &      & 0.30 & 0.35& 0.177 &  0.012& 0.201 &  0.011& 0.260 &  0.016& 0.186 &  0.030& 0.234 &  0.012\\ 
      &      & 0.35 & 0.40& 0.155 &  0.010& 0.171 &  0.009& 0.208 &  0.013& 0.141 &  0.021& 0.189 &  0.010\\ 
      &      & 0.40 & 0.45& 0.127 &  0.009& 0.157 &  0.009& 0.178 &  0.010& 0.179 &  0.030& 0.149 &  0.008\\ 
      &      & 0.45 & 0.50& 0.109 &  0.008& 0.124 &  0.008& 0.162 &  0.011& 0.149 &  0.022& 0.131 &  0.007\\ 
      &      & 0.50 & 0.60& 0.082 &  0.007& 0.096 &  0.007& 0.130 &  0.010& 0.117 &  0.020& 0.108 &  0.008\\ 
\hline
\end{tabular}
\end{center}
\end{table}

\begin{table}[hp!]
\begin{center}
\begin{tabular}{rrrr|r@{$\pm$}lr@{$\pm$}lr@{$\pm$}lr@{$\pm$}lr@{$\pm$}l}
\hline
$\theta_{\hbox{\small min}}$ &
$\theta_{\hbox{\small max}}$ &
$p_{\hbox{\small min}}$ &
$p_{\hbox{\small max}}$ &
\multicolumn{10}{c}{$d^2\sigma^{\pi^-}/(dpd\theta)$}
\\
(rad) & (rad) & (\GeVc) & (\GeVc) &
\multicolumn{10}{c}{($\barn/(\GeVc \cdot \rad)$)}
\\
  &  &  &
&\multicolumn{2}{c}{$ \bf{3 \ \GeVc}$}
&\multicolumn{2}{c}{$ \bf{5 \ \GeVc}$}
&\multicolumn{2}{c}{$ \bf{8 \ \GeVc}$}
&\multicolumn{2}{c}{$ \bf{12 \ \GeVc}$}
&\multicolumn{2}{c}{$ \bf{12.9 \ \GeVc}$}
\\
\hline
 1.15 & 1.35 & 0.10 & 0.15& 0.221 &  0.029& 0.222 &  0.027& 0.262 &  0.035& 0.191 &  0.066& 0.256 &  0.033\\ 
      &      & 0.15 & 0.20& 0.245 &  0.025& 0.283 &  0.021& 0.275 &  0.020& 0.396 &  0.061& 0.316 &  0.024\\ 
      &      & 0.20 & 0.25& 0.217 &  0.016& 0.253 &  0.015& 0.231 &  0.016& 0.288 &  0.046& 0.290 &  0.016\\ 
      &      & 0.25 & 0.30& 0.170 &  0.013& 0.200 &  0.012& 0.182 &  0.014& 0.213 &  0.036& 0.219 &  0.013\\ 
      &      & 0.30 & 0.35& 0.140 &  0.011& 0.162 &  0.010& 0.178 &  0.012& 0.151 &  0.026& 0.182 &  0.010\\ 
      &      & 0.35 & 0.40& 0.113 &  0.008& 0.130 &  0.007& 0.141 &  0.010& 0.109 &  0.021& 0.141 &  0.008\\ 
      &      & 0.40 & 0.45& 0.089 &  0.007& 0.108 &  0.006& 0.110 &  0.008& 0.077 &  0.016& 0.112 &  0.008\\ 
      &      & 0.45 & 0.50& 0.071 &  0.006& 0.082 &  0.007& 0.085 &  0.007& 0.057 &  0.014& 0.083 &  0.008\\ 
\hline  
 1.35 & 1.55 & 0.10 & 0.15& 0.213 &  0.031& 0.206 &  0.028& 0.251 &  0.035& 0.129 &  0.037& 0.229 &  0.030\\ 
      &      & 0.15 & 0.20& 0.274 &  0.021& 0.255 &  0.017& 0.278 &  0.020& 0.263 &  0.051& 0.238 &  0.017\\ 
      &      & 0.20 & 0.25& 0.190 &  0.016& 0.203 &  0.012& 0.210 &  0.015& 0.248 &  0.044& 0.209 &  0.014\\ 
      &      & 0.25 & 0.30& 0.144 &  0.011& 0.154 &  0.009& 0.166 &  0.013& 0.137 &  0.025& 0.189 &  0.012\\ 
      &      & 0.30 & 0.35& 0.113 &  0.009& 0.133 &  0.009& 0.133 &  0.010& 0.127 &  0.025& 0.145 &  0.012\\ 
      &      & 0.35 & 0.40& 0.085 &  0.007& 0.101 &  0.007& 0.100 &  0.008& 0.101 &  0.020& 0.100 &  0.008\\ 
      &      & 0.40 & 0.45& 0.066 &  0.006& 0.072 &  0.006& 0.085 &  0.007& 0.084 &  0.019& 0.072 &  0.006\\ 
      &      & 0.45 & 0.50& 0.052 &  0.005& 0.054 &  0.005& 0.066 &  0.007& 0.059 &  0.015& 0.056 &  0.005\\ 
\hline  
 1.55 & 1.75 & 0.10 & 0.15& 0.165 &  0.024& 0.191 &  0.020& 0.253 &  0.032& 0.290 &  0.067& 0.224 &  0.030\\ 
      &      & 0.15 & 0.20& 0.206 &  0.018& 0.208 &  0.014& 0.208 &  0.017& 0.254 &  0.050& 0.224 &  0.016\\ 
      &      & 0.20 & 0.25& 0.156 &  0.013& 0.155 &  0.010& 0.196 &  0.015& 0.101 &  0.022& 0.175 &  0.012\\ 
      &      & 0.25 & 0.30& 0.104 &  0.009& 0.124 &  0.009& 0.127 &  0.011& 0.141 &  0.031& 0.124 &  0.010\\ 
      &      & 0.30 & 0.35& 0.089 &  0.008& 0.093 &  0.007& 0.105 &  0.009& 0.110 &  0.025& 0.083 &  0.007\\ 
      &      & 0.35 & 0.40& 0.065 &  0.007& 0.069 &  0.005& 0.080 &  0.007& 0.069 &  0.018& 0.064 &  0.005\\ 
      &      & 0.40 & 0.45& 0.048 &  0.005& 0.054 &  0.004& 0.057 &  0.006& 0.045 &  0.014& 0.049 &  0.005\\ 
      &      & 0.45 & 0.50& 0.036 &  0.004& 0.042 &  0.004& 0.037 &  0.005& 0.031 &  0.011& 0.034 &  0.004\\ 
\hline  
 1.75 & 1.95 & 0.10 & 0.15& 0.171 &  0.020& 0.177 &  0.017& 0.158 &  0.023& 0.184 &  0.045& 0.194 &  0.023\\ 
      &      & 0.15 & 0.20& 0.192 &  0.015& 0.191 &  0.012& 0.200 &  0.017& 0.208 &  0.040& 0.178 &  0.013\\ 
      &      & 0.20 & 0.25& 0.141 &  0.013& 0.139 &  0.009& 0.122 &  0.011& 0.156 &  0.034& 0.131 &  0.010\\ 
      &      & 0.25 & 0.30& 0.074 &  0.009& 0.100 &  0.007& 0.084 &  0.008& 0.068 &  0.022& 0.097 &  0.008\\ 
      &      & 0.30 & 0.35& 0.056 &  0.006& 0.076 &  0.006& 0.067 &  0.007& 0.049 &  0.016& 0.068 &  0.008\\ 
      &      & 0.35 & 0.40& 0.042 &  0.005& 0.044 &  0.006& 0.048 &  0.006& 0.044 &  0.016& 0.045 &  0.004\\ 
      &      & 0.40 & 0.45& 0.026 &  0.004& 0.025 &  0.004& 0.035 &  0.005& 0.025 &  0.012& 0.035 &  0.004\\ 
      &      & 0.45 & 0.50& 0.017 &  0.003& 0.019 &  0.002& 0.023 &  0.003& 0.013 &  0.009& 0.022 &  0.003\\ 
\hline  
 1.95 & 2.15 & 0.10 & 0.15& 0.136 &  0.017& 0.127 &  0.013& 0.166 &  0.019& 0.125 &  0.034& 0.139 &  0.017\\ 
      &      & 0.15 & 0.20& 0.133 &  0.012& 0.143 &  0.011& 0.147 &  0.013& 0.145 &  0.033& 0.158 &  0.013\\ 
      &      & 0.20 & 0.25& 0.085 &  0.008& 0.100 &  0.008& 0.102 &  0.010& 0.115 &  0.031& 0.106 &  0.009\\ 
      &      & 0.25 & 0.30& 0.062 &  0.007& 0.071 &  0.006& 0.059 &  0.009& 0.047 &  0.018& 0.067 &  0.007\\ 
      &      & 0.30 & 0.35& 0.045 &  0.005& 0.048 &  0.005& 0.040 &  0.005& 0.026 &  0.011& 0.037 &  0.005\\ 
      &      & 0.35 & 0.40& 0.033 &  0.005& 0.032 &  0.003& 0.037 &  0.005& 0.043 &  0.020& 0.026 &  0.004\\ 
      &      & 0.40 & 0.45& 0.026 &  0.004& 0.025 &  0.003& 0.034 &  0.005& 0.023 &  0.013& 0.020 &  0.003\\ 
      &      & 0.45 & 0.50& 0.017 &  0.003& 0.019 &  0.003& 0.019 &  0.004& 0.013 &  0.010& 0.015 &  0.003\\ 
\end{tabular}
\end{center}
\end{table}

\clearpage
\begin{table}[hp!]
\begin{center}
  \caption{\label{tab:xsec-pimp-al}
    HARP results for the double-differential $\pi^+$ production
    cross-section in the laboratory system,
    $d^2\sigma^{\pi^+}/(dpd\theta)$ for $\pi^-$--Al interactions. Each row refers to a
    different $(p_{\hbox{\small min}} \le p<p_{\hbox{\small max}},
    \theta_{\hbox{\small min}} \le \theta<\theta_{\hbox{\small max}})$ bin,
    where $p$ and $\theta$ are the pion momentum and polar angle, respectively.
    The central value as well as the square-root of the diagonal elements
    of the covariance matrix are given.}
\vspace{2mm}
\begin{tabular}{rrrr|r@{$\pm$}lr@{$\pm$}lr@{$\pm$}lr@{$\pm$}l}
\hline
$\theta_{\hbox{\small min}}$ &
$\theta_{\hbox{\small max}}$ &
$p_{\hbox{\small min}}$ &
$p_{\hbox{\small max}}$ &
\multicolumn{8}{c}{$d^2\sigma^{\pi^+}/(dpd\theta)$}
\\
(rad) & (rad) & (\GeVc) & (\GeVc) &
\multicolumn{8}{c}{($\barn/(\GeVc \cdot \rad)$)}
\\
  &  &  &
&\multicolumn{2}{c}{$ \bf{3 \ \GeVc}$}
&\multicolumn{2}{c}{$ \bf{5 \ \GeVc}$}
&\multicolumn{2}{c}{$ \bf{8 \ \GeVc}$}
&\multicolumn{2}{c}{$ \bf{12 \ \GeVc}$}
\\
\hline  
 0.35 & 0.55 & 0.15 & 0.20& 0.217 &  0.034& 0.324 &  0.065& 0.326 &  0.056& 0.414 &  0.055\\ 
      &      & 0.20 & 0.25& 0.224 &  0.024& 0.373 &  0.037& 0.355 &  0.039& 0.489 &  0.048\\ 
      &      & 0.25 & 0.30& 0.288 &  0.025& 0.415 &  0.029& 0.449 &  0.037& 0.479 &  0.038\\ 
      &      & 0.30 & 0.35& 0.301 &  0.018& 0.406 &  0.030& 0.482 &  0.031& 0.573 &  0.029\\ 
      &      & 0.35 & 0.40& 0.293 &  0.016& 0.431 &  0.032& 0.498 &  0.033& 0.555 &  0.043\\ 
      &      & 0.40 & 0.45& 0.285 &  0.015& 0.486 &  0.032& 0.503 &  0.024& 0.545 &  0.029\\ 
      &      & 0.45 & 0.50& 0.299 &  0.021& 0.496 &  0.026& 0.537 &  0.031& 0.608 &  0.035\\ 
      &      & 0.50 & 0.60& 0.315 &  0.019& 0.474 &  0.030& 0.575 &  0.032& 0.591 &  0.030\\ 
      &      & 0.60 & 0.70& 0.281 &  0.028& 0.429 &  0.045& 0.518 &  0.047& 0.513 &  0.040\\ 
      &      & 0.70 & 0.80& 0.223 &  0.036& 0.300 &  0.050& 0.473 &  0.057& 0.379 &  0.061\\ 
\hline  
 0.55 & 0.75 & 0.10 & 0.15& 0.216 &  0.053& 0.213 &  0.066& 0.182 &  0.060& 0.277 &  0.055\\ 
      &      & 0.15 & 0.20& 0.282 &  0.025& 0.381 &  0.038& 0.321 &  0.033& 0.375 &  0.030\\ 
      &      & 0.20 & 0.25& 0.324 &  0.021& 0.403 &  0.028& 0.429 &  0.040& 0.481 &  0.028\\ 
      &      & 0.25 & 0.30& 0.335 &  0.027& 0.433 &  0.034& 0.488 &  0.032& 0.532 &  0.035\\ 
      &      & 0.30 & 0.35& 0.309 &  0.016& 0.422 &  0.023& 0.444 &  0.024& 0.489 &  0.025\\ 
      &      & 0.35 & 0.40& 0.303 &  0.017& 0.415 &  0.025& 0.472 &  0.030& 0.492 &  0.029\\ 
      &      & 0.40 & 0.45& 0.306 &  0.014& 0.398 &  0.019& 0.462 &  0.020& 0.487 &  0.020\\ 
      &      & 0.45 & 0.50& 0.297 &  0.014& 0.375 &  0.018& 0.446 &  0.020& 0.448 &  0.017\\ 
      &      & 0.50 & 0.60& 0.262 &  0.017& 0.319 &  0.024& 0.397 &  0.025& 0.356 &  0.025\\ 
      &      & 0.60 & 0.70& 0.204 &  0.027& 0.221 &  0.032& 0.298 &  0.038& 0.253 &  0.033\\ 
      &      & 0.70 & 0.80& 0.133 &  0.026& 0.154 &  0.032& 0.214 &  0.039& 0.165 &  0.030\\ 
\hline  
 0.75 & 0.95 & 0.10 & 0.15& 0.221 &  0.039& 0.244 &  0.044& 0.200 &  0.042& 0.279 &  0.038\\ 
      &      & 0.15 & 0.20& 0.299 &  0.023& 0.343 &  0.027& 0.351 &  0.026& 0.381 &  0.025\\ 
      &      & 0.20 & 0.25& 0.342 &  0.021& 0.409 &  0.031& 0.415 &  0.028& 0.413 &  0.035\\ 
      &      & 0.25 & 0.30& 0.322 &  0.017& 0.384 &  0.025& 0.397 &  0.021& 0.445 &  0.020\\ 
      &      & 0.30 & 0.35& 0.288 &  0.014& 0.332 &  0.020& 0.379 &  0.023& 0.399 &  0.018\\ 
      &      & 0.35 & 0.40& 0.261 &  0.012& 0.295 &  0.018& 0.339 &  0.015& 0.353 &  0.015\\ 
      &      & 0.40 & 0.45& 0.230 &  0.009& 0.295 &  0.015& 0.305 &  0.013& 0.321 &  0.014\\ 
      &      & 0.45 & 0.50& 0.201 &  0.009& 0.272 &  0.014& 0.279 &  0.014& 0.292 &  0.015\\ 
      &      & 0.50 & 0.60& 0.170 &  0.014& 0.208 &  0.019& 0.221 &  0.018& 0.198 &  0.020\\ 
      &      & 0.60 & 0.70& 0.105 &  0.017& 0.128 &  0.023& 0.140 &  0.022& 0.111 &  0.017\\ 
\hline  
 0.95 & 1.15 & 0.10 & 0.15& 0.204 &  0.032& 0.255 &  0.039& 0.200 &  0.033& 0.244 &  0.028\\ 
      &      & 0.15 & 0.20& 0.320 &  0.024& 0.339 &  0.026& 0.360 &  0.029& 0.401 &  0.027\\ 
      &      & 0.20 & 0.25& 0.303 &  0.015& 0.373 &  0.022& 0.361 &  0.019& 0.367 &  0.023\\ 
      &      & 0.25 & 0.30& 0.254 &  0.013& 0.306 &  0.018& 0.311 &  0.017& 0.349 &  0.019\\ 
      &      & 0.30 & 0.35& 0.219 &  0.011& 0.265 &  0.015& 0.270 &  0.013& 0.281 &  0.015\\ 
      &      & 0.35 & 0.40& 0.188 &  0.008& 0.219 &  0.012& 0.234 &  0.011& 0.235 &  0.011\\ 
      &      & 0.40 & 0.45& 0.157 &  0.008& 0.199 &  0.011& 0.197 &  0.008& 0.190 &  0.010\\ 
      &      & 0.45 & 0.50& 0.120 &  0.009& 0.154 &  0.011& 0.161 &  0.011& 0.144 &  0.011\\ 
      &      & 0.50 & 0.60& 0.083 &  0.010& 0.100 &  0.014& 0.114 &  0.012& 0.101 &  0.012\\ 
\hline
\end{tabular}
\end{center}
\end{table}

\begin{table}[hp!]
\begin{center}
\begin{tabular}{rrrr|r@{$\pm$}lr@{$\pm$}lr@{$\pm$}lr@{$\pm$}l}
\hline
$\theta_{\hbox{\small min}}$ &
$\theta_{\hbox{\small max}}$ &
$p_{\hbox{\small min}}$ &
$p_{\hbox{\small max}}$ &
\multicolumn{8}{c}{$d^2\sigma^{\pi^+}/(dpd\theta)$}
\\
(rad) & (rad) & (\GeVc) & (\GeVc) &
\multicolumn{8}{c}{(\barn/($\GeVc \cdot \rad$))}
\\
  &  &  &
&\multicolumn{2}{c}{$ \bf{3 \ \GeVc}$}
&\multicolumn{2}{c}{$ \bf{5 \ \GeVc}$}
&\multicolumn{2}{c}{$ \bf{8 \ \GeVc}$}
&\multicolumn{2}{c}{$ \bf{12 \ \GeVc}$}
\\
\hline
 1.15 & 1.35 & 0.10 & 0.15& 0.199 &  0.030& 0.244 &  0.035& 0.235 &  0.030& 0.258 &  0.034\\ 
      &      & 0.15 & 0.20& 0.312 &  0.019& 0.332 &  0.025& 0.347 &  0.026& 0.344 &  0.024\\ 
      &      & 0.20 & 0.25& 0.262 &  0.014& 0.307 &  0.019& 0.303 &  0.016& 0.322 &  0.017\\ 
      &      & 0.25 & 0.30& 0.211 &  0.011& 0.244 &  0.014& 0.230 &  0.013& 0.250 &  0.017\\ 
      &      & 0.30 & 0.35& 0.169 &  0.008& 0.177 &  0.010& 0.201 &  0.010& 0.180 &  0.011\\ 
      &      & 0.35 & 0.40& 0.138 &  0.008& 0.152 &  0.009& 0.156 &  0.008& 0.161 &  0.009\\ 
      &      & 0.40 & 0.45& 0.100 &  0.008& 0.127 &  0.008& 0.122 &  0.007& 0.127 &  0.009\\ 
      &      & 0.45 & 0.50& 0.077 &  0.007& 0.098 &  0.009& 0.091 &  0.008& 0.095 &  0.008\\ 
\hline  
 1.35 & 1.55 & 0.10 & 0.15& 0.216 &  0.030& 0.272 &  0.033& 0.233 &  0.029& 0.237 &  0.031\\ 
      &      & 0.15 & 0.20& 0.274 &  0.016& 0.270 &  0.017& 0.268 &  0.019& 0.328 &  0.018\\ 
      &      & 0.20 & 0.25& 0.218 &  0.011& 0.237 &  0.015& 0.264 &  0.014& 0.262 &  0.015\\ 
      &      & 0.25 & 0.30& 0.176 &  0.011& 0.185 &  0.014& 0.190 &  0.011& 0.190 &  0.014\\ 
      &      & 0.30 & 0.35& 0.121 &  0.009& 0.128 &  0.009& 0.142 &  0.009& 0.143 &  0.009\\ 
      &      & 0.35 & 0.40& 0.089 &  0.006& 0.103 &  0.007& 0.101 &  0.007& 0.103 &  0.008\\ 
      &      & 0.40 & 0.45& 0.069 &  0.005& 0.074 &  0.007& 0.070 &  0.006& 0.061 &  0.007\\ 
      &      & 0.45 & 0.50& 0.050 &  0.005& 0.049 &  0.006& 0.051 &  0.006& 0.052 &  0.006\\ 
\hline  
 1.55 & 1.75 & 0.10 & 0.15& 0.215 &  0.028& 0.220 &  0.027& 0.234 &  0.030& 0.216 &  0.028\\ 
      &      & 0.15 & 0.20& 0.251 &  0.016& 0.254 &  0.019& 0.252 &  0.014& 0.262 &  0.016\\ 
      &      & 0.20 & 0.25& 0.187 &  0.011& 0.227 &  0.017& 0.191 &  0.010& 0.206 &  0.012\\ 
      &      & 0.25 & 0.30& 0.125 &  0.008& 0.150 &  0.012& 0.136 &  0.009& 0.143 &  0.010\\ 
      &      & 0.30 & 0.35& 0.096 &  0.006& 0.118 &  0.008& 0.104 &  0.007& 0.102 &  0.008\\ 
      &      & 0.35 & 0.40& 0.066 &  0.005& 0.084 &  0.008& 0.076 &  0.006& 0.076 &  0.006\\ 
      &      & 0.40 & 0.45& 0.042 &  0.004& 0.050 &  0.006& 0.050 &  0.005& 0.042 &  0.005\\ 
      &      & 0.45 & 0.50& 0.026 &  0.003& 0.036 &  0.005& 0.030 &  0.005& 0.026 &  0.004\\ 
\hline  
 1.75 & 1.95 & 0.10 & 0.15& 0.188 &  0.022& 0.212 &  0.023& 0.180 &  0.020& 0.226 &  0.028\\ 
      &      & 0.15 & 0.20& 0.224 &  0.013& 0.235 &  0.016& 0.200 &  0.014& 0.218 &  0.014\\ 
      &      & 0.20 & 0.25& 0.156 &  0.011& 0.157 &  0.012& 0.169 &  0.009& 0.170 &  0.012\\ 
      &      & 0.25 & 0.30& 0.099 &  0.006& 0.101 &  0.008& 0.104 &  0.008& 0.107 &  0.008\\ 
      &      & 0.30 & 0.35& 0.071 &  0.005& 0.074 &  0.007& 0.069 &  0.005& 0.066 &  0.006\\ 
      &      & 0.35 & 0.40& 0.048 &  0.004& 0.046 &  0.005& 0.044 &  0.004& 0.043 &  0.004\\ 
      &      & 0.40 & 0.45& 0.029 &  0.005& 0.029 &  0.004& 0.028 &  0.004& 0.026 &  0.004\\ 
      &      & 0.45 & 0.50& 0.016 &  0.003& 0.021 &  0.004& 0.016 &  0.003& 0.018 &  0.002\\ 
\hline  
 1.95 & 2.15 & 0.10 & 0.15& 0.147 &  0.018& 0.170 &  0.022& 0.160 &  0.019& 0.168 &  0.024\\ 
      &      & 0.15 & 0.20& 0.150 &  0.009& 0.157 &  0.011& 0.169 &  0.010& 0.172 &  0.013\\ 
      &      & 0.20 & 0.25& 0.104 &  0.006& 0.117 &  0.010& 0.109 &  0.009& 0.135 &  0.012\\ 
      &      & 0.25 & 0.30& 0.073 &  0.005& 0.069 &  0.006& 0.063 &  0.005& 0.061 &  0.006\\ 
      &      & 0.30 & 0.35& 0.047 &  0.005& 0.045 &  0.005& 0.043 &  0.004& 0.044 &  0.004\\ 
      &      & 0.35 & 0.40& 0.028 &  0.003& 0.030 &  0.004& 0.027 &  0.003& 0.029 &  0.004\\ 
      &      & 0.40 & 0.45& 0.016 &  0.002& 0.021 &  0.003& 0.017 &  0.003& 0.018 &  0.002\\ 
      &      & 0.45 & 0.50& 0.011 &  0.002& 0.013 &  0.003& 0.010 &  0.002& 0.011 &  0.003\\ 

\end{tabular}
\end{center}
\end{table}

\begin{table}[hp!]
\begin{center}
  \caption{\label{tab:xsec-pimm-al}
    HARP results for the double-differential $\pi^-$ production
    cross-section in the laboratory system,
    $d^2\sigma^{\pi^-}/(dpd\theta)$ for $\pi^-$--Al interactions. Each row refers to a
    different $(p_{\hbox{\small min}} \le p<p_{\hbox{\small max}},
    \theta_{\hbox{\small min}} \le \theta<\theta_{\hbox{\small max}})$ bin,
    where $p$ and $\theta$ are the pion momentum and polar angle, respectively.
    The central value as well as the square-root of the diagonal elements
    of the covariance matrix are given.}
\vspace{2mm}
\begin{tabular}{rrrr|r@{$\pm$}lr@{$\pm$}lr@{$\pm$}lr@{$\pm$}l}
\hline
$\theta_{\hbox{\small min}}$ &
$\theta_{\hbox{\small max}}$ &
$p_{\hbox{\small min}}$ &
$p_{\hbox{\small max}}$ &
\multicolumn{8}{c}{$d^2\sigma^{\pi^-}/(dpd\theta)$}
\\
(rad) & (rad) & (\GeVc) & (\GeVc) &
\multicolumn{8}{c}{($\barn/(\GeVc \cdot \rad)$)}
\\
  &  &  &
&\multicolumn{2}{c}{$ \bf{3 \ \GeVc}$}
&\multicolumn{2}{c}{$ \bf{5 \ \GeVc}$}
&\multicolumn{2}{c}{$ \bf{8 \ \GeVc}$}
&\multicolumn{2}{c}{$ \bf{12 \ \GeVc}$}
\\
\hline  
 0.35 & 0.55 & 0.15 & 0.20& 0.301 &  0.044& 0.379 &  0.070& 0.357 &  0.070& 0.456 &  0.067\\ 
      &      & 0.20 & 0.25& 0.338 &  0.031& 0.466 &  0.042& 0.494 &  0.046& 0.560 &  0.044\\ 
      &      & 0.25 & 0.30& 0.393 &  0.026& 0.520 &  0.041& 0.565 &  0.041& 0.570 &  0.037\\ 
      &      & 0.30 & 0.35& 0.401 &  0.023& 0.502 &  0.030& 0.559 &  0.032& 0.555 &  0.025\\ 
      &      & 0.35 & 0.40& 0.375 &  0.022& 0.463 &  0.030& 0.602 &  0.041& 0.565 &  0.037\\ 
      &      & 0.40 & 0.45& 0.403 &  0.024& 0.517 &  0.041& 0.599 &  0.029& 0.555 &  0.030\\ 
      &      & 0.45 & 0.50& 0.409 &  0.022& 0.543 &  0.030& 0.596 &  0.030& 0.526 &  0.027\\ 
      &      & 0.50 & 0.60& 0.397 &  0.021& 0.550 &  0.030& 0.605 &  0.032& 0.560 &  0.027\\ 
      &      & 0.60 & 0.70& 0.387 &  0.031& 0.516 &  0.046& 0.599 &  0.047& 0.525 &  0.035\\ 
      &      & 0.70 & 0.80& 0.342 &  0.043& 0.435 &  0.052& 0.554 &  0.054& 0.462 &  0.049\\ 
\hline  
 0.55 & 0.75 & 0.10 & 0.15& 0.279 &  0.066& 0.356 &  0.086& 0.280 &  0.077& 0.352 &  0.067\\ 
      &      & 0.15 & 0.20& 0.377 &  0.029& 0.486 &  0.044& 0.428 &  0.037& 0.480 &  0.035\\ 
      &      & 0.20 & 0.25& 0.457 &  0.032& 0.577 &  0.039& 0.515 &  0.041& 0.513 &  0.030\\ 
      &      & 0.25 & 0.30& 0.475 &  0.026& 0.546 &  0.032& 0.550 &  0.029& 0.540 &  0.034\\ 
      &      & 0.30 & 0.35& 0.445 &  0.023& 0.470 &  0.027& 0.528 &  0.030& 0.500 &  0.027\\ 
      &      & 0.35 & 0.40& 0.435 &  0.024& 0.491 &  0.033& 0.527 &  0.028& 0.457 &  0.021\\ 
      &      & 0.40 & 0.45& 0.422 &  0.018& 0.507 &  0.025& 0.507 &  0.023& 0.434 &  0.022\\ 
      &      & 0.45 & 0.50& 0.408 &  0.018& 0.495 &  0.023& 0.484 &  0.021& 0.393 &  0.017\\ 
      &      & 0.50 & 0.60& 0.374 &  0.021& 0.427 &  0.025& 0.428 &  0.022& 0.346 &  0.019\\ 
      &      & 0.60 & 0.70& 0.327 &  0.026& 0.321 &  0.036& 0.384 &  0.030& 0.310 &  0.021\\ 
      &      & 0.70 & 0.80& 0.267 &  0.037& 0.260 &  0.036& 0.331 &  0.042& 0.272 &  0.033\\ 
\hline  
 0.75 & 0.95 & 0.10 & 0.15& 0.339 &  0.052& 0.328 &  0.051& 0.288 &  0.047& 0.297 &  0.038\\ 
      &      & 0.15 & 0.20& 0.463 &  0.028& 0.516 &  0.042& 0.439 &  0.033& 0.482 &  0.032\\ 
      &      & 0.20 & 0.25& 0.507 &  0.032& 0.548 &  0.032& 0.513 &  0.033& 0.464 &  0.024\\ 
      &      & 0.25 & 0.30& 0.469 &  0.023& 0.490 &  0.027& 0.487 &  0.025& 0.438 &  0.023\\ 
      &      & 0.30 & 0.35& 0.410 &  0.019& 0.429 &  0.021& 0.465 &  0.022& 0.411 &  0.019\\ 
      &      & 0.35 & 0.40& 0.367 &  0.016& 0.375 &  0.018& 0.382 &  0.017& 0.361 &  0.016\\ 
      &      & 0.40 & 0.45& 0.330 &  0.013& 0.370 &  0.020& 0.333 &  0.014& 0.305 &  0.013\\ 
      &      & 0.45 & 0.50& 0.301 &  0.011& 0.334 &  0.016& 0.311 &  0.014& 0.276 &  0.012\\ 
      &      & 0.50 & 0.60& 0.268 &  0.014& 0.278 &  0.018& 0.267 &  0.016& 0.236 &  0.012\\ 
      &      & 0.60 & 0.70& 0.217 &  0.021& 0.221 &  0.024& 0.210 &  0.023& 0.185 &  0.019\\ 
\hline  
 0.95 & 1.15 & 0.10 & 0.15& 0.362 &  0.043& 0.382 &  0.049& 0.328 &  0.044& 0.330 &  0.034\\ 
      &      & 0.15 & 0.20& 0.501 &  0.029& 0.507 &  0.030& 0.453 &  0.025& 0.458 &  0.032\\ 
      &      & 0.20 & 0.25& 0.450 &  0.023& 0.431 &  0.025& 0.451 &  0.025& 0.406 &  0.019\\ 
      &      & 0.25 & 0.30& 0.381 &  0.017& 0.421 &  0.028& 0.380 &  0.017& 0.356 &  0.018\\ 
      &      & 0.30 & 0.35& 0.328 &  0.015& 0.409 &  0.021& 0.337 &  0.015& 0.308 &  0.016\\ 
      &      & 0.35 & 0.40& 0.272 &  0.012& 0.320 &  0.018& 0.273 &  0.012& 0.252 &  0.012\\ 
      &      & 0.40 & 0.45& 0.238 &  0.011& 0.254 &  0.012& 0.223 &  0.012& 0.219 &  0.010\\ 
      &      & 0.45 & 0.50& 0.207 &  0.009& 0.218 &  0.011& 0.181 &  0.009& 0.194 &  0.011\\ 
      &      & 0.50 & 0.60& 0.171 &  0.010& 0.174 &  0.012& 0.150 &  0.009& 0.149 &  0.010\\ 
\hline
\end{tabular}
\end{center}
\end{table}

\begin{table}[hp!]
\begin{center}
\begin{tabular}{rrrr|r@{$\pm$}lr@{$\pm$}lr@{$\pm$}lr@{$\pm$}l}
\hline
$\theta_{\hbox{\small min}}$ &
$\theta_{\hbox{\small max}}$ &
$p_{\hbox{\small min}}$ &
$p_{\hbox{\small max}}$ &
\multicolumn{8}{c}{$d^2\sigma^{\pi^-}/(dpd\theta)$}
\\
(rad) & (rad) & (\GeVc) & (\GeVc) &
\multicolumn{8}{c}{(\barn/($\GeVc \cdot \rad$))}
\\
  &  &  &
&\multicolumn{2}{c}{$ \bf{3 \ \GeVc}$}
&\multicolumn{2}{c}{$ \bf{5 \ \GeVc}$}
&\multicolumn{2}{c}{$ \bf{8 \ \GeVc}$}
&\multicolumn{2}{c}{$ \bf{12 \ \GeVc}$}
\\
\hline
 1.15 & 1.35 & 0.10 & 0.15& 0.411 &  0.049& 0.408 &  0.049& 0.325 &  0.037& 0.317 &  0.033\\ 
      &      & 0.15 & 0.20& 0.484 &  0.024& 0.512 &  0.029& 0.441 &  0.026& 0.444 &  0.031\\ 
      &      & 0.20 & 0.25& 0.406 &  0.019& 0.455 &  0.026& 0.384 &  0.018& 0.360 &  0.019\\ 
      &      & 0.25 & 0.30& 0.306 &  0.014& 0.352 &  0.020& 0.316 &  0.015& 0.276 &  0.014\\ 
      &      & 0.30 & 0.35& 0.255 &  0.012& 0.253 &  0.015& 0.241 &  0.013& 0.216 &  0.013\\ 
      &      & 0.35 & 0.40& 0.209 &  0.009& 0.203 &  0.011& 0.183 &  0.009& 0.174 &  0.009\\ 
      &      & 0.40 & 0.45& 0.172 &  0.008& 0.159 &  0.009& 0.146 &  0.007& 0.146 &  0.007\\ 
      &      & 0.45 & 0.50& 0.135 &  0.008& 0.133 &  0.008& 0.124 &  0.007& 0.117 &  0.007\\ 
\hline  
 1.35 & 1.55 & 0.10 & 0.15& 0.419 &  0.044& 0.427 &  0.052& 0.314 &  0.032& 0.298 &  0.034\\ 
      &      & 0.15 & 0.20& 0.469 &  0.027& 0.456 &  0.026& 0.357 &  0.023& 0.383 &  0.023\\ 
      &      & 0.20 & 0.25& 0.366 &  0.019& 0.348 &  0.020& 0.315 &  0.017& 0.304 &  0.017\\ 
      &      & 0.25 & 0.30& 0.245 &  0.014& 0.268 &  0.017& 0.217 &  0.011& 0.203 &  0.011\\ 
      &      & 0.30 & 0.35& 0.185 &  0.010& 0.194 &  0.013& 0.177 &  0.010& 0.159 &  0.011\\ 
      &      & 0.35 & 0.40& 0.153 &  0.009& 0.142 &  0.010& 0.135 &  0.008& 0.123 &  0.008\\ 
      &      & 0.40 & 0.45& 0.118 &  0.006& 0.114 &  0.008& 0.100 &  0.007& 0.095 &  0.006\\ 
      &      & 0.45 & 0.50& 0.093 &  0.006& 0.087 &  0.007& 0.077 &  0.006& 0.081 &  0.006\\ 
\hline  
 1.55 & 1.75 & 0.10 & 0.15& 0.386 &  0.042& 0.370 &  0.043& 0.307 &  0.034& 0.318 &  0.033\\ 
      &      & 0.15 & 0.20& 0.398 &  0.021& 0.394 &  0.024& 0.348 &  0.018& 0.300 &  0.017\\ 
      &      & 0.20 & 0.25& 0.295 &  0.017& 0.304 &  0.019& 0.229 &  0.013& 0.241 &  0.014\\ 
      &      & 0.25 & 0.30& 0.200 &  0.012& 0.210 &  0.015& 0.162 &  0.009& 0.166 &  0.012\\ 
      &      & 0.30 & 0.35& 0.147 &  0.010& 0.134 &  0.011& 0.115 &  0.007& 0.113 &  0.008\\ 
      &      & 0.35 & 0.40& 0.111 &  0.007& 0.100 &  0.008& 0.085 &  0.006& 0.094 &  0.007\\ 
      &      & 0.40 & 0.45& 0.085 &  0.006& 0.075 &  0.007& 0.065 &  0.004& 0.072 &  0.005\\ 
      &      & 0.45 & 0.50& 0.065 &  0.006& 0.053 &  0.006& 0.052 &  0.004& 0.048 &  0.005\\ 
\hline  
 1.75 & 1.95 & 0.10 & 0.15& 0.352 &  0.034& 0.360 &  0.034& 0.242 &  0.026& 0.298 &  0.028\\ 
      &      & 0.15 & 0.20& 0.342 &  0.016& 0.318 &  0.019& 0.270 &  0.015& 0.258 &  0.014\\ 
      &      & 0.20 & 0.25& 0.244 &  0.015& 0.230 &  0.015& 0.201 &  0.011& 0.209 &  0.013\\ 
      &      & 0.25 & 0.30& 0.148 &  0.009& 0.125 &  0.013& 0.128 &  0.009& 0.121 &  0.010\\ 
      &      & 0.30 & 0.35& 0.103 &  0.008& 0.086 &  0.006& 0.090 &  0.006& 0.087 &  0.007\\ 
      &      & 0.35 & 0.40& 0.079 &  0.005& 0.077 &  0.006& 0.061 &  0.005& 0.068 &  0.005\\ 
      &      & 0.40 & 0.45& 0.061 &  0.005& 0.060 &  0.006& 0.044 &  0.004& 0.039 &  0.004\\ 
      &      & 0.45 & 0.50& 0.041 &  0.005& 0.040 &  0.005& 0.034 &  0.003& 0.030 &  0.003\\ 
\hline  
 1.95 & 2.15 & 0.10 & 0.15& 0.301 &  0.028& 0.252 &  0.024& 0.231 &  0.024& 0.237 &  0.029\\ 
      &      & 0.15 & 0.20& 0.289 &  0.014& 0.260 &  0.016& 0.227 &  0.013& 0.252 &  0.012\\ 
      &      & 0.20 & 0.25& 0.187 &  0.011& 0.204 &  0.012& 0.142 &  0.009& 0.160 &  0.010\\ 
      &      & 0.25 & 0.30& 0.114 &  0.008& 0.113 &  0.013& 0.098 &  0.007& 0.101 &  0.008\\ 
      &      & 0.30 & 0.35& 0.075 &  0.005& 0.064 &  0.006& 0.066 &  0.006& 0.061 &  0.006\\ 
      &      & 0.35 & 0.40& 0.054 &  0.005& 0.045 &  0.004& 0.040 &  0.004& 0.045 &  0.004\\ 
      &      & 0.40 & 0.45& 0.037 &  0.004& 0.035 &  0.004& 0.031 &  0.003& 0.029 &  0.004\\ 
      &      & 0.45 & 0.50& 0.027 &  0.003& 0.026 &  0.004& 0.022 &  0.003& 0.019 &  0.002\\ 

\end{tabular}
\end{center}
\end{table}

\clearpage
\begin{table}[hp!]
\begin{center}
  \caption{\label{tab:xsec-pipp-cu}
    HARP results for the double-differential $\pi^+$ production
    cross-section in the laboratory system,
    $d^2\sigma^{\pi^+}/(dpd\theta)$ for $\pi^+$--Cu interactions. Each row refers to a
    different $(p_{\hbox{\small min}} \le p<p_{\hbox{\small max}},
    \theta_{\hbox{\small min}} \le \theta<\theta_{\hbox{\small max}})$ bin,
    where $p$ and $\theta$ are the pion momentum and polar angle, respectively.
    The central value as well as the square-root of the diagonal elements
    of the covariance matrix are given.}
\vspace{2mm}
\begin{tabular}{rrrr|r@{$\pm$}lr@{$\pm$}lr@{$\pm$}lr@{$\pm$}l}
\hline
$\theta_{\hbox{\small min}}$ &
$\theta_{\hbox{\small max}}$ &
$p_{\hbox{\small min}}$ &
$p_{\hbox{\small max}}$ &
\multicolumn{8}{c}{$d^2\sigma^{\pi^+}/(dpd\theta)$}
\\
(rad) & (rad) & (\GeVc) & (\GeVc) &
\multicolumn{8}{c}{($\barn/(\GeVc \cdot \rad)$)}
\\
  &  &  &
&\multicolumn{2}{c}{$ \bf{3 \ \GeVc}$}
&\multicolumn{2}{c}{$ \bf{5 \ \GeVc}$}
&\multicolumn{2}{c}{$ \bf{8 \ \GeVc}$}
&\multicolumn{2}{c}{$ \bf{12 \ \GeVc}$}
\\
\hline  
 0.35 & 0.55 & 0.15 & 0.20& 0.44 &  0.08& 0.78 &  0.13& 0.82 &  0.13& 0.76 &  0.18\\ 
      &      & 0.20 & 0.25& 0.48 &  0.06& 0.87 &  0.07& 1.08 &  0.09& 1.00 &  0.15\\ 
      &      & 0.25 & 0.30& 0.51 &  0.05& 1.04 &  0.08& 1.26 &  0.08& 1.28 &  0.16\\ 
      &      & 0.30 & 0.35& 0.65 &  0.06& 1.05 &  0.07& 1.23 &  0.08& 1.63 &  0.20\\ 
      &      & 0.35 & 0.40& 0.68 &  0.06& 1.06 &  0.05& 1.38 &  0.11& 1.63 &  0.15\\ 
      &      & 0.40 & 0.45& 0.65 &  0.04& 1.09 &  0.07& 1.42 &  0.06& 1.35 &  0.13\\ 
      &      & 0.45 & 0.50& 0.67 &  0.06& 1.13 &  0.05& 1.32 &  0.07& 1.66 &  0.19\\ 
      &      & 0.50 & 0.60& 0.70 &  0.04& 1.06 &  0.05& 1.44 &  0.07& 1.48 &  0.12\\ 
      &      & 0.60 & 0.70& 0.62 &  0.06& 0.96 &  0.08& 1.34 &  0.12& 1.38 &  0.17\\ 
      &      & 0.70 & 0.80& 0.41 &  0.06& 0.75 &  0.10& 1.11 &  0.15& 0.98 &  0.15\\ 
\hline  
 0.55 & 0.75 & 0.10 & 0.15& 0.49 &  0.13& 0.54 &  0.17& 0.62 &  0.17& 0.66 &  0.26\\ 
      &      & 0.15 & 0.20& 0.67 &  0.07& 0.85 &  0.07& 0.93 &  0.08& 1.04 &  0.14\\ 
      &      & 0.20 & 0.25& 0.77 &  0.06& 1.05 &  0.08& 1.19 &  0.10& 1.26 &  0.20\\ 
      &      & 0.25 & 0.30& 0.76 &  0.07& 1.13 &  0.08& 1.33 &  0.09& 1.62 &  0.16\\ 
      &      & 0.30 & 0.35& 0.75 &  0.05& 1.08 &  0.07& 1.33 &  0.06& 1.24 &  0.14\\ 
      &      & 0.35 & 0.40& 0.78 &  0.05& 1.06 &  0.05& 1.19 &  0.05& 1.37 &  0.12\\ 
      &      & 0.40 & 0.45& 0.70 &  0.04& 1.02 &  0.05& 1.19 &  0.06& 1.26 &  0.11\\ 
      &      & 0.45 & 0.50& 0.61 &  0.04& 0.94 &  0.04& 1.10 &  0.05& 1.14 &  0.10\\ 
      &      & 0.50 & 0.60& 0.57 &  0.04& 0.85 &  0.05& 0.97 &  0.06& 0.99 &  0.09\\ 
      &      & 0.60 & 0.70& 0.42 &  0.06& 0.60 &  0.07& 0.74 &  0.08& 0.71 &  0.11\\ 
      &      & 0.70 & 0.80& 0.27 &  0.05& 0.38 &  0.06& 0.52 &  0.09& 0.47 &  0.10\\ 
\hline  
 0.75 & 0.95 & 0.10 & 0.15& 0.49 &  0.11& 0.62 &  0.14& 0.55 &  0.12& 0.45 &  0.15\\ 
      &      & 0.15 & 0.20& 0.90 &  0.08& 1.08 &  0.07& 1.03 &  0.07& 1.08 &  0.14\\ 
      &      & 0.20 & 0.25& 0.90 &  0.06& 1.11 &  0.06& 1.14 &  0.09& 1.07 &  0.12\\ 
      &      & 0.25 & 0.30& 0.85 &  0.06& 1.05 &  0.05& 1.13 &  0.07& 1.08 &  0.13\\ 
      &      & 0.30 & 0.35& 0.80 &  0.05& 0.90 &  0.04& 1.07 &  0.05& 0.95 &  0.09\\ 
      &      & 0.35 & 0.40& 0.67 &  0.04& 0.80 &  0.03& 0.92 &  0.04& 0.89 &  0.09\\ 
      &      & 0.40 & 0.45& 0.62 &  0.04& 0.74 &  0.03& 0.76 &  0.04& 0.88 &  0.09\\ 
      &      & 0.45 & 0.50& 0.53 &  0.04& 0.68 &  0.03& 0.66 &  0.03& 0.82 &  0.08\\ 
      &      & 0.50 & 0.60& 0.37 &  0.03& 0.50 &  0.04& 0.54 &  0.04& 0.58 &  0.08\\ 
      &      & 0.60 & 0.70& 0.26 &  0.03& 0.29 &  0.05& 0.34 &  0.05& 0.36 &  0.06\\ 
\hline  
 0.95 & 1.15 & 0.10 & 0.15& 0.60 &  0.11& 0.69 &  0.13& 0.70 &  0.12& 0.43 &  0.13\\ 
      &      & 0.15 & 0.20& 0.80 &  0.07& 1.06 &  0.07& 1.13 &  0.08& 1.11 &  0.15\\ 
      &      & 0.20 & 0.25& 0.79 &  0.05& 0.99 &  0.06& 1.01 &  0.05& 1.09 &  0.12\\ 
      &      & 0.25 & 0.30& 0.63 &  0.05& 0.90 &  0.05& 0.89 &  0.05& 0.84 &  0.09\\ 
      &      & 0.30 & 0.35& 0.56 &  0.04& 0.75 &  0.04& 0.80 &  0.04& 0.70 &  0.08\\ 
      &      & 0.35 & 0.40& 0.47 &  0.03& 0.58 &  0.03& 0.64 &  0.04& 0.68 &  0.07\\ 
      &      & 0.40 & 0.45& 0.42 &  0.03& 0.44 &  0.03& 0.49 &  0.03& 0.59 &  0.06\\ 
      &      & 0.45 & 0.50& 0.35 &  0.03& 0.37 &  0.02& 0.39 &  0.02& 0.44 &  0.06\\ 
      &      & 0.50 & 0.60& 0.24 &  0.03& 0.25 &  0.03& 0.28 &  0.03& 0.24 &  0.04\\ 
\hline
\end{tabular}
\end{center}
\end{table}

\begin{table}[hp!]
\begin{center}
\begin{tabular}{rrrr|r@{$\pm$}lr@{$\pm$}lr@{$\pm$}lr@{$\pm$}l}
\hline
$\theta_{\hbox{\small min}}$ &
$\theta_{\hbox{\small max}}$ &
$p_{\hbox{\small min}}$ &
$p_{\hbox{\small max}}$ &
\multicolumn{8}{c}{$d^2\sigma^{\pi^+}/(dpd\theta)$}
\\
(rad) & (rad) & (\GeVc) & (\GeVc) &
\multicolumn{8}{c}{(\barn/($\GeVc \cdot \rad$))}
\\
  &  &  &
&\multicolumn{2}{c}{$ \bf{3 \ \GeVc}$}
&\multicolumn{2}{c}{$ \bf{5 \ \GeVc}$}
&\multicolumn{2}{c}{$ \bf{8 \ \GeVc}$}
&\multicolumn{2}{c}{$ \bf{12 \ \GeVc}$}
\\
\hline
 1.15 & 1.35 & 0.10 & 0.15& 0.77 &  0.13& 0.72 &  0.14& 0.79 &  0.16& 0.79 &  0.20\\ 
      &      & 0.15 & 0.20& 0.85 &  0.07& 0.99 &  0.07& 1.08 &  0.08& 0.90 &  0.11\\ 
      &      & 0.20 & 0.25& 0.75 &  0.05& 0.85 &  0.05& 0.87 &  0.05& 0.73 &  0.09\\ 
      &      & 0.25 & 0.30& 0.54 &  0.04& 0.65 &  0.04& 0.66 &  0.04& 0.66 &  0.08\\ 
      &      & 0.30 & 0.35& 0.41 &  0.03& 0.51 &  0.03& 0.55 &  0.03& 0.62 &  0.07\\ 
      &      & 0.35 & 0.40& 0.31 &  0.02& 0.38 &  0.02& 0.42 &  0.03& 0.41 &  0.06\\ 
      &      & 0.40 & 0.45& 0.24 &  0.02& 0.29 &  0.02& 0.29 &  0.02& 0.31 &  0.04\\ 
      &      & 0.45 & 0.50& 0.18 &  0.02& 0.21 &  0.02& 0.22 &  0.02& 0.23 &  0.04\\ 
\hline  
 1.35 & 1.55 & 0.10 & 0.15& 0.80 &  0.18& 0.76 &  0.17& 0.76 &  0.20& 0.82 &  0.23\\ 
      &      & 0.15 & 0.20& 0.81 &  0.07& 0.97 &  0.08& 0.95 &  0.07& 0.94 &  0.12\\ 
      &      & 0.20 & 0.25& 0.66 &  0.05& 0.77 &  0.05& 0.74 &  0.05& 0.87 &  0.10\\ 
      &      & 0.25 & 0.30& 0.49 &  0.04& 0.48 &  0.04& 0.53 &  0.04& 0.50 &  0.07\\ 
      &      & 0.30 & 0.35& 0.30 &  0.03& 0.34 &  0.02& 0.41 &  0.03& 0.32 &  0.05\\ 
      &      & 0.35 & 0.40& 0.25 &  0.02& 0.26 &  0.02& 0.31 &  0.02& 0.21 &  0.04\\ 
      &      & 0.40 & 0.45& 0.18 &  0.02& 0.18 &  0.01& 0.21 &  0.01& 0.14 &  0.03\\ 
      &      & 0.45 & 0.50& 0.12 &  0.01& 0.11 &  0.01& 0.14 &  0.02& 0.09 &  0.02\\ 
\hline  
 1.55 & 1.75 & 0.10 & 0.15& 0.81 &  0.18& 0.75 &  0.17& 0.70 &  0.19& 0.63 &  0.16\\ 
      &      & 0.15 & 0.20& 0.74 &  0.06& 0.86 &  0.06& 0.83 &  0.06& 0.79 &  0.12\\ 
      &      & 0.20 & 0.25& 0.55 &  0.04& 0.61 &  0.04& 0.58 &  0.04& 0.75 &  0.10\\ 
      &      & 0.25 & 0.30& 0.38 &  0.04& 0.40 &  0.03& 0.41 &  0.03& 0.30 &  0.05\\ 
      &      & 0.30 & 0.35& 0.21 &  0.02& 0.26 &  0.02& 0.28 &  0.02& 0.27 &  0.04\\ 
      &      & 0.35 & 0.40& 0.16 &  0.01& 0.18 &  0.01& 0.20 &  0.02& 0.27 &  0.05\\ 
      &      & 0.40 & 0.45& 0.11 &  0.01& 0.13 &  0.01& 0.12 &  0.01& 0.16 &  0.03\\ 
      &      & 0.45 & 0.50& 0.08 &  0.01& 0.08 &  0.01& 0.07 &  0.01& 0.10 &  0.03\\ 
\hline  
 1.75 & 1.95 & 0.10 & 0.15& 0.71 &  0.13& 0.68 &  0.11& 0.68 &  0.12& 0.52 &  0.12\\ 
      &      & 0.15 & 0.20& 0.67 &  0.04& 0.71 &  0.04& 0.72 &  0.04& 0.61 &  0.09\\ 
      &      & 0.20 & 0.25& 0.41 &  0.04& 0.44 &  0.03& 0.47 &  0.03& 0.60 &  0.08\\ 
      &      & 0.25 & 0.30& 0.22 &  0.02& 0.27 &  0.02& 0.26 &  0.02& 0.32 &  0.06\\ 
      &      & 0.30 & 0.35& 0.16 &  0.02& 0.20 &  0.01& 0.15 &  0.01& 0.19 &  0.04\\ 
      &      & 0.35 & 0.40& 0.14 &  0.01& 0.13 &  0.01& 0.12 &  0.01& 0.12 &  0.03\\ 
      &      & 0.40 & 0.45& 0.10 &  0.01& 0.08 &  0.01& 0.08 &  0.01& 0.05 &  0.02\\ 
      &      & 0.45 & 0.50& 0.05 &  0.02& 0.05 &  0.01& 0.05 &  0.01& 0.03 &  0.01\\ 
\hline  
 1.95 & 2.15 & 0.10 & 0.15& 0.50 &  0.07& 0.57 &  0.08& 0.54 &  0.08& 0.57 &  0.14\\ 
      &      & 0.15 & 0.20& 0.49 &  0.04& 0.55 &  0.03& 0.56 &  0.03& 0.53 &  0.08\\ 
      &      & 0.20 & 0.25& 0.34 &  0.03& 0.34 &  0.02& 0.35 &  0.03& 0.35 &  0.08\\ 
      &      & 0.25 & 0.30& 0.18 &  0.02& 0.20 &  0.01& 0.18 &  0.02& 0.12 &  0.03\\ 
      &      & 0.30 & 0.35& 0.11 &  0.02& 0.13 &  0.01& 0.12 &  0.01& 0.06 &  0.02\\ 
      &      & 0.35 & 0.40& 0.06 &  0.01& 0.07 &  0.01& 0.07 &  0.01& 0.04 &  0.02\\ 
      &      & 0.40 & 0.45& 0.04 &  0.01& 0.04 &  0.01& 0.04 &  0.01& 0.04 &  0.02\\ 
      &      & 0.45 & 0.50& 0.02 &  0.01& 0.02 &  0.01& 0.03 &  0.01& 0.02 &  0.02\\ 

\end{tabular}
\end{center}
\end{table}

\begin{table}[hp!]
\begin{center}
  \caption{\label{tab:xsec-pipm-cu}
    HARP results for the double-differential $\pi^-$ production
    cross-section in the laboratory system,
    $d^2\sigma^{\pi^-}/(dpd\theta)$ for $\pi^+$--Cu interactions. Each row refers to a
    different $(p_{\hbox{\small min}} \le p<p_{\hbox{\small max}},
    \theta_{\hbox{\small min}} \le \theta<\theta_{\hbox{\small max}})$ bin,
    where $p$ and $\theta$ are the pion momentum and polar angle, respectively.
    The central value as well as the square-root of the diagonal elements
    of the covariance matrix are given.}
\vspace{2mm}
\begin{tabular}{rrrr|r@{$\pm$}lr@{$\pm$}lr@{$\pm$}lr@{$\pm$}l}
\hline
$\theta_{\hbox{\small min}}$ &
$\theta_{\hbox{\small max}}$ &
$p_{\hbox{\small min}}$ &
$p_{\hbox{\small max}}$ &
\multicolumn{8}{c}{$d^2\sigma^{\pi^-}/(dpd\theta)$}
\\
(rad) & (rad) & (\GeVc) & (\GeVc) &
\multicolumn{8}{c}{($\barn/(\GeVc \cdot \rad)$)}
\\
  &  &  &
&\multicolumn{2}{c}{$ \bf{3 \ \GeVc}$}
&\multicolumn{2}{c}{$ \bf{5 \ \GeVc}$}
&\multicolumn{2}{c}{$ \bf{8 \ \GeVc}$}
&\multicolumn{2}{c}{$ \bf{12 \ \GeVc}$}
\\
\hline  
 0.35 & 0.55 & 0.15 & 0.20& 0.35 &  0.08& 0.74 &  0.12& 0.75 &  0.12& 0.69 &  0.17\\ 
      &      & 0.20 & 0.25& 0.39 &  0.06& 0.76 &  0.07& 0.87 &  0.09& 0.81 &  0.15\\ 
      &      & 0.25 & 0.30& 0.55 &  0.06& 0.73 &  0.05& 1.02 &  0.06& 1.08 &  0.14\\ 
      &      & 0.30 & 0.35& 0.41 &  0.04& 0.71 &  0.05& 0.86 &  0.05& 1.10 &  0.13\\ 
      &      & 0.35 & 0.40& 0.38 &  0.03& 0.66 &  0.03& 0.86 &  0.05& 0.67 &  0.09\\ 
      &      & 0.40 & 0.45& 0.34 &  0.03& 0.58 &  0.03& 0.80 &  0.04& 0.76 &  0.13\\ 
      &      & 0.45 & 0.50& 0.32 &  0.03& 0.55 &  0.03& 0.75 &  0.04& 1.00 &  0.11\\ 
      &      & 0.50 & 0.60& 0.37 &  0.03& 0.54 &  0.03& 0.74 &  0.04& 0.78 &  0.08\\ 
      &      & 0.60 & 0.70& 0.30 &  0.03& 0.50 &  0.04& 0.69 &  0.05& 0.60 &  0.08\\ 
      &      & 0.70 & 0.80& 0.23 &  0.03& 0.41 &  0.05& 0.59 &  0.07& 0.50 &  0.09\\ 
\hline  
 0.55 & 0.75 & 0.10 & 0.15& 0.39 &  0.13& 0.67 &  0.18& 0.64 &  0.17& 0.67 &  0.23\\ 
      &      & 0.15 & 0.20& 0.54 &  0.07& 0.82 &  0.07& 0.83 &  0.08& 0.85 &  0.13\\ 
      &      & 0.20 & 0.25& 0.59 &  0.05& 0.76 &  0.05& 0.87 &  0.07& 0.81 &  0.13\\ 
      &      & 0.25 & 0.30& 0.59 &  0.05& 0.75 &  0.05& 0.92 &  0.05& 0.89 &  0.11\\ 
      &      & 0.30 & 0.35& 0.40 &  0.04& 0.63 &  0.03& 0.76 &  0.04& 0.95 &  0.11\\ 
      &      & 0.35 & 0.40& 0.36 &  0.03& 0.54 &  0.03& 0.72 &  0.04& 0.58 &  0.08\\ 
      &      & 0.40 & 0.45& 0.37 &  0.03& 0.50 &  0.02& 0.65 &  0.03& 0.48 &  0.06\\ 
      &      & 0.45 & 0.50& 0.35 &  0.03& 0.47 &  0.02& 0.60 &  0.03& 0.49 &  0.06\\ 
      &      & 0.50 & 0.60& 0.28 &  0.02& 0.43 &  0.02& 0.52 &  0.03& 0.53 &  0.06\\ 
      &      & 0.60 & 0.70& 0.24 &  0.02& 0.35 &  0.03& 0.45 &  0.03& 0.49 &  0.06\\ 
      &      & 0.70 & 0.80& 0.17 &  0.03& 0.27 &  0.03& 0.38 &  0.05& 0.42 &  0.07\\ 
\hline  
 0.75 & 0.95 & 0.10 & 0.15& 0.47 &  0.10& 0.54 &  0.10& 0.59 &  0.12& 0.72 &  0.19\\ 
      &      & 0.15 & 0.20& 0.60 &  0.06& 0.73 &  0.06& 0.80 &  0.06& 0.65 &  0.11\\ 
      &      & 0.20 & 0.25& 0.52 &  0.05& 0.67 &  0.04& 0.76 &  0.06& 0.92 &  0.17\\ 
      &      & 0.25 & 0.30& 0.48 &  0.05& 0.63 &  0.03& 0.75 &  0.04& 0.93 &  0.11\\ 
      &      & 0.30 & 0.35& 0.42 &  0.04& 0.54 &  0.03& 0.68 &  0.04& 0.70 &  0.08\\ 
      &      & 0.35 & 0.40& 0.30 &  0.02& 0.46 &  0.02& 0.58 &  0.03& 0.56 &  0.07\\ 
      &      & 0.40 & 0.45& 0.27 &  0.02& 0.41 &  0.02& 0.48 &  0.02& 0.47 &  0.06\\ 
      &      & 0.45 & 0.50& 0.26 &  0.02& 0.36 &  0.02& 0.40 &  0.02& 0.38 &  0.05\\ 
      &      & 0.50 & 0.60& 0.22 &  0.02& 0.30 &  0.02& 0.36 &  0.02& 0.34 &  0.04\\ 
      &      & 0.60 & 0.70& 0.15 &  0.02& 0.24 &  0.02& 0.30 &  0.03& 0.27 &  0.04\\ 
\hline  
 0.95 & 1.15 & 0.10 & 0.15& 0.49 &  0.09& 0.61 &  0.10& 0.60 &  0.11& 0.51 &  0.16\\ 
      &      & 0.15 & 0.20& 0.56 &  0.05& 0.72 &  0.05& 0.81 &  0.06& 0.83 &  0.11\\ 
      &      & 0.20 & 0.25& 0.38 &  0.03& 0.63 &  0.04& 0.72 &  0.04& 0.53 &  0.08\\ 
      &      & 0.25 & 0.30& 0.34 &  0.03& 0.55 &  0.03& 0.59 &  0.04& 0.33 &  0.05\\ 
      &      & 0.30 & 0.35& 0.27 &  0.02& 0.45 &  0.02& 0.51 &  0.03& 0.32 &  0.06\\ 
      &      & 0.35 & 0.40& 0.28 &  0.02& 0.35 &  0.02& 0.44 &  0.02& 0.44 &  0.06\\ 
      &      & 0.40 & 0.45& 0.21 &  0.02& 0.30 &  0.02& 0.35 &  0.02& 0.41 &  0.05\\ 
      &      & 0.45 & 0.50& 0.19 &  0.02& 0.25 &  0.01& 0.29 &  0.02& 0.31 &  0.04\\ 
      &      & 0.50 & 0.60& 0.14 &  0.01& 0.18 &  0.01& 0.24 &  0.01& 0.21 &  0.04\\ 
\hline
\end{tabular}
\end{center}
\end{table}

\begin{table}[hp!]
\begin{center}
\begin{tabular}{rrrr|r@{$\pm$}lr@{$\pm$}lr@{$\pm$}lr@{$\pm$}l}
\hline
$\theta_{\hbox{\small min}}$ &
$\theta_{\hbox{\small max}}$ &
$p_{\hbox{\small min}}$ &
$p_{\hbox{\small max}}$ &
\multicolumn{8}{c}{$d^2\sigma^{\pi^-}/(dpd\theta)$}
\\
(rad) & (rad) & (\GeVc) & (\GeVc) &
\multicolumn{8}{c}{(\barn/($\GeVc \cdot \rad$))}
\\
  &  &  &
&\multicolumn{2}{c}{$ \bf{3 \ \GeVc}$}
&\multicolumn{2}{c}{$ \bf{5 \ \GeVc}$}
&\multicolumn{2}{c}{$ \bf{8 \ \GeVc}$}
&\multicolumn{2}{c}{$ \bf{12 \ \GeVc}$}
\\
\hline
 1.15 & 1.35 & 0.10 & 0.15& 0.51 &  0.10& 0.64 &  0.12& 0.67 &  0.14& 0.47 &  0.15\\ 
      &      & 0.15 & 0.20& 0.59 &  0.06& 0.69 &  0.06& 0.83 &  0.06& 0.73 &  0.11\\ 
      &      & 0.20 & 0.25& 0.46 &  0.04& 0.56 &  0.03& 0.63 &  0.04& 0.63 &  0.09\\ 
      &      & 0.25 & 0.30& 0.36 &  0.03& 0.42 &  0.03& 0.45 &  0.03& 0.54 &  0.07\\ 
      &      & 0.30 & 0.35& 0.27 &  0.03& 0.33 &  0.02& 0.35 &  0.02& 0.41 &  0.06\\ 
      &      & 0.35 & 0.40& 0.19 &  0.02& 0.26 &  0.02& 0.28 &  0.02& 0.34 &  0.05\\ 
      &      & 0.40 & 0.45& 0.15 &  0.01& 0.20 &  0.01& 0.25 &  0.02& 0.30 &  0.04\\ 
      &      & 0.45 & 0.50& 0.12 &  0.01& 0.16 &  0.01& 0.19 &  0.01& 0.21 &  0.03\\ 
\hline  
 1.35 & 1.55 & 0.10 & 0.15& 0.65 &  0.14& 0.74 &  0.15& 0.67 &  0.13& 0.71 &  0.22\\ 
      &      & 0.15 & 0.20& 0.55 &  0.06& 0.65 &  0.06& 0.72 &  0.06& 0.73 &  0.11\\ 
      &      & 0.20 & 0.25& 0.39 &  0.04& 0.50 &  0.04& 0.57 &  0.04& 0.63 &  0.09\\ 
      &      & 0.25 & 0.30& 0.29 &  0.03& 0.39 &  0.03& 0.39 &  0.03& 0.44 &  0.06\\ 
      &      & 0.30 & 0.35& 0.20 &  0.02& 0.26 &  0.02& 0.28 &  0.02& 0.31 &  0.05\\ 
      &      & 0.35 & 0.40& 0.16 &  0.02& 0.18 &  0.01& 0.23 &  0.02& 0.20 &  0.03\\ 
      &      & 0.40 & 0.45& 0.11 &  0.01& 0.13 &  0.01& 0.18 &  0.01& 0.16 &  0.03\\ 
      &      & 0.45 & 0.50& 0.08 &  0.01& 0.09 &  0.01& 0.15 &  0.01& 0.13 &  0.03\\ 
\hline  
 1.55 & 1.75 & 0.10 & 0.15& 0.70 &  0.15& 0.71 &  0.14& 0.61 &  0.14& 0.52 &  0.18\\ 
      &      & 0.15 & 0.20& 0.54 &  0.05& 0.58 &  0.05& 0.59 &  0.05& 0.73 &  0.12\\ 
      &      & 0.20 & 0.25& 0.30 &  0.03& 0.39 &  0.03& 0.42 &  0.03& 0.54 &  0.08\\ 
      &      & 0.25 & 0.30& 0.22 &  0.02& 0.29 &  0.02& 0.27 &  0.02& 0.34 &  0.06\\ 
      &      & 0.30 & 0.35& 0.16 &  0.02& 0.20 &  0.02& 0.18 &  0.01& 0.22 &  0.04\\ 
      &      & 0.35 & 0.40& 0.11 &  0.01& 0.13 &  0.01& 0.14 &  0.01& 0.17 &  0.04\\ 
      &      & 0.40 & 0.45& 0.07 &  0.01& 0.09 &  0.01& 0.11 &  0.01& 0.12 &  0.03\\ 
      &      & 0.45 & 0.50& 0.05 &  0.01& 0.07 &  0.01& 0.08 &  0.01& 0.10 &  0.02\\ 
\hline  
 1.75 & 1.95 & 0.10 & 0.15& 0.39 &  0.07& 0.54 &  0.09& 0.51 &  0.10& 0.33 &  0.10\\ 
      &      & 0.15 & 0.20& 0.34 &  0.03& 0.42 &  0.03& 0.50 &  0.03& 0.44 &  0.07\\ 
      &      & 0.20 & 0.25& 0.28 &  0.03& 0.29 &  0.02& 0.34 &  0.02& 0.32 &  0.06\\ 
      &      & 0.25 & 0.30& 0.17 &  0.02& 0.19 &  0.02& 0.24 &  0.02& 0.18 &  0.04\\ 
      &      & 0.30 & 0.35& 0.10 &  0.01& 0.13 &  0.01& 0.14 &  0.01& 0.13 &  0.03\\ 
      &      & 0.35 & 0.40& 0.08 &  0.01& 0.09 &  0.01& 0.10 &  0.01& 0.13 &  0.03\\ 
      &      & 0.40 & 0.45& 0.06 &  0.01& 0.06 &  0.01& 0.09 &  0.01& 0.11 &  0.03\\ 
      &      & 0.45 & 0.50& 0.05 &  0.01& 0.05 &  0.01& 0.07 &  0.01& 0.09 &  0.03\\ 
\hline  
 1.95 & 2.15 & 0.10 & 0.15& 0.29 &  0.05& 0.44 &  0.06& 0.43 &  0.06& 0.53 &  0.14\\ 
      &      & 0.15 & 0.20& 0.32 &  0.03& 0.34 &  0.02& 0.38 &  0.02& 0.48 &  0.08\\ 
      &      & 0.20 & 0.25& 0.20 &  0.03& 0.23 &  0.01& 0.24 &  0.02& 0.25 &  0.06\\ 
      &      & 0.25 & 0.30& 0.13 &  0.02& 0.16 &  0.01& 0.15 &  0.02& 0.13 &  0.04\\ 
      &      & 0.30 & 0.35& 0.08 &  0.01& 0.09 &  0.01& 0.09 &  0.01& 0.06 &  0.02\\ 
      &      & 0.35 & 0.40& 0.06 &  0.01& 0.07 &  0.01& 0.07 &  0.01& 0.05 &  0.02\\ 
      &      & 0.40 & 0.45& 0.05 &  0.01& 0.05 &  0.01& 0.05 &  0.01& 0.04 &  0.02\\ 
      &      & 0.45 & 0.50& 0.03 &  0.01& 0.04 &  0.01& 0.03 &  0.01& 0.03 &  0.02\\ 

\end{tabular}
\end{center}
\end{table}
\clearpage
\begin{table}[hp!]
\begin{center}
  \caption{\label{tab:xsec-pimp-cu}
    HARP results for the double-differential $\pi^+$ production
    cross-section in the laboratory system,
    $d^2\sigma^{\pi^+}/(dpd\theta)$ for $\pi^-$--Cu interactions. Each row refers to a
    different $(p_{\hbox{\small min}} \le p<p_{\hbox{\small max}},
    \theta_{\hbox{\small min}} \le \theta<\theta_{\hbox{\small max}})$ bin,
    where $p$ and $\theta$ are the pion momentum and polar angle, respectively.
    The central value as well as the square-root of the diagonal elements
    of the covariance matrix are given.}
\vspace{2mm}
\begin{tabular}{rrrr|r@{$\pm$}lr@{$\pm$}lr@{$\pm$}lr@{$\pm$}l}
\hline
$\theta_{\hbox{\small min}}$ &
$\theta_{\hbox{\small max}}$ &
$p_{\hbox{\small min}}$ &
$p_{\hbox{\small max}}$ &
\multicolumn{8}{c}{$d^2\sigma^{\pi^+}/(dpd\theta)$}
\\
(rad) & (rad) & (\GeVc) & (\GeVc) &
\multicolumn{8}{c}{($\barn/(\GeVc \cdot \rad)$)}
\\
  &  &  &
&\multicolumn{2}{c}{$ \bf{3 \ \GeVc}$}
&\multicolumn{2}{c}{$ \bf{5 \ \GeVc}$}
&\multicolumn{2}{c}{$ \bf{8 \ \GeVc}$}
&\multicolumn{2}{c}{$ \bf{12 \ \GeVc}$}
\\
\hline  
 0.35 & 0.55 & 0.15 & 0.20& 0.34 &  0.09& 0.60 &  0.16& 0.53 &  0.17& 0.67 &  0.24\\ 
      &      & 0.20 & 0.25& 0.37 &  0.06& 0.65 &  0.10& 0.70 &  0.12& 0.89 &  0.16\\ 
      &      & 0.25 & 0.30& 0.46 &  0.05& 0.81 &  0.08& 0.90 &  0.09& 1.22 &  0.13\\ 
      &      & 0.30 & 0.35& 0.50 &  0.04& 0.85 &  0.06& 0.92 &  0.06& 1.10 &  0.09\\ 
      &      & 0.35 & 0.40& 0.46 &  0.03& 0.72 &  0.05& 0.91 &  0.07& 1.20 &  0.10\\ 
      &      & 0.40 & 0.45& 0.49 &  0.04& 0.76 &  0.06& 0.93 &  0.06& 1.25 &  0.09\\ 
      &      & 0.45 & 0.50& 0.50 &  0.03& 0.79 &  0.05& 0.99 &  0.06& 1.16 &  0.09\\ 
      &      & 0.50 & 0.60& 0.49 &  0.03& 0.78 &  0.05& 1.05 &  0.07& 1.21 &  0.09\\ 
      &      & 0.60 & 0.70& 0.43 &  0.05& 0.72 &  0.07& 1.07 &  0.10& 1.19 &  0.13\\ 
      &      & 0.70 & 0.80& 0.32 &  0.06& 0.58 &  0.09& 0.94 &  0.13& 1.27 &  0.14\\ 
\hline  
 0.55 & 0.75 & 0.10 & 0.15& 0.35 &  0.14& 0.51 &  0.20& 0.32 &  0.19& 0.36 &  0.21\\ 
      &      & 0.15 & 0.20& 0.52 &  0.07& 0.73 &  0.10& 0.64 &  0.10& 0.71 &  0.16\\ 
      &      & 0.20 & 0.25& 0.53 &  0.05& 0.81 &  0.09& 0.80 &  0.10& 1.13 &  0.14\\ 
      &      & 0.25 & 0.30& 0.59 &  0.05& 0.82 &  0.06& 0.95 &  0.07& 1.09 &  0.10\\ 
      &      & 0.30 & 0.35& 0.57 &  0.04& 0.82 &  0.05& 0.90 &  0.05& 1.08 &  0.07\\ 
      &      & 0.35 & 0.40& 0.55 &  0.03& 0.78 &  0.04& 0.83 &  0.04& 1.02 &  0.07\\ 
      &      & 0.40 & 0.45& 0.50 &  0.03& 0.71 &  0.04& 0.84 &  0.05& 1.06 &  0.07\\ 
      &      & 0.45 & 0.50& 0.47 &  0.03& 0.70 &  0.04& 0.85 &  0.05& 1.07 &  0.06\\ 
      &      & 0.50 & 0.60& 0.41 &  0.03& 0.66 &  0.05& 0.80 &  0.06& 0.91 &  0.07\\ 
      &      & 0.60 & 0.70& 0.28 &  0.04& 0.49 &  0.07& 0.60 &  0.08& 0.64 &  0.09\\ 
      &      & 0.70 & 0.80& 0.18 &  0.04& 0.33 &  0.07& 0.41 &  0.08& 0.52 &  0.09\\ 
\hline  
 0.75 & 0.95 & 0.10 & 0.15& 0.41 &  0.12& 0.57 &  0.15& 0.37 &  0.15& 0.35 &  0.15\\ 
      &      & 0.15 & 0.20& 0.63 &  0.06& 0.82 &  0.07& 0.74 &  0.08& 0.72 &  0.10\\ 
      &      & 0.20 & 0.25& 0.60 &  0.06& 0.84 &  0.08& 0.85 &  0.08& 0.91 &  0.10\\ 
      &      & 0.25 & 0.30& 0.53 &  0.04& 0.74 &  0.05& 0.80 &  0.05& 1.02 &  0.08\\ 
      &      & 0.30 & 0.35& 0.51 &  0.04& 0.64 &  0.04& 0.75 &  0.04& 0.90 &  0.06\\ 
      &      & 0.35 & 0.40& 0.46 &  0.03& 0.58 &  0.03& 0.66 &  0.03& 0.84 &  0.05\\ 
      &      & 0.40 & 0.45& 0.38 &  0.02& 0.55 &  0.03& 0.59 &  0.03& 0.83 &  0.05\\ 
      &      & 0.45 & 0.50& 0.32 &  0.02& 0.50 &  0.03& 0.54 &  0.03& 0.67 &  0.05\\ 
      &      & 0.50 & 0.60& 0.25 &  0.02& 0.38 &  0.04& 0.43 &  0.04& 0.39 &  0.04\\ 
      &      & 0.60 & 0.70& 0.16 &  0.03& 0.25 &  0.04& 0.29 &  0.04& 0.26 &  0.03\\ 
\hline  
 0.95 & 1.15 & 0.10 & 0.15& 0.40 &  0.10& 0.55 &  0.12& 0.39 &  0.10& 0.32 &  0.09\\ 
      &      & 0.15 & 0.20& 0.65 &  0.06& 0.73 &  0.07& 0.77 &  0.08& 0.79 &  0.10\\ 
      &      & 0.20 & 0.25& 0.58 &  0.04& 0.72 &  0.06& 0.77 &  0.06& 0.88 &  0.10\\ 
      &      & 0.25 & 0.30& 0.45 &  0.03& 0.63 &  0.04& 0.63 &  0.04& 0.81 &  0.07\\ 
      &      & 0.30 & 0.35& 0.39 &  0.02& 0.53 &  0.03& 0.56 &  0.03& 0.63 &  0.04\\ 
      &      & 0.35 & 0.40& 0.31 &  0.02& 0.44 &  0.03& 0.47 &  0.02& 0.59 &  0.04\\ 
      &      & 0.40 & 0.45& 0.25 &  0.01& 0.37 &  0.03& 0.40 &  0.02& 0.48 &  0.04\\ 
      &      & 0.45 & 0.50& 0.20 &  0.02& 0.29 &  0.02& 0.33 &  0.02& 0.33 &  0.04\\ 
      &      & 0.50 & 0.60& 0.15 &  0.02& 0.18 &  0.02& 0.23 &  0.02& 0.22 &  0.02\\ 
\hline
\end{tabular}
\end{center}
\end{table}

\begin{table}[hp!]
\begin{center}
\begin{tabular}{rrrr|r@{$\pm$}lr@{$\pm$}lr@{$\pm$}lr@{$\pm$}l}
\hline
$\theta_{\hbox{\small min}}$ &
$\theta_{\hbox{\small max}}$ &
$p_{\hbox{\small min}}$ &
$p_{\hbox{\small max}}$ &
\multicolumn{8}{c}{$d^2\sigma^{\pi^+}/(dpd\theta)$}
\\
(rad) & (rad) & (\GeVc) & (\GeVc) &
\multicolumn{8}{c}{(\barn/($\GeVc \cdot \rad$))}
\\
  &  &  &
&\multicolumn{2}{c}{$ \bf{3 \ \GeVc}$}
&\multicolumn{2}{c}{$ \bf{5 \ \GeVc}$}
&\multicolumn{2}{c}{$ \bf{8 \ \GeVc}$}
&\multicolumn{2}{c}{$ \bf{12 \ \GeVc}$}
\\
\hline
 1.15 & 1.35 & 0.10 & 0.15& 0.47 &  0.12& 0.52 &  0.12& 0.44 &  0.11& 0.39 &  0.09\\ 
      &      & 0.15 & 0.20& 0.59 &  0.06& 0.72 &  0.07& 0.66 &  0.08& 0.70 &  0.08\\ 
      &      & 0.20 & 0.25& 0.51 &  0.04& 0.59 &  0.04& 0.60 &  0.05& 0.77 &  0.06\\ 
      &      & 0.25 & 0.30& 0.38 &  0.03& 0.48 &  0.04& 0.52 &  0.03& 0.58 &  0.04\\ 
      &      & 0.30 & 0.35& 0.27 &  0.02& 0.41 &  0.03& 0.43 &  0.02& 0.45 &  0.04\\ 
      &      & 0.35 & 0.40& 0.22 &  0.02& 0.33 &  0.02& 0.34 &  0.02& 0.41 &  0.03\\ 
      &      & 0.40 & 0.45& 0.17 &  0.02& 0.24 &  0.02& 0.25 &  0.02& 0.22 &  0.02\\ 
      &      & 0.45 & 0.50& 0.14 &  0.01& 0.17 &  0.02& 0.18 &  0.02& 0.17 &  0.02\\ 
\hline  
 1.35 & 1.55 & 0.10 & 0.15& 0.49 &  0.14& 0.55 &  0.14& 0.54 &  0.13& 0.51 &  0.14\\ 
      &      & 0.15 & 0.20& 0.58 &  0.06& 0.66 &  0.07& 0.60 &  0.07& 0.70 &  0.09\\ 
      &      & 0.20 & 0.25& 0.44 &  0.04& 0.52 &  0.04& 0.56 &  0.04& 0.57 &  0.05\\ 
      &      & 0.25 & 0.30& 0.32 &  0.03& 0.37 &  0.03& 0.36 &  0.03& 0.41 &  0.03\\ 
      &      & 0.30 & 0.35& 0.23 &  0.02& 0.28 &  0.02& 0.27 &  0.02& 0.27 &  0.02\\ 
      &      & 0.35 & 0.40& 0.16 &  0.01& 0.21 &  0.01& 0.21 &  0.01& 0.22 &  0.03\\ 
      &      & 0.40 & 0.45& 0.11 &  0.01& 0.16 &  0.01& 0.15 &  0.01& 0.16 &  0.02\\ 
      &      & 0.45 & 0.50& 0.08 &  0.01& 0.11 &  0.01& 0.10 &  0.01& 0.09 &  0.02\\ 
\hline  
 1.55 & 1.75 & 0.10 & 0.15& 0.44 &  0.12& 0.50 &  0.13& 0.49 &  0.12& 0.47 &  0.11\\ 
      &      & 0.15 & 0.20& 0.52 &  0.05& 0.62 &  0.06& 0.49 &  0.05& 0.60 &  0.07\\ 
      &      & 0.20 & 0.25& 0.36 &  0.03& 0.44 &  0.03& 0.44 &  0.03& 0.55 &  0.06\\ 
      &      & 0.25 & 0.30& 0.26 &  0.02& 0.28 &  0.02& 0.28 &  0.02& 0.31 &  0.02\\ 
      &      & 0.30 & 0.35& 0.18 &  0.02& 0.21 &  0.02& 0.21 &  0.01& 0.21 &  0.02\\ 
      &      & 0.35 & 0.40& 0.13 &  0.01& 0.15 &  0.01& 0.14 &  0.01& 0.13 &  0.01\\ 
      &      & 0.40 & 0.45& 0.09 &  0.01& 0.10 &  0.01& 0.10 &  0.01& 0.11 &  0.01\\ 
      &      & 0.45 & 0.50& 0.06 &  0.01& 0.07 &  0.01& 0.07 &  0.01& 0.05 &  0.01\\ 
\hline  
 1.75 & 1.95 & 0.10 & 0.15& 0.40 &  0.08& 0.47 &  0.09& 0.48 &  0.09& 0.40 &  0.08\\ 
      &      & 0.15 & 0.20& 0.45 &  0.03& 0.48 &  0.04& 0.44 &  0.03& 0.44 &  0.04\\ 
      &      & 0.20 & 0.25& 0.27 &  0.02& 0.32 &  0.02& 0.32 &  0.02& 0.34 &  0.03\\ 
      &      & 0.25 & 0.30& 0.16 &  0.01& 0.21 &  0.02& 0.21 &  0.01& 0.20 &  0.02\\ 
      &      & 0.30 & 0.35& 0.12 &  0.01& 0.14 &  0.01& 0.16 &  0.01& 0.13 &  0.02\\ 
      &      & 0.35 & 0.40& 0.08 &  0.01& 0.09 &  0.01& 0.11 &  0.01& 0.07 &  0.01\\ 
      &      & 0.40 & 0.45& 0.05 &  0.01& 0.06 &  0.01& 0.06 &  0.01& 0.05 &  0.01\\ 
      &      & 0.45 & 0.50& 0.03 &  0.01& 0.04 &  0.01& 0.04 &  0.01& 0.02 &  0.01\\ 
\hline  
 1.95 & 2.15 & 0.10 & 0.15& 0.34 &  0.06& 0.39 &  0.07& 0.35 &  0.05& 0.32 &  0.08\\ 
      &      & 0.15 & 0.20& 0.34 &  0.02& 0.38 &  0.03& 0.35 &  0.02& 0.35 &  0.04\\ 
      &      & 0.20 & 0.25& 0.20 &  0.02& 0.25 &  0.02& 0.22 &  0.02& 0.23 &  0.03\\ 
      &      & 0.25 & 0.30& 0.12 &  0.01& 0.16 &  0.01& 0.13 &  0.01& 0.12 &  0.02\\ 
      &      & 0.30 & 0.35& 0.06 &  0.01& 0.10 &  0.01& 0.09 &  0.01& 0.08 &  0.01\\ 
      &      & 0.35 & 0.40& 0.05 &  0.01& 0.06 &  0.01& 0.06 &  0.01& 0.04 &  0.01\\ 
      &      & 0.40 & 0.45& 0.04 &  0.01& 0.04 &  0.01& 0.04 &  0.01& 0.02 &  0.01\\ 
      &      & 0.45 & 0.50& 0.03 &  0.01& 0.02 &  0.01& 0.02 &  0.01& 0.01 &  0.01\\ 

\end{tabular}
\end{center}
\end{table}

\begin{table}[hp!]
\begin{center}
  \caption{\label{tab:xsec-pimm-cu}
    HARP results for the double-differential $\pi^-$ production
    cross-section in the laboratory system,
    $d^2\sigma^{\pi^-}/(dpd\theta)$ for $\pi^-$--Cu interactions. Each row refers to a
    different $(p_{\hbox{\small min}} \le p<p_{\hbox{\small max}},
    \theta_{\hbox{\small min}} \le \theta<\theta_{\hbox{\small max}})$ bin,
    where $p$ and $\theta$ are the pion momentum and polar angle, respectively.
    The central value as well as the square-root of the diagonal elements
    of the covariance matrix are given.}
\vspace{2mm}
\begin{tabular}{rrrr|r@{$\pm$}lr@{$\pm$}lr@{$\pm$}lr@{$\pm$}l}
\hline
$\theta_{\hbox{\small min}}$ &
$\theta_{\hbox{\small max}}$ &
$p_{\hbox{\small min}}$ &
$p_{\hbox{\small max}}$ &
\multicolumn{8}{c}{$d^2\sigma^{\pi^-}/(dpd\theta)$}
\\
(rad) & (rad) & (\GeVc) & (\GeVc) &
\multicolumn{8}{c}{($\barn/(\GeVc \cdot \rad)$)}
\\
  &  &  &
&\multicolumn{2}{c}{$ \bf{3 \ \GeVc}$}
&\multicolumn{2}{c}{$ \bf{5 \ \GeVc}$}
&\multicolumn{2}{c}{$ \bf{8 \ \GeVc}$}
&\multicolumn{2}{c}{$ \bf{12 \ \GeVc}$}
\\
\hline  
 0.35 & 0.55 & 0.15 & 0.20& 0.56 &  0.11& 0.74 &  0.18& 0.68 &  0.22& 0.86 &  0.25\\ 
      &      & 0.20 & 0.25& 0.60 &  0.07& 0.93 &  0.12& 0.87 &  0.13& 1.07 &  0.18\\ 
      &      & 0.25 & 0.30& 0.69 &  0.06& 1.06 &  0.09& 1.05 &  0.10& 1.30 &  0.11\\ 
      &      & 0.30 & 0.35& 0.65 &  0.05& 0.99 &  0.07& 1.06 &  0.08& 1.44 &  0.11\\ 
      &      & 0.35 & 0.40& 0.61 &  0.04& 0.90 &  0.07& 1.09 &  0.07& 1.10 &  0.08\\ 
      &      & 0.40 & 0.45& 0.65 &  0.05& 0.97 &  0.07& 1.05 &  0.06& 1.18 &  0.12\\ 
      &      & 0.45 & 0.50& 0.66 &  0.04& 0.92 &  0.06& 1.09 &  0.07& 1.34 &  0.09\\ 
      &      & 0.50 & 0.60& 0.65 &  0.04& 0.96 &  0.07& 1.10 &  0.07& 1.18 &  0.08\\ 
      &      & 0.60 & 0.70& 0.61 &  0.05& 0.98 &  0.09& 1.10 &  0.09& 1.26 &  0.10\\ 
      &      & 0.70 & 0.80& 0.53 &  0.07& 0.86 &  0.11& 1.01 &  0.12& 1.03 &  0.15\\ 
\hline  
 0.55 & 0.75 & 0.10 & 0.15& 0.60 &  0.17& 0.70 &  0.24& 0.47 &  0.22& 0.52 &  0.24\\ 
      &      & 0.15 & 0.20& 0.75 &  0.08& 1.00 &  0.12& 0.79 &  0.12& 0.84 &  0.16\\ 
      &      & 0.20 & 0.25& 0.81 &  0.07& 1.05 &  0.09& 1.09 &  0.10& 1.17 &  0.13\\ 
      &      & 0.25 & 0.30& 0.85 &  0.06& 1.10 &  0.08& 1.04 &  0.07& 1.13 &  0.09\\ 
      &      & 0.30 & 0.35& 0.76 &  0.05& 0.96 &  0.06& 1.02 &  0.06& 1.25 &  0.09\\ 
      &      & 0.35 & 0.40& 0.68 &  0.04& 0.86 &  0.05& 0.94 &  0.05& 1.11 &  0.06\\ 
      &      & 0.40 & 0.45& 0.67 &  0.04& 0.89 &  0.06& 0.86 &  0.05& 0.96 &  0.05\\ 
      &      & 0.45 & 0.50& 0.62 &  0.03& 0.87 &  0.05& 0.83 &  0.04& 0.99 &  0.06\\ 
      &      & 0.50 & 0.60& 0.56 &  0.03& 0.79 &  0.05& 0.80 &  0.04& 0.82 &  0.06\\ 
      &      & 0.60 & 0.70& 0.49 &  0.04& 0.70 &  0.06& 0.74 &  0.06& 0.68 &  0.07\\ 
      &      & 0.70 & 0.80& 0.42 &  0.06& 0.58 &  0.08& 0.63 &  0.09& 0.54 &  0.07\\ 
\hline  
 0.75 & 0.95 & 0.10 & 0.15& 0.75 &  0.16& 0.83 &  0.19& 0.54 &  0.16& 0.63 &  0.19\\ 
      &      & 0.15 & 0.20& 0.95 &  0.08& 1.08 &  0.10& 0.94 &  0.09& 0.99 &  0.11\\ 
      &      & 0.20 & 0.25& 0.85 &  0.06& 1.05 &  0.08& 0.97 &  0.07& 1.11 &  0.10\\ 
      &      & 0.25 & 0.30& 0.75 &  0.05& 0.98 &  0.06& 0.94 &  0.06& 1.10 &  0.07\\ 
      &      & 0.30 & 0.35& 0.64 &  0.04& 0.87 &  0.05& 0.83 &  0.05& 0.86 &  0.06\\ 
      &      & 0.35 & 0.40& 0.58 &  0.03& 0.75 &  0.04& 0.76 &  0.04& 0.68 &  0.04\\ 
      &      & 0.40 & 0.45& 0.55 &  0.03& 0.69 &  0.04& 0.69 &  0.04& 0.63 &  0.05\\ 
      &      & 0.45 & 0.50& 0.49 &  0.02& 0.63 &  0.04& 0.58 &  0.03& 0.60 &  0.04\\ 
      &      & 0.50 & 0.60& 0.43 &  0.03& 0.53 &  0.04& 0.50 &  0.03& 0.47 &  0.04\\ 
      &      & 0.60 & 0.70& 0.35 &  0.04& 0.44 &  0.04& 0.42 &  0.04& 0.35 &  0.04\\ 
\hline  
 0.95 & 1.15 & 0.10 & 0.15& 0.84 &  0.16& 0.97 &  0.18& 0.73 &  0.16& 0.69 &  0.15\\ 
      &      & 0.15 & 0.20& 0.95 &  0.08& 1.11 &  0.09& 0.99 &  0.08& 0.93 &  0.10\\ 
      &      & 0.20 & 0.25& 0.78 &  0.06& 1.01 &  0.07& 0.85 &  0.06& 0.82 &  0.07\\ 
      &      & 0.25 & 0.30& 0.72 &  0.04& 0.81 &  0.05& 0.77 &  0.05& 0.74 &  0.04\\ 
      &      & 0.30 & 0.35& 0.57 &  0.03& 0.66 &  0.04& 0.67 &  0.04& 0.58 &  0.04\\ 
      &      & 0.35 & 0.40& 0.49 &  0.03& 0.57 &  0.03& 0.52 &  0.03& 0.56 &  0.04\\ 
      &      & 0.40 & 0.45& 0.40 &  0.02& 0.49 &  0.03& 0.41 &  0.02& 0.44 &  0.04\\ 
      &      & 0.45 & 0.50& 0.34 &  0.02& 0.41 &  0.03& 0.35 &  0.02& 0.36 &  0.05\\ 
      &      & 0.50 & 0.60& 0.28 &  0.02& 0.33 &  0.02& 0.29 &  0.02& 0.28 &  0.03\\ 
\hline
\end{tabular}
\end{center}
\end{table}

\begin{table}[hp!]
\begin{center}
\begin{tabular}{rrrr|r@{$\pm$}lr@{$\pm$}lr@{$\pm$}lr@{$\pm$}l}
\hline
$\theta_{\hbox{\small min}}$ &
$\theta_{\hbox{\small max}}$ &
$p_{\hbox{\small min}}$ &
$p_{\hbox{\small max}}$ &
\multicolumn{8}{c}{$d^2\sigma^{\pi^-}/(dpd\theta)$}
\\
(rad) & (rad) & (\GeVc) & (\GeVc) &
\multicolumn{8}{c}{(\barn/($\GeVc \cdot \rad$))}
\\
  &  &  &
&\multicolumn{2}{c}{$ \bf{3 \ \GeVc}$}
&\multicolumn{2}{c}{$ \bf{5 \ \GeVc}$}
&\multicolumn{2}{c}{$ \bf{8 \ \GeVc}$}
&\multicolumn{2}{c}{$ \bf{12 \ \GeVc}$}
\\
\hline
 1.15 & 1.35 & 0.10 & 0.15& 0.87 &  0.18& 1.05 &  0.21& 0.80 &  0.17& 0.76 &  0.15\\ 
      &      & 0.15 & 0.20& 0.99 &  0.09& 1.10 &  0.10& 0.90 &  0.08& 0.89 &  0.09\\ 
      &      & 0.20 & 0.25& 0.73 &  0.05& 0.80 &  0.06& 0.77 &  0.05& 0.75 &  0.06\\ 
      &      & 0.25 & 0.30& 0.55 &  0.04& 0.63 &  0.04& 0.59 &  0.04& 0.61 &  0.05\\ 
      &      & 0.30 & 0.35& 0.42 &  0.03& 0.51 &  0.03& 0.46 &  0.03& 0.46 &  0.03\\ 
      &      & 0.35 & 0.40& 0.34 &  0.02& 0.40 &  0.03& 0.37 &  0.02& 0.36 &  0.03\\ 
      &      & 0.40 & 0.45& 0.29 &  0.02& 0.32 &  0.02& 0.29 &  0.02& 0.27 &  0.03\\ 
      &      & 0.45 & 0.50& 0.23 &  0.02& 0.26 &  0.02& 0.24 &  0.02& 0.21 &  0.03\\ 
\hline  
 1.35 & 1.55 & 0.10 & 0.15& 0.91 &  0.21& 0.96 &  0.22& 0.79 &  0.19& 0.75 &  0.19\\ 
      &      & 0.15 & 0.20& 0.91 &  0.09& 1.00 &  0.10& 0.83 &  0.08& 0.74 &  0.08\\ 
      &      & 0.20 & 0.25& 0.61 &  0.05& 0.73 &  0.05& 0.69 &  0.05& 0.61 &  0.06\\ 
      &      & 0.25 & 0.30& 0.44 &  0.03& 0.53 &  0.04& 0.48 &  0.04& 0.54 &  0.06\\ 
      &      & 0.30 & 0.35& 0.31 &  0.02& 0.35 &  0.03& 0.33 &  0.02& 0.34 &  0.03\\ 
      &      & 0.35 & 0.40& 0.25 &  0.02& 0.26 &  0.02& 0.25 &  0.02& 0.21 &  0.03\\ 
      &      & 0.40 & 0.45& 0.21 &  0.02& 0.21 &  0.02& 0.20 &  0.01& 0.15 &  0.02\\ 
      &      & 0.45 & 0.50& 0.16 &  0.01& 0.16 &  0.01& 0.16 &  0.01& 0.10 &  0.01\\ 
\hline  
 1.55 & 1.75 & 0.10 & 0.15& 0.85 &  0.18& 0.93 &  0.21& 0.77 &  0.18& 0.48 &  0.13\\ 
      &      & 0.15 & 0.20& 0.73 &  0.07& 0.80 &  0.07& 0.71 &  0.06& 0.59 &  0.06\\ 
      &      & 0.20 & 0.25& 0.50 &  0.04& 0.56 &  0.04& 0.49 &  0.04& 0.48 &  0.04\\ 
      &      & 0.25 & 0.30& 0.37 &  0.03& 0.40 &  0.04& 0.34 &  0.03& 0.30 &  0.02\\ 
      &      & 0.30 & 0.35& 0.26 &  0.02& 0.27 &  0.02& 0.24 &  0.02& 0.23 &  0.02\\ 
      &      & 0.35 & 0.40& 0.19 &  0.01& 0.19 &  0.02& 0.18 &  0.01& 0.15 &  0.02\\ 
      &      & 0.40 & 0.45& 0.15 &  0.01& 0.13 &  0.01& 0.14 &  0.01& 0.10 &  0.02\\ 
      &      & 0.45 & 0.50& 0.10 &  0.01& 0.10 &  0.01& 0.10 &  0.01& 0.07 &  0.01\\ 
\hline  
 1.75 & 1.95 & 0.10 & 0.15& 0.74 &  0.13& 0.77 &  0.13& 0.62 &  0.12& 0.43 &  0.09\\ 
      &      & 0.15 & 0.20& 0.65 &  0.05& 0.65 &  0.05& 0.61 &  0.04& 0.57 &  0.06\\ 
      &      & 0.20 & 0.25& 0.45 &  0.04& 0.44 &  0.03& 0.37 &  0.03& 0.37 &  0.04\\ 
      &      & 0.25 & 0.30& 0.28 &  0.02& 0.28 &  0.02& 0.24 &  0.02& 0.20 &  0.02\\ 
      &      & 0.30 & 0.35& 0.17 &  0.02& 0.18 &  0.01& 0.17 &  0.01& 0.15 &  0.02\\ 
      &      & 0.35 & 0.40& 0.12 &  0.01& 0.15 &  0.01& 0.13 &  0.01& 0.08 &  0.01\\ 
      &      & 0.40 & 0.45& 0.09 &  0.01& 0.11 &  0.01& 0.10 &  0.01& 0.07 &  0.01\\ 
      &      & 0.45 & 0.50& 0.06 &  0.01& 0.07 &  0.01& 0.07 &  0.01& 0.06 &  0.01\\ 
\hline  
 1.95 & 2.15 & 0.10 & 0.15& 0.65 &  0.10& 0.62 &  0.09& 0.50 &  0.08& 0.43 &  0.06\\ 
      &      & 0.15 & 0.20& 0.52 &  0.04& 0.50 &  0.03& 0.48 &  0.03& 0.46 &  0.04\\ 
      &      & 0.20 & 0.25& 0.33 &  0.02& 0.32 &  0.02& 0.30 &  0.02& 0.29 &  0.03\\ 
      &      & 0.25 & 0.30& 0.20 &  0.02& 0.21 &  0.02& 0.16 &  0.01& 0.16 &  0.02\\ 
      &      & 0.30 & 0.35& 0.13 &  0.01& 0.14 &  0.01& 0.11 &  0.01& 0.09 &  0.01\\ 
      &      & 0.35 & 0.40& 0.09 &  0.01& 0.10 &  0.01& 0.08 &  0.01& 0.04 &  0.01\\ 
      &      & 0.40 & 0.45& 0.06 &  0.01& 0.08 &  0.01& 0.06 &  0.01& 0.03 &  0.01\\ 
      &      & 0.45 & 0.50& 0.05 &  0.01& 0.05 &  0.01& 0.05 &  0.01& 0.03 &  0.01\\ 

\end{tabular}
\end{center}
\end{table}
\clearpage
\begin{table}[hp!]
\begin{center}
  \caption{\label{tab:xsec-pipp-sn}
    HARP results for the double-differential $\pi^+$ production
    cross-section in the laboratory system,
    $d^2\sigma^{\pi^+}/(dpd\theta)$ for $\pi^+$--Sn interactions. Each row refers to a
    different $(p_{\hbox{\small min}} \le p<p_{\hbox{\small max}},
    \theta_{\hbox{\small min}} \le \theta<\theta_{\hbox{\small max}})$ bin,
    where $p$ and $\theta$ are the pion momentum and polar angle, respectively.
    The central value as well as the square-root of the diagonal elements
    of the covariance matrix are given.}
\vspace{2mm}
\begin{tabular}{rrrr|r@{$\pm$}lr@{$\pm$}lr@{$\pm$}lr@{$\pm$}l}
\hline
$\theta_{\hbox{\small min}}$ &
$\theta_{\hbox{\small max}}$ &
$p_{\hbox{\small min}}$ &
$p_{\hbox{\small max}}$ &
\multicolumn{8}{c}{$d^2\sigma^{\pi^+}/(dpd\theta)$}
\\
(rad) & (rad) & (\GeVc) & (\GeVc) &
\multicolumn{8}{c}{($\barn/(\GeVc \cdot \rad)$)}
\\
  &  &  &
&\multicolumn{2}{c}{$ \bf{3 \ \GeVc}$}
&\multicolumn{2}{c}{$ \bf{5 \ \GeVc}$}
&\multicolumn{2}{c}{$ \bf{8 \ \GeVc}$}
&\multicolumn{2}{c}{$ \bf{12 \ \GeVc}$}
\\
\hline  
 0.35 & 0.55 & 0.15 & 0.20& 0.65 &  0.20& 1.08 &  0.19& 1.43 &  0.21& 1.78 &  0.33\\ 
      &      & 0.20 & 0.25& 0.81 &  0.15& 1.15 &  0.11& 1.83 &  0.15& 1.99 &  0.23\\ 
      &      & 0.25 & 0.30& 0.90 &  0.07& 1.53 &  0.13& 1.89 &  0.12& 2.70 &  0.25\\ 
      &      & 0.30 & 0.35& 0.86 &  0.06& 1.43 &  0.07& 1.90 &  0.12& 2.34 &  0.16\\ 
      &      & 0.35 & 0.40& 0.83 &  0.05& 1.52 &  0.10& 1.89 &  0.11& 2.45 &  0.18\\ 
      &      & 0.40 & 0.45& 0.90 &  0.07& 1.54 &  0.07& 1.84 &  0.08& 2.30 &  0.14\\ 
      &      & 0.45 & 0.50& 0.97 &  0.05& 1.51 &  0.06& 1.89 &  0.11& 2.44 &  0.21\\ 
      &      & 0.50 & 0.60& 0.91 &  0.05& 1.47 &  0.07& 1.90 &  0.10& 2.54 &  0.17\\ 
      &      & 0.60 & 0.70& 0.75 &  0.07& 1.33 &  0.11& 1.74 &  0.16& 2.44 &  0.22\\ 
      &      & 0.70 & 0.80& 0.61 &  0.08& 0.94 &  0.12& 1.33 &  0.23& 2.05 &  0.28\\ 
\hline  
 0.55 & 0.75 & 0.10 & 0.15& 0.65 &  0.20& 1.08 &  0.23& 1.11 &  0.24& 1.16 &  0.31\\ 
      &      & 0.15 & 0.20& 0.95 &  0.12& 1.46 &  0.12& 1.81 &  0.13& 1.42 &  0.18\\ 
      &      & 0.20 & 0.25& 1.00 &  0.09& 1.61 &  0.10& 1.84 &  0.12& 2.15 &  0.23\\ 
      &      & 0.25 & 0.30& 1.08 &  0.07& 1.58 &  0.09& 1.99 &  0.15& 2.04 &  0.20\\ 
      &      & 0.30 & 0.35& 1.02 &  0.07& 1.52 &  0.09& 1.90 &  0.08& 2.22 &  0.17\\ 
      &      & 0.35 & 0.40& 0.99 &  0.05& 1.52 &  0.07& 1.85 &  0.12& 2.01 &  0.12\\ 
      &      & 0.40 & 0.45& 0.96 &  0.06& 1.39 &  0.06& 1.74 &  0.07& 1.95 &  0.12\\ 
      &      & 0.45 & 0.50& 0.92 &  0.05& 1.28 &  0.05& 1.62 &  0.07& 1.78 &  0.11\\ 
      &      & 0.50 & 0.60& 0.77 &  0.06& 1.12 &  0.07& 1.38 &  0.08& 1.66 &  0.12\\ 
      &      & 0.60 & 0.70& 0.58 &  0.06& 0.78 &  0.09& 0.98 &  0.12& 1.42 &  0.14\\ 
      &      & 0.70 & 0.80& 0.42 &  0.07& 0.53 &  0.10& 0.64 &  0.11& 1.02 &  0.16\\ 
\hline  
 0.75 & 0.95 & 0.10 & 0.15& 0.69 &  0.16& 1.01 &  0.16& 1.12 &  0.18& 1.05 &  0.23\\ 
      &      & 0.15 & 0.20& 1.10 &  0.11& 1.59 &  0.11& 1.76 &  0.12& 1.86 &  0.18\\ 
      &      & 0.20 & 0.25& 1.20 &  0.07& 1.58 &  0.09& 1.94 &  0.13& 2.15 &  0.19\\ 
      &      & 0.25 & 0.30& 0.98 &  0.06& 1.54 &  0.07& 1.71 &  0.09& 1.98 &  0.14\\ 
      &      & 0.30 & 0.35& 0.86 &  0.05& 1.33 &  0.06& 1.53 &  0.08& 1.85 &  0.13\\ 
      &      & 0.35 & 0.40& 0.81 &  0.05& 1.13 &  0.05& 1.43 &  0.07& 1.28 &  0.09\\ 
      &      & 0.40 & 0.45& 0.73 &  0.04& 1.07 &  0.04& 1.14 &  0.06& 1.21 &  0.08\\ 
      &      & 0.45 & 0.50& 0.67 &  0.03& 0.91 &  0.05& 0.98 &  0.05& 1.15 &  0.08\\ 
      &      & 0.50 & 0.60& 0.52 &  0.04& 0.67 &  0.06& 0.76 &  0.06& 0.98 &  0.09\\ 
      &      & 0.60 & 0.70& 0.33 &  0.05& 0.42 &  0.06& 0.46 &  0.07& 0.55 &  0.10\\ 
\hline  
 0.95 & 1.15 & 0.10 & 0.15& 0.73 &  0.14& 1.04 &  0.14& 1.14 &  0.15& 1.04 &  0.18\\ 
      &      & 0.15 & 0.20& 1.09 &  0.09& 1.42 &  0.08& 1.64 &  0.11& 1.71 &  0.16\\ 
      &      & 0.20 & 0.25& 0.98 &  0.06& 1.37 &  0.07& 1.62 &  0.11& 1.67 &  0.13\\ 
      &      & 0.25 & 0.30& 0.97 &  0.06& 1.21 &  0.06& 1.32 &  0.07& 1.37 &  0.11\\ 
      &      & 0.30 & 0.35& 0.74 &  0.04& 1.00 &  0.06& 1.15 &  0.06& 1.37 &  0.11\\ 
      &      & 0.35 & 0.40& 0.63 &  0.04& 0.82 &  0.04& 0.98 &  0.05& 1.09 &  0.08\\ 
      &      & 0.40 & 0.45& 0.56 &  0.04& 0.71 &  0.03& 0.86 &  0.05& 0.88 &  0.08\\ 
      &      & 0.45 & 0.50& 0.49 &  0.04& 0.55 &  0.04& 0.61 &  0.05& 0.71 &  0.06\\ 
      &      & 0.50 & 0.60& 0.31 &  0.03& 0.36 &  0.04& 0.38 &  0.04& 0.49 &  0.06\\ 
\hline
\end{tabular}
\end{center}
\end{table}

\begin{table}[hp!]
\begin{center}
\begin{tabular}{rrrr|r@{$\pm$}lr@{$\pm$}lr@{$\pm$}lr@{$\pm$}l}
\hline
$\theta_{\hbox{\small min}}$ &
$\theta_{\hbox{\small max}}$ &
$p_{\hbox{\small min}}$ &
$p_{\hbox{\small max}}$ &
\multicolumn{8}{c}{$d^2\sigma^{\pi^+}/(dpd\theta)$}
\\
(rad) & (rad) & (\GeVc) & (\GeVc) &
\multicolumn{8}{c}{(\barn/($\GeVc \cdot \rad$))}
\\
  &  &  &
&\multicolumn{2}{c}{$ \bf{3 \ \GeVc}$}
&\multicolumn{2}{c}{$ \bf{5 \ \GeVc}$}
&\multicolumn{2}{c}{$ \bf{8 \ \GeVc}$}
&\multicolumn{2}{c}{$ \bf{12 \ \GeVc}$}
\\
\hline
 1.15 & 1.35 & 0.10 & 0.15& 0.81 &  0.15& 1.12 &  0.16& 1.09 &  0.17& 1.26 &  0.20\\ 
      &      & 0.15 & 0.20& 1.19 &  0.08& 1.49 &  0.08& 1.60 &  0.11& 1.59 &  0.18\\ 
      &      & 0.20 & 0.25& 0.97 &  0.06& 1.21 &  0.06& 1.33 &  0.08& 1.47 &  0.12\\ 
      &      & 0.25 & 0.30& 0.79 &  0.05& 0.87 &  0.05& 1.04 &  0.06& 1.04 &  0.09\\ 
      &      & 0.30 & 0.35& 0.56 &  0.04& 0.70 &  0.04& 0.89 &  0.05& 0.99 &  0.09\\ 
      &      & 0.35 & 0.40& 0.44 &  0.02& 0.57 &  0.03& 0.68 &  0.04& 0.76 &  0.07\\ 
      &      & 0.40 & 0.45& 0.35 &  0.02& 0.45 &  0.03& 0.50 &  0.04& 0.54 &  0.06\\ 
      &      & 0.45 & 0.50& 0.28 &  0.02& 0.36 &  0.03& 0.34 &  0.04& 0.39 &  0.04\\ 
\hline  
 1.35 & 1.55 & 0.10 & 0.15& 1.01 &  0.18& 1.12 &  0.19& 1.27 &  0.22& 1.32 &  0.26\\ 
      &      & 0.15 & 0.20& 1.20 &  0.09& 1.52 &  0.12& 1.61 &  0.12& 1.69 &  0.19\\ 
      &      & 0.20 & 0.25& 0.98 &  0.07& 1.14 &  0.08& 1.35 &  0.10& 1.30 &  0.12\\ 
      &      & 0.25 & 0.30& 0.60 &  0.04& 0.76 &  0.05& 0.82 &  0.07& 0.94 &  0.09\\ 
      &      & 0.30 & 0.35& 0.50 &  0.03& 0.61 &  0.04& 0.60 &  0.04& 0.59 &  0.07\\ 
      &      & 0.35 & 0.40& 0.37 &  0.03& 0.46 &  0.03& 0.51 &  0.03& 0.52 &  0.06\\ 
      &      & 0.40 & 0.45& 0.27 &  0.02& 0.29 &  0.03& 0.33 &  0.03& 0.46 &  0.06\\ 
      &      & 0.45 & 0.50& 0.18 &  0.02& 0.20 &  0.02& 0.22 &  0.03& 0.26 &  0.05\\ 
\hline  
 1.55 & 1.75 & 0.10 & 0.15& 1.03 &  0.17& 1.15 &  0.19& 1.29 &  0.21& 1.08 &  0.20\\ 
      &      & 0.15 & 0.20& 1.19 &  0.09& 1.45 &  0.11& 1.46 &  0.11& 1.63 &  0.17\\ 
      &      & 0.20 & 0.25& 0.81 &  0.06& 0.95 &  0.07& 1.08 &  0.08& 0.97 &  0.10\\ 
      &      & 0.25 & 0.30& 0.50 &  0.05& 0.62 &  0.05& 0.66 &  0.05& 0.76 &  0.08\\ 
      &      & 0.30 & 0.35& 0.35 &  0.03& 0.42 &  0.03& 0.46 &  0.04& 0.45 &  0.05\\ 
      &      & 0.35 & 0.40& 0.27 &  0.02& 0.32 &  0.03& 0.28 &  0.02& 0.32 &  0.04\\ 
      &      & 0.40 & 0.45& 0.19 &  0.02& 0.18 &  0.02& 0.19 &  0.02& 0.23 &  0.03\\ 
      &      & 0.45 & 0.50& 0.13 &  0.02& 0.12 &  0.02& 0.13 &  0.02& 0.13 &  0.03\\ 
\hline  
 1.75 & 1.95 & 0.10 & 0.15& 0.81 &  0.10& 0.89 &  0.11& 0.97 &  0.12& 0.88 &  0.16\\ 
      &      & 0.15 & 0.20& 0.90 &  0.05& 1.02 &  0.05& 1.00 &  0.06& 1.05 &  0.10\\ 
      &      & 0.20 & 0.25& 0.55 &  0.05& 0.65 &  0.04& 0.69 &  0.05& 0.77 &  0.08\\ 
      &      & 0.25 & 0.30& 0.32 &  0.03& 0.38 &  0.03& 0.43 &  0.03& 0.52 &  0.06\\ 
      &      & 0.30 & 0.35& 0.21 &  0.02& 0.27 &  0.02& 0.31 &  0.03& 0.27 &  0.04\\ 
      &      & 0.35 & 0.40& 0.16 &  0.01& 0.16 &  0.02& 0.17 &  0.02& 0.19 &  0.03\\ 
      &      & 0.40 & 0.45& 0.12 &  0.01& 0.11 &  0.01& 0.10 &  0.01& 0.13 &  0.03\\ 
      &      & 0.45 & 0.50& 0.07 &  0.01& 0.06 &  0.01& 0.07 &  0.01& 0.07 &  0.02\\ 
\hline  
 1.95 & 2.15 & 0.10 & 0.15& 0.62 &  0.08& 0.76 &  0.08& 0.78 &  0.09& 0.71 &  0.11\\ 
      &      & 0.15 & 0.20& 0.80 &  0.04& 0.86 &  0.05& 0.85 &  0.05& 0.64 &  0.07\\ 
      &      & 0.20 & 0.25& 0.51 &  0.04& 0.54 &  0.04& 0.56 &  0.04& 0.50 &  0.07\\ 
      &      & 0.25 & 0.30& 0.27 &  0.03& 0.30 &  0.02& 0.31 &  0.02& 0.32 &  0.05\\ 
      &      & 0.30 & 0.35& 0.15 &  0.01& 0.21 &  0.01& 0.19 &  0.02& 0.23 &  0.04\\ 
      &      & 0.35 & 0.40& 0.11 &  0.01& 0.14 &  0.01& 0.12 &  0.02& 0.17 &  0.03\\ 
      &      & 0.40 & 0.45& 0.07 &  0.01& 0.08 &  0.01& 0.06 &  0.01& 0.11 &  0.03\\ 
      &      & 0.45 & 0.50& 0.04 &  0.01& 0.05 &  0.01& 0.03 &  0.01& 0.05 &  0.02\\ 

\end{tabular}
\end{center}
\end{table}

\begin{table}[hp!]
\begin{center}
  \caption{\label{tab:xsec-pipm-sn}
    HARP results for the double-differential $\pi^-$ production
    cross-section in the laboratory system,
    $d^2\sigma^{\pi^-}/(dpd\theta)$ for $\pi^+$--Sn interactions. Each row refers to a
    different $(p_{\hbox{\small min}} \le p<p_{\hbox{\small max}},
    \theta_{\hbox{\small min}} \le \theta<\theta_{\hbox{\small max}})$ bin,
    where $p$ and $\theta$ are the pion momentum and polar angle, respectively.
    The central value as well as the square-root of the diagonal elements
    of the covariance matrix are given.}
\vspace{2mm}
\begin{tabular}{rrrr|r@{$\pm$}lr@{$\pm$}lr@{$\pm$}lr@{$\pm$}l}
\hline
$\theta_{\hbox{\small min}}$ &
$\theta_{\hbox{\small max}}$ &
$p_{\hbox{\small min}}$ &
$p_{\hbox{\small max}}$ &
\multicolumn{8}{c}{$d^2\sigma^{\pi^-}/(dpd\theta)$}
\\
(rad) & (rad) & (\GeVc) & (\GeVc) &
\multicolumn{8}{c}{($\barn/(\GeVc \cdot \rad)$)}
\\
  &  &  &
&\multicolumn{2}{c}{$ \bf{3 \ \GeVc}$}
&\multicolumn{2}{c}{$ \bf{5 \ \GeVc}$}
&\multicolumn{2}{c}{$ \bf{8 \ \GeVc}$}
&\multicolumn{2}{c}{$ \bf{12 \ \GeVc}$}
\\
\hline  
 0.35 & 0.55 & 0.15 & 0.20& 0.68 &  0.21& 1.24 &  0.17& 1.52 &  0.20& 1.47 &  0.30\\ 
      &      & 0.20 & 0.25& 0.65 &  0.12& 1.07 &  0.12& 1.55 &  0.13& 1.85 &  0.21\\ 
      &      & 0.25 & 0.30& 0.58 &  0.07& 1.17 &  0.08& 1.68 &  0.13& 1.70 &  0.21\\ 
      &      & 0.30 & 0.35& 0.61 &  0.05& 1.11 &  0.06& 1.55 &  0.08& 1.93 &  0.15\\ 
      &      & 0.35 & 0.40& 0.59 &  0.04& 1.00 &  0.05& 1.39 &  0.08& 1.52 &  0.12\\ 
      &      & 0.40 & 0.45& 0.54 &  0.04& 0.86 &  0.04& 1.33 &  0.06& 1.49 &  0.15\\ 
      &      & 0.45 & 0.50& 0.48 &  0.03& 0.82 &  0.04& 1.12 &  0.05& 1.51 &  0.11\\ 
      &      & 0.50 & 0.60& 0.48 &  0.03& 0.76 &  0.04& 1.09 &  0.06& 1.40 &  0.10\\ 
      &      & 0.60 & 0.70& 0.43 &  0.04& 0.65 &  0.05& 1.04 &  0.08& 1.34 &  0.14\\ 
      &      & 0.70 & 0.80& 0.39 &  0.04& 0.56 &  0.06& 0.81 &  0.10& 1.12 &  0.13\\ 
\hline  
 0.55 & 0.75 & 0.10 & 0.15& 0.75 &  0.22& 1.01 &  0.23& 1.32 &  0.26& 1.56 &  0.34\\ 
      &      & 0.15 & 0.20& 0.94 &  0.12& 1.23 &  0.11& 1.69 &  0.12& 1.51 &  0.20\\ 
      &      & 0.20 & 0.25& 0.80 &  0.06& 1.27 &  0.09& 1.54 &  0.09& 1.95 &  0.18\\ 
      &      & 0.25 & 0.30& 0.64 &  0.05& 1.11 &  0.06& 1.32 &  0.08& 1.51 &  0.14\\ 
      &      & 0.30 & 0.35& 0.70 &  0.05& 0.99 &  0.05& 1.32 &  0.08& 1.44 &  0.12\\ 
      &      & 0.35 & 0.40& 0.58 &  0.04& 0.89 &  0.04& 1.36 &  0.07& 1.44 &  0.13\\ 
      &      & 0.40 & 0.45& 0.50 &  0.03& 0.77 &  0.03& 1.12 &  0.06& 1.13 &  0.11\\ 
      &      & 0.45 & 0.50& 0.44 &  0.03& 0.69 &  0.03& 0.98 &  0.04& 1.06 &  0.08\\ 
      &      & 0.50 & 0.60& 0.39 &  0.02& 0.66 &  0.03& 0.86 &  0.05& 1.09 &  0.08\\ 
      &      & 0.60 & 0.70& 0.35 &  0.03& 0.56 &  0.05& 0.65 &  0.06& 0.94 &  0.09\\ 
      &      & 0.70 & 0.80& 0.27 &  0.04& 0.40 &  0.05& 0.53 &  0.06& 0.80 &  0.11\\ 
\hline  
 0.75 & 0.95 & 0.10 & 0.15& 0.69 &  0.14& 0.90 &  0.14& 1.04 &  0.15& 1.06 &  0.21\\ 
      &      & 0.15 & 0.20& 0.90 &  0.09& 1.28 &  0.08& 1.62 &  0.11& 1.90 &  0.20\\ 
      &      & 0.20 & 0.25& 0.76 &  0.06& 1.08 &  0.06& 1.46 &  0.09& 1.53 &  0.14\\ 
      &      & 0.25 & 0.30& 0.61 &  0.04& 0.95 &  0.05& 1.24 &  0.06& 1.43 &  0.15\\ 
      &      & 0.30 & 0.35& 0.57 &  0.04& 0.83 &  0.04& 1.12 &  0.06& 1.32 &  0.11\\ 
      &      & 0.35 & 0.40& 0.46 &  0.03& 0.68 &  0.03& 0.97 &  0.05& 1.04 &  0.08\\ 
      &      & 0.40 & 0.45& 0.43 &  0.02& 0.62 &  0.03& 0.82 &  0.04& 0.97 &  0.08\\ 
      &      & 0.45 & 0.50& 0.38 &  0.02& 0.56 &  0.02& 0.67 &  0.03& 0.80 &  0.08\\ 
      &      & 0.50 & 0.60& 0.31 &  0.02& 0.47 &  0.03& 0.58 &  0.03& 0.60 &  0.05\\ 
      &      & 0.60 & 0.70& 0.23 &  0.03& 0.35 &  0.03& 0.43 &  0.04& 0.54 &  0.05\\ 
\hline  
 0.95 & 1.15 & 0.10 & 0.15& 0.69 &  0.13& 0.82 &  0.12& 1.09 &  0.12& 0.82 &  0.14\\ 
      &      & 0.15 & 0.20& 0.85 &  0.07& 1.14 &  0.07& 1.40 &  0.09& 1.41 &  0.16\\ 
      &      & 0.20 & 0.25& 0.71 &  0.05& 1.00 &  0.06& 1.15 &  0.06& 1.52 &  0.13\\ 
      &      & 0.25 & 0.30& 0.61 &  0.04& 0.78 &  0.04& 0.96 &  0.05& 0.96 &  0.10\\ 
      &      & 0.30 & 0.35& 0.49 &  0.03& 0.66 &  0.03& 0.83 &  0.05& 0.79 &  0.07\\ 
      &      & 0.35 & 0.40& 0.37 &  0.02& 0.59 &  0.03& 0.72 &  0.04& 0.69 &  0.06\\ 
      &      & 0.40 & 0.45& 0.33 &  0.02& 0.49 &  0.02& 0.58 &  0.03& 0.71 &  0.07\\ 
      &      & 0.45 & 0.50& 0.26 &  0.02& 0.41 &  0.02& 0.49 &  0.03& 0.69 &  0.06\\ 
      &      & 0.50 & 0.60& 0.19 &  0.02& 0.32 &  0.02& 0.38 &  0.02& 0.49 &  0.05\\ 
\hline
\end{tabular}
\end{center}
\end{table}

\begin{table}[hp!]
\begin{center}
\begin{tabular}{rrrr|r@{$\pm$}lr@{$\pm$}lr@{$\pm$}lr@{$\pm$}l}
\hline
$\theta_{\hbox{\small min}}$ &
$\theta_{\hbox{\small max}}$ &
$p_{\hbox{\small min}}$ &
$p_{\hbox{\small max}}$ &
\multicolumn{8}{c}{$d^2\sigma^{\pi^-}/(dpd\theta)$}
\\
(rad) & (rad) & (\GeVc) & (\GeVc) &
\multicolumn{8}{c}{(\barn/($\GeVc \cdot \rad$))}
\\
  &  &  &
&\multicolumn{2}{c}{$ \bf{3 \ \GeVc}$}
&\multicolumn{2}{c}{$ \bf{5 \ \GeVc}$}
&\multicolumn{2}{c}{$ \bf{8 \ \GeVc}$}
&\multicolumn{2}{c}{$ \bf{12 \ \GeVc}$}
\\
\hline
 1.15 & 1.35 & 0.10 & 0.15& 0.66 &  0.12& 0.79 &  0.11& 1.13 &  0.14& 1.23 &  0.19\\ 
      &      & 0.15 & 0.20& 0.78 &  0.06& 1.01 &  0.08& 1.24 &  0.08& 1.17 &  0.12\\ 
      &      & 0.20 & 0.25& 0.55 &  0.04& 0.86 &  0.05& 0.99 &  0.06& 1.20 &  0.13\\ 
      &      & 0.25 & 0.30& 0.46 &  0.03& 0.65 &  0.04& 0.89 &  0.06& 0.89 &  0.09\\ 
      &      & 0.30 & 0.35& 0.43 &  0.03& 0.50 &  0.03& 0.71 &  0.05& 0.66 &  0.06\\ 
      &      & 0.35 & 0.40& 0.33 &  0.02& 0.42 &  0.02& 0.59 &  0.04& 0.51 &  0.05\\ 
      &      & 0.40 & 0.45& 0.22 &  0.02& 0.35 &  0.02& 0.43 &  0.03& 0.41 &  0.04\\ 
      &      & 0.45 & 0.50& 0.18 &  0.01& 0.26 &  0.02& 0.30 &  0.03& 0.31 &  0.04\\ 
\hline  
 1.35 & 1.55 & 0.10 & 0.15& 0.72 &  0.12& 0.92 &  0.15& 1.23 &  0.20& 1.56 &  0.27\\ 
      &      & 0.15 & 0.20& 0.80 &  0.07& 1.11 &  0.09& 1.24 &  0.09& 1.25 &  0.14\\ 
      &      & 0.20 & 0.25& 0.48 &  0.04& 0.91 &  0.06& 0.92 &  0.06& 1.08 &  0.11\\ 
      &      & 0.25 & 0.30& 0.43 &  0.03& 0.63 &  0.05& 0.69 &  0.05& 0.85 &  0.09\\ 
      &      & 0.30 & 0.35& 0.30 &  0.03& 0.44 &  0.03& 0.49 &  0.04& 0.56 &  0.06\\ 
      &      & 0.35 & 0.40& 0.23 &  0.02& 0.33 &  0.02& 0.36 &  0.02& 0.40 &  0.04\\ 
      &      & 0.40 & 0.45& 0.18 &  0.01& 0.26 &  0.02& 0.28 &  0.02& 0.34 &  0.04\\ 
      &      & 0.45 & 0.50& 0.15 &  0.01& 0.20 &  0.01& 0.22 &  0.02& 0.27 &  0.03\\ 
\hline  
 1.55 & 1.75 & 0.10 & 0.15& 0.70 &  0.11& 0.94 &  0.14& 1.03 &  0.16& 1.37 &  0.26\\ 
      &      & 0.15 & 0.20& 0.76 &  0.06& 1.08 &  0.07& 1.27 &  0.09& 1.26 &  0.13\\ 
      &      & 0.20 & 0.25& 0.52 &  0.04& 0.66 &  0.05& 0.76 &  0.06& 0.86 &  0.09\\ 
      &      & 0.25 & 0.30& 0.34 &  0.03& 0.45 &  0.03& 0.50 &  0.04& 0.62 &  0.07\\ 
      &      & 0.30 & 0.35& 0.23 &  0.02& 0.34 &  0.03& 0.41 &  0.03& 0.49 &  0.06\\ 
      &      & 0.35 & 0.40& 0.17 &  0.01& 0.22 &  0.02& 0.35 &  0.03& 0.27 &  0.04\\ 
      &      & 0.40 & 0.45& 0.12 &  0.01& 0.15 &  0.01& 0.26 &  0.02& 0.21 &  0.03\\ 
      &      & 0.45 & 0.50& 0.09 &  0.01& 0.13 &  0.01& 0.17 &  0.02& 0.18 &  0.03\\ 
\hline  
 1.75 & 1.95 & 0.10 & 0.15& 0.58 &  0.07& 0.73 &  0.08& 0.79 &  0.10& 0.80 &  0.14\\ 
      &      & 0.15 & 0.20& 0.54 &  0.04& 0.69 &  0.04& 0.85 &  0.05& 0.91 &  0.09\\ 
      &      & 0.20 & 0.25& 0.36 &  0.03& 0.49 &  0.03& 0.55 &  0.04& 0.66 &  0.07\\ 
      &      & 0.25 & 0.30& 0.24 &  0.03& 0.32 &  0.02& 0.34 &  0.03& 0.43 &  0.06\\ 
      &      & 0.30 & 0.35& 0.17 &  0.02& 0.20 &  0.02& 0.27 &  0.02& 0.26 &  0.04\\ 
      &      & 0.35 & 0.40& 0.15 &  0.01& 0.13 &  0.01& 0.20 &  0.02& 0.13 &  0.03\\ 
      &      & 0.40 & 0.45& 0.10 &  0.01& 0.10 &  0.01& 0.15 &  0.01& 0.11 &  0.02\\ 
      &      & 0.45 & 0.50& 0.07 &  0.01& 0.09 &  0.01& 0.10 &  0.01& 0.10 &  0.02\\ 
\hline  
 1.95 & 2.15 & 0.10 & 0.15& 0.46 &  0.05& 0.64 &  0.05& 0.65 &  0.06& 0.62 &  0.11\\ 
      &      & 0.15 & 0.20& 0.45 &  0.03& 0.63 &  0.04& 0.59 &  0.04& 0.72 &  0.09\\ 
      &      & 0.20 & 0.25& 0.29 &  0.03& 0.39 &  0.03& 0.43 &  0.03& 0.49 &  0.06\\ 
      &      & 0.25 & 0.30& 0.18 &  0.02& 0.19 &  0.02& 0.22 &  0.02& 0.30 &  0.05\\ 
      &      & 0.30 & 0.35& 0.14 &  0.01& 0.15 &  0.01& 0.15 &  0.02& 0.16 &  0.04\\ 
      &      & 0.35 & 0.40& 0.09 &  0.01& 0.13 &  0.01& 0.11 &  0.01& 0.08 &  0.02\\ 
      &      & 0.40 & 0.45& 0.07 &  0.01& 0.08 &  0.01& 0.08 &  0.01& 0.12 &  0.03\\ 
      &      & 0.45 & 0.50& 0.05 &  0.01& 0.06 &  0.01& 0.07 &  0.01& 0.07 &  0.02\\ 

\end{tabular}
\end{center}
\end{table}
\clearpage
\begin{table}[hp!]
\begin{center}
  \caption{\label{tab:xsec-pimp-sn}
    HARP results for the double-differential $\pi^+$ production
    cross-section in the laboratory system,
    $d^2\sigma^{\pi^+}/(dpd\theta)$ for $\pi^-$--Sn interactions. Each row refers to a
    different $(p_{\hbox{\small min}} \le p<p_{\hbox{\small max}},
    \theta_{\hbox{\small min}} \le \theta<\theta_{\hbox{\small max}})$ bin,
    where $p$ and $\theta$ are the pion momentum and polar angle, respectively.
    The central value as well as the square-root of the diagonal elements
    of the covariance matrix are given.}
\vspace{2mm}
\begin{tabular}{rrrr|r@{$\pm$}lr@{$\pm$}lr@{$\pm$}lr@{$\pm$}l}
\hline
$\theta_{\hbox{\small min}}$ &
$\theta_{\hbox{\small max}}$ &
$p_{\hbox{\small min}}$ &
$p_{\hbox{\small max}}$ &
\multicolumn{8}{c}{$d^2\sigma^{\pi^+}/(dpd\theta)$}
\\
(rad) & (rad) & (\GeVc) & (\GeVc) &
\multicolumn{8}{c}{($\barn/(\GeVc \cdot \rad)$)}
\\
  &  &  &
&\multicolumn{2}{c}{$ \bf{3 \ \GeVc}$}
&\multicolumn{2}{c}{$ \bf{5 \ \GeVc}$}
&\multicolumn{2}{c}{$ \bf{8 \ \GeVc}$}
&\multicolumn{2}{c}{$ \bf{12 \ \GeVc}$}
\\
\hline  
 0.35 & 0.55 & 0.15 & 0.20& 0.45 &  0.12& 1.12 &  0.20& 1.12 &  0.24& 1.68 &  0.25\\ 
      &      & 0.20 & 0.25& 0.56 &  0.09& 1.03 &  0.12& 1.29 &  0.17& 1.62 &  0.17\\ 
      &      & 0.25 & 0.30& 0.67 &  0.06& 1.11 &  0.08& 1.48 &  0.12& 1.82 &  0.14\\ 
      &      & 0.30 & 0.35& 0.58 &  0.04& 1.10 &  0.07& 1.56 &  0.09& 1.97 &  0.11\\ 
      &      & 0.35 & 0.40& 0.57 &  0.04& 1.00 &  0.05& 1.42 &  0.07& 1.90 &  0.09\\ 
      &      & 0.40 & 0.45& 0.59 &  0.04& 1.04 &  0.07& 1.34 &  0.07& 1.81 &  0.10\\ 
      &      & 0.45 & 0.50& 0.59 &  0.03& 1.10 &  0.06& 1.41 &  0.08& 1.83 &  0.08\\ 
      &      & 0.50 & 0.60& 0.61 &  0.04& 1.08 &  0.07& 1.53 &  0.09& 1.81 &  0.11\\ 
      &      & 0.60 & 0.70& 0.52 &  0.06& 1.01 &  0.10& 1.47 &  0.14& 1.68 &  0.16\\ 
      &      & 0.70 & 0.80& 0.34 &  0.06& 0.76 &  0.13& 1.21 &  0.16& 1.33 &  0.20\\ 
\hline  
 0.55 & 0.75 & 0.10 & 0.15& 0.70 &  0.17& 0.90 &  0.22& 0.79 &  0.27& 1.30 &  0.32\\ 
      &      & 0.15 & 0.20& 0.70 &  0.09& 1.13 &  0.13& 1.12 &  0.14& 1.63 &  0.15\\ 
      &      & 0.20 & 0.25& 0.77 &  0.06& 1.16 &  0.09& 1.46 &  0.12& 1.72 &  0.11\\ 
      &      & 0.25 & 0.30& 0.76 &  0.05& 1.10 &  0.08& 1.33 &  0.09& 1.78 &  0.12\\ 
      &      & 0.30 & 0.35& 0.63 &  0.03& 1.07 &  0.06& 1.39 &  0.07& 1.72 &  0.08\\ 
      &      & 0.35 & 0.40& 0.65 &  0.05& 1.06 &  0.05& 1.36 &  0.06& 1.60 &  0.08\\ 
      &      & 0.40 & 0.45& 0.64 &  0.03& 1.01 &  0.04& 1.26 &  0.06& 1.61 &  0.08\\ 
      &      & 0.45 & 0.50& 0.57 &  0.03& 0.94 &  0.04& 1.26 &  0.06& 1.54 &  0.07\\ 
      &      & 0.50 & 0.60& 0.51 &  0.04& 0.86 &  0.06& 1.13 &  0.08& 1.30 &  0.08\\ 
      &      & 0.60 & 0.70& 0.39 &  0.05& 0.64 &  0.09& 0.89 &  0.10& 0.96 &  0.12\\ 
      &      & 0.70 & 0.80& 0.26 &  0.05& 0.43 &  0.09& 0.63 &  0.12& 0.70 &  0.12\\ 
\hline  
 0.75 & 0.95 & 0.10 & 0.15& 0.57 &  0.12& 0.82 &  0.17& 0.77 &  0.18& 1.11 &  0.22\\ 
      &      & 0.15 & 0.20& 0.75 &  0.07& 1.16 &  0.09& 1.35 &  0.14& 1.66 &  0.12\\ 
      &      & 0.20 & 0.25& 0.78 &  0.05& 1.18 &  0.07& 1.43 &  0.09& 1.72 &  0.10\\ 
      &      & 0.25 & 0.30& 0.71 &  0.04& 1.12 &  0.06& 1.26 &  0.06& 1.59 &  0.07\\ 
      &      & 0.30 & 0.35& 0.58 &  0.03& 0.87 &  0.05& 1.19 &  0.06& 1.40 &  0.06\\ 
      &      & 0.35 & 0.40& 0.55 &  0.03& 0.74 &  0.04& 1.19 &  0.06& 1.24 &  0.06\\ 
      &      & 0.40 & 0.45& 0.47 &  0.03& 0.73 &  0.04& 1.05 &  0.05& 1.10 &  0.05\\ 
      &      & 0.45 & 0.50& 0.42 &  0.03& 0.65 &  0.04& 0.87 &  0.05& 0.94 &  0.05\\ 
      &      & 0.50 & 0.60& 0.32 &  0.03& 0.49 &  0.04& 0.68 &  0.05& 0.75 &  0.06\\ 
      &      & 0.60 & 0.70& 0.21 &  0.03& 0.32 &  0.05& 0.50 &  0.08& 0.50 &  0.07\\ 
\hline  
 0.95 & 1.15 & 0.10 & 0.15& 0.54 &  0.10& 0.75 &  0.14& 0.73 &  0.16& 1.05 &  0.16\\ 
      &      & 0.15 & 0.20& 0.78 &  0.06& 1.17 &  0.08& 1.29 &  0.10& 1.50 &  0.09\\ 
      &      & 0.20 & 0.25& 0.68 &  0.05& 0.93 &  0.05& 1.22 &  0.06& 1.49 &  0.08\\ 
      &      & 0.25 & 0.30& 0.56 &  0.03& 0.85 &  0.05& 1.06 &  0.06& 1.22 &  0.06\\ 
      &      & 0.30 & 0.35& 0.44 &  0.02& 0.71 &  0.04& 0.88 &  0.04& 0.98 &  0.06\\ 
      &      & 0.35 & 0.40& 0.39 &  0.02& 0.60 &  0.03& 0.83 &  0.04& 0.83 &  0.04\\ 
      &      & 0.40 & 0.45& 0.33 &  0.02& 0.53 &  0.03& 0.65 &  0.04& 0.72 &  0.04\\ 
      &      & 0.45 & 0.50& 0.27 &  0.02& 0.41 &  0.03& 0.51 &  0.04& 0.58 &  0.04\\ 
      &      & 0.50 & 0.60& 0.20 &  0.02& 0.29 &  0.03& 0.35 &  0.03& 0.40 &  0.04\\ 
\hline
\end{tabular}
\end{center}
\end{table}

\begin{table}[hp!]
\begin{center}
\begin{tabular}{rrrr|r@{$\pm$}lr@{$\pm$}lr@{$\pm$}lr@{$\pm$}l}
\hline
$\theta_{\hbox{\small min}}$ &
$\theta_{\hbox{\small max}}$ &
$p_{\hbox{\small min}}$ &
$p_{\hbox{\small max}}$ &
\multicolumn{8}{c}{$d^2\sigma^{\pi^+}/(dpd\theta)$}
\\
(rad) & (rad) & (\GeVc) & (\GeVc) &
\multicolumn{8}{c}{(\barn/($\GeVc \cdot \rad$))}
\\
  &  &  &
&\multicolumn{2}{c}{$ \bf{3 \ \GeVc}$}
&\multicolumn{2}{c}{$ \bf{5 \ \GeVc}$}
&\multicolumn{2}{c}{$ \bf{8 \ \GeVc}$}
&\multicolumn{2}{c}{$ \bf{12 \ \GeVc}$}
\\
\hline
 1.15 & 1.35 & 0.10 & 0.15& 0.59 &  0.11& 0.72 &  0.14& 0.66 &  0.15& 1.00 &  0.16\\ 
      &      & 0.15 & 0.20& 0.78 &  0.06& 0.98 &  0.08& 1.08 &  0.10& 1.32 &  0.09\\ 
      &      & 0.20 & 0.25& 0.68 &  0.04& 0.82 &  0.04& 1.06 &  0.07& 1.24 &  0.07\\ 
      &      & 0.25 & 0.30& 0.49 &  0.03& 0.68 &  0.04& 0.84 &  0.05& 0.92 &  0.05\\ 
      &      & 0.30 & 0.35& 0.37 &  0.02& 0.56 &  0.03& 0.63 &  0.04& 0.72 &  0.04\\ 
      &      & 0.35 & 0.40& 0.28 &  0.02& 0.45 &  0.02& 0.52 &  0.03& 0.60 &  0.03\\ 
      &      & 0.40 & 0.45& 0.23 &  0.02& 0.34 &  0.02& 0.44 &  0.03& 0.44 &  0.03\\ 
      &      & 0.45 & 0.50& 0.18 &  0.01& 0.25 &  0.02& 0.29 &  0.03& 0.35 &  0.03\\ 
\hline  
 1.35 & 1.55 & 0.10 & 0.15& 0.63 &  0.12& 0.82 &  0.16& 0.77 &  0.16& 1.07 &  0.18\\ 
      &      & 0.15 & 0.20& 0.80 &  0.07& 1.07 &  0.09& 1.09 &  0.10& 1.22 &  0.10\\ 
      &      & 0.20 & 0.25& 0.59 &  0.04& 0.84 &  0.05& 0.89 &  0.05& 1.02 &  0.08\\ 
      &      & 0.25 & 0.30& 0.45 &  0.03& 0.60 &  0.04& 0.67 &  0.04& 0.72 &  0.05\\ 
      &      & 0.30 & 0.35& 0.35 &  0.03& 0.43 &  0.03& 0.50 &  0.03& 0.50 &  0.03\\ 
      &      & 0.35 & 0.40& 0.23 &  0.02& 0.31 &  0.02& 0.37 &  0.03& 0.37 &  0.02\\ 
      &      & 0.40 & 0.45& 0.16 &  0.01& 0.23 &  0.02& 0.29 &  0.02& 0.26 &  0.02\\ 
      &      & 0.45 & 0.50& 0.11 &  0.01& 0.14 &  0.02& 0.18 &  0.02& 0.19 &  0.02\\ 
\hline  
 1.55 & 1.75 & 0.10 & 0.15& 0.65 &  0.12& 0.89 &  0.16& 0.83 &  0.17& 1.03 &  0.16\\ 
      &      & 0.15 & 0.20& 0.73 &  0.05& 0.96 &  0.07& 1.06 &  0.08& 1.23 &  0.09\\ 
      &      & 0.20 & 0.25& 0.51 &  0.04& 0.64 &  0.04& 0.75 &  0.06& 0.86 &  0.06\\ 
      &      & 0.25 & 0.30& 0.35 &  0.03& 0.43 &  0.03& 0.52 &  0.04& 0.57 &  0.04\\ 
      &      & 0.30 & 0.35& 0.26 &  0.02& 0.33 &  0.02& 0.37 &  0.03& 0.40 &  0.03\\ 
      &      & 0.35 & 0.40& 0.18 &  0.01& 0.25 &  0.02& 0.25 &  0.02& 0.28 &  0.02\\ 
      &      & 0.40 & 0.45& 0.13 &  0.01& 0.18 &  0.02& 0.16 &  0.02& 0.19 &  0.01\\ 
      &      & 0.45 & 0.50& 0.07 &  0.01& 0.10 &  0.01& 0.09 &  0.01& 0.13 &  0.02\\ 
\hline  
 1.75 & 1.95 & 0.10 & 0.15& 0.56 &  0.07& 0.68 &  0.09& 0.56 &  0.08& 0.80 &  0.10\\ 
      &      & 0.15 & 0.20& 0.56 &  0.04& 0.68 &  0.04& 0.71 &  0.05& 0.84 &  0.05\\ 
      &      & 0.20 & 0.25& 0.42 &  0.03& 0.48 &  0.03& 0.55 &  0.04& 0.55 &  0.05\\ 
      &      & 0.25 & 0.30& 0.25 &  0.02& 0.28 &  0.02& 0.33 &  0.04& 0.33 &  0.03\\ 
      &      & 0.30 & 0.35& 0.16 &  0.02& 0.20 &  0.02& 0.20 &  0.02& 0.25 &  0.02\\ 
      &      & 0.35 & 0.40& 0.12 &  0.01& 0.16 &  0.01& 0.14 &  0.01& 0.17 &  0.01\\ 
      &      & 0.40 & 0.45& 0.08 &  0.01& 0.11 &  0.01& 0.09 &  0.01& 0.11 &  0.01\\ 
      &      & 0.45 & 0.50& 0.04 &  0.01& 0.07 &  0.01& 0.06 &  0.01& 0.07 &  0.01\\ 
\hline  
 1.95 & 2.15 & 0.10 & 0.15& 0.39 &  0.05& 0.50 &  0.05& 0.45 &  0.06& 0.57 &  0.06\\ 
      &      & 0.15 & 0.20& 0.43 &  0.03& 0.52 &  0.03& 0.61 &  0.03& 0.63 &  0.03\\ 
      &      & 0.20 & 0.25& 0.28 &  0.02& 0.39 &  0.02& 0.39 &  0.03& 0.41 &  0.03\\ 
      &      & 0.25 & 0.30& 0.17 &  0.02& 0.24 &  0.02& 0.22 &  0.02& 0.26 &  0.02\\ 
      &      & 0.30 & 0.35& 0.11 &  0.01& 0.13 &  0.01& 0.13 &  0.01& 0.15 &  0.01\\ 
      &      & 0.35 & 0.40& 0.08 &  0.01& 0.10 &  0.01& 0.10 &  0.01& 0.10 &  0.01\\ 
      &      & 0.40 & 0.45& 0.06 &  0.01& 0.07 &  0.01& 0.07 &  0.01& 0.06 &  0.01\\ 
      &      & 0.45 & 0.50& 0.04 &  0.01& 0.04 &  0.01& 0.04 &  0.01& 0.04 &  0.01\\ 

\end{tabular}
\end{center}
\end{table}

\begin{table}[hp!]
\begin{center}
  \caption{\label{tab:xsec-pimm-sn}
    HARP results for the double-differential $\pi^-$ production
    cross-section in the laboratory system,
    $d^2\sigma^{\pi^-}/(dpd\theta)$ for $\pi^-$--Sn interactions. Each row refers to a
    different $(p_{\hbox{\small min}} \le p<p_{\hbox{\small max}},
    \theta_{\hbox{\small min}} \le \theta<\theta_{\hbox{\small max}})$ bin,
    where $p$ and $\theta$ are the pion momentum and polar angle, respectively.
    The central value as well as the square-root of the diagonal elements
    of the covariance matrix are given.}
\vspace{2mm}
\begin{tabular}{rrrr|r@{$\pm$}lr@{$\pm$}lr@{$\pm$}lr@{$\pm$}l}
\hline
$\theta_{\hbox{\small min}}$ &
$\theta_{\hbox{\small max}}$ &
$p_{\hbox{\small min}}$ &
$p_{\hbox{\small max}}$ &
\multicolumn{8}{c}{$d^2\sigma^{\pi^-}/(dpd\theta)$}
\\
(rad) & (rad) & (\GeVc) & (\GeVc) &
\multicolumn{8}{c}{($\barn/(\GeVc \cdot \rad)$)}
\\
  &  &  &
&\multicolumn{2}{c}{$ \bf{3 \ \GeVc}$}
&\multicolumn{2}{c}{$ \bf{5 \ \GeVc}$}
&\multicolumn{2}{c}{$ \bf{8 \ \GeVc}$}
&\multicolumn{2}{c}{$ \bf{12 \ \GeVc}$}
\\
\hline  
 0.35 & 0.55 & 0.15 & 0.20& 0.70 &  0.16& 1.27 &  0.22& 1.40 &  0.28& 1.98 &  0.30\\ 
      &      & 0.20 & 0.25& 0.85 &  0.10& 1.50 &  0.14& 1.67 &  0.17& 2.10 &  0.17\\ 
      &      & 0.25 & 0.30& 0.90 &  0.07& 1.46 &  0.08& 1.91 &  0.15& 2.22 &  0.13\\ 
      &      & 0.30 & 0.35& 0.89 &  0.05& 1.43 &  0.10& 1.78 &  0.09& 2.03 &  0.10\\ 
      &      & 0.35 & 0.40& 0.86 &  0.05& 1.42 &  0.07& 1.60 &  0.07& 1.98 &  0.10\\ 
      &      & 0.40 & 0.45& 0.84 &  0.05& 1.25 &  0.06& 1.61 &  0.08& 1.98 &  0.10\\ 
      &      & 0.45 & 0.50& 0.85 &  0.04& 1.25 &  0.06& 1.63 &  0.08& 1.87 &  0.08\\ 
      &      & 0.50 & 0.60& 0.82 &  0.05& 1.30 &  0.07& 1.67 &  0.08& 1.88 &  0.10\\ 
      &      & 0.60 & 0.70& 0.79 &  0.06& 1.27 &  0.10& 1.66 &  0.13& 1.84 &  0.15\\ 
      &      & 0.70 & 0.80& 0.70 &  0.08& 1.13 &  0.13& 1.52 &  0.18& 1.60 &  0.17\\ 
\hline  
 0.55 & 0.75 & 0.10 & 0.15& 0.82 &  0.20& 1.22 &  0.26& 1.16 &  0.29& 1.59 &  0.32\\ 
      &      & 0.15 & 0.20& 1.22 &  0.10& 1.55 &  0.13& 1.57 &  0.19& 1.91 &  0.17\\ 
      &      & 0.20 & 0.25& 1.16 &  0.08& 1.55 &  0.11& 1.80 &  0.12& 2.16 &  0.13\\ 
      &      & 0.25 & 0.30& 1.07 &  0.05& 1.58 &  0.08& 1.80 &  0.10& 2.04 &  0.11\\ 
      &      & 0.30 & 0.35& 0.97 &  0.05& 1.41 &  0.07& 1.62 &  0.08& 1.84 &  0.08\\ 
      &      & 0.35 & 0.40& 0.91 &  0.04& 1.30 &  0.06& 1.54 &  0.07& 1.70 &  0.08\\ 
      &      & 0.40 & 0.45& 0.86 &  0.04& 1.27 &  0.05& 1.49 &  0.07& 1.57 &  0.07\\ 
      &      & 0.45 & 0.50& 0.85 &  0.04& 1.21 &  0.05& 1.41 &  0.06& 1.45 &  0.06\\ 
      &      & 0.50 & 0.60& 0.82 &  0.04& 1.07 &  0.06& 1.27 &  0.07& 1.35 &  0.06\\ 
      &      & 0.60 & 0.70& 0.66 &  0.06& 0.93 &  0.08& 1.12 &  0.09& 1.16 &  0.10\\ 
      &      & 0.70 & 0.80& 0.52 &  0.07& 0.78 &  0.11& 0.94 &  0.12& 0.93 &  0.14\\ 
\hline  
 0.75 & 0.95 & 0.10 & 0.15& 0.91 &  0.13& 1.26 &  0.19& 1.17 &  0.20& 1.46 &  0.21\\ 
      &      & 0.15 & 0.20& 1.32 &  0.08& 1.86 &  0.12& 1.74 &  0.12& 2.12 &  0.13\\ 
      &      & 0.20 & 0.25& 1.15 &  0.07& 1.49 &  0.07& 1.86 &  0.13& 1.97 &  0.11\\ 
      &      & 0.25 & 0.30& 1.01 &  0.05& 1.43 &  0.08& 1.65 &  0.07& 1.75 &  0.08\\ 
      &      & 0.30 & 0.35& 0.89 &  0.05& 1.26 &  0.05& 1.44 &  0.06& 1.43 &  0.06\\ 
      &      & 0.35 & 0.40& 0.80 &  0.04& 1.12 &  0.05& 1.19 &  0.05& 1.29 &  0.06\\ 
      &      & 0.40 & 0.45& 0.70 &  0.03& 1.04 &  0.05& 1.06 &  0.05& 1.14 &  0.04\\ 
      &      & 0.45 & 0.50& 0.63 &  0.03& 0.93 &  0.04& 0.96 &  0.04& 1.03 &  0.04\\ 
      &      & 0.50 & 0.60& 0.58 &  0.03& 0.74 &  0.05& 0.81 &  0.05& 0.91 &  0.05\\ 
      &      & 0.60 & 0.70& 0.48 &  0.05& 0.59 &  0.06& 0.65 &  0.06& 0.71 &  0.07\\ 
\hline  
 0.95 & 1.15 & 0.10 & 0.15& 1.08 &  0.12& 1.28 &  0.15& 1.26 &  0.17& 1.41 &  0.18\\ 
      &      & 0.15 & 0.20& 1.28 &  0.07& 1.66 &  0.09& 1.57 &  0.09& 1.82 &  0.10\\ 
      &      & 0.20 & 0.25& 1.09 &  0.06& 1.31 &  0.07& 1.45 &  0.09& 1.70 &  0.10\\ 
      &      & 0.25 & 0.30& 0.87 &  0.04& 1.29 &  0.07& 1.26 &  0.06& 1.42 &  0.06\\ 
      &      & 0.30 & 0.35& 0.71 &  0.03& 0.96 &  0.05& 1.03 &  0.05& 1.10 &  0.05\\ 
      &      & 0.35 & 0.40& 0.60 &  0.03& 0.81 &  0.03& 0.88 &  0.04& 0.92 &  0.04\\ 
      &      & 0.40 & 0.45& 0.52 &  0.02& 0.71 &  0.03& 0.79 &  0.04& 0.79 &  0.04\\ 
      &      & 0.45 & 0.50& 0.47 &  0.02& 0.60 &  0.03& 0.66 &  0.03& 0.67 &  0.03\\ 
      &      & 0.50 & 0.60& 0.40 &  0.02& 0.50 &  0.03& 0.51 &  0.04& 0.55 &  0.04\\ 
\hline
\end{tabular}
\end{center}
\end{table}

\begin{table}[hp!]
\begin{center}
\begin{tabular}{rrrr|r@{$\pm$}lr@{$\pm$}lr@{$\pm$}lr@{$\pm$}l}
\hline
$\theta_{\hbox{\small min}}$ &
$\theta_{\hbox{\small max}}$ &
$p_{\hbox{\small min}}$ &
$p_{\hbox{\small max}}$ &
\multicolumn{8}{c}{$d^2\sigma^{\pi^-}/(dpd\theta)$}
\\
(rad) & (rad) & (\GeVc) & (\GeVc) &
\multicolumn{8}{c}{(\barn/($\GeVc \cdot \rad$))}
\\
  &  &  &
&\multicolumn{2}{c}{$ \bf{3 \ \GeVc}$}
&\multicolumn{2}{c}{$ \bf{5 \ \GeVc}$}
&\multicolumn{2}{c}{$ \bf{8 \ \GeVc}$}
&\multicolumn{2}{c}{$ \bf{12 \ \GeVc}$}
\\
\hline
 1.15 & 1.35 & 0.10 & 0.15& 1.22 &  0.14& 1.37 &  0.18& 1.26 &  0.18& 1.47 &  0.20\\ 
      &      & 0.15 & 0.20& 1.27 &  0.07& 1.62 &  0.10& 1.59 &  0.10& 1.73 &  0.10\\ 
      &      & 0.20 & 0.25& 0.93 &  0.05& 1.27 &  0.07& 1.23 &  0.07& 1.35 &  0.07\\ 
      &      & 0.25 & 0.30& 0.76 &  0.04& 0.99 &  0.06& 0.98 &  0.05& 1.08 &  0.05\\ 
      &      & 0.30 & 0.35& 0.59 &  0.03& 0.71 &  0.04& 0.77 &  0.04& 0.88 &  0.05\\ 
      &      & 0.35 & 0.40& 0.46 &  0.03& 0.57 &  0.03& 0.58 &  0.03& 0.73 &  0.04\\ 
      &      & 0.40 & 0.45& 0.38 &  0.02& 0.46 &  0.02& 0.47 &  0.02& 0.57 &  0.04\\ 
      &      & 0.45 & 0.50& 0.34 &  0.02& 0.38 &  0.02& 0.41 &  0.02& 0.43 &  0.03\\ 
\hline  
 1.35 & 1.55 & 0.10 & 0.15& 1.29 &  0.20& 1.52 &  0.22& 1.40 &  0.22& 1.68 &  0.24\\ 
      &      & 0.15 & 0.20& 1.36 &  0.09& 1.70 &  0.11& 1.61 &  0.11& 1.74 &  0.12\\ 
      &      & 0.20 & 0.25& 0.86 &  0.06& 1.16 &  0.07& 1.16 &  0.07& 1.19 &  0.07\\ 
      &      & 0.25 & 0.30& 0.64 &  0.04& 0.83 &  0.06& 0.85 &  0.05& 0.84 &  0.06\\ 
      &      & 0.30 & 0.35& 0.49 &  0.03& 0.59 &  0.04& 0.66 &  0.04& 0.63 &  0.04\\ 
      &      & 0.35 & 0.40& 0.38 &  0.02& 0.46 &  0.03& 0.45 &  0.03& 0.49 &  0.03\\ 
      &      & 0.40 & 0.45& 0.29 &  0.02& 0.35 &  0.02& 0.35 &  0.02& 0.37 &  0.02\\ 
      &      & 0.45 & 0.50& 0.21 &  0.02& 0.26 &  0.02& 0.31 &  0.02& 0.30 &  0.02\\ 
\hline  
 1.55 & 1.75 & 0.10 & 0.15& 1.27 &  0.18& 1.56 &  0.25& 1.20 &  0.19& 1.52 &  0.21\\ 
      &      & 0.15 & 0.20& 1.25 &  0.08& 1.50 &  0.09& 1.34 &  0.09& 1.56 &  0.10\\ 
      &      & 0.20 & 0.25& 0.77 &  0.05& 0.89 &  0.06& 0.93 &  0.06& 1.06 &  0.07\\ 
      &      & 0.25 & 0.30& 0.51 &  0.04& 0.59 &  0.04& 0.66 &  0.05& 0.67 &  0.05\\ 
      &      & 0.30 & 0.35& 0.37 &  0.03& 0.42 &  0.03& 0.45 &  0.04& 0.48 &  0.03\\ 
      &      & 0.35 & 0.40& 0.31 &  0.02& 0.32 &  0.02& 0.35 &  0.02& 0.36 &  0.03\\ 
      &      & 0.40 & 0.45& 0.22 &  0.02& 0.25 &  0.02& 0.25 &  0.02& 0.27 &  0.02\\ 
      &      & 0.45 & 0.50& 0.15 &  0.01& 0.17 &  0.02& 0.20 &  0.02& 0.19 &  0.01\\ 
\hline  
 1.75 & 1.95 & 0.10 & 0.15& 1.02 &  0.11& 1.10 &  0.12& 0.96 &  0.11& 1.13 &  0.12\\ 
      &      & 0.15 & 0.20& 0.90 &  0.05& 1.04 &  0.05& 0.95 &  0.04& 1.12 &  0.05\\ 
      &      & 0.20 & 0.25& 0.52 &  0.04& 0.67 &  0.04& 0.64 &  0.04& 0.68 &  0.04\\ 
      &      & 0.25 & 0.30& 0.33 &  0.02& 0.42 &  0.03& 0.41 &  0.03& 0.45 &  0.04\\ 
      &      & 0.30 & 0.35& 0.25 &  0.02& 0.31 &  0.02& 0.27 &  0.02& 0.28 &  0.02\\ 
      &      & 0.35 & 0.40& 0.19 &  0.01& 0.25 &  0.02& 0.21 &  0.01& 0.23 &  0.01\\ 
      &      & 0.40 & 0.45& 0.15 &  0.01& 0.17 &  0.02& 0.16 &  0.01& 0.18 &  0.01\\ 
      &      & 0.45 & 0.50& 0.11 &  0.01& 0.11 &  0.01& 0.13 &  0.01& 0.13 &  0.01\\ 
\hline  
 1.95 & 2.15 & 0.10 & 0.15& 0.82 &  0.07& 0.93 &  0.07& 0.90 &  0.07& 0.91 &  0.07\\ 
      &      & 0.15 & 0.20& 0.73 &  0.04& 0.88 &  0.04& 0.77 &  0.04& 0.81 &  0.05\\ 
      &      & 0.20 & 0.25& 0.40 &  0.03& 0.47 &  0.03& 0.48 &  0.03& 0.52 &  0.03\\ 
      &      & 0.25 & 0.30& 0.29 &  0.02& 0.29 &  0.02& 0.32 &  0.02& 0.30 &  0.03\\ 
      &      & 0.30 & 0.35& 0.20 &  0.02& 0.22 &  0.01& 0.21 &  0.02& 0.19 &  0.01\\ 
      &      & 0.35 & 0.40& 0.14 &  0.01& 0.17 &  0.01& 0.16 &  0.01& 0.16 &  0.01\\ 
      &      & 0.40 & 0.45& 0.11 &  0.01& 0.11 &  0.01& 0.12 &  0.01& 0.11 &  0.01\\ 
      &      & 0.45 & 0.50& 0.08 &  0.01& 0.07 &  0.01& 0.09 &  0.01& 0.08 &  0.01\\ 

\end{tabular}
\end{center}
\end{table}
\clearpage
\begin{table}[hp!]
\begin{center}
  \caption{\label{tab:xsec-pipp-ta}
    HARP results for the double-differential $\pi^+$ production
    cross-section in the laboratory system,
    $d^2\sigma^{\pi^+}/(dpd\theta)$ for $\pi^+$--Ta interactions. Each row refers to a
    different $(p_{\hbox{\small min}} \le p<p_{\hbox{\small max}},
    \theta_{\hbox{\small min}} \le \theta<\theta_{\hbox{\small max}})$ bin,
    where $p$ and $\theta$ are the pion momentum and polar angle, respectively.
    The central value as well as the square-root of the diagonal elements
    of the covariance matrix are given.}
\vspace{2mm}
\begin{tabular}{rrrr|r@{$\pm$}lr@{$\pm$}lr@{$\pm$}lr@{$\pm$}l}
\hline
$\theta_{\hbox{\small min}}$ &
$\theta_{\hbox{\small max}}$ &
$p_{\hbox{\small min}}$ &
$p_{\hbox{\small max}}$ &
\multicolumn{8}{c}{$d^2\sigma^{\pi^+}/(dpd\theta)$}
\\
(rad) & (rad) & (\GeVc) & (\GeVc) &
\multicolumn{8}{c}{($\barn/(\GeVc \cdot \rad)$)}
\\
  &  &  &
&\multicolumn{2}{c}{$ \bf{3 \ \GeVc}$}
&\multicolumn{2}{c}{$ \bf{5 \ \GeVc}$}
&\multicolumn{2}{c}{$ \bf{8 \ \GeVc}$}
&\multicolumn{2}{c}{$ \bf{12 \ \GeVc}$}
\\
\hline  
 0.35 & 0.55 & 0.15 & 0.20& 0.47 &  0.25& 0.99 &  0.48& 1.28 &  0.56& 1.55 &  0.70\\ 
      &      & 0.20 & 0.25& 0.63 &  0.17& 1.31 &  0.34& 1.73 &  0.37& 1.65 &  0.46\\ 
      &      & 0.25 & 0.30& 0.87 &  0.10& 1.57 &  0.19& 2.38 &  0.23& 3.05 &  0.50\\ 
      &      & 0.30 & 0.35& 1.01 &  0.09& 1.71 &  0.12& 2.60 &  0.20& 3.16 &  0.28\\ 
      &      & 0.35 & 0.40& 1.03 &  0.06& 1.76 &  0.10& 2.41 &  0.11& 3.00 &  0.27\\ 
      &      & 0.40 & 0.45& 1.08 &  0.07& 1.74 &  0.09& 2.41 &  0.12& 2.88 &  0.26\\ 
      &      & 0.45 & 0.50& 1.09 &  0.06& 1.77 &  0.09& 2.53 &  0.18& 2.98 &  0.22\\ 
      &      & 0.50 & 0.60& 0.94 &  0.06& 1.77 &  0.08& 2.74 &  0.15& 3.09 &  0.24\\ 
      &      & 0.60 & 0.70& 0.80 &  0.08& 1.51 &  0.14& 2.33 &  0.25& 2.66 &  0.31\\ 
      &      & 0.70 & 0.80& 0.66 &  0.09& 1.11 &  0.16& 1.78 &  0.24& 2.03 &  0.31\\ 
\hline  
 0.55 & 0.75 & 0.10 & 0.15& 0.45 &  0.29& 0.65 &  0.42& 0.60 &  0.44& 0.75 &  0.48\\ 
      &      & 0.15 & 0.20& 0.99 &  0.23& 1.47 &  0.39& 1.63 &  0.39& 2.15 &  0.60\\ 
      &      & 0.20 & 0.25& 1.02 &  0.13& 1.90 &  0.20& 2.36 &  0.26& 3.16 &  0.36\\ 
      &      & 0.25 & 0.30& 1.20 &  0.10& 1.83 &  0.11& 2.62 &  0.15& 2.69 &  0.25\\ 
      &      & 0.30 & 0.35& 1.03 &  0.07& 1.71 &  0.10& 2.67 &  0.14& 3.12 &  0.37\\ 
      &      & 0.35 & 0.40& 1.01 &  0.08& 1.76 &  0.09& 2.31 &  0.11& 3.12 &  0.24\\ 
      &      & 0.40 & 0.45& 1.01 &  0.05& 1.68 &  0.07& 2.18 &  0.14& 2.43 &  0.22\\ 
      &      & 0.45 & 0.50& 0.93 &  0.05& 1.57 &  0.07& 2.19 &  0.11& 2.11 &  0.18\\ 
      &      & 0.50 & 0.60& 0.78 &  0.06& 1.36 &  0.08& 1.97 &  0.13& 2.28 &  0.20\\ 
      &      & 0.60 & 0.70& 0.58 &  0.07& 0.96 &  0.11& 1.45 &  0.17& 1.44 &  0.26\\ 
      &      & 0.70 & 0.80& 0.37 &  0.07& 0.64 &  0.11& 0.93 &  0.16& 0.89 &  0.20\\ 
\hline  
 0.75 & 0.95 & 0.10 & 0.15& 0.72 &  0.29& 1.01 &  0.39& 1.02 &  0.43& 1.35 &  0.50\\ 
      &      & 0.15 & 0.20& 1.26 &  0.15& 1.84 &  0.23& 2.30 &  0.21& 2.37 &  0.37\\ 
      &      & 0.20 & 0.25& 1.34 &  0.11& 1.96 &  0.13& 2.49 &  0.18& 3.15 &  0.27\\ 
      &      & 0.25 & 0.30& 1.28 &  0.08& 1.79 &  0.10& 2.31 &  0.12& 2.66 &  0.26\\ 
      &      & 0.30 & 0.35& 1.09 &  0.06& 1.66 &  0.08& 2.18 &  0.11& 2.81 &  0.24\\ 
      &      & 0.35 & 0.40& 0.99 &  0.06& 1.48 &  0.07& 1.90 &  0.09& 2.48 &  0.20\\ 
      &      & 0.40 & 0.45& 0.86 &  0.05& 1.27 &  0.06& 1.68 &  0.08& 2.05 &  0.18\\ 
      &      & 0.45 & 0.50& 0.71 &  0.04& 1.11 &  0.05& 1.45 &  0.08& 1.70 &  0.16\\ 
      &      & 0.50 & 0.60& 0.56 &  0.04& 0.85 &  0.06& 1.09 &  0.08& 1.12 &  0.15\\ 
      &      & 0.60 & 0.70& 0.36 &  0.05& 0.50 &  0.08& 0.67 &  0.09& 0.60 &  0.12\\ 
\hline  
 0.95 & 1.15 & 0.10 & 0.15& 0.89 &  0.24& 1.38 &  0.31& 1.56 &  0.33& 1.79 &  0.39\\ 
      &      & 0.15 & 0.20& 1.29 &  0.13& 1.98 &  0.15& 2.26 &  0.16& 2.26 &  0.24\\ 
      &      & 0.20 & 0.25& 1.21 &  0.08& 1.81 &  0.08& 2.34 &  0.14& 2.19 &  0.20\\ 
      &      & 0.25 & 0.30& 1.10 &  0.07& 1.46 &  0.08& 1.83 &  0.10& 2.32 &  0.23\\ 
      &      & 0.30 & 0.35& 0.88 &  0.06& 1.18 &  0.06& 1.54 &  0.09& 1.84 &  0.20\\ 
      &      & 0.35 & 0.40& 0.73 &  0.04& 1.03 &  0.05& 1.36 &  0.08& 1.21 &  0.13\\ 
      &      & 0.40 & 0.45& 0.62 &  0.04& 0.86 &  0.04& 1.12 &  0.07& 0.89 &  0.11\\ 
      &      & 0.45 & 0.50& 0.46 &  0.03& 0.69 &  0.04& 0.87 &  0.06& 0.73 &  0.08\\ 
      &      & 0.50 & 0.60& 0.36 &  0.03& 0.47 &  0.05& 0.57 &  0.05& 0.55 &  0.08\\ 
\hline
\end{tabular}
\end{center}
\end{table}

\begin{table}[hp!]
\begin{center}
\begin{tabular}{rrrr|r@{$\pm$}lr@{$\pm$}lr@{$\pm$}lr@{$\pm$}l}
\hline
$\theta_{\hbox{\small min}}$ &
$\theta_{\hbox{\small max}}$ &
$p_{\hbox{\small min}}$ &
$p_{\hbox{\small max}}$ &
\multicolumn{8}{c}{$d^2\sigma^{\pi^+}/(dpd\theta)$}
\\
(rad) & (rad) & (\GeVc) & (\GeVc) &
\multicolumn{8}{c}{(\barn/($\GeVc \cdot \rad$))}
\\
  &  &  &
&\multicolumn{2}{c}{$ \bf{3 \ \GeVc}$}
&\multicolumn{2}{c}{$ \bf{5 \ \GeVc}$}
&\multicolumn{2}{c}{$ \bf{8 \ \GeVc}$}
&\multicolumn{2}{c}{$ \bf{12 \ \GeVc}$}
\\
\hline
 1.15 & 1.35 & 0.10 & 0.15& 1.05 &  0.23& 1.45 &  0.28& 1.83 &  0.33& 2.02 &  0.42\\ 
      &      & 0.15 & 0.20& 1.44 &  0.14& 1.86 &  0.15& 2.22 &  0.20& 2.13 &  0.26\\ 
      &      & 0.20 & 0.25& 1.31 &  0.07& 1.70 &  0.10& 2.00 &  0.13& 2.10 &  0.21\\ 
      &      & 0.25 & 0.30& 0.87 &  0.06& 1.21 &  0.07& 1.49 &  0.10& 1.88 &  0.18\\ 
      &      & 0.30 & 0.35& 0.69 &  0.05& 0.94 &  0.05& 1.09 &  0.07& 1.04 &  0.17\\ 
      &      & 0.35 & 0.40& 0.60 &  0.04& 0.74 &  0.04& 0.91 &  0.05& 0.75 &  0.08\\ 
      &      & 0.40 & 0.45& 0.48 &  0.03& 0.58 &  0.04& 0.71 &  0.04& 0.69 &  0.09\\ 
      &      & 0.45 & 0.50& 0.34 &  0.03& 0.43 &  0.04& 0.53 &  0.04& 0.59 &  0.08\\ 
\hline  
 1.35 & 1.55 & 0.10 & 0.15& 1.00 &  0.24& 1.46 &  0.31& 1.87 &  0.43& 2.22 &  0.48\\ 
      &      & 0.15 & 0.20& 1.43 &  0.17& 1.78 &  0.18& 2.16 &  0.24& 2.66 &  0.30\\ 
      &      & 0.20 & 0.25& 1.23 &  0.09& 1.61 &  0.12& 1.83 &  0.12& 2.21 &  0.20\\ 
      &      & 0.25 & 0.30& 0.80 &  0.06& 1.02 &  0.08& 1.14 &  0.08& 1.24 &  0.15\\ 
      &      & 0.30 & 0.35& 0.57 &  0.04& 0.66 &  0.05& 0.84 &  0.06& 0.79 &  0.09\\ 
      &      & 0.35 & 0.40& 0.44 &  0.03& 0.52 &  0.03& 0.63 &  0.04& 0.66 &  0.08\\ 
      &      & 0.40 & 0.45& 0.33 &  0.03& 0.39 &  0.03& 0.47 &  0.03& 0.56 &  0.08\\ 
      &      & 0.45 & 0.50& 0.21 &  0.02& 0.26 &  0.03& 0.35 &  0.03& 0.46 &  0.08\\ 
\hline  
 1.55 & 1.75 & 0.10 & 0.15& 1.11 &  0.26& 1.39 &  0.30& 1.76 &  0.43& 2.10 &  0.53\\ 
      &      & 0.15 & 0.20& 1.38 &  0.13& 1.73 &  0.17& 1.92 &  0.20& 2.38 &  0.32\\ 
      &      & 0.20 & 0.25& 1.06 &  0.08& 1.41 &  0.09& 1.51 &  0.11& 1.87 &  0.21\\ 
      &      & 0.25 & 0.30& 0.68 &  0.06& 0.82 &  0.07& 0.87 &  0.06& 1.01 &  0.12\\ 
      &      & 0.30 & 0.35& 0.49 &  0.03& 0.54 &  0.04& 0.63 &  0.04& 0.68 &  0.11\\ 
      &      & 0.35 & 0.40& 0.35 &  0.03& 0.38 &  0.03& 0.45 &  0.04& 0.35 &  0.07\\ 
      &      & 0.40 & 0.45& 0.22 &  0.02& 0.26 &  0.02& 0.28 &  0.03& 0.20 &  0.04\\ 
      &      & 0.45 & 0.50& 0.14 &  0.02& 0.17 &  0.02& 0.18 &  0.02& 0.14 &  0.03\\ 
\hline  
 1.75 & 1.95 & 0.10 & 0.15& 1.12 &  0.20& 1.34 &  0.22& 1.44 &  0.22& 1.64 &  0.36\\ 
      &      & 0.15 & 0.20& 1.18 &  0.08& 1.51 &  0.09& 1.55 &  0.11& 1.99 &  0.20\\ 
      &      & 0.20 & 0.25& 0.87 &  0.06& 1.00 &  0.08& 1.07 &  0.07& 1.26 &  0.18\\ 
      &      & 0.25 & 0.30& 0.52 &  0.05& 0.51 &  0.04& 0.60 &  0.05& 0.52 &  0.09\\ 
      &      & 0.30 & 0.35& 0.32 &  0.02& 0.35 &  0.02& 0.39 &  0.03& 0.32 &  0.06\\ 
      &      & 0.35 & 0.40& 0.24 &  0.02& 0.26 &  0.02& 0.26 &  0.02& 0.29 &  0.06\\ 
      &      & 0.40 & 0.45& 0.18 &  0.02& 0.15 &  0.02& 0.17 &  0.02& 0.17 &  0.05\\ 
      &      & 0.45 & 0.50& 0.10 &  0.02& 0.08 &  0.01& 0.10 &  0.02& 0.10 &  0.03\\ 
\hline  
 1.95 & 2.15 & 0.10 & 0.15& 0.98 &  0.16& 1.17 &  0.17& 1.09 &  0.16& 0.82 &  0.20\\ 
      &      & 0.15 & 0.20& 1.05 &  0.05& 1.13 &  0.06& 1.24 &  0.06& 1.49 &  0.19\\ 
      &      & 0.20 & 0.25& 0.61 &  0.04& 0.70 &  0.05& 0.72 &  0.05& 0.83 &  0.15\\ 
      &      & 0.25 & 0.30& 0.36 &  0.03& 0.37 &  0.03& 0.45 &  0.04& 0.39 &  0.08\\ 
      &      & 0.30 & 0.35& 0.23 &  0.02& 0.24 &  0.02& 0.30 &  0.02& 0.24 &  0.06\\ 
      &      & 0.35 & 0.40& 0.16 &  0.02& 0.19 &  0.01& 0.19 &  0.03& 0.15 &  0.05\\ 
      &      & 0.40 & 0.45& 0.10 &  0.01& 0.12 &  0.02& 0.10 &  0.02& 0.09 &  0.04\\ 
      &      & 0.45 & 0.50& 0.06 &  0.01& 0.06 &  0.01& 0.05 &  0.01& 0.06 &  0.03\\ 

\end{tabular}
\end{center}
\end{table}

\begin{table}[hp!]
\begin{center}
  \caption{\label{tab:xsec-pipm-ta}
    HARP results for the double-differential $\pi^-$ production
    cross-section in the laboratory system,
    $d^2\sigma^{\pi^-}/(dpd\theta)$ for $\pi^+$--Ta interactions. Each row refers to a
    different $(p_{\hbox{\small min}} \le p<p_{\hbox{\small max}},
    \theta_{\hbox{\small min}} \le \theta<\theta_{\hbox{\small max}})$ bin,
    where $p$ and $\theta$ are the pion momentum and polar angle, respectively.
    The central value as well as the square-root of the diagonal elements
    of the covariance matrix are given.}
\vspace{2mm}
\begin{tabular}{rrrr|r@{$\pm$}lr@{$\pm$}lr@{$\pm$}lr@{$\pm$}l}
\hline
$\theta_{\hbox{\small min}}$ &
$\theta_{\hbox{\small max}}$ &
$p_{\hbox{\small min}}$ &
$p_{\hbox{\small max}}$ &
\multicolumn{8}{c}{$d^2\sigma^{\pi^-}/(dpd\theta)$}
\\
(rad) & (rad) & (\GeVc) & (\GeVc) &
\multicolumn{8}{c}{($\barn/(\GeVc \cdot \rad)$)}
\\
  &  &  &
&\multicolumn{2}{c}{$ \bf{3 \ \GeVc}$}
&\multicolumn{2}{c}{$ \bf{5 \ \GeVc}$}
&\multicolumn{2}{c}{$ \bf{8 \ \GeVc}$}
&\multicolumn{2}{c}{$ \bf{12 \ \GeVc}$}
\\
\hline  
 0.35 & 0.55 & 0.15 & 0.20& 0.64 &  0.26& 0.96 &  0.49& 1.31 &  0.54& 1.83 &  0.77\\ 
      &      & 0.20 & 0.25& 0.77 &  0.17& 1.28 &  0.32& 1.65 &  0.37& 1.97 &  0.50\\ 
      &      & 0.25 & 0.30& 0.71 &  0.10& 1.52 &  0.19& 2.08 &  0.19& 1.87 &  0.33\\ 
      &      & 0.30 & 0.35& 0.57 &  0.05& 1.51 &  0.09& 2.13 &  0.13& 2.41 &  0.27\\ 
      &      & 0.35 & 0.40& 0.65 &  0.06& 1.20 &  0.08& 1.70 &  0.10& 1.65 &  0.19\\ 
      &      & 0.40 & 0.45& 0.64 &  0.04& 1.02 &  0.05& 1.57 &  0.09& 1.38 &  0.15\\ 
      &      & 0.45 & 0.50& 0.63 &  0.04& 0.98 &  0.05& 1.57 &  0.08& 1.46 &  0.18\\ 
      &      & 0.50 & 0.60& 0.59 &  0.04& 0.96 &  0.05& 1.44 &  0.08& 1.67 &  0.17\\ 
      &      & 0.60 & 0.70& 0.53 &  0.05& 0.91 &  0.07& 1.36 &  0.10& 1.52 &  0.20\\ 
      &      & 0.70 & 0.80& 0.41 &  0.06& 0.67 &  0.09& 1.20 &  0.15& 1.15 &  0.19\\ 
\hline  
 0.55 & 0.75 & 0.10 & 0.15& 0.71 &  0.33& 0.91 &  0.51& 0.95 &  0.55& 1.34 &  0.65\\ 
      &      & 0.15 & 0.20& 0.91 &  0.22& 1.43 &  0.37& 1.69 &  0.38& 2.20 &  0.54\\ 
      &      & 0.20 & 0.25& 0.95 &  0.10& 1.49 &  0.15& 2.20 &  0.17& 2.39 &  0.32\\ 
      &      & 0.25 & 0.30& 0.84 &  0.07& 1.34 &  0.09& 1.85 &  0.11& 2.20 &  0.24\\ 
      &      & 0.30 & 0.35& 0.80 &  0.05& 1.23 &  0.07& 1.61 &  0.09& 1.54 &  0.17\\ 
      &      & 0.35 & 0.40& 0.68 &  0.04& 1.10 &  0.05& 1.56 &  0.09& 1.59 &  0.17\\ 
      &      & 0.40 & 0.45& 0.70 &  0.05& 1.00 &  0.05& 1.51 &  0.08& 1.50 &  0.14\\ 
      &      & 0.45 & 0.50& 0.65 &  0.04& 0.94 &  0.04& 1.32 &  0.07& 1.54 &  0.16\\ 
      &      & 0.50 & 0.60& 0.55 &  0.03& 0.84 &  0.04& 1.19 &  0.06& 1.39 &  0.14\\ 
      &      & 0.60 & 0.70& 0.42 &  0.04& 0.70 &  0.06& 0.98 &  0.08& 1.14 &  0.13\\ 
      &      & 0.70 & 0.80& 0.30 &  0.05& 0.49 &  0.07& 0.76 &  0.10& 0.94 &  0.14\\ 
\hline  
 0.75 & 0.95 & 0.10 & 0.15& 0.96 &  0.28& 1.29 &  0.39& 1.46 &  0.44& 1.78 &  0.54\\ 
      &      & 0.15 & 0.20& 1.16 &  0.14& 1.70 &  0.19& 1.96 &  0.20& 1.93 &  0.29\\ 
      &      & 0.20 & 0.25& 1.03 &  0.08& 1.47 &  0.09& 2.02 &  0.12& 1.91 &  0.22\\ 
      &      & 0.25 & 0.30& 0.83 &  0.05& 1.24 &  0.06& 1.59 &  0.09& 1.82 &  0.20\\ 
      &      & 0.30 & 0.35& 0.75 &  0.05& 1.06 &  0.05& 1.50 &  0.08& 1.80 &  0.20\\ 
      &      & 0.35 & 0.40& 0.68 &  0.04& 0.89 &  0.04& 1.28 &  0.06& 1.65 &  0.16\\ 
      &      & 0.40 & 0.45& 0.58 &  0.03& 0.76 &  0.03& 1.16 &  0.06& 1.29 &  0.13\\ 
      &      & 0.45 & 0.50& 0.49 &  0.03& 0.69 &  0.03& 1.03 &  0.05& 1.15 &  0.11\\ 
      &      & 0.50 & 0.60& 0.39 &  0.03& 0.60 &  0.03& 0.81 &  0.05& 0.87 &  0.10\\ 
      &      & 0.60 & 0.70& 0.28 &  0.03& 0.47 &  0.04& 0.64 &  0.05& 0.70 &  0.09\\ 
\hline  
 0.95 & 1.15 & 0.10 & 0.15& 1.19 &  0.23& 1.62 &  0.30& 2.02 &  0.34& 2.33 &  0.45\\ 
      &      & 0.15 & 0.20& 1.17 &  0.12& 1.74 &  0.14& 2.02 &  0.15& 2.28 &  0.23\\ 
      &      & 0.20 & 0.25& 0.81 &  0.06& 1.32 &  0.08& 1.81 &  0.12& 2.31 &  0.23\\ 
      &      & 0.25 & 0.30& 0.67 &  0.05& 1.11 &  0.06& 1.52 &  0.09& 1.67 &  0.18\\ 
      &      & 0.30 & 0.35& 0.59 &  0.04& 0.92 &  0.05& 1.19 &  0.07& 1.29 &  0.14\\ 
      &      & 0.35 & 0.40& 0.48 &  0.03& 0.76 &  0.04& 0.93 &  0.05& 1.22 &  0.13\\ 
      &      & 0.40 & 0.45& 0.38 &  0.02& 0.61 &  0.03& 0.82 &  0.04& 0.94 &  0.11\\ 
      &      & 0.45 & 0.50& 0.34 &  0.02& 0.52 &  0.02& 0.69 &  0.04& 0.64 &  0.08\\ 
      &      & 0.50 & 0.60& 0.26 &  0.02& 0.42 &  0.03& 0.54 &  0.04& 0.43 &  0.06\\ 
\hline
\end{tabular}
\end{center}
\end{table}

\begin{table}[hp!]
\begin{center}
\begin{tabular}{rrrr|r@{$\pm$}lr@{$\pm$}lr@{$\pm$}lr@{$\pm$}l}
\hline
$\theta_{\hbox{\small min}}$ &
$\theta_{\hbox{\small max}}$ &
$p_{\hbox{\small min}}$ &
$p_{\hbox{\small max}}$ &
\multicolumn{8}{c}{$d^2\sigma^{\pi^-}/(dpd\theta)$}
\\
(rad) & (rad) & (\GeVc) & (\GeVc) &
\multicolumn{8}{c}{(\barn/($\GeVc \cdot \rad$))}
\\
  &  &  &
&\multicolumn{2}{c}{$ \bf{3 \ \GeVc}$}
&\multicolumn{2}{c}{$ \bf{5 \ \GeVc}$}
&\multicolumn{2}{c}{$ \bf{8 \ \GeVc}$}
&\multicolumn{2}{c}{$ \bf{12 \ \GeVc}$}
\\
\hline
 1.15 & 1.35 & 0.10 & 0.15& 1.38 &  0.24& 1.78 &  0.30& 2.29 &  0.39& 2.55 &  0.48\\ 
      &      & 0.15 & 0.20& 1.15 &  0.12& 1.63 &  0.13& 2.02 &  0.15& 2.20 &  0.25\\ 
      &      & 0.20 & 0.25& 0.80 &  0.06& 1.25 &  0.08& 1.52 &  0.10& 2.07 &  0.20\\ 
      &      & 0.25 & 0.30& 0.61 &  0.04& 0.88 &  0.06& 1.11 &  0.07& 1.52 &  0.18\\ 
      &      & 0.30 & 0.35& 0.46 &  0.03& 0.69 &  0.04& 0.96 &  0.06& 0.95 &  0.11\\ 
      &      & 0.35 & 0.40& 0.37 &  0.03& 0.63 &  0.04& 0.76 &  0.05& 0.64 &  0.08\\ 
      &      & 0.40 & 0.45& 0.29 &  0.02& 0.50 &  0.03& 0.58 &  0.04& 0.51 &  0.07\\ 
      &      & 0.45 & 0.50& 0.23 &  0.02& 0.38 &  0.03& 0.48 &  0.03& 0.50 &  0.07\\ 
\hline  
 1.35 & 1.55 & 0.10 & 0.15& 1.32 &  0.25& 1.78 &  0.32& 2.63 &  0.50& 2.93 &  0.76\\ 
      &      & 0.15 & 0.20& 1.09 &  0.13& 1.47 &  0.15& 1.98 &  0.18& 2.03 &  0.24\\ 
      &      & 0.20 & 0.25& 0.77 &  0.06& 1.15 &  0.09& 1.30 &  0.10& 1.78 &  0.18\\ 
      &      & 0.25 & 0.30& 0.47 &  0.04& 0.77 &  0.06& 0.87 &  0.07& 1.27 &  0.17\\ 
      &      & 0.30 & 0.35& 0.39 &  0.03& 0.50 &  0.04& 0.69 &  0.05& 0.62 &  0.09\\ 
      &      & 0.35 & 0.40& 0.32 &  0.02& 0.41 &  0.03& 0.56 &  0.04& 0.51 &  0.08\\ 
      &      & 0.40 & 0.45& 0.25 &  0.02& 0.34 &  0.02& 0.42 &  0.03& 0.46 &  0.07\\ 
      &      & 0.45 & 0.50& 0.17 &  0.02& 0.25 &  0.02& 0.32 &  0.03& 0.36 &  0.06\\ 
\hline  
 1.55 & 1.75 & 0.10 & 0.15& 1.28 &  0.26& 1.66 &  0.32& 1.99 &  0.41& 2.90 &  0.55\\ 
      &      & 0.15 & 0.20& 1.00 &  0.11& 1.29 &  0.12& 1.70 &  0.18& 1.79 &  0.20\\ 
      &      & 0.20 & 0.25& 0.68 &  0.06& 0.91 &  0.07& 1.07 &  0.09& 1.33 &  0.15\\ 
      &      & 0.25 & 0.30& 0.36 &  0.03& 0.61 &  0.05& 0.73 &  0.06& 0.89 &  0.12\\ 
      &      & 0.30 & 0.35& 0.28 &  0.03& 0.41 &  0.03& 0.57 &  0.04& 0.47 &  0.08\\ 
      &      & 0.35 & 0.40& 0.24 &  0.02& 0.30 &  0.02& 0.41 &  0.03& 0.30 &  0.05\\ 
      &      & 0.40 & 0.45& 0.18 &  0.02& 0.22 &  0.01& 0.29 &  0.02& 0.25 &  0.06\\ 
      &      & 0.45 & 0.50& 0.12 &  0.01& 0.16 &  0.01& 0.20 &  0.02& 0.19 &  0.04\\ 
\hline  
 1.75 & 1.95 & 0.10 & 0.15& 1.13 &  0.19& 1.48 &  0.21& 1.55 &  0.24& 2.36 &  0.42\\ 
      &      & 0.15 & 0.20& 0.84 &  0.06& 1.03 &  0.07& 1.33 &  0.09& 1.41 &  0.16\\ 
      &      & 0.20 & 0.25& 0.54 &  0.04& 0.65 &  0.04& 0.85 &  0.05& 0.83 &  0.11\\ 
      &      & 0.25 & 0.30& 0.30 &  0.03& 0.44 &  0.03& 0.56 &  0.05& 0.47 &  0.09\\ 
      &      & 0.30 & 0.35& 0.21 &  0.02& 0.31 &  0.02& 0.33 &  0.03& 0.27 &  0.06\\ 
      &      & 0.35 & 0.40& 0.19 &  0.02& 0.22 &  0.02& 0.26 &  0.02& 0.22 &  0.05\\ 
      &      & 0.40 & 0.45& 0.14 &  0.01& 0.18 &  0.01& 0.19 &  0.02& 0.22 &  0.05\\ 
      &      & 0.45 & 0.50& 0.09 &  0.01& 0.13 &  0.01& 0.14 &  0.01& 0.12 &  0.04\\ 
\hline  
 1.95 & 2.15 & 0.10 & 0.15& 0.92 &  0.13& 1.26 &  0.16& 1.36 &  0.18& 1.72 &  0.35\\ 
      &      & 0.15 & 0.20& 0.67 &  0.04& 0.89 &  0.05& 1.04 &  0.06& 1.30 &  0.16\\ 
      &      & 0.20 & 0.25& 0.42 &  0.03& 0.49 &  0.04& 0.60 &  0.04& 0.65 &  0.12\\ 
      &      & 0.25 & 0.30& 0.26 &  0.03& 0.29 &  0.02& 0.42 &  0.04& 0.22 &  0.05\\ 
      &      & 0.30 & 0.35& 0.17 &  0.02& 0.20 &  0.02& 0.25 &  0.02& 0.22 &  0.06\\ 
      &      & 0.35 & 0.40& 0.14 &  0.01& 0.14 &  0.01& 0.19 &  0.02& 0.16 &  0.05\\ 
      &      & 0.40 & 0.45& 0.11 &  0.01& 0.11 &  0.01& 0.15 &  0.02& 0.14 &  0.05\\ 
      &      & 0.45 & 0.50& 0.07 &  0.01& 0.08 &  0.01& 0.12 &  0.02& 0.11 &  0.04\\ 

\end{tabular}
\end{center}
\end{table}
\clearpage
\begin{table}[hp!]
\begin{center}
  \caption{\label{tab:xsec-pimp-ta}
    HARP results for the double-differential $\pi^+$ production
    cross-section in the laboratory system,
    $d^2\sigma^{\pi^+}/(dpd\theta)$ for $\pi^-$--Ta interactions. Each row refers to a
    different $(p_{\hbox{\small min}} \le p<p_{\hbox{\small max}},
    \theta_{\hbox{\small min}} \le \theta<\theta_{\hbox{\small max}})$ bin,
    where $p$ and $\theta$ are the pion momentum and polar angle, respectively.
    The central value as well as the square-root of the diagonal elements
    of the covariance matrix are given.}
\vspace{2mm}
\begin{tabular}{rrrr|r@{$\pm$}lr@{$\pm$}lr@{$\pm$}lr@{$\pm$}l}
\hline
$\theta_{\hbox{\small min}}$ &
$\theta_{\hbox{\small max}}$ &
$p_{\hbox{\small min}}$ &
$p_{\hbox{\small max}}$ &
\multicolumn{8}{c}{$d^2\sigma^{\pi^+}/(dpd\theta)$}
\\
(rad) & (rad) & (\GeVc) & (\GeVc) &
\multicolumn{8}{c}{($\barn/(\GeVc \cdot \rad)$)}
\\
  &  &  &
&\multicolumn{2}{c}{$ \bf{3 \ \GeVc}$}
&\multicolumn{2}{c}{$ \bf{5 \ \GeVc}$}
&\multicolumn{2}{c}{$ \bf{8 \ \GeVc}$}
&\multicolumn{2}{c}{$ \bf{12 \ \GeVc}$}
\\
\hline  
 0.35 & 0.55 & 0.15 & 0.20& 0.47 &  0.24& 1.19 &  0.43& 1.23 &  0.51& 2.14 &  0.64\\ 
      &      & 0.20 & 0.25& 0.62 &  0.17& 1.24 &  0.29& 1.28 &  0.36& 2.48 &  0.39\\ 
      &      & 0.25 & 0.30& 0.69 &  0.10& 1.51 &  0.14& 1.91 &  0.20& 2.78 &  0.21\\ 
      &      & 0.30 & 0.35& 0.75 &  0.08& 1.56 &  0.11& 1.94 &  0.12& 2.78 &  0.17\\ 
      &      & 0.35 & 0.40& 0.67 &  0.06& 1.40 &  0.07& 1.93 &  0.11& 2.60 &  0.13\\ 
      &      & 0.40 & 0.45& 0.62 &  0.05& 1.34 &  0.07& 1.77 &  0.08& 2.59 &  0.11\\ 
      &      & 0.45 & 0.50& 0.58 &  0.05& 1.35 &  0.08& 1.88 &  0.10& 2.58 &  0.11\\ 
      &      & 0.50 & 0.60& 0.69 &  0.07& 1.42 &  0.08& 1.78 &  0.09& 2.64 &  0.13\\ 
      &      & 0.60 & 0.70& 0.58 &  0.07& 1.20 &  0.13& 1.59 &  0.14& 2.53 &  0.21\\ 
      &      & 0.70 & 0.80& 0.42 &  0.08& 0.96 &  0.15& 1.24 &  0.19& 2.14 &  0.29\\ 
\hline  
 0.55 & 0.75 & 0.10 & 0.15& 0.52 &  0.32& 0.92 &  0.47& 0.83 &  0.49& 1.66 &  0.73\\ 
      &      & 0.15 & 0.20& 0.77 &  0.22& 1.43 &  0.29& 1.33 &  0.32& 2.51 &  0.37\\ 
      &      & 0.20 & 0.25& 0.90 &  0.10& 1.66 &  0.14& 1.94 &  0.20& 2.61 &  0.21\\ 
      &      & 0.25 & 0.30& 0.72 &  0.08& 1.61 &  0.09& 1.95 &  0.12& 2.66 &  0.17\\ 
      &      & 0.30 & 0.35& 0.81 &  0.07& 1.46 &  0.08& 1.95 &  0.13& 2.58 &  0.11\\ 
      &      & 0.35 & 0.40& 0.78 &  0.06& 1.39 &  0.07& 1.79 &  0.08& 2.51 &  0.14\\ 
      &      & 0.40 & 0.45& 0.71 &  0.06& 1.22 &  0.06& 1.58 &  0.07& 2.42 &  0.11\\ 
      &      & 0.45 & 0.50& 0.69 &  0.06& 1.10 &  0.06& 1.55 &  0.07& 2.15 &  0.10\\ 
      &      & 0.50 & 0.60& 0.54 &  0.06& 1.02 &  0.07& 1.33 &  0.09& 1.99 &  0.12\\ 
      &      & 0.60 & 0.70& 0.33 &  0.06& 0.77 &  0.10& 1.00 &  0.12& 1.48 &  0.19\\ 
      &      & 0.70 & 0.80& 0.22 &  0.04& 0.53 &  0.10& 0.65 &  0.12& 0.96 &  0.20\\ 
\hline  
 0.75 & 0.95 & 0.10 & 0.15& 0.47 &  0.26& 1.05 &  0.35& 0.91 &  0.35& 1.67 &  0.44\\ 
      &      & 0.15 & 0.20& 0.99 &  0.15& 1.64 &  0.16& 1.70 &  0.21& 2.45 &  0.18\\ 
      &      & 0.20 & 0.25& 0.87 &  0.09& 1.51 &  0.10& 2.03 &  0.12& 2.65 &  0.14\\ 
      &      & 0.25 & 0.30& 0.90 &  0.08& 1.31 &  0.07& 1.90 &  0.10& 2.37 &  0.12\\ 
      &      & 0.30 & 0.35& 0.66 &  0.05& 1.23 &  0.07& 1.58 &  0.07& 2.01 &  0.11\\ 
      &      & 0.35 & 0.40& 0.65 &  0.06& 1.10 &  0.06& 1.36 &  0.07& 1.81 &  0.10\\ 
      &      & 0.40 & 0.45& 0.56 &  0.05& 1.04 &  0.05& 1.27 &  0.06& 1.71 &  0.08\\ 
      &      & 0.45 & 0.50& 0.47 &  0.04& 0.96 &  0.05& 1.12 &  0.06& 1.49 &  0.08\\ 
      &      & 0.50 & 0.60& 0.38 &  0.04& 0.70 &  0.06& 0.86 &  0.06& 1.13 &  0.08\\ 
      &      & 0.60 & 0.70& 0.25 &  0.04& 0.47 &  0.07& 0.55 &  0.08& 0.72 &  0.11\\ 
\hline  
 0.95 & 1.15 & 0.10 & 0.15& 0.56 &  0.20& 1.21 &  0.26& 1.19 &  0.25& 1.61 &  0.29\\ 
      &      & 0.15 & 0.20& 1.14 &  0.15& 1.83 &  0.13& 1.81 &  0.15& 2.30 &  0.14\\ 
      &      & 0.20 & 0.25& 0.84 &  0.08& 1.60 &  0.09& 1.82 &  0.10& 2.28 &  0.11\\ 
      &      & 0.25 & 0.30& 0.62 &  0.05& 1.24 &  0.07& 1.58 &  0.09& 1.84 &  0.11\\ 
      &      & 0.30 & 0.35& 0.56 &  0.05& 1.02 &  0.06& 1.33 &  0.07& 1.55 &  0.10\\ 
      &      & 0.35 & 0.40& 0.45 &  0.04& 0.81 &  0.05& 1.07 &  0.05& 1.29 &  0.06\\ 
      &      & 0.40 & 0.45& 0.41 &  0.04& 0.67 &  0.04& 0.84 &  0.05& 1.10 &  0.06\\ 
      &      & 0.45 & 0.50& 0.37 &  0.04& 0.55 &  0.04& 0.64 &  0.05& 0.88 &  0.06\\ 
      &      & 0.50 & 0.60& 0.25 &  0.03& 0.38 &  0.04& 0.41 &  0.04& 0.61 &  0.06\\ 
\hline
\end{tabular}
\end{center}
\end{table}

\begin{table}[hp!]
\begin{center}
\begin{tabular}{rrrr|r@{$\pm$}lr@{$\pm$}lr@{$\pm$}lr@{$\pm$}l}
\hline
$\theta_{\hbox{\small min}}$ &
$\theta_{\hbox{\small max}}$ &
$p_{\hbox{\small min}}$ &
$p_{\hbox{\small max}}$ &
\multicolumn{8}{c}{$d^2\sigma^{\pi^+}/(dpd\theta)$}
\\
(rad) & (rad) & (\GeVc) & (\GeVc) &
\multicolumn{8}{c}{(\barn/($\GeVc \cdot \rad$))}
\\
  &  &  &
&\multicolumn{2}{c}{$ \bf{3 \ \GeVc}$}
&\multicolumn{2}{c}{$ \bf{5 \ \GeVc}$}
&\multicolumn{2}{c}{$ \bf{8 \ \GeVc}$}
&\multicolumn{2}{c}{$ \bf{12 \ \GeVc}$}
\\
\hline
 1.15 & 1.35 & 0.10 & 0.15& 0.68 &  0.18& 1.22 &  0.23& 1.31 &  0.25& 1.71 &  0.31\\ 
      &      & 0.15 & 0.20& 0.92 &  0.12& 1.57 &  0.13& 1.72 &  0.13& 2.19 &  0.16\\ 
      &      & 0.20 & 0.25& 0.86 &  0.07& 1.39 &  0.08& 1.55 &  0.09& 1.88 &  0.10\\ 
      &      & 0.25 & 0.30& 0.57 &  0.05& 1.01 &  0.06& 1.28 &  0.07& 1.37 &  0.07\\ 
      &      & 0.30 & 0.35& 0.53 &  0.04& 0.74 &  0.05& 0.93 &  0.06& 1.11 &  0.07\\ 
      &      & 0.35 & 0.40& 0.42 &  0.04& 0.60 &  0.04& 0.71 &  0.04& 0.90 &  0.05\\ 
      &      & 0.40 & 0.45& 0.30 &  0.03& 0.46 &  0.03& 0.56 &  0.03& 0.66 &  0.04\\ 
      &      & 0.45 & 0.50& 0.21 &  0.02& 0.35 &  0.03& 0.42 &  0.04& 0.45 &  0.04\\ 
\hline  
 1.35 & 1.55 & 0.10 & 0.15& 0.67 &  0.18& 1.04 &  0.24& 1.10 &  0.26& 1.63 &  0.34\\ 
      &      & 0.15 & 0.20& 0.88 &  0.13& 1.35 &  0.15& 1.56 &  0.18& 1.96 &  0.19\\ 
      &      & 0.20 & 0.25& 0.78 &  0.08& 1.23 &  0.08& 1.38 &  0.10& 1.60 &  0.11\\ 
      &      & 0.25 & 0.30& 0.47 &  0.04& 0.81 &  0.06& 0.94 &  0.07& 1.06 &  0.07\\ 
      &      & 0.30 & 0.35& 0.38 &  0.03& 0.55 &  0.04& 0.68 &  0.05& 0.78 &  0.05\\ 
      &      & 0.35 & 0.40& 0.29 &  0.03& 0.40 &  0.03& 0.49 &  0.03& 0.58 &  0.04\\ 
      &      & 0.40 & 0.45& 0.20 &  0.02& 0.31 &  0.02& 0.35 &  0.03& 0.43 &  0.03\\ 
      &      & 0.45 & 0.50& 0.14 &  0.02& 0.23 &  0.02& 0.23 &  0.02& 0.28 &  0.03\\ 
\hline  
 1.55 & 1.75 & 0.10 & 0.15& 0.74 &  0.16& 1.02 &  0.22& 1.19 &  0.25& 1.47 &  0.34\\ 
      &      & 0.15 & 0.20& 0.91 &  0.12& 1.16 &  0.12& 1.39 &  0.13& 1.65 &  0.16\\ 
      &      & 0.20 & 0.25& 0.76 &  0.07& 1.03 &  0.08& 1.22 &  0.08& 1.35 &  0.09\\ 
      &      & 0.25 & 0.30& 0.51 &  0.05& 0.69 &  0.05& 0.73 &  0.05& 0.85 &  0.06\\ 
      &      & 0.30 & 0.35& 0.37 &  0.04& 0.40 &  0.03& 0.51 &  0.04& 0.58 &  0.04\\ 
      &      & 0.35 & 0.40& 0.29 &  0.03& 0.31 &  0.02& 0.35 &  0.03& 0.40 &  0.03\\ 
      &      & 0.40 & 0.45& 0.20 &  0.02& 0.22 &  0.02& 0.23 &  0.02& 0.28 &  0.02\\ 
      &      & 0.45 & 0.50& 0.13 &  0.02& 0.16 &  0.02& 0.14 &  0.02& 0.18 &  0.02\\ 
\hline  
 1.75 & 1.95 & 0.10 & 0.15& 0.73 &  0.16& 0.98 &  0.16& 1.20 &  0.19& 1.33 &  0.22\\ 
      &      & 0.15 & 0.20& 0.83 &  0.07& 1.04 &  0.07& 1.17 &  0.07& 1.41 &  0.08\\ 
      &      & 0.20 & 0.25& 0.56 &  0.05& 0.82 &  0.05& 0.83 &  0.05& 0.97 &  0.06\\ 
      &      & 0.25 & 0.30& 0.38 &  0.04& 0.46 &  0.04& 0.47 &  0.04& 0.56 &  0.04\\ 
      &      & 0.30 & 0.35& 0.26 &  0.03& 0.28 &  0.02& 0.32 &  0.02& 0.34 &  0.03\\ 
      &      & 0.35 & 0.40& 0.18 &  0.02& 0.19 &  0.02& 0.21 &  0.02& 0.22 &  0.02\\ 
      &      & 0.40 & 0.45& 0.11 &  0.02& 0.12 &  0.01& 0.14 &  0.02& 0.13 &  0.01\\ 
      &      & 0.45 & 0.50& 0.06 &  0.01& 0.09 &  0.01& 0.08 &  0.01& 0.10 &  0.01\\ 
\hline  
 1.95 & 2.15 & 0.10 & 0.15& 0.67 &  0.12& 0.78 &  0.13& 0.93 &  0.14& 1.09 &  0.16\\ 
      &      & 0.15 & 0.20& 0.58 &  0.05& 0.83 &  0.05& 0.86 &  0.04& 1.10 &  0.05\\ 
      &      & 0.20 & 0.25& 0.36 &  0.04& 0.55 &  0.04& 0.59 &  0.04& 0.69 &  0.04\\ 
      &      & 0.25 & 0.30& 0.30 &  0.03& 0.27 &  0.03& 0.31 &  0.03& 0.40 &  0.03\\ 
      &      & 0.30 & 0.35& 0.15 &  0.03& 0.17 &  0.01& 0.18 &  0.02& 0.25 &  0.02\\ 
      &      & 0.35 & 0.40& 0.08 &  0.01& 0.13 &  0.01& 0.12 &  0.01& 0.16 &  0.02\\ 
      &      & 0.40 & 0.45& 0.04 &  0.01& 0.08 &  0.01& 0.10 &  0.01& 0.10 &  0.01\\ 
      &      & 0.45 & 0.50& 0.02 &  0.01& 0.05 &  0.01& 0.05 &  0.01& 0.06 &  0.01\\ 

\end{tabular}
\end{center}
\end{table}

\begin{table}[hp!]
\begin{center}
  \caption{\label{tab:xsec-pimm-ta}
    HARP results for the double-differential $\pi^-$ production
    cross-section in the laboratory system,
    $d^2\sigma^{\pi^-}/(dpd\theta)$ for $\pi^-$--Ta interactions. Each row refers to a
    different $(p_{\hbox{\small min}} \le p<p_{\hbox{\small max}},
    \theta_{\hbox{\small min}} \le \theta<\theta_{\hbox{\small max}})$ bin,
    where $p$ and $\theta$ are the pion momentum and polar angle, respectively.
    The central value as well as the square-root of the diagonal elements
    of the covariance matrix are given.}
\vspace{2mm}
\begin{tabular}{rrrr|r@{$\pm$}lr@{$\pm$}lr@{$\pm$}lr@{$\pm$}l}
\hline
$\theta_{\hbox{\small min}}$ &
$\theta_{\hbox{\small max}}$ &
$p_{\hbox{\small min}}$ &
$p_{\hbox{\small max}}$ &
\multicolumn{8}{c}{$d^2\sigma^{\pi^-}/(dpd\theta)$}
\\
(rad) & (rad) & (\GeVc) & (\GeVc) &
\multicolumn{8}{c}{($\barn/(\GeVc \cdot \rad)$)}
\\
  &  &  &
&\multicolumn{2}{c}{$ \bf{3 \ \GeVc}$}
&\multicolumn{2}{c}{$ \bf{5 \ \GeVc}$}
&\multicolumn{2}{c}{$ \bf{8 \ \GeVc}$}
&\multicolumn{2}{c}{$ \bf{12 \ \GeVc}$}
\\
\hline  
 0.35 & 0.55 & 0.15 & 0.20& 0.96 &  0.30& 1.68 &  0.46& 1.71 &  0.57& 2.70 &  0.69\\ 
      &      & 0.20 & 0.25& 0.94 &  0.17& 1.93 &  0.27& 2.02 &  0.35& 3.05 &  0.38\\ 
      &      & 0.25 & 0.30& 1.11 &  0.14& 2.00 &  0.15& 2.40 &  0.22& 3.38 &  0.24\\ 
      &      & 0.30 & 0.35& 1.25 &  0.10& 2.01 &  0.12& 2.43 &  0.15& 3.27 &  0.17\\ 
      &      & 0.35 & 0.40& 0.88 &  0.07& 1.76 &  0.09& 2.29 &  0.11& 2.98 &  0.13\\ 
      &      & 0.40 & 0.45& 1.00 &  0.12& 1.57 &  0.09& 2.16 &  0.10& 2.89 &  0.14\\ 
      &      & 0.45 & 0.50& 1.15 &  0.09& 1.61 &  0.09& 2.14 &  0.09& 2.77 &  0.12\\ 
      &      & 0.50 & 0.60& 0.99 &  0.07& 1.64 &  0.09& 2.04 &  0.10& 2.65 &  0.14\\ 
      &      & 0.60 & 0.70& 0.97 &  0.09& 1.58 &  0.13& 1.90 &  0.13& 2.67 &  0.19\\ 
      &      & 0.70 & 0.80& 0.91 &  0.11& 1.41 &  0.15& 1.62 &  0.19& 2.37 &  0.25\\ 
\hline  
 0.55 & 0.75 & 0.10 & 0.15& 0.98 &  0.40& 1.69 &  0.57& 1.25 &  0.54& 2.44 &  0.76\\ 
      &      & 0.15 & 0.20& 1.41 &  0.21& 2.30 &  0.29& 2.11 &  0.33& 3.14 &  0.43\\ 
      &      & 0.20 & 0.25& 1.44 &  0.14& 2.12 &  0.15& 2.44 &  0.19& 3.31 &  0.23\\ 
      &      & 0.25 & 0.30& 1.52 &  0.12& 1.96 &  0.12& 2.48 &  0.14& 3.08 &  0.17\\ 
      &      & 0.30 & 0.35& 1.25 &  0.09& 1.93 &  0.10& 2.33 &  0.14& 2.82 &  0.13\\ 
      &      & 0.35 & 0.40& 1.19 &  0.08& 1.64 &  0.08& 2.25 &  0.10& 2.52 &  0.11\\ 
      &      & 0.40 & 0.45& 1.11 &  0.08& 1.58 &  0.08& 1.98 &  0.08& 2.27 &  0.09\\ 
      &      & 0.45 & 0.50& 1.12 &  0.07& 1.52 &  0.07& 1.78 &  0.07& 2.17 &  0.10\\ 
      &      & 0.50 & 0.60& 1.08 &  0.07& 1.43 &  0.07& 1.58 &  0.08& 1.97 &  0.10\\ 
      &      & 0.60 & 0.70& 0.87 &  0.10& 1.21 &  0.12& 1.31 &  0.11& 1.68 &  0.15\\ 
      &      & 0.70 & 0.80& 0.68 &  0.10& 0.95 &  0.14& 0.99 &  0.13& 1.41 &  0.19\\ 
\hline  
 0.75 & 0.95 & 0.10 & 0.15& 1.42 &  0.35& 1.99 &  0.42& 1.85 &  0.43& 2.51 &  0.52\\ 
      &      & 0.15 & 0.20& 1.77 &  0.16& 2.39 &  0.18& 2.63 &  0.21& 3.40 &  0.21\\ 
      &      & 0.20 & 0.25& 1.54 &  0.12& 2.12 &  0.13& 2.52 &  0.15& 2.96 &  0.15\\ 
      &      & 0.25 & 0.30& 1.32 &  0.10& 1.95 &  0.10& 2.20 &  0.11& 2.65 &  0.13\\ 
      &      & 0.30 & 0.35& 1.12 &  0.07& 1.65 &  0.08& 1.92 &  0.09& 2.26 &  0.10\\ 
      &      & 0.35 & 0.40& 1.00 &  0.07& 1.39 &  0.07& 1.71 &  0.08& 1.96 &  0.08\\ 
      &      & 0.40 & 0.45& 0.94 &  0.06& 1.24 &  0.06& 1.45 &  0.07& 1.67 &  0.07\\ 
      &      & 0.45 & 0.50& 0.85 &  0.05& 1.12 &  0.05& 1.21 &  0.06& 1.54 &  0.06\\ 
      &      & 0.50 & 0.60& 0.69 &  0.05& 1.00 &  0.05& 0.95 &  0.05& 1.29 &  0.07\\ 
      &      & 0.60 & 0.70& 0.56 &  0.06& 0.83 &  0.08& 0.76 &  0.06& 1.05 &  0.10\\ 
\hline  
 0.95 & 1.15 & 0.10 & 0.15& 1.76 &  0.32& 2.46 &  0.39& 2.39 &  0.39& 2.86 &  0.44\\ 
      &      & 0.15 & 0.20& 1.98 &  0.15& 2.38 &  0.15& 2.85 &  0.17& 3.27 &  0.17\\ 
      &      & 0.20 & 0.25& 1.67 &  0.11& 2.09 &  0.12& 2.40 &  0.12& 2.52 &  0.12\\ 
      &      & 0.25 & 0.30& 1.27 &  0.09& 1.66 &  0.10& 1.91 &  0.10& 2.13 &  0.11\\ 
      &      & 0.30 & 0.35& 1.00 &  0.08& 1.28 &  0.07& 1.48 &  0.08& 1.67 &  0.09\\ 
      &      & 0.35 & 0.40& 0.77 &  0.05& 1.05 &  0.05& 1.19 &  0.06& 1.37 &  0.06\\ 
      &      & 0.40 & 0.45& 0.73 &  0.05& 0.91 &  0.04& 1.03 &  0.05& 1.18 &  0.05\\ 
      &      & 0.45 & 0.50& 0.65 &  0.04& 0.81 &  0.04& 0.87 &  0.04& 1.04 &  0.04\\ 
      &      & 0.50 & 0.60& 0.51 &  0.04& 0.63 &  0.04& 0.67 &  0.04& 0.82 &  0.05\\ 
\hline
\end{tabular}
\end{center}
\end{table}

\begin{table}[hp!]
\begin{center}
\begin{tabular}{rrrr|r@{$\pm$}lr@{$\pm$}lr@{$\pm$}lr@{$\pm$}l}
\hline
$\theta_{\hbox{\small min}}$ &
$\theta_{\hbox{\small max}}$ &
$p_{\hbox{\small min}}$ &
$p_{\hbox{\small max}}$ &
\multicolumn{8}{c}{$d^2\sigma^{\pi^-}/(dpd\theta)$}
\\
(rad) & (rad) & (\GeVc) & (\GeVc) &
\multicolumn{8}{c}{(\barn/($\GeVc \cdot \rad$))}
\\
  &  &  &
&\multicolumn{2}{c}{$ \bf{3 \ \GeVc}$}
&\multicolumn{2}{c}{$ \bf{5 \ \GeVc}$}
&\multicolumn{2}{c}{$ \bf{8 \ \GeVc}$}
&\multicolumn{2}{c}{$ \bf{12 \ \GeVc}$}
\\
\hline
 1.15 & 1.35 & 0.10 & 0.15& 2.07 &  0.35& 2.73 &  0.43& 2.58 &  0.47& 3.04 &  0.52\\ 
      &      & 0.15 & 0.20& 1.74 &  0.16& 2.42 &  0.16& 2.63 &  0.19& 2.96 &  0.19\\ 
      &      & 0.20 & 0.25& 1.18 &  0.10& 1.75 &  0.11& 2.00 &  0.12& 2.30 &  0.12\\ 
      &      & 0.25 & 0.30& 0.99 &  0.08& 1.35 &  0.08& 1.54 &  0.09& 1.69 &  0.10\\ 
      &      & 0.30 & 0.35& 0.86 &  0.07& 1.02 &  0.07& 1.12 &  0.07& 1.26 &  0.08\\ 
      &      & 0.35 & 0.40& 0.65 &  0.05& 0.77 &  0.04& 0.88 &  0.05& 0.97 &  0.05\\ 
      &      & 0.40 & 0.45& 0.51 &  0.04& 0.60 &  0.03& 0.70 &  0.04& 0.81 &  0.04\\ 
      &      & 0.45 & 0.50& 0.41 &  0.03& 0.51 &  0.03& 0.58 &  0.03& 0.67 &  0.03\\ 
\hline  
 1.35 & 1.55 & 0.10 & 0.15& 2.35 &  0.50& 2.29 &  0.48& 2.57 &  0.49& 2.91 &  0.61\\ 
      &      & 0.15 & 0.20& 1.59 &  0.16& 2.11 &  0.20& 2.29 &  0.22& 2.58 &  0.23\\ 
      &      & 0.20 & 0.25& 0.99 &  0.09& 1.36 &  0.11& 1.69 &  0.12& 1.86 &  0.13\\ 
      &      & 0.25 & 0.30& 0.65 &  0.06& 0.94 &  0.08& 1.16 &  0.09& 1.28 &  0.10\\ 
      &      & 0.30 & 0.35& 0.64 &  0.06& 0.73 &  0.05& 0.80 &  0.06& 0.91 &  0.07\\ 
      &      & 0.35 & 0.40& 0.55 &  0.05& 0.52 &  0.04& 0.63 &  0.04& 0.68 &  0.04\\ 
      &      & 0.40 & 0.45& 0.43 &  0.04& 0.42 &  0.03& 0.50 &  0.03& 0.53 &  0.03\\ 
      &      & 0.45 & 0.50& 0.32 &  0.03& 0.36 &  0.02& 0.37 &  0.03& 0.42 &  0.03\\ 
\hline  
 1.55 & 1.75 & 0.10 & 0.15& 1.94 &  0.36& 2.06 &  0.42& 2.58 &  0.50& 2.50 &  0.54\\ 
      &      & 0.15 & 0.20& 1.36 &  0.16& 1.84 &  0.17& 1.96 &  0.17& 2.18 &  0.20\\ 
      &      & 0.20 & 0.25& 0.88 &  0.09& 1.14 &  0.09& 1.39 &  0.10& 1.44 &  0.11\\ 
      &      & 0.25 & 0.30& 0.53 &  0.05& 0.74 &  0.06& 0.89 &  0.07& 0.98 &  0.07\\ 
      &      & 0.30 & 0.35& 0.44 &  0.04& 0.51 &  0.04& 0.62 &  0.05& 0.67 &  0.05\\ 
      &      & 0.35 & 0.40& 0.39 &  0.04& 0.38 &  0.03& 0.43 &  0.03& 0.48 &  0.03\\ 
      &      & 0.40 & 0.45& 0.28 &  0.03& 0.30 &  0.02& 0.32 &  0.02& 0.37 &  0.02\\ 
      &      & 0.45 & 0.50& 0.20 &  0.02& 0.23 &  0.02& 0.23 &  0.02& 0.27 &  0.02\\ 
\hline  
 1.75 & 1.95 & 0.10 & 0.15& 1.65 &  0.26& 1.91 &  0.28& 2.02 &  0.29& 2.21 &  0.33\\ 
      &      & 0.15 & 0.20& 1.36 &  0.10& 1.54 &  0.10& 1.56 &  0.10& 1.81 &  0.10\\ 
      &      & 0.20 & 0.25& 0.84 &  0.07& 0.93 &  0.06& 1.01 &  0.06& 1.10 &  0.07\\ 
      &      & 0.25 & 0.30& 0.47 &  0.05& 0.61 &  0.05& 0.57 &  0.05& 0.67 &  0.05\\ 
      &      & 0.30 & 0.35& 0.36 &  0.04& 0.40 &  0.03& 0.42 &  0.03& 0.43 &  0.03\\ 
      &      & 0.35 & 0.40& 0.24 &  0.03& 0.28 &  0.02& 0.32 &  0.02& 0.31 &  0.02\\ 
      &      & 0.40 & 0.45& 0.15 &  0.02& 0.22 &  0.02& 0.23 &  0.02& 0.24 &  0.02\\ 
      &      & 0.45 & 0.50& 0.11 &  0.01& 0.17 &  0.02& 0.17 &  0.01& 0.19 &  0.02\\ 
\hline  
 1.95 & 2.15 & 0.10 & 0.15& 1.64 &  0.23& 1.85 &  0.24& 1.70 &  0.24& 2.03 &  0.27\\ 
      &      & 0.15 & 0.20& 1.06 &  0.08& 1.20 &  0.08& 1.26 &  0.07& 1.44 &  0.08\\ 
      &      & 0.20 & 0.25& 0.50 &  0.06& 0.74 &  0.05& 0.65 &  0.05& 0.76 &  0.04\\ 
      &      & 0.25 & 0.30& 0.33 &  0.03& 0.36 &  0.04& 0.38 &  0.03& 0.51 &  0.04\\ 
      &      & 0.30 & 0.35& 0.33 &  0.04& 0.26 &  0.02& 0.28 &  0.02& 0.30 &  0.03\\ 
      &      & 0.35 & 0.40& 0.21 &  0.03& 0.22 &  0.02& 0.23 &  0.02& 0.21 &  0.02\\ 
      &      & 0.40 & 0.45& 0.14 &  0.02& 0.15 &  0.02& 0.16 &  0.01& 0.15 &  0.01\\ 
      &      & 0.45 & 0.50& 0.10 &  0.02& 0.11 &  0.01& 0.12 &  0.01& 0.12 &  0.01\\ 

\end{tabular}
\end{center}
\end{table}
\clearpage
\begin{table}[hp!]
\begin{center}
  \caption{\label{tab:xsec-pipp-pb}
    HARP results for the double-differential $\pi^+$ production
    cross-section in the laboratory system,
    $d^2\sigma^{\pi^+}/(dpd\theta)$ for $\pi^+$--Pb interactions. Each row refers to a
    different $(p_{\hbox{\small min}} \le p<p_{\hbox{\small max}},
    \theta_{\hbox{\small min}} \le \theta<\theta_{\hbox{\small max}})$ bin,
    where $p$ and $\theta$ are the pion momentum and polar angle, respectively.
    The central value as well as the square-root of the diagonal elements
    of the covariance matrix are given.}
\vspace{2mm}
\begin{tabular}{rrrr|r@{$\pm$}lr@{$\pm$}lr@{$\pm$}lr@{$\pm$}l}
\hline
$\theta_{\hbox{\small min}}$ &
$\theta_{\hbox{\small max}}$ &
$p_{\hbox{\small min}}$ &
$p_{\hbox{\small max}}$ &
\multicolumn{8}{c}{$d^2\sigma^{\pi^+}/(dpd\theta)$}
\\
(rad) & (rad) & (\GeVc) & (\GeVc) &
\multicolumn{8}{c}{($\barn/(\GeVc \cdot \rad)$)}
\\
  &  &  &
&\multicolumn{2}{c}{$ \bf{3 \ \GeVc}$}
&\multicolumn{2}{c}{$ \bf{5 \ \GeVc}$}
&\multicolumn{2}{c}{$ \bf{8 \ \GeVc}$}
&\multicolumn{2}{c}{$ \bf{12 \ \GeVc}$}
\\
\hline  
 0.35 & 0.55 & 0.15 & 0.20& 0.52 &  0.21& 1.38 &  0.35& 1.60 &  0.38& 1.76 &  0.58\\ 
      &      & 0.20 & 0.25& 0.88 &  0.15& 1.59 &  0.21& 2.25 &  0.25& 3.00 &  0.57\\ 
      &      & 0.25 & 0.30& 0.85 &  0.08& 2.04 &  0.19& 2.80 &  0.19& 3.51 &  0.56\\ 
      &      & 0.30 & 0.35& 1.01 &  0.10& 2.01 &  0.12& 2.63 &  0.17& 2.68 &  0.40\\ 
      &      & 0.35 & 0.40& 1.01 &  0.07& 1.82 &  0.10& 2.62 &  0.12& 3.03 &  0.43\\ 
      &      & 0.40 & 0.45& 1.03 &  0.08& 1.79 &  0.12& 2.71 &  0.16& 3.24 &  0.51\\ 
      &      & 0.45 & 0.50& 1.08 &  0.07& 1.94 &  0.11& 2.62 &  0.11& 3.28 &  0.40\\ 
      &      & 0.50 & 0.60& 1.00 &  0.07& 1.87 &  0.10& 2.65 &  0.13& 3.14 &  0.37\\ 
      &      & 0.60 & 0.70& 0.77 &  0.08& 1.65 &  0.14& 2.44 &  0.22& 2.58 &  0.40\\ 
      &      & 0.70 & 0.80& 0.55 &  0.08& 1.31 &  0.19& 1.69 &  0.26& 1.91 &  0.34\\ 
\hline  
 0.55 & 0.75 & 0.10 & 0.15& 0.65 &  0.29& 0.92 &  0.37& 1.04 &  0.42& 1.33 &  0.69\\ 
      &      & 0.15 & 0.20& 1.05 &  0.16& 1.55 &  0.22& 2.19 &  0.25& 1.90 &  0.42\\ 
      &      & 0.20 & 0.25& 1.24 &  0.14& 2.15 &  0.17& 2.74 &  0.20& 2.82 &  0.57\\ 
      &      & 0.25 & 0.30& 1.28 &  0.08& 2.06 &  0.13& 2.86 &  0.18& 3.19 &  0.44\\ 
      &      & 0.30 & 0.35& 1.22 &  0.09& 2.01 &  0.10& 2.80 &  0.16& 2.93 &  0.44\\ 
      &      & 0.35 & 0.40& 1.23 &  0.07& 2.00 &  0.14& 2.50 &  0.12& 2.66 &  0.37\\ 
      &      & 0.40 & 0.45& 1.10 &  0.06& 1.91 &  0.10& 2.48 &  0.15& 2.88 &  0.43\\ 
      &      & 0.45 & 0.50& 1.06 &  0.07& 1.85 &  0.09& 2.32 &  0.11& 2.87 &  0.38\\ 
      &      & 0.50 & 0.60& 0.85 &  0.07& 1.60 &  0.11& 1.84 &  0.13& 1.75 &  0.30\\ 
      &      & 0.60 & 0.70& 0.55 &  0.07& 0.99 &  0.14& 1.30 &  0.16& 1.47 &  0.23\\ 
      &      & 0.70 & 0.80& 0.34 &  0.07& 0.62 &  0.12& 0.83 &  0.17& 1.23 &  0.25\\ 
\hline  
 0.75 & 0.95 & 0.10 & 0.15& 1.01 &  0.23& 1.17 &  0.31& 1.55 &  0.37& 2.10 &  0.65\\ 
      &      & 0.15 & 0.20& 1.39 &  0.11& 1.96 &  0.15& 2.34 &  0.15& 2.86 &  0.46\\ 
      &      & 0.20 & 0.25& 1.46 &  0.11& 2.08 &  0.17& 2.57 &  0.17& 3.05 &  0.43\\ 
      &      & 0.25 & 0.30& 1.20 &  0.08& 1.92 &  0.15& 2.23 &  0.13& 2.89 &  0.41\\ 
      &      & 0.30 & 0.35& 1.35 &  0.08& 1.78 &  0.10& 2.21 &  0.12& 2.25 &  0.33\\ 
      &      & 0.35 & 0.40& 1.00 &  0.06& 1.67 &  0.09& 2.14 &  0.10& 2.72 &  0.41\\ 
      &      & 0.40 & 0.45& 0.89 &  0.05& 1.48 &  0.07& 1.83 &  0.08& 2.06 &  0.40\\ 
      &      & 0.45 & 0.50& 0.84 &  0.05& 1.27 &  0.07& 1.61 &  0.08& 0.95 &  0.24\\ 
      &      & 0.50 & 0.60& 0.61 &  0.05& 0.91 &  0.08& 1.14 &  0.11& 0.72 &  0.12\\ 
      &      & 0.60 & 0.70& 0.39 &  0.05& 0.51 &  0.08& 0.64 &  0.10& 0.75 &  0.15\\ 
\hline  
 0.95 & 1.15 & 0.10 & 0.15& 1.08 &  0.21& 1.45 &  0.29& 1.65 &  0.29& 1.52 &  0.46\\ 
      &      & 0.15 & 0.20& 1.54 &  0.10& 2.13 &  0.13& 2.43 &  0.19& 2.44 &  0.38\\ 
      &      & 0.20 & 0.25& 1.37 &  0.09& 1.99 &  0.12& 2.35 &  0.14& 2.75 &  0.42\\ 
      &      & 0.25 & 0.30& 1.09 &  0.07& 1.61 &  0.12& 1.86 &  0.12& 2.81 &  0.38\\ 
      &      & 0.30 & 0.35& 0.87 &  0.06& 1.33 &  0.09& 1.59 &  0.11& 1.92 &  0.31\\ 
      &      & 0.35 & 0.40& 0.68 &  0.05& 1.14 &  0.07& 1.40 &  0.07& 1.23 &  0.21\\ 
      &      & 0.40 & 0.45& 0.65 &  0.04& 0.93 &  0.05& 1.16 &  0.06& 0.86 &  0.15\\ 
      &      & 0.45 & 0.50& 0.50 &  0.04& 0.75 &  0.06& 0.90 &  0.05& 0.88 &  0.15\\ 
      &      & 0.50 & 0.60& 0.31 &  0.04& 0.45 &  0.05& 0.62 &  0.06& 0.87 &  0.17\\ 
\hline
\end{tabular}
\end{center}
\end{table}

\begin{table}[hp!]
\begin{center}
\begin{tabular}{rrrr|r@{$\pm$}lr@{$\pm$}lr@{$\pm$}lr@{$\pm$}l}
\hline
$\theta_{\hbox{\small min}}$ &
$\theta_{\hbox{\small max}}$ &
$p_{\hbox{\small min}}$ &
$p_{\hbox{\small max}}$ &
\multicolumn{8}{c}{$d^2\sigma^{\pi^+}/(dpd\theta)$}
\\
(rad) & (rad) & (\GeVc) & (\GeVc) &
\multicolumn{8}{c}{(\barn/($\GeVc \cdot \rad$))}
\\
  &  &  &
&\multicolumn{2}{c}{$ \bf{3 \ \GeVc}$}
&\multicolumn{2}{c}{$ \bf{5 \ \GeVc}$}
&\multicolumn{2}{c}{$ \bf{8 \ \GeVc}$}
&\multicolumn{2}{c}{$ \bf{12 \ \GeVc}$}
\\
\hline
 1.15 & 1.35 & 0.10 & 0.15& 1.10 &  0.20& 1.64 &  0.30& 1.84 &  0.34& 1.56 &  0.45\\ 
      &      & 0.15 & 0.20& 1.59 &  0.13& 2.19 &  0.14& 2.65 &  0.17& 2.31 &  0.36\\ 
      &      & 0.20 & 0.25& 1.30 &  0.09& 1.96 &  0.12& 2.27 &  0.12& 3.22 &  0.47\\ 
      &      & 0.25 & 0.30& 0.86 &  0.07& 1.33 &  0.09& 1.75 &  0.10& 1.80 &  0.28\\ 
      &      & 0.30 & 0.35& 0.72 &  0.05& 1.02 &  0.07& 1.27 &  0.09& 1.70 &  0.26\\ 
      &      & 0.35 & 0.40& 0.58 &  0.04& 0.84 &  0.05& 1.03 &  0.06& 1.25 &  0.22\\ 
      &      & 0.40 & 0.45& 0.47 &  0.03& 0.67 &  0.05& 0.76 &  0.05& 0.77 &  0.16\\ 
      &      & 0.45 & 0.50& 0.36 &  0.03& 0.48 &  0.05& 0.56 &  0.04& 0.66 &  0.14\\ 
\hline  
 1.35 & 1.55 & 0.10 & 0.15& 1.16 &  0.22& 1.57 &  0.32& 1.77 &  0.40& 1.37 &  0.47\\ 
      &      & 0.15 & 0.20& 1.56 &  0.13& 2.10 &  0.18& 2.39 &  0.20& 2.42 &  0.39\\ 
      &      & 0.20 & 0.25& 1.23 &  0.08& 1.76 &  0.10& 1.92 &  0.12& 1.56 &  0.27\\ 
      &      & 0.25 & 0.30& 0.81 &  0.06& 1.13 &  0.08& 1.22 &  0.08& 1.50 &  0.28\\ 
      &      & 0.30 & 0.35& 0.59 &  0.04& 0.76 &  0.07& 0.84 &  0.06& 1.20 &  0.24\\ 
      &      & 0.35 & 0.40& 0.46 &  0.03& 0.57 &  0.04& 0.62 &  0.04& 0.59 &  0.15\\ 
      &      & 0.40 & 0.45& 0.30 &  0.02& 0.41 &  0.03& 0.46 &  0.03& 0.47 &  0.12\\ 
      &      & 0.45 & 0.50& 0.23 &  0.02& 0.29 &  0.03& 0.31 &  0.03& 0.39 &  0.12\\ 
\hline  
 1.55 & 1.75 & 0.10 & 0.15& 1.24 &  0.26& 1.47 &  0.30& 1.70 &  0.33& 1.10 &  0.39\\ 
      &      & 0.15 & 0.20& 1.49 &  0.10& 1.87 &  0.15& 2.13 &  0.16& 2.03 &  0.37\\ 
      &      & 0.20 & 0.25& 1.06 &  0.08& 1.43 &  0.09& 1.68 &  0.11& 1.71 &  0.32\\ 
      &      & 0.25 & 0.30& 0.71 &  0.05& 0.84 &  0.06& 0.92 &  0.07& 0.76 &  0.18\\ 
      &      & 0.30 & 0.35& 0.51 &  0.04& 0.58 &  0.04& 0.62 &  0.05& 0.48 &  0.12\\ 
      &      & 0.35 & 0.40& 0.31 &  0.03& 0.41 &  0.03& 0.46 &  0.03& 0.45 &  0.13\\ 
      &      & 0.40 & 0.45& 0.19 &  0.02& 0.28 &  0.03& 0.30 &  0.03& 0.46 &  0.15\\ 
      &      & 0.45 & 0.50& 0.14 &  0.02& 0.18 &  0.02& 0.20 &  0.02& 0.37 &  0.15\\ 
\hline  
 1.75 & 1.95 & 0.10 & 0.15& 1.15 &  0.20& 1.55 &  0.24& 1.69 &  0.24& 1.03 &  0.32\\ 
      &      & 0.15 & 0.20& 1.32 &  0.07& 1.59 &  0.09& 1.80 &  0.10& 1.61 &  0.28\\ 
      &      & 0.20 & 0.25& 0.87 &  0.06& 1.03 &  0.08& 1.18 &  0.07& 1.22 &  0.25\\ 
      &      & 0.25 & 0.30& 0.52 &  0.04& 0.59 &  0.05& 0.67 &  0.05& 0.61 &  0.16\\ 
      &      & 0.30 & 0.35& 0.34 &  0.03& 0.41 &  0.03& 0.42 &  0.04& 0.57 &  0.17\\ 
      &      & 0.35 & 0.40& 0.21 &  0.03& 0.32 &  0.03& 0.25 &  0.03& 0.35 &  0.13\\ 
      &      & 0.40 & 0.45& 0.13 &  0.02& 0.19 &  0.03& 0.15 &  0.03& 0.20 &  0.10\\ 
      &      & 0.45 & 0.50& 0.08 &  0.01& 0.11 &  0.02& 0.08 &  0.01& 0.11 &  0.08\\ 
\hline  
 1.95 & 2.15 & 0.10 & 0.15& 0.77 &  0.13& 1.31 &  0.20& 1.29 &  0.19& 0.91 &  0.32\\ 
      &      & 0.15 & 0.20& 1.01 &  0.07& 1.39 &  0.07& 1.39 &  0.07& 1.21 &  0.25\\ 
      &      & 0.20 & 0.25& 0.68 &  0.06& 0.77 &  0.07& 0.79 &  0.07& 1.09 &  0.25\\ 
      &      & 0.25 & 0.30& 0.32 &  0.03& 0.39 &  0.04& 0.45 &  0.04& 0.74 &  0.22\\ 
      &      & 0.30 & 0.35& 0.24 &  0.02& 0.28 &  0.02& 0.30 &  0.03& 0.32 &  0.14\\ 
      &      & 0.35 & 0.40& 0.16 &  0.02& 0.20 &  0.02& 0.17 &  0.02& 0.15 &  0.09\\ 
      &      & 0.40 & 0.45& 0.12 &  0.02& 0.15 &  0.02& 0.10 &  0.01& 0.15 &  0.10\\ 
      &      & 0.45 & 0.50& 0.07 &  0.01& 0.09 &  0.02& 0.07 &  0.01& 0.14 &  0.12\\ 

\end{tabular}
\end{center}
\end{table}

\begin{table}[hp!]
\begin{center}
  \caption{\label{tab:xsec-pipm-pb}
    HARP results for the double-differential $\pi^-$ production
    cross-section in the laboratory system,
    $d^2\sigma^{\pi^-}/(dpd\theta)$ for $\pi^+$--Pb interactions. Each row refers to a
    different $(p_{\hbox{\small min}} \le p<p_{\hbox{\small max}},
    \theta_{\hbox{\small min}} \le \theta<\theta_{\hbox{\small max}})$ bin,
    where $p$ and $\theta$ are the pion momentum and polar angle, respectively.
    The central value as well as the square-root of the diagonal elements
    of the covariance matrix are given.}
\vspace{2mm}
\begin{tabular}{rrrr|r@{$\pm$}lr@{$\pm$}lr@{$\pm$}lr@{$\pm$}l}
\hline
$\theta_{\hbox{\small min}}$ &
$\theta_{\hbox{\small max}}$ &
$p_{\hbox{\small min}}$ &
$p_{\hbox{\small max}}$ &
\multicolumn{8}{c}{$d^2\sigma^{\pi^-}/(dpd\theta)$}
\\
(rad) & (rad) & (\GeVc) & (\GeVc) &
\multicolumn{8}{c}{($\barn/(\GeVc \cdot \rad)$)}
\\
  &  &  &
&\multicolumn{2}{c}{$ \bf{3 \ \GeVc}$}
&\multicolumn{2}{c}{$ \bf{5 \ \GeVc}$}
&\multicolumn{2}{c}{$ \bf{8 \ \GeVc}$}
&\multicolumn{2}{c}{$ \bf{12 \ \GeVc}$}
\\
\hline  
 0.35 & 0.55 & 0.15 & 0.20& 0.75 &  0.22& 1.44 &  0.35& 2.03 &  0.42& 2.93 &  0.73\\ 
      &      & 0.20 & 0.25& 0.77 &  0.13& 1.44 &  0.20& 2.26 &  0.23& 2.63 &  0.53\\ 
      &      & 0.25 & 0.30& 0.91 &  0.10& 1.59 &  0.15& 2.09 &  0.16& 2.19 &  0.43\\ 
      &      & 0.30 & 0.35& 0.86 &  0.07& 1.54 &  0.10& 2.14 &  0.12& 2.14 &  0.39\\ 
      &      & 0.35 & 0.40& 0.72 &  0.05& 1.38 &  0.08& 1.90 &  0.12& 2.22 &  0.35\\ 
      &      & 0.40 & 0.45& 0.69 &  0.05& 1.26 &  0.07& 2.02 &  0.11& 2.44 &  0.44\\ 
      &      & 0.45 & 0.50& 0.66 &  0.05& 1.12 &  0.06& 1.68 &  0.10& 2.18 &  0.35\\ 
      &      & 0.50 & 0.60& 0.50 &  0.04& 1.13 &  0.06& 1.49 &  0.08& 1.70 &  0.26\\ 
      &      & 0.60 & 0.70& 0.45 &  0.04& 0.93 &  0.08& 1.41 &  0.10& 1.51 &  0.30\\ 
      &      & 0.70 & 0.80& 0.36 &  0.05& 0.84 &  0.08& 1.24 &  0.14& 0.91 &  0.22\\ 
\hline  
 0.55 & 0.75 & 0.10 & 0.15& 1.03 &  0.29& 1.37 &  0.43& 1.49 &  0.49& 1.63 &  0.66\\ 
      &      & 0.15 & 0.20& 1.14 &  0.14& 1.74 &  0.21& 2.28 &  0.24& 1.74 &  0.39\\ 
      &      & 0.20 & 0.25& 0.95 &  0.09& 1.69 &  0.12& 2.68 &  0.15& 2.61 &  0.46\\ 
      &      & 0.25 & 0.30& 0.89 &  0.08& 1.62 &  0.10& 1.93 &  0.11& 2.12 &  0.37\\ 
      &      & 0.30 & 0.35& 0.78 &  0.06& 1.39 &  0.08& 1.81 &  0.12& 1.87 &  0.31\\ 
      &      & 0.35 & 0.40& 0.77 &  0.06& 1.11 &  0.06& 1.67 &  0.08& 1.83 &  0.30\\ 
      &      & 0.40 & 0.45& 0.70 &  0.05& 1.13 &  0.07& 1.74 &  0.10& 1.60 &  0.26\\ 
      &      & 0.45 & 0.50& 0.61 &  0.04& 1.10 &  0.06& 1.60 &  0.08& 1.58 &  0.27\\ 
      &      & 0.50 & 0.60& 0.53 &  0.04& 0.98 &  0.06& 1.23 &  0.07& 1.37 &  0.22\\ 
      &      & 0.60 & 0.70& 0.41 &  0.04& 0.69 &  0.07& 1.03 &  0.08& 1.14 &  0.22\\ 
      &      & 0.70 & 0.80& 0.28 &  0.04& 0.52 &  0.07& 0.84 &  0.10& 0.74 &  0.20\\ 
\hline  
 0.75 & 0.95 & 0.10 & 0.15& 1.17 &  0.23& 1.55 &  0.30& 2.02 &  0.40& 1.22 &  0.49\\ 
      &      & 0.15 & 0.20& 1.15 &  0.11& 1.96 &  0.14& 2.44 &  0.16& 3.67 &  0.58\\ 
      &      & 0.20 & 0.25& 1.08 &  0.09& 1.75 &  0.11& 2.21 &  0.13& 2.00 &  0.37\\ 
      &      & 0.25 & 0.30& 0.95 &  0.06& 1.37 &  0.08& 1.98 &  0.10& 1.67 &  0.31\\ 
      &      & 0.30 & 0.35& 0.76 &  0.05& 1.21 &  0.08& 1.66 &  0.08& 1.90 &  0.31\\ 
      &      & 0.35 & 0.40& 0.60 &  0.04& 1.13 &  0.06& 1.39 &  0.07& 1.66 &  0.27\\ 
      &      & 0.40 & 0.45& 0.55 &  0.04& 0.97 &  0.05& 1.27 &  0.07& 1.32 &  0.21\\ 
      &      & 0.45 & 0.50& 0.49 &  0.03& 0.83 &  0.04& 1.07 &  0.06& 1.34 &  0.23\\ 
      &      & 0.50 & 0.60& 0.39 &  0.03& 0.65 &  0.04& 0.80 &  0.05& 1.35 &  0.21\\ 
      &      & 0.60 & 0.70& 0.27 &  0.03& 0.48 &  0.04& 0.62 &  0.06& 0.98 &  0.19\\ 
\hline  
 0.95 & 1.15 & 0.10 & 0.15& 1.46 &  0.21& 1.61 &  0.26& 2.49 &  0.36& 1.76 &  0.46\\ 
      &      & 0.15 & 0.20& 1.22 &  0.09& 1.82 &  0.13& 2.48 &  0.14& 3.03 &  0.50\\ 
      &      & 0.20 & 0.25& 0.92 &  0.07& 1.53 &  0.10& 1.95 &  0.11& 2.55 &  0.38\\ 
      &      & 0.25 & 0.30& 0.76 &  0.05& 1.19 &  0.08& 1.63 &  0.09& 2.41 &  0.37\\ 
      &      & 0.30 & 0.35& 0.59 &  0.04& 1.03 &  0.07& 1.32 &  0.08& 2.08 &  0.33\\ 
      &      & 0.35 & 0.40& 0.51 &  0.03& 0.88 &  0.05& 1.11 &  0.06& 1.53 &  0.25\\ 
      &      & 0.40 & 0.45& 0.44 &  0.03& 0.74 &  0.04& 0.93 &  0.05& 1.33 &  0.23\\ 
      &      & 0.45 & 0.50& 0.36 &  0.02& 0.62 &  0.04& 0.80 &  0.04& 1.06 &  0.20\\ 
      &      & 0.50 & 0.60& 0.28 &  0.02& 0.44 &  0.04& 0.59 &  0.04& 0.68 &  0.14\\ 
\hline
\end{tabular}
\end{center}
\end{table}

\begin{table}[hp!]
\begin{center}
\begin{tabular}{rrrr|r@{$\pm$}lr@{$\pm$}lr@{$\pm$}lr@{$\pm$}l}
\hline
$\theta_{\hbox{\small min}}$ &
$\theta_{\hbox{\small max}}$ &
$p_{\hbox{\small min}}$ &
$p_{\hbox{\small max}}$ &
\multicolumn{8}{c}{$d^2\sigma^{\pi^-}/(dpd\theta)$}
\\
(rad) & (rad) & (\GeVc) & (\GeVc) &
\multicolumn{8}{c}{(\barn/($\GeVc \cdot \rad$))}
\\
  &  &  &
&\multicolumn{2}{c}{$ \bf{3 \ \GeVc}$}
&\multicolumn{2}{c}{$ \bf{5 \ \GeVc}$}
&\multicolumn{2}{c}{$ \bf{8 \ \GeVc}$}
&\multicolumn{2}{c}{$ \bf{12 \ \GeVc}$}
\\
\hline
 1.15 & 1.35 & 0.10 & 0.15& 1.42 &  0.23& 1.71 &  0.29& 2.47 &  0.39& 2.84 &  0.60\\ 
      &      & 0.15 & 0.20& 1.26 &  0.09& 1.91 &  0.12& 2.37 &  0.16& 2.57 &  0.42\\ 
      &      & 0.20 & 0.25& 0.84 &  0.06& 1.32 &  0.09& 1.83 &  0.11& 1.87 &  0.30\\ 
      &      & 0.25 & 0.30& 0.62 &  0.05& 0.99 &  0.07& 1.42 &  0.09& 1.56 &  0.27\\ 
      &      & 0.30 & 0.35& 0.50 &  0.04& 0.88 &  0.06& 1.05 &  0.07& 1.06 &  0.23\\ 
      &      & 0.35 & 0.40& 0.38 &  0.03& 0.72 &  0.05& 0.80 &  0.05& 0.64 &  0.13\\ 
      &      & 0.40 & 0.45& 0.30 &  0.02& 0.52 &  0.04& 0.61 &  0.04& 0.71 &  0.16\\ 
      &      & 0.45 & 0.50& 0.25 &  0.02& 0.40 &  0.03& 0.48 &  0.03& 0.74 &  0.17\\ 
\hline  
 1.35 & 1.55 & 0.10 & 0.15& 1.50 &  0.30& 1.97 &  0.34& 2.43 &  0.46& 4.26 &  0.94\\ 
      &      & 0.15 & 0.20& 1.25 &  0.08& 1.81 &  0.11& 2.16 &  0.15& 2.73 &  0.37\\ 
      &      & 0.20 & 0.25& 0.71 &  0.06& 1.18 &  0.08& 1.55 &  0.09& 2.08 &  0.34\\ 
      &      & 0.25 & 0.30& 0.55 &  0.05& 0.81 &  0.06& 1.15 &  0.08& 1.29 &  0.23\\ 
      &      & 0.30 & 0.35& 0.46 &  0.04& 0.67 &  0.05& 0.76 &  0.05& 0.86 &  0.19\\ 
      &      & 0.35 & 0.40& 0.31 &  0.03& 0.54 &  0.04& 0.60 &  0.04& 0.53 &  0.12\\ 
      &      & 0.40 & 0.45& 0.21 &  0.02& 0.37 &  0.03& 0.48 &  0.03& 0.51 &  0.13\\ 
      &      & 0.45 & 0.50& 0.18 &  0.02& 0.25 &  0.03& 0.35 &  0.03& 0.38 &  0.10\\ 
\hline  
 1.55 & 1.75 & 0.10 & 0.15& 1.38 &  0.24& 1.86 &  0.29& 2.15 &  0.40& 3.10 &  0.73\\ 
      &      & 0.15 & 0.20& 1.07 &  0.07& 1.42 &  0.10& 1.86 &  0.11& 2.19 &  0.33\\ 
      &      & 0.20 & 0.25& 0.59 &  0.05& 0.96 &  0.07& 1.17 &  0.08& 1.73 &  0.30\\ 
      &      & 0.25 & 0.30& 0.46 &  0.04& 0.62 &  0.05& 0.77 &  0.07& 1.20 &  0.26\\ 
      &      & 0.30 & 0.35& 0.33 &  0.03& 0.52 &  0.04& 0.52 &  0.04& 0.61 &  0.16\\ 
      &      & 0.35 & 0.40& 0.22 &  0.02& 0.41 &  0.03& 0.41 &  0.03& 0.54 &  0.15\\ 
      &      & 0.40 & 0.45& 0.16 &  0.02& 0.32 &  0.03& 0.29 &  0.02& 0.45 &  0.14\\ 
      &      & 0.45 & 0.50& 0.12 &  0.01& 0.24 &  0.02& 0.21 &  0.02& 0.29 &  0.11\\ 
\hline  
 1.75 & 1.95 & 0.10 & 0.15& 1.17 &  0.15& 1.55 &  0.21& 1.82 &  0.26& 2.05 &  0.51\\ 
      &      & 0.15 & 0.20& 0.81 &  0.06& 1.08 &  0.07& 1.44 &  0.08& 1.34 &  0.25\\ 
      &      & 0.20 & 0.25& 0.48 &  0.04& 0.76 &  0.05& 0.91 &  0.06& 0.72 &  0.17\\ 
      &      & 0.25 & 0.30& 0.30 &  0.03& 0.45 &  0.04& 0.60 &  0.05& 0.37 &  0.12\\ 
      &      & 0.30 & 0.35& 0.18 &  0.02& 0.34 &  0.03& 0.41 &  0.03& 0.25 &  0.10\\ 
      &      & 0.35 & 0.40& 0.16 &  0.02& 0.25 &  0.02& 0.30 &  0.02& 0.22 &  0.10\\ 
      &      & 0.40 & 0.45& 0.12 &  0.01& 0.18 &  0.02& 0.22 &  0.02& 0.15 &  0.08\\ 
      &      & 0.45 & 0.50& 0.09 &  0.01& 0.13 &  0.01& 0.16 &  0.02& 0.12 &  0.07\\ 
\hline  
 1.95 & 2.15 & 0.10 & 0.15& 1.03 &  0.14& 1.36 &  0.20& 1.63 &  0.19& 2.02 &  0.55\\ 
      &      & 0.15 & 0.20& 0.63 &  0.05& 0.91 &  0.06& 1.16 &  0.06& 1.41 &  0.31\\ 
      &      & 0.20 & 0.25& 0.42 &  0.04& 0.56 &  0.04& 0.70 &  0.05& 0.83 &  0.24\\ 
      &      & 0.25 & 0.30& 0.27 &  0.03& 0.38 &  0.03& 0.42 &  0.04& 0.26 &  0.15\\ 
      &      & 0.30 & 0.35& 0.21 &  0.03& 0.26 &  0.03& 0.25 &  0.03& 0.15 &  0.10\\ 
      &      & 0.35 & 0.40& 0.16 &  0.03& 0.18 &  0.02& 0.15 &  0.02& 0.18 &  0.13\\ 
      &      & 0.40 & 0.45& 0.10 &  0.02& 0.12 &  0.01& 0.11 &  0.01& 0.09 &  0.08\\ 
      &      & 0.45 & 0.50& 0.07 &  0.01& 0.08 &  0.01& 0.09 &  0.01& 0.07 &  0.07\\ 

\end{tabular}
\end{center}
\end{table}
\clearpage
\begin{table}[hp!]
\begin{center}
  \caption{\label{tab:xsec-pimp-pb}
    HARP results for the double-differential $\pi^+$ production
    cross-section in the laboratory system,
    $d^2\sigma^{\pi^+}/(dpd\theta)$ for $\pi^-$--Pb interactions. Each row refers to a
    different $(p_{\hbox{\small min}} \le p<p_{\hbox{\small max}},
    \theta_{\hbox{\small min}} \le \theta<\theta_{\hbox{\small max}})$ bin,
    where $p$ and $\theta$ are the pion momentum and polar angle, respectively.
    The central value as well as the square-root of the diagonal elements
    of the covariance matrix are given.}
\vspace{2mm}
\begin{tabular}{rrrr|r@{$\pm$}lr@{$\pm$}lr@{$\pm$}lr@{$\pm$}l}
\hline
$\theta_{\hbox{\small min}}$ &
$\theta_{\hbox{\small max}}$ &
$p_{\hbox{\small min}}$ &
$p_{\hbox{\small max}}$ &
\multicolumn{8}{c}{$d^2\sigma^{\pi^+}/(dpd\theta)$}
\\
(rad) & (rad) & (\GeVc) & (\GeVc) &
\multicolumn{8}{c}{($\barn/(\GeVc \cdot \rad)$)}
\\
  &  &  &
&\multicolumn{2}{c}{$ \bf{3 \ \GeVc}$}
&\multicolumn{2}{c}{$ \bf{5 \ \GeVc}$}
&\multicolumn{2}{c}{$ \bf{8 \ \GeVc}$}
&\multicolumn{2}{c}{$ \bf{12 \ \GeVc}$}
\\
\hline  
 0.35 & 0.55 & 0.15 & 0.20& 0.46 &  0.28& 1.35 &  0.47& 0.91 &  0.56& 1.36 &  0.79\\ 
      &      & 0.20 & 0.25& 0.67 &  0.22& 1.63 &  0.27& 1.65 &  0.44& 2.16 &  0.63\\ 
      &      & 0.25 & 0.30& 0.64 &  0.12& 1.68 &  0.16& 2.04 &  0.20& 2.84 &  0.28\\ 
      &      & 0.30 & 0.35& 0.71 &  0.07& 1.55 &  0.09& 2.37 &  0.18& 2.95 &  0.17\\ 
      &      & 0.35 & 0.40& 0.71 &  0.08& 1.52 &  0.08& 2.19 &  0.09& 2.88 &  0.16\\ 
      &      & 0.40 & 0.45& 0.65 &  0.06& 1.35 &  0.07& 2.10 &  0.11& 2.92 &  0.13\\ 
      &      & 0.45 & 0.50& 0.61 &  0.06& 1.45 &  0.09& 2.25 &  0.10& 2.81 &  0.12\\ 
      &      & 0.50 & 0.60& 0.68 &  0.06& 1.48 &  0.09& 2.30 &  0.13& 2.91 &  0.17\\ 
      &      & 0.60 & 0.70& 0.59 &  0.08& 1.41 &  0.13& 2.01 &  0.21& 2.83 &  0.24\\ 
      &      & 0.70 & 0.80& 0.42 &  0.08& 1.08 &  0.18& 1.53 &  0.22& 2.29 &  0.33\\ 
\hline  
 0.55 & 0.75 & 0.10 & 0.15& 0.30 &  0.24& 1.11 &  0.51& 0.59 &  0.47& 0.77 &  0.60\\ 
      &      & 0.15 & 0.20& 0.62 &  0.29& 1.51 &  0.30& 1.39 &  0.48& 2.12 &  0.81\\ 
      &      & 0.20 & 0.25& 1.00 &  0.12& 1.60 &  0.14& 2.26 &  0.22& 2.78 &  0.28\\ 
      &      & 0.25 & 0.30& 0.96 &  0.09& 1.66 &  0.12& 2.24 &  0.13& 2.66 &  0.19\\ 
      &      & 0.30 & 0.35& 0.81 &  0.07& 1.59 &  0.08& 2.20 &  0.14& 2.88 &  0.16\\ 
      &      & 0.35 & 0.40& 0.68 &  0.06& 1.48 &  0.08& 2.02 &  0.09& 2.67 &  0.11\\ 
      &      & 0.40 & 0.45& 0.62 &  0.06& 1.37 &  0.07& 1.89 &  0.09& 2.47 &  0.10\\ 
      &      & 0.45 & 0.50& 0.66 &  0.06& 1.34 &  0.07& 1.75 &  0.08& 2.30 &  0.10\\ 
      &      & 0.50 & 0.60& 0.67 &  0.06& 1.14 &  0.09& 1.62 &  0.10& 2.09 &  0.12\\ 
      &      & 0.60 & 0.70& 0.42 &  0.08& 0.83 &  0.10& 1.22 &  0.15& 1.61 &  0.19\\ 
      &      & 0.70 & 0.80& 0.23 &  0.05& 0.58 &  0.10& 0.78 &  0.16& 1.12 &  0.20\\ 
\hline  
 0.75 & 0.95 & 0.10 & 0.15& 0.49 &  0.30& 1.07 &  0.34& 0.88 &  0.43& 0.97 &  0.58\\ 
      &      & 0.15 & 0.20& 0.82 &  0.22& 1.85 &  0.16& 1.92 &  0.24& 2.35 &  0.35\\ 
      &      & 0.20 & 0.25& 0.94 &  0.10& 1.61 &  0.10& 2.15 &  0.15& 2.67 &  0.16\\ 
      &      & 0.25 & 0.30& 0.96 &  0.08& 1.39 &  0.07& 1.96 &  0.12& 2.43 &  0.12\\ 
      &      & 0.30 & 0.35& 0.75 &  0.06& 1.31 &  0.08& 1.79 &  0.09& 2.29 &  0.11\\ 
      &      & 0.35 & 0.40& 0.72 &  0.06& 1.23 &  0.06& 1.55 &  0.08& 2.05 &  0.09\\ 
      &      & 0.40 & 0.45& 0.60 &  0.05& 1.06 &  0.05& 1.40 &  0.07& 1.78 &  0.08\\ 
      &      & 0.45 & 0.50& 0.49 &  0.04& 0.96 &  0.05& 1.26 &  0.05& 1.59 &  0.08\\ 
      &      & 0.50 & 0.60& 0.38 &  0.04& 0.74 &  0.06& 1.00 &  0.09& 1.26 &  0.09\\ 
      &      & 0.60 & 0.70& 0.23 &  0.04& 0.48 &  0.07& 0.63 &  0.10& 0.83 &  0.12\\ 
\hline  
 0.95 & 1.15 & 0.10 & 0.15& 0.67 &  0.28& 1.21 &  0.27& 1.11 &  0.29& 1.22 &  0.39\\ 
      &      & 0.15 & 0.20& 0.93 &  0.14& 1.90 &  0.12& 2.01 &  0.18& 2.20 &  0.19\\ 
      &      & 0.20 & 0.25& 0.85 &  0.08& 1.61 &  0.09& 2.04 &  0.14& 2.30 &  0.12\\ 
      &      & 0.25 & 0.30& 0.81 &  0.06& 1.27 &  0.07& 1.56 &  0.11& 1.97 &  0.10\\ 
      &      & 0.30 & 0.35& 0.58 &  0.05& 1.11 &  0.06& 1.28 &  0.06& 1.67 &  0.08\\ 
      &      & 0.35 & 0.40& 0.46 &  0.04& 0.93 &  0.05& 1.16 &  0.06& 1.33 &  0.06\\ 
      &      & 0.40 & 0.45& 0.40 &  0.04& 0.76 &  0.04& 0.99 &  0.04& 1.11 &  0.06\\ 
      &      & 0.45 & 0.50& 0.32 &  0.03& 0.61 &  0.04& 0.82 &  0.05& 0.89 &  0.06\\ 
      &      & 0.50 & 0.60& 0.21 &  0.03& 0.42 &  0.04& 0.53 &  0.06& 0.65 &  0.06\\ 
\hline
\end{tabular}
\end{center}
\end{table}

\begin{table}[hp!]
\begin{center}
\begin{tabular}{rrrr|r@{$\pm$}lr@{$\pm$}lr@{$\pm$}lr@{$\pm$}l}
\hline
$\theta_{\hbox{\small min}}$ &
$\theta_{\hbox{\small max}}$ &
$p_{\hbox{\small min}}$ &
$p_{\hbox{\small max}}$ &
\multicolumn{8}{c}{$d^2\sigma^{\pi^+}/(dpd\theta)$}
\\
(rad) & (rad) & (\GeVc) & (\GeVc) &
\multicolumn{8}{c}{(\barn/($\GeVc \cdot \rad$))}
\\
  &  &  &
&\multicolumn{2}{c}{$ \bf{3 \ \GeVc}$}
&\multicolumn{2}{c}{$ \bf{5 \ \GeVc}$}
&\multicolumn{2}{c}{$ \bf{8 \ \GeVc}$}
&\multicolumn{2}{c}{$ \bf{12 \ \GeVc}$}
\\
\hline
 1.15 & 1.35 & 0.10 & 0.15& 0.76 &  0.23& 1.30 &  0.26& 1.30 &  0.25& 1.42 &  0.26\\ 
      &      & 0.15 & 0.20& 1.08 &  0.13& 1.75 &  0.10& 1.93 &  0.16& 1.98 &  0.17\\ 
      &      & 0.20 & 0.25& 0.92 &  0.07& 1.43 &  0.08& 1.86 &  0.10& 2.00 &  0.11\\ 
      &      & 0.25 & 0.30& 0.63 &  0.05& 1.04 &  0.07& 1.27 &  0.08& 1.56 &  0.09\\ 
      &      & 0.30 & 0.35& 0.48 &  0.04& 0.82 &  0.05& 0.99 &  0.05& 1.21 &  0.07\\ 
      &      & 0.35 & 0.40& 0.34 &  0.03& 0.62 &  0.04& 0.81 &  0.04& 0.95 &  0.05\\ 
      &      & 0.40 & 0.45& 0.27 &  0.02& 0.42 &  0.03& 0.62 &  0.04& 0.73 &  0.04\\ 
      &      & 0.45 & 0.50& 0.22 &  0.02& 0.32 &  0.03& 0.49 &  0.04& 0.54 &  0.04\\ 
\hline  
 1.35 & 1.55 & 0.10 & 0.15& 0.81 &  0.19& 1.28 &  0.25& 1.44 &  0.27& 1.48 &  0.29\\ 
      &      & 0.15 & 0.20& 0.94 &  0.11& 1.60 &  0.13& 1.81 &  0.13& 1.95 &  0.15\\ 
      &      & 0.20 & 0.25& 0.87 &  0.07& 1.33 &  0.08& 1.47 &  0.09& 1.80 &  0.10\\ 
      &      & 0.25 & 0.30& 0.56 &  0.05& 0.90 &  0.06& 0.99 &  0.07& 1.23 &  0.08\\ 
      &      & 0.30 & 0.35& 0.37 &  0.04& 0.65 &  0.04& 0.70 &  0.05& 0.91 &  0.05\\ 
      &      & 0.35 & 0.40& 0.27 &  0.03& 0.47 &  0.04& 0.54 &  0.04& 0.64 &  0.04\\ 
      &      & 0.40 & 0.45& 0.19 &  0.02& 0.34 &  0.03& 0.38 &  0.03& 0.47 &  0.04\\ 
      &      & 0.45 & 0.50& 0.14 &  0.02& 0.23 &  0.02& 0.25 &  0.03& 0.31 &  0.03\\ 
\hline  
 1.55 & 1.75 & 0.10 & 0.15& 0.76 &  0.17& 1.27 &  0.23& 1.34 &  0.26& 1.47 &  0.27\\ 
      &      & 0.15 & 0.20& 0.87 &  0.10& 1.44 &  0.10& 1.61 &  0.12& 1.82 &  0.12\\ 
      &      & 0.20 & 0.25& 0.67 &  0.06& 1.11 &  0.07& 1.20 &  0.07& 1.48 &  0.08\\ 
      &      & 0.25 & 0.30& 0.50 &  0.05& 0.67 &  0.05& 0.75 &  0.06& 0.90 &  0.06\\ 
      &      & 0.30 & 0.35& 0.32 &  0.04& 0.46 &  0.03& 0.53 &  0.04& 0.65 &  0.04\\ 
      &      & 0.35 & 0.40& 0.23 &  0.03& 0.35 &  0.02& 0.38 &  0.03& 0.45 &  0.03\\ 
      &      & 0.40 & 0.45& 0.18 &  0.02& 0.24 &  0.02& 0.27 &  0.03& 0.30 &  0.02\\ 
      &      & 0.45 & 0.50& 0.11 &  0.02& 0.16 &  0.02& 0.17 &  0.02& 0.21 &  0.02\\ 
\hline  
 1.75 & 1.95 & 0.10 & 0.15& 0.68 &  0.13& 1.24 &  0.19& 1.13 &  0.19& 1.35 &  0.22\\ 
      &      & 0.15 & 0.20& 0.91 &  0.08& 1.21 &  0.06& 1.32 &  0.07& 1.46 &  0.08\\ 
      &      & 0.20 & 0.25& 0.50 &  0.05& 0.80 &  0.05& 0.86 &  0.06& 1.08 &  0.05\\ 
      &      & 0.25 & 0.30& 0.31 &  0.03& 0.49 &  0.03& 0.54 &  0.04& 0.60 &  0.05\\ 
      &      & 0.30 & 0.35& 0.21 &  0.03& 0.32 &  0.02& 0.36 &  0.03& 0.41 &  0.02\\ 
      &      & 0.35 & 0.40& 0.16 &  0.02& 0.23 &  0.02& 0.22 &  0.02& 0.29 &  0.03\\ 
      &      & 0.40 & 0.45& 0.11 &  0.02& 0.14 &  0.02& 0.16 &  0.01& 0.17 &  0.02\\ 
      &      & 0.45 & 0.50& 0.07 &  0.01& 0.08 &  0.01& 0.11 &  0.01& 0.10 &  0.02\\ 
\hline  
 1.95 & 2.15 & 0.10 & 0.15& 0.51 &  0.12& 0.96 &  0.15& 0.96 &  0.15& 1.06 &  0.15\\ 
      &      & 0.15 & 0.20& 0.73 &  0.06& 0.94 &  0.05& 0.96 &  0.05& 1.09 &  0.06\\ 
      &      & 0.20 & 0.25& 0.49 &  0.05& 0.69 &  0.04& 0.60 &  0.04& 0.76 &  0.05\\ 
      &      & 0.25 & 0.30& 0.27 &  0.04& 0.40 &  0.03& 0.36 &  0.03& 0.40 &  0.04\\ 
      &      & 0.30 & 0.35& 0.15 &  0.02& 0.22 &  0.02& 0.20 &  0.02& 0.26 &  0.02\\ 
      &      & 0.35 & 0.40& 0.11 &  0.02& 0.14 &  0.01& 0.13 &  0.02& 0.19 &  0.01\\ 
      &      & 0.40 & 0.45& 0.09 &  0.02& 0.09 &  0.01& 0.10 &  0.01& 0.12 &  0.02\\ 
      &      & 0.45 & 0.50& 0.05 &  0.01& 0.05 &  0.01& 0.06 &  0.01& 0.07 &  0.01\\ 

\end{tabular}
\end{center}
\end{table}

\begin{table}[hp!]
\begin{center}
  \caption{\label{tab:xsec-pimm-pb}
    HARP results for the double-differential $\pi^-$ production
    cross-section in the laboratory system,
    $d^2\sigma^{\pi^-}/(dpd\theta)$ for $\pi^-$--Pb interactions. Each row refers to a
    different $(p_{\hbox{\small min}} \le p<p_{\hbox{\small max}},
    \theta_{\hbox{\small min}} \le \theta<\theta_{\hbox{\small max}})$ bin,
    where $p$ and $\theta$ are the pion momentum and polar angle, respectively.
    The central value as well as the square-root of the diagonal elements
    of the covariance matrix are given.}
\vspace{2mm}
\begin{tabular}{rrrr|r@{$\pm$}lr@{$\pm$}lr@{$\pm$}lr@{$\pm$}l}
\hline
$\theta_{\hbox{\small min}}$ &
$\theta_{\hbox{\small max}}$ &
$p_{\hbox{\small min}}$ &
$p_{\hbox{\small max}}$ &
\multicolumn{8}{c}{$d^2\sigma^{\pi^-}/(dpd\theta)$}
\\
(rad) & (rad) & (\GeVc) & (\GeVc) &
\multicolumn{8}{c}{($\barn/(\GeVc \cdot \rad)$)}
\\
  &  &  &
&\multicolumn{2}{c}{$ \bf{3 \ \GeVc}$}
&\multicolumn{2}{c}{$ \bf{5 \ \GeVc}$}
&\multicolumn{2}{c}{$ \bf{8 \ \GeVc}$}
&\multicolumn{2}{c}{$ \bf{12 \ \GeVc}$}
\\
\hline  
 0.35 & 0.55 & 0.15 & 0.20& 0.89 &  0.31& 1.87 &  0.50& 1.66 &  0.69& 1.99 &  0.97\\ 
      &      & 0.20 & 0.25& 0.82 &  0.22& 2.04 &  0.30& 2.38 &  0.38& 2.90 &  0.60\\ 
      &      & 0.25 & 0.30& 1.24 &  0.14& 2.18 &  0.17& 2.47 &  0.19& 3.41 &  0.32\\ 
      &      & 0.30 & 0.35& 1.24 &  0.11& 2.19 &  0.12& 2.50 &  0.15& 3.43 &  0.19\\ 
      &      & 0.35 & 0.40& 1.13 &  0.09& 1.91 &  0.09& 2.38 &  0.11& 2.96 &  0.14\\ 
      &      & 0.40 & 0.45& 1.11 &  0.09& 1.81 &  0.10& 2.25 &  0.12& 2.83 &  0.17\\ 
      &      & 0.45 & 0.50& 1.20 &  0.10& 1.81 &  0.09& 2.17 &  0.10& 2.86 &  0.13\\ 
      &      & 0.50 & 0.60& 1.22 &  0.09& 1.77 &  0.10& 2.24 &  0.12& 2.79 &  0.14\\ 
      &      & 0.60 & 0.70& 1.00 &  0.10& 1.66 &  0.13& 2.06 &  0.17& 2.80 &  0.21\\ 
      &      & 0.70 & 0.80& 0.85 &  0.12& 1.50 &  0.15& 1.84 &  0.19& 2.48 &  0.30\\ 
\hline  
 0.55 & 0.75 & 0.10 & 0.15& 1.12 &  0.42& 1.69 &  0.61& 1.32 &  0.63& 1.46 &  0.86\\ 
      &      & 0.15 & 0.20& 1.31 &  0.25& 2.52 &  0.30& 2.24 &  0.41& 2.93 &  0.72\\ 
      &      & 0.20 & 0.25& 1.55 &  0.15& 2.45 &  0.16& 2.76 &  0.22& 3.56 &  0.28\\ 
      &      & 0.25 & 0.30& 1.61 &  0.12& 2.41 &  0.14& 2.58 &  0.13& 3.11 &  0.20\\ 
      &      & 0.30 & 0.35& 1.37 &  0.09& 2.05 &  0.10& 2.34 &  0.13& 2.96 &  0.14\\ 
      &      & 0.35 & 0.40& 1.17 &  0.08& 1.81 &  0.09& 2.24 &  0.10& 2.57 &  0.11\\ 
      &      & 0.40 & 0.45& 1.10 &  0.07& 1.70 &  0.08& 2.05 &  0.09& 2.40 &  0.10\\ 
      &      & 0.45 & 0.50& 1.09 &  0.07& 1.59 &  0.07& 1.96 &  0.08& 2.19 &  0.09\\ 
      &      & 0.50 & 0.60& 1.06 &  0.07& 1.49 &  0.08& 1.75 &  0.09& 2.06 &  0.11\\ 
      &      & 0.60 & 0.70& 0.82 &  0.10& 1.24 &  0.11& 1.41 &  0.13& 1.83 &  0.16\\ 
      &      & 0.70 & 0.80& 0.63 &  0.10& 1.06 &  0.13& 1.15 &  0.16& 1.51 &  0.19\\ 
\hline  
 0.75 & 0.95 & 0.10 & 0.15& 1.67 &  0.39& 1.97 &  0.46& 2.00 &  0.51& 1.97 &  0.68\\ 
      &      & 0.15 & 0.20& 1.75 &  0.18& 2.83 &  0.18& 2.70 &  0.21& 3.21 &  0.28\\ 
      &      & 0.20 & 0.25& 1.70 &  0.14& 2.32 &  0.13& 2.62 &  0.17& 3.18 &  0.18\\ 
      &      & 0.25 & 0.30& 1.53 &  0.10& 2.03 &  0.11& 2.39 &  0.12& 2.81 &  0.13\\ 
      &      & 0.30 & 0.35& 1.17 &  0.08& 1.72 &  0.08& 1.99 &  0.09& 2.41 &  0.11\\ 
      &      & 0.35 & 0.40& 1.09 &  0.08& 1.58 &  0.07& 1.79 &  0.08& 2.09 &  0.09\\ 
      &      & 0.40 & 0.45& 1.04 &  0.07& 1.44 &  0.07& 1.54 &  0.06& 1.81 &  0.08\\ 
      &      & 0.45 & 0.50& 0.99 &  0.06& 1.24 &  0.06& 1.35 &  0.06& 1.56 &  0.07\\ 
      &      & 0.50 & 0.60& 0.85 &  0.07& 1.06 &  0.06& 1.14 &  0.06& 1.33 &  0.07\\ 
      &      & 0.60 & 0.70& 0.62 &  0.08& 0.88 &  0.08& 0.89 &  0.09& 1.10 &  0.09\\ 
\hline  
 0.95 & 1.15 & 0.10 & 0.15& 2.08 &  0.32& 2.37 &  0.41& 2.62 &  0.44& 2.61 &  0.46\\ 
      &      & 0.15 & 0.20& 1.69 &  0.15& 2.84 &  0.15& 2.89 &  0.16& 3.23 &  0.18\\ 
      &      & 0.20 & 0.25& 1.57 &  0.12& 2.32 &  0.13& 2.38 &  0.14& 2.66 &  0.14\\ 
      &      & 0.25 & 0.30& 1.26 &  0.10& 1.83 &  0.09& 2.02 &  0.10& 2.28 &  0.11\\ 
      &      & 0.30 & 0.35& 0.93 &  0.07& 1.38 &  0.08& 1.59 &  0.08& 1.84 &  0.09\\ 
      &      & 0.35 & 0.40& 0.80 &  0.06& 1.14 &  0.06& 1.34 &  0.07& 1.44 &  0.07\\ 
      &      & 0.40 & 0.45& 0.68 &  0.05& 0.99 &  0.05& 1.10 &  0.05& 1.19 &  0.06\\ 
      &      & 0.45 & 0.50& 0.61 &  0.04& 0.86 &  0.04& 0.95 &  0.04& 1.03 &  0.04\\ 
      &      & 0.50 & 0.60& 0.53 &  0.04& 0.70 &  0.04& 0.76 &  0.05& 0.84 &  0.05\\ 
\hline 
\end{tabular}
\end{center}
\end{table}

\begin{table}[hp!]
\begin{center}
\begin{tabular}{rrrr|r@{$\pm$}lr@{$\pm$}lr@{$\pm$}lr@{$\pm$}l}
\hline
$\theta_{\hbox{\small min}}$ &
$\theta_{\hbox{\small max}}$ &
$p_{\hbox{\small min}}$ &
$p_{\hbox{\small max}}$ &
\multicolumn{8}{c}{$d^2\sigma^{\pi^-}/(dpd\theta)$}
\\
(rad) & (rad) & (\GeVc) & (\GeVc) &
\multicolumn{8}{c}{(\barn/($\GeVc \cdot \rad$))}
\\
  &  &  &
&\multicolumn{2}{c}{$ \bf{3 \ \GeVc}$}
&\multicolumn{2}{c}{$ \bf{5 \ \GeVc}$}
&\multicolumn{2}{c}{$ \bf{8 \ \GeVc}$}
&\multicolumn{2}{c}{$ \bf{12 \ \GeVc}$}
\\
\hline
 1.15 & 1.35 & 0.10 & 0.15& 2.17 &  0.36& 2.72 &  0.42& 2.74 &  0.44& 2.92 &  0.46\\ 
      &      & 0.15 & 0.20& 1.76 &  0.13& 2.76 &  0.16& 2.74 &  0.17& 3.05 &  0.19\\ 
      &      & 0.20 & 0.25& 1.40 &  0.10& 1.91 &  0.11& 2.17 &  0.11& 2.23 &  0.13\\ 
      &      & 0.25 & 0.30& 0.98 &  0.08& 1.44 &  0.08& 1.63 &  0.09& 1.78 &  0.10\\ 
      &      & 0.30 & 0.35& 0.74 &  0.06& 1.07 &  0.07& 1.20 &  0.07& 1.43 &  0.08\\ 
      &      & 0.35 & 0.40& 0.66 &  0.05& 0.82 &  0.05& 0.93 &  0.05& 1.13 &  0.06\\ 
      &      & 0.40 & 0.45& 0.54 &  0.04& 0.65 &  0.03& 0.75 &  0.04& 0.88 &  0.05\\ 
      &      & 0.45 & 0.50& 0.41 &  0.03& 0.55 &  0.03& 0.62 &  0.03& 0.69 &  0.04\\ 
\hline  
 1.35 & 1.55 & 0.10 & 0.15& 2.15 &  0.39& 2.81 &  0.49& 2.59 &  0.47& 2.86 &  0.52\\ 
      &      & 0.15 & 0.20& 1.58 &  0.15& 2.50 &  0.17& 2.54 &  0.18& 2.89 &  0.17\\ 
      &      & 0.20 & 0.25& 1.30 &  0.10& 1.62 &  0.10& 1.76 &  0.11& 1.88 &  0.12\\ 
      &      & 0.25 & 0.30& 0.99 &  0.08& 1.18 &  0.07& 1.30 &  0.08& 1.47 &  0.09\\ 
      &      & 0.30 & 0.35& 0.73 &  0.06& 0.93 &  0.06& 0.97 &  0.07& 1.06 &  0.07\\ 
      &      & 0.35 & 0.40& 0.59 &  0.05& 0.68 &  0.04& 0.69 &  0.05& 0.76 &  0.05\\ 
      &      & 0.40 & 0.45& 0.44 &  0.04& 0.52 &  0.03& 0.53 &  0.03& 0.54 &  0.04\\ 
      &      & 0.45 & 0.50& 0.31 &  0.03& 0.38 &  0.03& 0.41 &  0.03& 0.42 &  0.03\\ 
\hline  
 1.55 & 1.75 & 0.10 & 0.15& 1.83 &  0.30& 2.74 &  0.48& 2.51 &  0.44& 2.60 &  0.47\\ 
      &      & 0.15 & 0.20& 1.53 &  0.13& 2.15 &  0.13& 2.28 &  0.14& 2.51 &  0.15\\ 
      &      & 0.20 & 0.25& 1.07 &  0.08& 1.32 &  0.09& 1.45 &  0.09& 1.56 &  0.09\\ 
      &      & 0.25 & 0.30& 0.67 &  0.07& 0.79 &  0.06& 0.98 &  0.07& 1.11 &  0.07\\ 
      &      & 0.30 & 0.35& 0.46 &  0.04& 0.61 &  0.04& 0.68 &  0.05& 0.76 &  0.05\\ 
      &      & 0.35 & 0.40& 0.40 &  0.04& 0.49 &  0.03& 0.47 &  0.03& 0.54 &  0.03\\ 
      &      & 0.40 & 0.45& 0.30 &  0.03& 0.36 &  0.03& 0.39 &  0.02& 0.41 &  0.03\\ 
      &      & 0.45 & 0.50& 0.21 &  0.03& 0.26 &  0.02& 0.29 &  0.02& 0.30 &  0.02\\ 
\hline  
 1.75 & 1.95 & 0.10 & 0.15& 1.71 &  0.25& 2.40 &  0.33& 2.23 &  0.35& 2.40 &  0.34\\ 
      &      & 0.15 & 0.20& 1.45 &  0.09& 1.75 &  0.09& 1.86 &  0.09& 1.94 &  0.10\\ 
      &      & 0.20 & 0.25& 0.82 &  0.07& 1.05 &  0.06& 1.15 &  0.06& 1.20 &  0.06\\ 
      &      & 0.25 & 0.30& 0.56 &  0.05& 0.66 &  0.05& 0.71 &  0.05& 0.77 &  0.06\\ 
      &      & 0.30 & 0.35& 0.32 &  0.04& 0.45 &  0.03& 0.45 &  0.04& 0.53 &  0.03\\ 
      &      & 0.35 & 0.40& 0.25 &  0.03& 0.35 &  0.02& 0.32 &  0.02& 0.41 &  0.02\\ 
      &      & 0.40 & 0.45& 0.24 &  0.03& 0.25 &  0.02& 0.27 &  0.03& 0.29 &  0.02\\ 
      &      & 0.45 & 0.50& 0.20 &  0.03& 0.18 &  0.02& 0.19 &  0.02& 0.21 &  0.02\\ 
\hline  
 1.95 & 2.15 & 0.10 & 0.15& 1.58 &  0.19& 2.07 &  0.26& 1.81 &  0.25& 2.22 &  0.28\\ 
      &      & 0.15 & 0.20& 1.06 &  0.09& 1.40 &  0.07& 1.37 &  0.07& 1.47 &  0.07\\ 
      &      & 0.20 & 0.25& 0.62 &  0.07& 0.79 &  0.05& 0.83 &  0.04& 0.82 &  0.04\\ 
      &      & 0.25 & 0.30& 0.40 &  0.04& 0.48 &  0.03& 0.52 &  0.04& 0.54 &  0.04\\ 
      &      & 0.30 & 0.35& 0.28 &  0.03& 0.32 &  0.03& 0.35 &  0.03& 0.34 &  0.03\\ 
      &      & 0.35 & 0.40& 0.21 &  0.03& 0.24 &  0.02& 0.24 &  0.02& 0.24 &  0.02\\ 
      &      & 0.40 & 0.45& 0.16 &  0.02& 0.16 &  0.02& 0.18 &  0.02& 0.16 &  0.02\\ 
      &      & 0.45 & 0.50& 0.14 &  0.02& 0.10 &  0.02& 0.13 &  0.01& 0.11 &  0.02\\ 

\end{tabular}
\end{center}
\end{table}
\clearpage


\end{appendix}
\fi

\end{document}